\documentclass[a4paper,12pt]{book}

\usepackage[english]{babel}
\usepackage[utf8]{inputenc}
\usepackage{cite}
\usepackage{booktabs}
\usepackage{graphics, rotating, amsfonts, amssymb}
\usepackage{amsmath}

\let\openright=\cleardoublepage

\usepackage[pdftex,colorlinks]{hyperref}
\hypersetup{pdftitle=General relativity in higher dimensions}
\hypersetup{pdfauthor=Tom\'a\v{s} M\'alek}

\allowdisplaybreaks[1]

\def \half {\frac{1}{2}}
\def \itilde {\tilde{\imath}}
\def \jtilde {\tilde{\jmath}}
\def \d {\mathrm{d}}
\def \D {\mathrm{D}}
\def \T {\mbox{\ensuremath{\bigtriangleup}}}
\def \H {\mathcal{H}}
\def \K {\mathcal{K}}
\def \bn {\boldsymbol{n}}
\def \bk {\boldsymbol{k}}
\def \bl {\boldsymbol{\ell}}
\def \bm {\boldsymbol{m}}

\newcommand{\M}[2] {\mbox{\ensuremath{\stackrel{#1}{M}_{#2}}}}
\newcommand{\hM}[2] {\mbox{\ensuremath{\stackrel{#1}{\hat M}_{#2}}}}
\newcommand{\bM}[2] {\mbox{\ensuremath{\stackrel{#1}{\bar{M}}_{#2}}}}

\newtheorem{proposition}{Proposition}

\newtheorem{corollary}[proposition]{Corollary}

\newcommand{\pp}{{\it pp\,}-}

\newcommand{\sgn}{\operatorname{sgn}}

\newcommand\uv[1]{\quotedblbase #1\textquotedblleft}

\usepackage{fancyhdr}
\fancypagestyle{plain}{%
   \fancyhead{} 
}

\begin{document}
\frontmatter
\thispagestyle{empty}
\begin{center}
{\Large Charles University in~Prague\\
Faculty of Mathematics and Physics\\}
\vspace{1.8cm}
{\bf{\Huge{DOCTORAL THESIS}}}\\
\vspace{1cm}
\scalebox{0.45}{\includegraphics{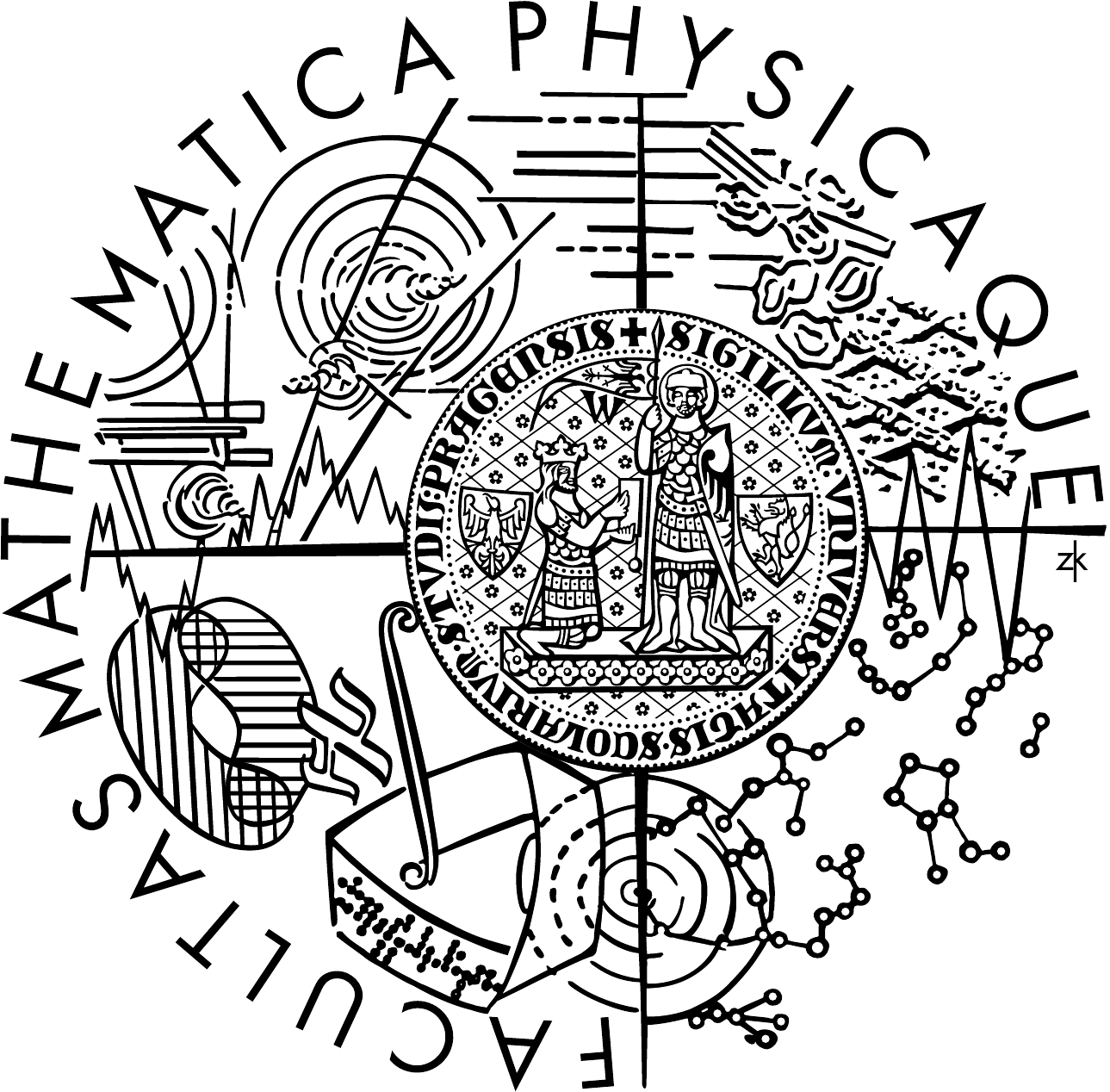}}\\
\vspace{1.8cm}
{\LARGE Tom\'a\v{s} M\'alek}\\
\vspace{1cm}
{\bf{\huge General Relativity in Higher Dimensions}}\\
\vfill
{\Large{Institute of Theoretical Physics}\\}
\vspace{0.42cm}
{\Large Supervisor: {Mgr. Vojt\v ech Pravda, Ph.D.}\\}
\vspace{0.35cm}
{\Large Branch of study: {F–1 Theoretical physics, astronomy and astrophysics}\\}
\vspace{1.5cm}
{\Large Prague 2012}
\end{center}

\addtolength{\textheight}{-20mm}

\newpage
\thispagestyle{empty}
\openright


\noindent
First of all, I would like to thank my supervisor Vojt\v{e}ch Pravda for
constant guidance, valuable comments and helpful suggestions he provided
throughout my doctoral studies. I am grateful to Alena Pravdov\'a
and Jan Nov\'ak for
careful reading the manuscript and pointing out numerous misprints.

I also truly appreciate my girlfriend Pavla for the endless patience she showed
especially during the completion of this thesis.\\

Finally, I acknowledge the financial support by the projects SVV 261301
and SVV 263301 of the Charles University in Prague.

\newpage
\openright


\vglue 0pt plus 1fill

\noindent
I declare that I carried out this doctoral thesis independently, and only with the cited
sources, literature and other professional sources.

\medskip\noindent
I understand that my work relates to the rights and obligations under the Act No.
121/2000 Coll., the Copyright Act, as amended, in particular the fact that the Charles
University in Prague has the right to conclude a license agreement on the use of this
work as a school work pursuant to Section 60 paragraph 1 of the Copyright Act.

\vspace{20mm}

\hbox{\hbox to 0.5\hsize{%
Prague, 12 January 2012
\hss}\hbox to 0.5\hsize{%
Tomáš Málek
\hss}}

\vspace{20mm}
\newpage


\vbox to 0.5\vsize{
\setlength\parindent{0mm}
\setlength\parskip{5mm}

N\'azev pr\'ace:
Obecn\'a relativita ve vy\v{s}\v{s}\'ich dimenz\'ich

Autor:
Tom\'a\v{s} M\'alek

\'Ustav:
\'Ustav teoretick\'e fyziky

Vedouc\'i diserta\v{c}n\'i pr\'ace:
Mgr. Vojt\v{e}ch Pravda, PhD., Matematick\'y \'ustav\\Akademie v\v{e}d \v{C}R, vvi.

Abstrakt:
V první části této práce analyzujeme Kerrovy--Schildovy a rozšířené
Kerrovy--Schildovy metriky v kontextu vícerozměrné obecné relativity. Pomocí
zo\-bec\-ně\-ní Newmanova--Penroseova formalizmu a algebraické
kla\-si\-fi\-ka\-ce Weylova ten\-so\-ru, založené na existenci a násobnosti jeho
vlastních nulových směrů, do vyšších dimenzí jsou studovány geometrické
vlastnosti Kerrových--Schildových kongruencí, určeny kompatibilní algebraické
typy a v expandujících případech diskutována přítomnost singularit. Uvedeme také
známá přesná řešení, která lze převést na Kerrův--Schildův tvar metriky a
zkonstruujeme nová řešení pomocí Brinkmannova \uv{warp produktu}.
V druhé části této práce uvažujeme vliv kvantových korekcí sestávajících se z kvadratických invariantů křivosti
na Einsteinovu--Hilbertovu akci a studujeme přesná řešení těchto kvadratických
teorií gravitace v libovolné dimenzi. Nalezneme třídy Einsteinových prostoročasů
a prostoročasů s nulovým zářením splňující vakuové polní rovnice a uvedeme příklady těchto metrik.

Klíčová slova:
algebraická klasifikace, gravitace ve vyšších dimenzích,
Kerrovy--Schildovy metriky, kvadratická teorie gravitace
\vss}
\vskip 1cm
\nobreak
\vbox to 0.49\vsize{
\setlength\parindent{0mm}
\setlength\parskip{5mm}

Title:
General relativity in higher dimensions

Author:
Tom\'a\v{s} M\'alek

Institute:
Institute of Theoretical Physics

Supervisor:
Mgr. Vojt\v{e}ch Pravda, PhD., Institute of Mathematics of the\\
Academy of Sciences of the Czech Republic

Abstract:
In the first part of this thesis, Kerr--Schild metrics and extended Kerr--Schild
metrics are analyzed in the context of higher dimensional general relativity.
Employing the higher dimensional generalizations of the Newman--Penrose
formalism and the algebraic classification of spacetimes based on the existence
and multiplicity of Weyl aligned null directions, we establish various
geometrical properties of the Kerr--Schild congruences, determine compatible
Weyl types and in the expanding case discuss the presence of curvature
singularities. We also present known exact solutions admitting these
Kerr--Schild forms and construct some new ones using the Brinkmann warp product.
In the second part, the influence of quantum corrections consisting of quadratic
curvature invariants on the Einstein--Hilbert action is considered and exact
vacuum solutions of these quadratic gravities are studied in arbitrary
dimension. We investigate classes of Einstein spacetimes and spacetimes with a
null radiation term in the Ricci tensor satisfying the vacuum field equations of
quadratic gravity and provide examples of these metrics.

\vss}
\newpage

\tableofcontents

\mainmatter

\openright
\setcounter{page}{1}

\chapter{Introduction}

Almost a century ago, it was Einstein's great insight that gravity as a
universal force could be described by a curvature of spacetime consisting of one
time and three spatial dimensions that has led him to formulate the famous
Einstein field equations of general relativity. Although since then the validity
of general relativity has been confirmed by many experiments, it breaks down at
the Planck scale and is expected to emerge as a low energy limit of a full
theory of quantum gravity, whatever that is.

In recent years, growing interest in higher dimensional general relativity and
black hole solutions within this theory \cite{EmparanReall2008} has been
influenced by several fields including string theory which contains general
relativity and consistency of which requires an appropriate number of extra
dimensions. Let us also mention higher dimensional supergravity theories, the
AdS/CFT correspondence relating string theory in an $n$-dimensional anti-de
Sitter bulk spacetime with conformal field theory on the lower dimensional
boundary and various brane-world scenarios considering that our four-dimensional
universe lies on a brane embedded in a higher dimensional spacetime.

However, motivations for studying higher dimensional general relativity also
come from this theory itself since it turns out that it exhibits much more
richer dynamics then in the four-dimensional case. Let us point out,
for instance, the existence of black rings or other black objects with various
non-spherical horizon topologies leading to the violation of the
four-dimensional black hole uniqueness theorem in higher dimensions.

In $n$ dimensions, general relativity can be described by the Einstein--Hilbert
action with a Lagrangian $\mathcal{L}_{\text{matter}}$ of matter fields
appearing in the theory
\begin{equation}
  S = \int \d^n x \, \sqrt{-g} \left( \frac{1}{\kappa} \left( R - 2 \Lambda \right)
    + \mathcal{L}_{\text{matter}} \right),
  \label{intro:Eistein-Hilbert:action}
\end{equation}
where $\Lambda$ is a cosmological constant and $\kappa = \frac{8 \pi G}{c^4}$ is
Einstein's constant given by Newton's gravitational constant $G$ and the speed
of light $c$ so that the non-relativistic limit in four dimensions yields
Newton's gravity. Using geometrized units $G = c = 1$, the variation of the
action \eqref{intro:Eistein-Hilbert:action} with respect to the metric leads to
the Einstein field equations
\begin{equation}
  R_{ab} - \half R g_{ab} + \Lambda g_{ab} = 8\pi T_{ab},
  \label{intro:EFEs}
\end{equation}
where the energy--momentum tensor $T_{ab}$ is given by the variation of the
matter field Lagrangian $\mathcal{L}_\text{matter}$
\begin{equation}
  T_{ab} = \frac{2}{\sqrt{-g}} \frac{\delta (\sqrt{-g} \mathcal{L}_\text{matter})}{\delta g^{ab}}.
\end{equation}
From the trace of \eqref{intro:EFEs}, one may express the Ricci scalar as
\begin{equation}
  R = \frac{2n\Lambda}{n-2} - \frac{16\pi}{n-2} T^c_{\phantom{c}c}
  \label{R}
\end{equation}
and then substituting back, the Einstein field equations can be rewritten to the
equivalent form
\begin{equation}
  R_{ab} = \frac{2 \Lambda}{n-2} g_{ab} + 8\pi T_{ab} - \frac{8\pi}{n-2} T^c_{\phantom{c}c} g_{ab}.
  \label{intro:EFEs:traceout}
\end{equation}

In this thesis, we are mainly focused on Einstein spacetimes for which the Ricci
tensor is proportional to the metric and with regard to
\eqref{intro:EFEs:traceout} we define them by
\begin{equation}
  R_{ab} = \frac{2 \Lambda}{n - 2} g_{ab}.
\end{equation}
Occasionally, we also consider spacetimes with null radiation whose
energy--momentum tensor $T_{ab}$ is of the form
\begin{equation}
  T_{ab} = \frac{\Phi}{8 \pi} \ell_a \ell_b,
\end{equation}
where $\bl$ is a null vector.

The increasing attention to higher dimensional general relativity has also led
to the effort to generalize methods successfully applied in four dimensions such
as the Newman--Penrose formalism and algebraic classification briefly reviewed
in the following sections.

\section{Newman--Penrose formalism}
\label{sec:intro:NP}

In this thesis, we frequently employ the higher dimensional generalization of
the Newman--Penrose (NP) formalism developed in
\cite{PravdaPravdovaColeyMilson2004, OrtaggioPravdaPravdova2007} along with the
algebraic classification of the Weyl tensor based on the existence of Weyl
aligned null directions (WANDs) and their multiplicity outlined in the following
section \ref{sec:intro:algclass}. In this section, let us mention the basic
concepts of the NP formalism and list some necessary definitions and relations.

In four dimensions, the NP formalism provides a useful tool that has been
successfully applied to construct exact solutions or prove theorems of general
relativity. Although the number of equations is greater than in the standard
coordinate approach with the Einstein field equations, these differential
equations are only of the first order and some are redundant. Moreover, the
advantages of the NP formalism arise in connection with the algebraic
classification if one assumes a special algebraic type or if one works in a
frame reflecting some symmetry. It then leads to considerable simplifications
since many of the components vanish.

First of all, let us mention the convention for indices. Throughout the thesis
we mainly use two types of them. Latin lower case indices $a, b, \ldots$ going
from 0 to $n-1$ mostly denoting the vector components and Latin lower case
indices $i, j, \ldots$ that range from 2 to $n-1$ mostly numbering the spacelike
frame vectors. Especially in chapter \ref{sec:xKS}, we also employ indices
$\itilde$, $\jtilde$ denoted by tilde running from 3 to $n - 2$ to indicate that
$\bm^{(\itilde)}$ does not include $\bm^{(2)}$. Only in exceptional cases when
we construct a metric in $(n + 1)$ dimensions, the indices $\itilde$, $\jtilde$
range from 2 to $n$ as will be properly emphasized.

A complex null tetrad consisting of two real null vectors $\bl$, $\bn$ and two
complex conjugate null vectors $\bm$, $\bar\bm$ plays a key role in the
four-dimensional NP formalism. However, such a complex frame cannot be
constructed in odd dimensions and thus, in $n$-dimensional spacetimes, it is
convenient to introduce a real null frame consisting of two null vectors
$\bn \equiv \bm^{(0)}$, $\bl \equiv \bm^{(1)}$ and $n - 2$ spacelike orthonormal
vectors $\bm^{(i)}$ obeying
\begin{align}
  n^a n_a = \ell^a \ell_a = n^a m^{(i)}_a = \ell^a m^{(i)}_a = 0, \qquad
  n^a \ell_a = 1, \qquad
  m^{(i) a} m^{(j)}_a = \delta_{ij}.
  \label{intro:frame:constraints}
\end{align}
Although only two of the frame vectors $\bm^{(a)}$ are null, we often refer to
such a frame \eqref{intro:frame:constraints} as a null frame. Due to the
constraints \eqref{intro:frame:constraints} the metric can be expressed in terms
of the frame vectors as
\begin{equation}
  g_{ab} = 2 n_{(a} \ell_{b)} + \delta_{ij} m^{(i)}_a m^{(j)}_b.
  \label{intro:frame:metric}
\end{equation}

The relations \eqref{intro:frame:constraints} remain valid and consequently the
form of the metric \eqref{intro:frame:metric} is preserved if one performs
Lorentz transformations of the frame. An arbitrary action of the Lorentz group
on the vectors $\bm^{(a)}$ can be described in terms of null rotations with
$\bn$ or $\bl$ fixed, boosts in the plane spanned by $\bn$ and $\bl$ and spatial
rotations as follows \cite{PravdaPravdovaColeyMilson2004}. Under null rotations
with $\bl$ fixed the frame is transformed as
\begin{equation}
  \hat \bl=\bl, \qquad
  \hat \bn =\bn + z_i \bm^{(i)} - \half z^2 \bl, \qquad
  {\hat \bm}^{(i)} = \bm^{(i)} - z_i \bl,
  \label{intro:frame:nullrot}
\end{equation}
where $z_i$ are real functions and $z^2 \equiv z_iz^i$, whereas null rotations
with $\bn$ fixed are obtained just by interchanging $\bl \leftrightarrow \bn$.
Boosts of the frame can be expressed as
\begin{equation}
  \hat \bl = \lambda \bl, \qquad
  \hat \bn = \lambda^{-1} \bn, \qquad \hat \bm^{(i)} = \bm^{(i)},
  \label{intro:frame:boosts}
\end{equation}
where $\lambda$ is a real function. Spatial rotations are generated by
$(n - 2) \times (n - 2)$ orthogonal matrices $X^{i}_{\phantom{i}j}$
\begin{equation}
  \hat \bl =  \bl, \qquad \hat \bn = \bn, \qquad
  \hat \bm^{(i)} =  X^{i}_{\ j} \bm^{(j)}.
  \label{intro:frame:spins}
\end{equation}

The Ricci rotation coefficients $L_{ab}$, $N_{ab}$ and $\M{i}{ab}$ are defined
as frame components of covariant derivatives of the frame vectors
\begin{equation}
  \ell_{a;b} = L_{cd} \, m_a^{(c)} m_b^{(d)}, \qquad
  n_{a;b} = N_{cd} \, m_a^{(c)} m_b^{(d)}, \qquad
  m_{a;b}^{(i)} = \M i {cd} \, m_a^{(c)} m_b^{(d)}.
  \label{intro:Riccicoeff}
\end{equation}
From \eqref{intro:frame:constraints} it follows that some of the Ricci rotation
coefficients are related to others or even vanish
\begin{equation}
  \begin{aligned}
  L_{0a} &= 0, \qquad
  L_{1a} = - N_{0a}, \qquad
  L_{ia} = - \M{i}{0a}, \\
  N_{1a} &= 0, \qquad
  N_{ia} = - \M{i}{1a}, \qquad
  \M{i}{ja} = - \M{j}{ia}
  \end{aligned}
  \label{intro:Riccicoeff:constraints}
\end{equation}
and therefore we consider only $L_{10}$, $L_{11}$, $L_{1i}$, $L_{i0}$, $L_{i1}$,
$L_{ij}$, $N_{i0}$, $N_{i1}$, $N_{ij}$, $\M{i}{j0}$, $\M{i}{j1}$ and $\M{i}{jk}$
as independent. Note that, in $n$ dimensions, the number of independent
components is $n^2 (n - 1) / 2$. In four dimensions, these Ricci rotation
coefficients correspond to the twelve complex spin coefficients denoted by Greek
letters.

However, in certain special cases, the number of independent coefficients can be
further reduced by an appropriate choice of the frame vectors. For instance,
since
\begin{equation}
  \ell_{a;b} \ell^b = L_{10} \ell_a + L_{i0} m_a^{(i)},
\end{equation}
the coefficients $L_{i0}$ vanish if the frame vector $\bl$ is geodetic, i.e.\
$\ell_{a;b} \ell^b \propto \ell_a$. Moreover, if $\bl$ is also affinely
parametrized then $\ell_{a;b} \ell^b = 0$ and consequently $L_{10} = 0$.
In the case that $\bl$ is geodetic and affinely parametrized, one may still
perform boosts and spins to transform the frame vectors $\bn$, $\bm^{(i)}$ to be
parallelly transported along the geodesics $\bl$ and thus $N_{i0}$ and
$\M{i}{j0}$ vanish. Alternatively, in Kundt spacetimes, appropriate boost and
spins always lead to
\begin{equation}
  L_{[1i]} = 0, \qquad
  L_{12} \neq 0, \qquad
  L_{1\itilde} = 0,
\end{equation}
as it is shown in section \ref{sec:KS:nonexpanding}.

The directional derivatives along the frame vectors are denoted as
\begin{equation}
  \D \equiv \ell^a \nabla_a, \qquad
  \T \equiv n^a \nabla_a, \qquad
  \delta_i \equiv m_{(i)}^a \nabla_a
  \label{intro:frame:derivatives}
\end{equation}
and one can straightforwardly show that the commutators of these derivatives
satisfy \cite{ColeyMilsonPravdaPravdova2004}
\begin{equation}
  \begin{aligned}
  \T \D - \D \T &= L_{11} \D + L_{10} \T + L_{i1} \delta_i - N_{i0} \delta_i, \\
  \delta_i \D - \D \delta_i &= (L_{1i} + N_{i0}) \D + L_{i0} \T + (L_{ji} - \M{i}{j0}) \delta_j, \\
  \delta_i \T - \T \delta_i  &= N_{i1} \D + (L_{i1} - L_{1i}) \T + (N_{ji} - \M{i}{j1}) \delta_j, \\
  \delta_i \delta_j - \delta_j \delta_i &= (N_{ij} - N_{ji}) \D + (L_{ij} - L_{ji}) \T
    + (\M{j}{ki} - \M{i}{kj}) \delta_k.
  \end{aligned}
  \label{intro:commutators}
\end{equation}

Lorentz transformations of the frame act on the Ricci rotation coefficients in
the following way \cite{OrtaggioPravdaPravdova2007}. Under null rotations with
$\bl$ fixed \eqref{intro:frame:nullrot}, the Ricci rotation coefficients
transform as
\begin{align}
  \hat L_{11} &= L_{11} + z_i (L_{1i} + L_{i1}) + z_i z_j L_{ij} - \half z^2 L_{10} - \half z^2 z_i L_{i0}, \notag \\
  \hat L_{10} &= L_{10} + z_i L_{i0} , \qquad 
  \hat L_{1i} = L_{1i} - z_i L_{10} + z_j L_{ji} - z_i z_j L_{j0}, \notag \phantom{\half} \\
  \hat L_{i1} &= L_{i1} + z_j L_{ij} - \half z^2 L_{i0}, \qquad 
  \hat L_{i0} = L_{i0}, \qquad
  \hat L_{ij} = L_{ij} - z_j L_{i0}, \notag \\
  \hat N_{i1} &= N_{i1} + z_j N_{ij} + z_i L_{11} + z_j \M{j}{i1} - \half z^2 (N_{i0} + L_{i1})
    + z_i z_j(L_{1j} + L_{j1}) \notag \\
    &\qquad + z_j z_k \M{j}{ik} - \half z^2(z_i L_{10} + z_j L_{ij} + z_j \M{j}{i0}) + z_i z_j z_k L_{jk} \notag \\
    &\qquad + \half z^2 \left(-z_i z_j L_{j0} + \half z^2 L_{i0} \right) + \T z_i + z_j \delta_j z_i - \half z^2 \D z_i,
  \label{intro:Riccicoeff:nullrot} \\
  \hat N_{i0} &= N_{i0} + z_i L_{10} + z_j \M{j}{i0} + z_i z_j L_{j0} - \half z^2 L_{i0} + \D z_i , \notag \\
  \hat N_{ij} &= N_{ij} + z_i L_{1j} - z_j N_{i0} + z_k \M{k}{ij} - z_j (z_i L_{10} + z_k \M{k}{i0}) + z_i z_k L_{kj} \phantom{\half} \notag \\
    &\qquad - \half z^2 L_{ij} - z_i z_j z_k L_{k0} + \half z^2 z_j L_{i0} + \delta_j z_i - z_j \D z_i , \nonumber \\
  \hM{i}{j1} &= \M{i}{j1} + 2 z_{[j} L_{i]1} + z_k \M{i}{jk} + 2 z_k z_{[j} L_{i]k} - \half z^2 \M{i}{j0} - z^2 z_{[j} L_{i]0} , \notag \\
  \hM{i}{j0} &= \M{i}{j0} + 2 z_{[j} L_{i]0}, \qquad \hM{i}{jk} = \M{i}{jk} + 2 z_{[j} L_{i]k} - z_k \M{i}{j0} + 2 z_k z_{[i} L_{j]0} , \nonumber
\end{align}
whereas null rotations with $\bn$ fixed are obtained by interchanging
$L\leftrightarrow N$ and $0\leftrightarrow 1$. Under boosts
\eqref{intro:frame:boosts}, we obtain \cite{OrtaggioPravdaPravdova2007}
\begin{equation}
  \begin{aligned}
  \hat L_{11} &= \lambda^{-1} L_{11} + \lambda^{-2} \T\lambda, \qquad
  \hat L_{10} = \lambda L_{10} + \D\lambda, \qquad
  \hat L_{i1} = L_{i1}, \\
  \hat L_{1i} &= L_{1i} + \lambda^{-1} \delta_i \lambda, \qquad
  \hat L_{i0} = \lambda^2 L_{i0}, \qquad
  \hat L_{ij} = \lambda L_{ij}, \phantom{\hM{i}{ii}} \\
  \hat N_{i1} &= \lambda^{-2} N_{i1}, \qquad
  \hat N_{i0} = N_{i0}, \qquad
  \hat N_{ij} = \lambda^{-1} N_{ij}, \phantom{\hM{i}{ii}} \\
  \hM{i}{j1} &= \lambda^{-1} \M{i}{j1}, \qquad \hM{i}{j0} = \lambda \M{i}{j0}, \qquad \hM{i}{jk} = \M{i}{jk}.
  \end{aligned}
  \label{intro:Riccicoeff:boosts}
\end{equation}
Finally, spatial rotations \eqref{intro:frame:spins} transform the Ricci
rotation coefficients as \cite{OrtaggioPravdaPravdova2007}
\begin{align}
  \hat L_{11} &= L_{11}, \qquad
  \hat L_{10} = L_{10}, \qquad
  \hat L_{1i} = X^{i}_{\ j} L_{1j}, \notag \\
  \hat L_{i1} &= X^{i}_{\ j}L_{j1}, \qquad
  \hat L_{i0} = X^{i}_{\ j} L_{j0}, \qquad  
  \hat L_{ij} = X^{i}_{\ k} X^{j}_{\ l} L_{kl}, \phantom{\hM{i}{ii}} \notag \\
  \hat N_{i0} &= X^{i}_{\ j} N_{j0}, \qquad
  \hM{i}{j1} = X^{i}_{\ k} X^{j}_{\ l} \M{k}{l1} + X^{j}_{\ k} \T X^{i}_{\ k}, \label{intro:Riccicoeff:spins} \\
  \hat N_{ij} &= X^{i}_{\ k} X^{j}_{\ l} N_{kl}, \qquad
  \hM{i}{j0} = X^{i}_{\ k} X^{j}_{\ l} \M{k}{l0} + X^{j}_{\ k} \D X^{i}_{\ k}, \notag \\
  \hat N_{i1} &= X^{i}_{\ j} N_{j1}, \qquad
  \hM{i}{jk} = X^{i}_{\ l} X^{j}_{\ m} X^{k}_{\ n} \M{l}{mn} + X^{j}_{\ m} X^{k}_{\ n} \delta_n X^{i}_{\ m}. \notag 
\end{align}

For geodetic $\bl$, it follows from \eqref{intro:Riccicoeff:nullrot} that
the optical matrix $L_{ij}$ is invariant under null rotations with $\bl$ fixed.
Thus, $L_{ij}$ has a special geometrical meaning and can be split into three
quantities, its trace $\theta$, traceless symmetric part $\sigma_{ij}$ and
anti-symmetric part $A_{ij}$
\begin{equation}
  S_{ij} \equiv L_{(ij)} = \sigma_{ij} + \theta \delta_{ij}, \qquad
  A_{ij} \equiv L_{[ij]},
  \label{intro:optical_matrix:decomposition}
\end{equation}
that are related to the expansion, shear and twist of the geodetic null
congruence $\bl$, respectively. The corresponding expansion, shear and twist
scalars are defined as
\begin{equation}
  \theta \equiv \frac{1}{n-2} L_{ii}, \qquad
  \sigma^2 \equiv \sigma_{ij} \sigma_{ij}, \qquad
  \omega^2 \equiv A_{ij} A_{ij}.
  \label{intro:optical_scalars}
\end{equation}
If the geodetic null congruence $\bl$ is also affinely parametrized, the
following identities could be useful 
\begin{equation}
  \ell^a_{\phantom{a};a} = L_{ii}, \qquad
  \ell_{a;b} \, \ell^{a;b} = L_{ij} L_{ij}, \qquad
  \ell_{a;b} \, \ell^{b;a} = L_{ij} L_{ji}
\end{equation}
and therefore
\begin{equation}
  \theta = \frac{1}{n - 2} \ell^a_{\phantom{a};a}, \qquad
  \sigma^2 = \ell_{(a;b)} \, \ell^{a;b} - \frac{1}{n - 2} (\ell^a_{\phantom{a};a})^2, \qquad
  \omega^2 = \ell_{[a;b]} \, \ell^{a;b}.
  \label{intro:optical_scalars:ell}
\end{equation}

The NP formalism can be completed expressing the frame components of the Ricci
identities
\begin{equation}
  v_{a;bc} - v_{a;cb} = R^d_{\phantom{d}abc} v_d,
\end{equation}
applied on all frame vectors $v^a = \ell^a, n^a, m^a_{(2)}, \dots, m^a_{(n-1)}$,
and the frame components of the Bianchi identities
\begin{equation}
  R_{ab[cd;e]} = 0.
\end{equation}
However, we do not list them all since, in this way, one obtains many lengthy
expressions which can be found in \cite{OrtaggioPravdaPravdova2007} and
\cite{PravdaPravdovaColeyMilson2004}, respectively. Here we present only one of
the possible contractions of the Ricci identities with the frame vectors, namely
\begin{equation}
  (\ell_{a;bc} - \ell_{a;cb}) \, m_{(i)}^a \, \ell^b \, m_{(j)}^c =
    R^d_{\phantom{d}abc} \, \ell_d \, m_{(i)}^a \, \ell^b \, m_{(j)}^c,
\end{equation}
leading to \cite{OrtaggioPravdaPravdova2007}
\begin{align}
  \D L_{ij} - \delta_j L_{i0} &= L_{10} L_{ij}
    - L_{i0} (2 L_{1j} + N_{j0})
    - L_{i1} L_{j0}
    + 2 L_{k[0|} \M{k}{i|j]} \notag \\
    &\qquad - L_{ik} (L_{kj} + \M{k}{j0})
    - C_{0i0j}
    - \frac{1}{n-2} R_{00} \delta_{ij}.
  \label{intro:Ricciid:11g}
\end{align}
If the null vector field $\bl$ is geodetic and affinely parametrized then
\eqref{intro:Ricciid:11g} reduces to the Sachs equation that for spacetimes of
algebraic type I or more special, i.e.\ $C_{0i0j} = 0$ see table
\ref{tab:Weyl:types}, with $R_{00} = 0$ including e.g.\ Einstein spaces takes
the simple form
\begin{equation}
  \D L_{ij} = - L_{ik} L_{kj}.
  \label{intro:Sachs_equation}
\end{equation}
In chapter \ref{sec:KS}, this equation allows us to express an explicit form of
the optical matrix $L_{ij}$ for expanding Einstein generalized Kerr--Schild
spacetimes.

\section{Algebraic classification of the Weyl tensor}
\label{sec:intro:algclass}

In this section, we outline one of the basic tools used in the following
chapters, namely the algebraic classification of the Weyl tensor in higher
dimensions based on the existence of preferred null directions and their
multiplicity. Further details can be found in \cite{Coleyetal2004,
MilsonColeyPravdaPravdova2004,Coley2007}, see also \cite{Reall2011} for an
introductory review.

In $n > 3$ dimensions, the Weyl tensor is defined as
\begin{equation}
  C_{abcd} = R_{abcd}
    - \frac{2}{n-2} \left( g_{a[c} R_{d]b} - g_{b[c} R_{d]a} \right)
    + \frac{2}{(n-1)(n-2)} R g_{a[c} g_{d]b}
  \label{intro:Weyl:definition}
\end{equation}
and inherits all the symmetries of the Riemann tensor, moreover, it is
completely traceless. Obviously, the Weyl tensor contains all information about
the curvature in Ricci-flat spacetimes and therefore it is considered as the
part of the Riemann tensor describing pure gravitational field. Note also that
two spacetimes related by a conformal transformation have the same Weyl tensors.

One of the equivalent approaches to algebraic classification of the Weyl tensor
in four dimensions is based on the properties of the principal null directions
(PNDs). Every four-dimensional spacetime admits exactly four discrete PNDs and
the Petrov type is determined by their multiplicity. Other equivalent
classifications can be formulated in terms of two-forms, spinors or scalar
invariants, see e.g. \cite{StephaniKramer2003}.

Only the algebraic classification of the Weyl tensor generalizing the
four-dimensional Petrov classification based on the existence of preferred null
directions developed in \cite{Coleyetal2004} has been successfully formulated
in arbitrary dimension. Note also that the spinorial approach in five dimensions
established in \cite{DeSmet2002} is not equivalent to the null directions
approach neither to the spinorial classification in four dimensions
\cite{Godazgar2010}.

First, let us introduce the following definitions. If some quantity $q$
transforms under boosts \eqref{intro:frame:boosts} as
\begin{equation}
  \hat{q} = \lambda^w q
\end{equation}
we say that $q$ has a boost weight $w$. Therefore, the null frame vectors $\bl$,
$\bn$ have boost weight 1 and $-1$, respectively, and the spacelike frame
vectors $\bm^{(i)}$ are of boost weight zero. It then follows that every index
$0$ contributes to the boost weight of the given frame component by one, whereas
each index $1$ decreases the boost weight by one, for instance, the components
$C_{010i}$ have boost weight 1. It is also convenient to define the following
operation reflecting the symmetries of the Riemann tensor
\begin{equation}
  T_{\{pqrs\}} \equiv \half \left( T_{[ab][cd]} + T_{[cd][ab]} \right),
  \label{intro:Riemann:symmetries}
\end{equation}
which allows us to express the frame components of the Weyl tensor more
compactly. Such symmetries immediately imply that the maximal boost weight of
the Weyl tensor components is 2 and the minimal boost weight is $-2$.

Next, using \eqref{intro:Riemann:symmetries}, we decompose the Weyl tensor into
the frame components and sort them according to their boost weight
\begin{align}
  C_{abcd} &= \overbrace{4 C_{0i0j} \, n_{\{a} m^{(i)}_b n_c m^{(j)}_{d\}}}^\text{boost weight 2} \notag \\
    &\qquad + \overbrace{8 C_{010i} \, n_{\{a} \ell_b n_c m^{(i)}_{d\}}
      + 4 C_{0ijk} \, n_{\{a} m^{(i)}_b m^{(j)}_c m^{(k)}_{d\}}}^\text{boost weight 1} \notag \\
    &\qquad\!\begin{aligned}&+ 4 C_{0101} \, n_{\{a} \ell_b n_c \ell_{d\}}
      + 4 C_{01ij} \, n_{\{a} \ell_b m^{(i)}_c m^{(j)}_{d\}}  \\
      &+ 8 C_{0i1j} \, n_{\{a} m^{(i)}_b \ell_c m^{(j)}_{d\}}
      + C_{ijkl} \, m^{(i)}_{\{a} m^{(j)}_b m^{(k)}_c m^{(l)}_{d\}}\end{aligned} \Bigg\} \text{\scriptsize boost weight 0}
  \label{intro:Weyl:decomposition} \\
    &\qquad + \underbrace{8 C_{101i} \, \ell_{\{a} n_b \ell_c m^{(i)}_{d\}}
      + 4 C_{1ijk} \, \ell_{\{a} m^{(i)}_b m^{(j)}_c m^{(k)}_{d\}}}_\text{boost weight $-1$} \notag \\
    &\qquad + \underbrace{4 C_{1i1j} \, \ell_{\{a} m^{(i)}_b \ell_c m^{(j)}_{d\}}}_\text{boost weight $-2$}. \notag
\end{align}
However, some of these components are redundant due to the remaining symmetries
$C_{a[bcd]}$ and tracelessness of the Weyl tensor leading to the additional
relations 
\begin{equation}
  \begin{aligned}
  &C_{0i0i} = 0, \qquad
  C_{010j} = C_{0iji}, \qquad
  C_{0[ijk]} = 0, \qquad \phantom{\half} \\
  &C_{0101} = C_{0i1i}, \qquad
  C_{i[jkl]} = 0, \qquad
  C_{0i1j} = -\half C_{ikjk} + \half C_{01ij}, \\
  &C_{011j} = - C_{1iji}, \qquad
  C_{1[ijk]} = 0, \qquad
  C_{1i1i} = 0, \phantom{\half}
  \end{aligned}
  \label{intro:Weyl:symmetries}
\end{equation}
which reduce number of independent frame components of the Weyl tensor.

Choosing the vector $\bl$ such that as many as possible leading terms in
\eqref{intro:Weyl:definition} vanish, the highest boost weight of the remaining
components determines the primary algebraic type and we say that $\bl$ is a Weyl
aligned null direction (WAND). Effectively, one may choose an arbitrary null
frame and if none of the null frame vectors is just a desired WAND perform null
rotations with $\bn$ fixed \eqref{intro:frame:nullrot} to find $\bl$ that
transforms away the highest possible number of the components.

If there is no WAND $\bl$, i.e.\ there is no frame with all boost weight 2
components $C_{0i0j}$ vanishing, the spacetime is of general Weyl type G. If all
boost weight 2 components vanish and some of the boost weight 1 components are
non-zero the spacetime is of type I. Types II, III and N are determined by
vanishing components of all corresponding boost weights up to 1, 0, or $-1$,
respectively, and $\bl$ is then referred to as a multiple WAND. Type O denotes
conformally flat spacetimes with the Weyl tensor completely vanishing. Using the
relations \eqref{intro:Weyl:symmetries}, the conditions on the frame components
of the Weyl tensor determining primary algebraic type are summarized in table
\ref{tab:Weyl:types}.
\begin{table}
  \caption{The conditions on the frame components of the Weyl tensor determining
  primary algebraic type of a spacetime.}
  \begin{center}
  \begin{tabular}{cl}
    \toprule
    Weyl type & vanishing components of the Weyl tensor \\
    \midrule
    G & $\exists i, j: C_{0i0j} \neq 0$ \\
    I & $C_{0i0j} = 0$ \\
    II & $C_{0i0j} = C_{0ijk} = 0$ \\
    III & $C_{0i0j} = C_{0ijk} = C_{ijkl} = C_{01ij} = 0$ \\
    N & $C_{0i0j} = C_{0ijk} = C_{ijkl} = C_{01ij} = C_{1ijk} = 0$ \\
   \bottomrule
  \end{tabular}
  \end{center}
  \label{tab:Weyl:types}
\end{table}

Note that the Bel--Deveber criteria generalized to higher dimensions in
\cite{Ortaggio2009} are equivalent conditions to those ones given in table
\ref{tab:Weyl:types} involving only the null vector $\bl$ and thus construction
of complete null frame in not required.

Although in \cite{Coleyetal2004} the notion algebraically special spacetimes
means spacetimes of Weyl type I or more special, in subsequent papers and also
in this thesis, spacetimes are denoted as algebraically special if they admit a
multiple WAND, i.e.\ spacetimes of Weyl type II or more special similarly as in
the four dimensional case.

Having established the primary type, similarly we introduce secondary types.
Using null rotations with the WAND $\bl$ fixed \eqref{intro:frame:nullrot} one
may find a WAND $\bn$ such that as many as possible trailing terms in
\eqref{intro:Weyl:definition} vanish. The lowest boost weight of the remaining
components determine the secondary algebraic type. If $\bn$ is a simple WAND we
denote by I$_i$, II$_i$ and III$_i$ the subtypes of the corresponding types I,
II and III, respectively. Type D is defined as a subtype of the primary type II
with $\bn$ being also a multiple WAND and thus only the boost weight zero
components of the Weyl tensor are non-vanishing.

Note that an additional subclasses can be defined, for instance,
type III(a) is a subclass of type III with $C_{101i} = 0$ and thus
the Weyl tensor takes the form
\begin{equation}
  C_{abcd} = 4 C_{1ijk} \, \ell_{\{a} m^{(i)}_b m^{(j)}_c m^{(k)}_{d\}}
    + 4 C_{1i1j} \, \ell_{\{a} m^{(i)}_b \ell_c m^{(j)}_{d\}}.
\end{equation}
In chapter \ref{sec:QuadGr}, we introduce other subclasses of type III,
namely types III(A) and III(B) depending on whether the quantity
$\half C_{1ijk} C_{1ijk} - C_{101i} C_{101i}$ is non-vanishing or vanishing,
respectively. Obviously, type III(a) is a subclass of type III(A).

Finally, let us recall the differences of the algebraic classification based on
the existence of preferred null directions and their multiplicity in four and
higher dimensions. In four dimensions, every spacetime admits exactly four
discrete PNDs. The most general algebraic type is type I and if two or more PNDs
coincide we say that a spacetime is algebraically special. On the other hand, in
higher dimensions, a spacetime admits no, a finite number or a continuous family
of WANDs and thus new general type G arises. Types I, II and III in four
dimensions correspond to types I$_i$, II$_i$ and III$_i$ in higher dimensions,
respectively.

\section{Goldberg--Sachs theorem}

Let us briefly comment on the validity of the Goldberg--Sachs theorem in higher
dimensions. In four dimensional general relativity, the Goldberg--Sachs theorem
states that an Einstein spacetime is algebraically special if and only if it
admits a congruence of non-shearing null geodesics. Such congruence then
corresponds to the PND. In four dimensions, this theorem is useful when
searching for new exact solutions and, for instance, it has led to the discovery
of the Kerr black hole. However, the statement of the Goldberg--Sachs theorem
cannot be straightforwardly generalized to higher dimensions.

It has been shown that a multiple WAND of Weyl type III and N Ricci-flat
spacetimes is geodetic \cite{PravdaPravdovaColeyMilson2004}. This result holds
also in the case of Einstein spacetimes. A multiple WAND of ``generic'' type II
and D Einstein spacetimes is also geodetic, nevertheless there exit type D
spacetimes that admit a non-geodetic multiple WAND
\cite{PravdaPravdovaOrtaggio2007}. However, it was shown in
\cite{DurkeeReall2009} that an Einstein spacetime admitting a non-geodetic
multiple WAND also admits a geodetic multiple WAND.

It also turns out that there exist shearing multiple WANDs. An example of such
spacetime, namely the Kerr--(anti-)de Sitter black hole, is discussed in section
\ref{sec:KS:Kerr-(A)dS}. This fact suggests that the shear-free condition should
be weakened. Let us point out, for instance, the result of chapter \ref{sec:KS}
that the optical matrix $L_{ij}$ of expanding Einstein generalized Kerr--Schild
spacetimes takes the block-diagonal form \eqref{KS:expanding:Lij:blockdiagonal}
with $2 \times 2$ blocks \eqref{KS:expanding:Lij:block} and therefore there is
no shear in any planes spanned by pairs of the spacelike frame vectors
corresponding to the $2 \times 2$ blocks of the optical matrix. Note also that
the optical matrices of all type N Einstein spacetimes are of this form with
just one $2 \times 2$ block \cite{PravdaPravdovaColeyMilson2004}.

\section{Brinkmann warp product}
\label{sec:intro:warp}

Solving the Einstein field equations is a rather complicated task, especially
in higher dimensions. It would be convenient to have a method for generating new
solutions from the already existing ones. The Brinkmann warp product
\cite{Brinkmann1925} is one of such methods allowing construction of
$n$-dimensional Einstein spacetimes from known $(n-1)$-dimensional Ricci-flat or
Einstein metrics.

Although gravity in higher dimensions exhibits much richer dynamics due to the
existence of extended black objects with a same mass and angular momentum but
with different horizon topologies such as black strings, black rings, black
Saturns, etc., forbidden by the no-hair theorem in four dimensions, some of the
known four-dimensional exact solutions still have no higher dimensional analogue
which may be obtained using the Brinkmann warp product.

For instance, a higher dimensional C-metric is unknown unlike the four
dimensional case and moreover it cannot belong to the Robinson--Trautman class
\cite{PodolskyOrtaggio2006} of spacetimes admitting expanding, non-shearing and
non-twisting geodetic null congruence. However, an example of five-dimensional
C-metric is presented in \cite{OrtaggioPravdaPravdova2010b} using the Brinkmann
warp product.

We employ the Brinkmann warp product in section \ref{sec:KS:nonexpanding:warp}
not only to generate solutions with one extra dimension but also in order to
introduce a cosmological constant to Ricci-flat solutions. Thus we construct
examples of higher dimensional type N Einstein Kundt spacetimes from higher
dimensional type N Ricci-flat Kundt metrics belonging to the class of spacetimes
with vanishing scalar invariants and from four-dimensional type N Einstein Kundt
metrics. In section \ref{sec:KS:expanding:warp}, we also present a few examples
of expanding warped metrics such as a rotating black string constructed from the
higher dimensional Kerr--(A)dS black hole. Taking the Minkowski and
Schwarzschild metric as a seed one may also obtain Randall--Sundrum brane model
and Chamblin--Hawking--Reall black hole on a brane, respectively. Unfortunately,
application of the Brinkmann warp product has certain limits. As we mention
below, the sign of cosmological constant of the warped metric is not entirely
arbitrary and, in some cases, a naked singularity may be introduced to the
spacetime.

It has been shown in \cite{Brinkmann1925} that using an $(n-1)$-dimensional
Einstein metric as a seed $\d\tilde s^2$, we can construct an $n$-dimensional
Einstein metric $\d s^2$
\begin{equation}
  \d s^2 = \frac{1}{f(z)} \d z^2 + f(z) \d \tilde s^2
  \label{intro:warp:warpedmetric} 
\end{equation}
with the warp factor $f(z)$ given by
\begin{equation}
  f(z) = - \lambda z^2 + 2 d z + b,
  \label{intro:warp:factor}
\end{equation}
where the cosmological constant $\Lambda$ of the $n$-dimensional warped Einstein
metric is introduced via $\lambda = \frac{2\Lambda}{(n-1)(n-2)}$ and $b$, $d$
are constant parameters subject to
\begin{equation}
  \tilde R = (n-1)(n-2) (\lambda b + d^2),
  \label{intro:warp:seedricci}
\end{equation}
where $\tilde R$ is the Ricci scalar of the $(n-1)$-dimensional seed metric
$\d\tilde s^2$. Note that in the case $\tilde R = R = 0$, the warp product
reduces to the trivial direct product of a seed metric with a one-dimensional
flat space $\d z^2$.

Since we consider only Lorentzian metrics the warp factor $f(z)$ has to be
positive. Note that the Ricci scalar $\tilde R$ of the seed is proportional to
the discriminant of the quadratic equation $f(z) = 0$ and the Ricci scalar
$R = \frac{2 n \Lambda}{n-2} = n (n-1) \lambda$ of the warped metric is
proportional to the quadratic term of $f(z) = 0$ with the opposite sign.
Therefore, it is obvious \cite{OrtaggioPravdaPravdova2010b} that only the
combinations of signs of the Ricci scalars $R$ and $\tilde R$ listed in table
\ref{tab:warp:allowedsigns} are allowed.

\begin{table}
  \caption{Allowed combinations of signs of the Ricci scalar $\tilde R$
  corresponding to the seed metric $\d\tilde{s}^2$ and of the Ricci scalar $R$
  corresponding to the warped metric $\d s^2$ \eqref{intro:warp:warpedmetric}.}
  \begin{center}
    \begin{tabular}{cccc}
      \toprule
      & $R < 0$ & $R = 0$ & $R > 0$ \\
      \midrule
      $\tilde R < 0$ & \checkmark  & $\times$ & $\times$ \\
      $\tilde R = 0$ & \checkmark  & \checkmark    & $\times$ \\
      $\tilde R > 0$ & \checkmark  & \checkmark    & \checkmark \\
      \bottomrule
    \end{tabular}
  \end{center}
  \label{tab:warp:allowedsigns}
\end{table}

Several useful statements about Weyl types of the warped metrics have been given
in \cite{OrtaggioPravdaPravdova2010b}. It turns out that if the Weyl tensor of
the seed metric $\d \tilde s^2$ is algebraically special, i.e.\ of Weyl type II
or more special, then the warped metric $\d s^2$ is of the same Weyl type. On
the other hand, seed spacetimes of the general Weyl type G lead to warped
spacetimes of types G, I$_i$ or D and finally a seed of the Weyl type I yields
warped metrics of types I or I$_i$. It was also shown in
\cite{OrtaggioPravdaPravdova2010b} that the Brinkmann warp product introduces a
curvature or parallelly propagated singularity to warped spacetimes $\d s^2$ at
any point where the warp factor vanishes $f(z) = 0$. Only in the cases when both
metrics $\d \tilde{s}^2$ and $\d s^2$ are Ricci-flat $\tilde R = 0$, $R = 0$ or
Einstein with negative cosmological constants $\tilde R < 0$, $R < 0$ the warp
factor $f(z)$ does not admit roots and therefore such spacetimes are free from
this kind of singularities.

Let us conclude this brief summary of properties of the Brinkmann warp product
by listing a few expressions that are employed in sections
\ref{sec:KS:nonexpanding:warp} and \ref{sec:KS:expanding:warp}. The warped
metric \eqref{intro:warp:warpedmetric} can be rewritten by an appropriate
coordinate transformation to a different form that could be more convenient in
certain cases. Two such forms are given in \cite{OrtaggioPravdaPravdova2010b}
and both depend on the combination of the signs of the Ricci scalars $\tilde R$
and $R$. Using one of the possible transformations, one may put the metric
\eqref{intro:warp:warpedmetric} to the form conformal to a direct product
\begin{align}
  \lambda>0: \qquad & \d s^2 = \cosh^{-2}(\sqrt{\lambda}x) (\d x^2 + \d\tilde s^2) & \tilde R &> 0,
  \label{intro:warp:conformal++} \\
  \lambda=0: \qquad & \d s^2 = \d x^2 + \d \tilde s^2 & \tilde R &= 0,
  \label{intro:warp:conformal00} \\
  & \d s^2 = 2 e^{2x} ( \d x^2 + \d \tilde s^2 ) & \tilde R &> 0,
  \label{intro:warp:conformal0+} \\
  \lambda<0: \qquad & \d s^2 = \cos^{-2}(\sqrt{-\lambda}x) (\d x^2 + \d\tilde s^2) & \tilde R &< 0,
  \label{intro:warp:conformal--} \\
  & \d s^2 = (-\lambda x^2)^{-1} (\d x^2 + \d\tilde s^2) & \tilde R &= 0,
  \label{intro:warp:conformal-0} \\
  & \d s^2 = \sinh^{-2}(\sqrt{-\lambda}x) (\d x^2 + \d\tilde s^2) & \tilde R &> 0,
  \label{intro:warp:conformal-+}
\end{align}
where $\tilde R$ and $\lambda$ are related by $|\tilde R|=(n-1)(n-2) |\lambda|$
apart from \eqref{intro:warp:conformal0+} where $\tilde R = (n-1)(n-2)$. A
different form can be obtained from \eqref{intro:warp:warpedmetric} employing
the transformation $\d z^2 f(z)^{-1} = \d y^2$ that leads to metrics with a flat
extra dimension
\begin{align}
  \lambda>0: \qquad & \d s^2 = \d y^2 + \cos^2(\sqrt{\lambda}y) \, \d\tilde s^2 & \tilde R &> 0,
  \label{intro:warp:kk++} \\
  \lambda=0: \qquad & \d s^2 = \d y^2 + y^2 \, \d \tilde s^2 & \tilde R &> 0,
  \label{intro:warp:kk0+} \\
  \lambda<0: \qquad & \d s^2 = \d y^2 + \cosh^2(\sqrt{-\lambda}y) \, \d\tilde s^2 & \tilde R &< 0,
  \label{intro:warp:kk--} \\
  & \d s^2 = \d y^2 + e^{2 \sqrt{-\lambda} y} \, \d\tilde s^2 & \tilde R &= 0,
  \label{intro:warp:kk-0} \\
  & \d s^2 = \d y^2 + \sinh^{2}(\sqrt{-\lambda}y) \, \d\tilde s^2 \qquad & \tilde R &> 0.
  \label{intro:warp:kk-+}
\end{align}
The case $R = 0$, $\tilde R = 0$ when the Brinkmann warp product reduces just to
a direct product is same as \eqref{intro:warp:conformal00}.

\section{New results of this thesis}

The Kerr--Schild ansatz has turned out to be an effective tool for finding exact
solutions of general relativity in four dimensions. In chapter \ref{sec:KS}, we
study higher dimensional generalized Kerr--Schild (GKS) metrics having an
(anti-)de Sitter background. We obtain the necessary and sufficient condition
under which the null Kerr-Schild vector field is geodetic. It is shown that
non-expanding Einstein GKS spacetimes are of Weyl type N and belong to the Kundt
class. Using the Brinkmann warp product, some new explicit non-expanding
solutions are constructed. In the case of expanding Einstein GKS spacetimes, the
compatible Weyl types are D or II and the corresponding optical matrices take a
special block-diagonal form satisfying the so-called optical constraint. This
allows us to determine the dependence of various geometric quantities on the
affine parameter $r$ along the geodetic Kerr--Schild congruence and to discuss
presence of curvature singularities at the origin $r = 0$. We also express the
optical matrix of the five-dimensional Kerr--(anti-)de Sitter black hole in
order to compare this important example of expanding GKS spacetime with our
general results, namely the forms of the optical matrices and the Kerr--Schild
scalar functions and presence of singularities.

In a similar way, we analyze the extended Kerr--Schild (xKS) ansatz in chapter
\ref{sec:xKS}. It is a further extension of the GKS ansatz where, in addition to
the null Kerr--Schild vector, a spacelike vector field appears in the metric. We
are motivated by the known fact that a straightforward generalization of the KS
form of the Kerr--Newman black hole to higher dimensions has failed and,
moreover, that the CCLP solution of charged rotating black hole in
five-dimensional minimal gauged supergravity takes the xKS form. In contrast to the GKS
case, we obtain in general only the necessary condition under which the
Kerr--Schild vector is geodetic. However, it is shown that this condition
becomes sufficient if we appropriately restrict the geometry of the null and
spacelike vectors appearing in the metric. It turns out that xKS spacetimes with
a geodetic Kerr--Schild vector field are of Weyl type I or more special. In the
case of Kundt xKS spacetimes, the compatible Weyl types are further restricted
depending on the form of the energy--momentum tensor. A few examples of such
spacetimes are also briefly discussed. For an example of an expanding xKS
spacetime, namely the CCLP black hole, we express the optical matrix which
interestingly satisfies the optical constraint obtained in the case of expanding
GKS spacetimes.

In chapter \ref{sec:QuadGr}, we focus on quadratic gravity (QG) in arbitrary
dimension, i.e.\ a generalization of the Einstein theory with a Lagrangian
containing all possible polynomial curvature invariants as quantum corrections
up to the second order in the Riemann tensor. We show that all higher
dimensional Einstein spacetimes of the Weyl type N with an appropriately chosen
effective cosmological constant $\Lambda$ depending on the particular parameters
of the theory are exact solutions to QG. We refer to explicitly known metrics
within this class and construct some new ones using the Brinkmann warp product.
In the case of type III Einstein spacetimes, it is shown that the field
equations of QG impose an additional constraint on the Weyl tensor and some
examples of such type III solutions are given. It turns out that not all
spacetimes with vanishing scalar invariants (VSI) solve QG since type III
\pp waves do not satisfy this constraint. We also study a wider class of
spacetimes admitting a null radiation term in the Ricci tensor aligned with a
WAND. We show that such spacetimes of type N and certain subclass of type III
solve the source-free field equations of QG and, in contrast to the Einstein
case, the optical properties of the null geodetic congruence are restricted so
that these solutions belong to the Kundt class. Examples of such type N metrics
are also given explicitly.

\chapter{Kerr--Schild spacetimes}
\label{sec:KS}

This chapter is mainly based on the original results published in
\cite{MalekPravda2010,MalekPravda2011b}. First, we include a cosmological
constant to the Kerr--Schild ansatz in an appropriate way. Without any additional
assumptions, we show under which conditions the Kerr--Schild vector $\bk$ is
geodetic and consequently a multiple WAND. For Einstein spacetimes which then
have $\bk$ necessarily geodetic, the Einstein field equations naturally lead to
the splitting of our analysis to the non-expanding and expanding case. It will
be shown that non-expanding Einstein Kerr--Schild spacetimes are only of Weyl
type N, whereas the expanding spacetimes are of Weyl types II or D. In the
expanding case, we determine the $r$-dependence of the optical matrix and of the
boost weight zero components of the Weyl tensor. This then allows us to discuss
the presence of curvature singularities. Some known examples of both
non-expanding and expanding Kerr--Schild spacetimes will be presented and a few
new solutions will be also given using the Brinkmann warp product.

The complexity of the Einstein field equations, which are a system of quasilinear
partial differential equations of the second order for an unknown metric, has led
to the development of many approaches that simplify solving of these equations.
This is motivated by the fact that a direct attack on the equations is hopeless,
especially in the case of higher dimensions.
Since most of the known exact solutions of four and higher dimensional general
gravity are algebraically special, one can use an appropriate formalism, for
instance the Newman--Penrose formalism briefly summarized in section
\ref{sec:intro:NP}, and assume a special algebraic type of the spacetime under
consideration to reduce the number and simplify the form of the field equations.
In fact, this method will be employed in chapter \ref{sec:QuadGr} in the context
of quadratic theory of gravity.

Another approach is to reduce the number of unknown independent metric components
by considering an appropriate special form of the metric. For instance, this form
may follow from an assumption of some kind of spacetime symmetries as in the
case of the discovery of the static Schwarzschild black hole using spherical
symmetry.
Alternatively, one may directly propose a convenient form of the unknown metric
in order to simplify subsequent calculations. An important example representing
this approach, which has been successfully applied for finding exact solutions,
is the Kerr--Schild (KS) ansatz
\begin{equation}
  g_{ab} = \eta_{ab} - 2 \H k_a k_b,
  \label{KS:KSansatz}
\end{equation}
where $\H$ is a scalar function and $\bk$ is a null vector with respect to both
the Minkowski background metric $\eta_{ab}$ and full metric $g_{ab}$.

The KS ansatz proposed by Kerr and Schild in 1965 \cite{KerrSchild1965} has led
to the rediscovery of the four-dimensional rotating solution known as the Kerr
black hole \cite{Kerr1963}. The reason why they considered metrics in the KS
form \eqref{KS:KSansatz} is that the inverse metric is simply given by
$g^{ab} = \eta^{ab} + 2 \H k^a k^b$ which also means that the full metric
corresponds exactly to its linear approximation around the flat background.

Twenty years later, Myers and Perry managed to generalize the KS form of the
Kerr solution so that they obtained a metric of a rotating black hole in
arbitrary dimension \cite{MyersPerry1986}. The situation differs significantly
from the four-dimensional case where the black hole rotates in one rotation
plane. Since in $n$ dimensions, the black hole may rotate arbitrarily in $p$
independent rotation planes, where $p$ is given by
\begin{equation}
  p \equiv \left\lfloor \frac{n-1}{2} \right\rfloor.
\end{equation}
This exhibits the fact that the rotation group $SO(n-1)$ has a Cartan subgroup
$U(1)^p$.

Despite the simplicity of the Kerr--Schild ansatz \eqref{KS:KSansatz}, which
makes analytic calculations tractable, this class of metrics contains physically
interesting solutions such as the above-mentioned Kerr black hole and the
Myers--Perry black hole, both being of Weyl type D, but it also contains
radiative spacetimes of Weyl type N represented, for instance, by \pp waves.

Let us mention that various non-vacuum exact solutions in four dimensions
also admit the Kerr--Schild form, see e.g.\ \cite{StephaniKramer2003},
and some of them can be straightforwardly generalized to higher dimensions.

Vaidya's radiating star \cite{Vaidya1951} and the corresponding higher
dimensional analogue \cite{IyerVishveshwara1989} are slight modifications
of the spherically symmetric Schwarzschild black hole where the constant mass
parameter appearing in the Kerr--Schild function $\H$ is replaced by a function
$m = m(u)$ depending on the retarded time $u$. As a consequence, the additional
term in the Ricci tensor corresponding to a null radiation appears due to the
change of mass $m(u)$.

Kinnersley's photon rocket \cite{Kinnersley1969} and its higher dimensional
counterparts \cite{GursesSarioglu2002} are other examples of non-vacuum
solutions of the Einstein field equations describing the gravitational field of
an accelerating object anisotropically emitting null radiation.

A well-known example of electro-vacuum solutions, the Kerr--Newman black hole,
which has not yet been successfully generalized to higher dimensions as opposite
to its static limit, the Reissner--Nordst\"om black hole, can be also cast to the
Kerr--Schild form. In this case, the Maxwell field is aligned with the
geodetic null congruence $\bk$, i.e.\ the vector potential is proportional to
the Kerr--Schild vector. 

All the above examples of spacetimes admitting the Kerr--Schild form are
solutions of the Einstein field equations without cosmological constant. General
properties of such Ricci-flat Kerr--Schild metrics have been studied recently in
\cite{OrtaggioPravdaPravdova2008}. The aim of this chapter is to introduce
cosmological constant to the Kerr--Schild ansatz and generalize the results from
the Ricci-flat case to Einstein spaces. Moreover, we will also briefly discuss
solutions with possible cosmological constant which also contain aligned matter
fields.

One of the simplest higher dimensional Ricci-flat solutions admitting
Kerr--Schild form is the Schwarzschild--Tangherlini black hole
\cite{Tangherlini1963}. In the standard Schwarzschild coordinates, cosmological
constant $\Lambda$ enters the metric in a simple way, i.e.\ the
(A)dS--Schwarzschild--Tangherlini metric \cite{GibbonsHartnoll2002} takes the
form
\begin{equation}
  \d s^2 = - U(r) \, \d t^2 + U(r)^{-1} \, \d r^2 + r^2 \, \d \Omega^2_{(n-2)},
  \label{KS:AdS-S-T}
\end{equation}
where 
\begin{align}
  U(r) &= 1 - \lambda r^2 - \left( \frac{2m}{r} \right)^{n-3}, \label{KS:AdS-S-T:U} \\
  \lambda &= \frac{2 \Lambda}{(n-1)(n-2)}
\end{align}
and $\d \Omega_{(n-2)}^2$ is a metric on the $(n-2)$-dimensional sphere $S^{n-2}$
of unit radius
\begin{equation}
  \d \Omega_1^2 = \d \varphi^2 , \qquad
  \d \Omega_{i+1}^2 = \d \theta_i^2 + \sin^2 \theta_i \, \d \Omega_i^2 \quad
  (i \in \mathbb{N}),
\end{equation}
with the standard angular coordinates $\varphi \in \langle 0,2\pi \rangle$,
$\theta_i \in \langle 0, \pi \rangle$. Using the transformation
\begin{equation}
  \d t' = \d t - \frac{1 - U(r)}{U(r)} \, \d r,
\end{equation}  
the metric \eqref{KS:AdS-S-T} can be rewritten as
\begin{equation}
  \d s^2 = - \d t'^2 + \d r^2 + r^2 \, \d \Omega^2_{(n-2)}
    + \left[ \lambda r^2 + \left( \frac{2m}{r} \right)^{n-3} \right] ( \d t' - \d r )^2.
  \label{KS:AdS-S-T:KS}
\end{equation}
The first three terms in \eqref{KS:AdS-S-T:KS} correspond to the Minkowski metric
in the spherical coordinates and the last term is a multiple of a null vector,
therefore, the (A)dS--Schwarzschild--Tangherlini metric \eqref{KS:AdS-S-T:KS}
takes the Kerr--Schild form with a flat background \eqref{KS:KSansatz}.

Note that, setting $m=0$ in \eqref{KS:AdS-S-T:KS}, one directly obtains the
(Anti-)de Sitter metric in the Kerr--Schild form with a flat background
\begin{equation}
  \d s^2 = - \d t'^2 + \d r^2 + r^2 \, \d \Omega^2_{(n-2)}
    + \frac{2\Lambda}{(n-1)(n-2)} r^2 ( \d t' - \d r )^2.
  \label{KS:AdS:KS}
\end{equation}

The metric \eqref{KS:AdS-S-T} can be also expressed in another form. One may
start with the higher dimensional Kerr--(A)dS metric
\cite{GibbonsLuPagePope2004} and set all the rotation parameters to zero. Thus,
we arrive at the (A)dS--Schwarzschild--Tangherlini metric as a static limit in
the form
\begin{equation}
  \d s^2 = - (1 - \lambda r^2) \, \d \tilde t^2
    + \frac{\d r^2}{1 - \lambda r^2} 
    + r^2 \, \d \Omega^2_{(n-2)}
    + \frac{2 m}{r^{n-1}} \left(\d \tilde t 
    - \frac{\d r}{1 - \lambda r^2} \right)^2.
  \label{KS:AdS-S-T:GKS}
\end{equation}
The first three terms now represent the (Anti-)de Sitter metric in the
spherical coordinates and the last term is again a multiple of a null vector.
Hence, the metric \eqref{KS:AdS-S-T:GKS} takes the generalized Kerr--Schild
(GKS) form
\begin{equation}
  g_{ab} = \bar{g}_{ab} - 2 \H k_a k_b,
  \label{KS:GKSmetric}
\end{equation}
where the background metric $\bar{g}_{ab}$ corresponds to an (A)dS spacetime,
$\mathcal{H}$ is a scalar function and the Kerr--Schild vector $\bk$ is null with
respect to both the full metric $g_{ab}$ and background metric $\bar{g}_{ab}$.

Therefore, the (A)dS--Schwarzschild--Tangherlini metric can be cast to both the
KS form \eqref{KS:KSansatz} with the flat background and the GKS form
\eqref{KS:GKSmetric} with an (anti-)de Sitter background. It implies that the
cosmological constant $\Lambda$ can be included either to the scalar function
$\mathcal{H}$ in the KS form or to the background metric $\bar{g}_{ab}$ in the
GKS form. However, these two possibilities how the cosmological constant enters
the metric occur only in exceptional cases where the Kerr--Schild vector field
$\bk$ in the KS form of the (A)dS--Schwarzschild--Tangherlini metric
\eqref{KS:AdS-S-T:KS} and in the KS form of the (A)dS metric \eqref{KS:AdS:KS}
has the same geometrical properties. Namely, if $\bk$ is non-twisting and
shear-free with the same expansion. It follows that Robinson--Trautman solutions
with a non-vanishing cosmological constant admitting the KS form can be
transformed to the GKS form as has been shown in the case of the higher
dimensional Vaidya metric in section \ref{sec:KS:expanding:nullrad}. However,
in general, Einstein GKS spacetimes do not admit the KS form with a flat
background and therefore we adopt the GKS form as a generalization of the
original KS ansatz to the cases of spacetimes with a non-vanishing cosmological
constant.

In fact, it was shown first by Carter in 1968 \cite{Carter1968} that his solution
of four-dimensional rotating black hole with a non-vanishing cosmological constant
$\Lambda$ can be cast to the GKS form \eqref{KS:GKSmetric}.
Later, Hawking et al.\ in 1999 \cite{HawkingHunterTaylor1999} generalized Carter's
solution to five dimensions, but without detailed derivation. Moreover, their
solution is not given in the GKS form.
Finally, Gibbons et al.\ in 2004 \cite{GibbonsLuPagePope2004} inspired by the
previous works employed the GKS form and special ellipsoidal coordinates which
allowed them to construct a solution representing a rotating black hole with a
cosmological constant in arbitrary dimension.

Throughout the thesis, we will assume that the $n$-dimensional background
(anti-)de Sitter metric with a cosmological constant $\Lambda$ takes the
conformally flat form
\begin{equation}
  \bar{g}_{ab} = \Omega \eta_{ab},
  \label{KS:bgmetric}
\end{equation}
where the conformal factor $\Omega$ is given by
\begin{equation}
\begin{split}
  \Omega_\text{AdS} &= \frac{(n-2)(n-1)}{2\Lambda t^2}, \\
  \Omega_\text{dS} &= - \frac{(n-2)(n-1)}{2\Lambda {x_1}^2},
  \end{split}
  \label{KS:Omega(A)dS}
\end{equation}
respectively, and the Minkowski metric $\eta_{ab}$ is in the canonical form
\begin{equation}
  \eta_{ab} = -\d t^2 + \d x_1^2 + \ldots + \d x_{n-1}^2,
  \label{KS:flatmetric}
\end{equation}
which is convenient for the following calculations.

The vacuum Einstein field equations for the conformally flat metric
\eqref{KS:bgmetric} imply that the conformal factor $\Omega$ satisfies
\begin{equation}
  \frac{\Omega_{,ab}}{\Omega} = \frac{3}{2} \frac{\Omega_{,a} \Omega_{,b}}{\Omega^2}, \qquad
  - \frac{1}{4} \frac{\Omega_{,a} \Omega_{,b}}{\Omega^2} \bar{g}^{ab} = \frac{2}{(n-2)(n-1)} {\Lambda},
  \label{KS:bgEFEs}
\end{equation}
with both possible signs of the cosmological constant $\Lambda$. Note that the
Minkowski limit $\Lambda = 0$ can be obtained by setting $\Omega = 1$.

\section{General Kerr--Schild vector field}
\label{sec:KS:general_k}

We will study general properties of the GKS metric \eqref{KS:GKSmetric} with an
(anti-)de Sitter background $\bar{g}_{ab}$ where $\H$ is a scalar function and
the Kerr--Schild vector $\bk$ is null with respect to the full metric. However,
if $\bk$ is null with respect to the full metric $g_{ab}$ then it follows that
it is also null with respect to the background metric $\bar{g}_{ab}$ and vice
versa. Consequently, the inverse metric to \eqref{KS:GKSmetric} takes the
simple form
\begin{equation}
  g^{ab} = \bar{g}^{ab} + 2 \H k^a k^b,
  \label{KS:inverseGKSmetric}
\end{equation}
where $\bar{g}^{ab} = \Omega^{-1} \eta^{ab}$. This implies that one may use both
metrics for raising or lowering index of the Kerr--Schild vector $\bk$
\begin{equation}
  k_a \equiv g_{ab} k^b = \bar{g}_{ab} k^b, \qquad
  k^a \equiv g^{ab} k_b = \bar{g}^{ab} k_b.
  \label{KS:raiselower_k}
\end{equation}
Our choice of the canonical form for the background metric $\bar{g}_{ab}$
\eqref{KS:bgmetric}, \eqref{KS:flatmetric} allows us to express Christoffel
symbols
\begin{equation}
  \begin{split}
  \Gamma^a_{bc} &= - \left( \H k^a k_b \right)_{,c}
    - \left( \H k^a k_c \right)_{,b}
    + g^{as} \left( \H k_b k_c \right)_{,s} \\
    &\qquad + \half \frac{\Omega_{,c}}{\Omega} \delta^a_b
    + \half \frac{\Omega_{,b}}{\Omega} \delta^a_c
    - \half \frac{\Omega_{,s}}{\Omega} g^{as} \bar{g}_{bc}.
  \end{split}
  \label{KS:Christoffel}
\end{equation}

The first crucial step in our study of properties of the GKS metric
\eqref{KS:GKSmetric} is to show under which conditions the Einstein field
equations imply that the KS vector field $\bk$ is geodetic.

Obviously, due to the form of the GKS ansatz \eqref{KS:GKSmetric} it will be
convenient to employ the higher dimensional Newman--Penrose frame formalism,
briefly summarized in section \ref{sec:intro:NP}, and naturally identify the
Kerr--Schild vector $\bk$ with the null frame vector $\bl$
\eqref{intro:frame:constraints}. From now on, we will denote both vectors as
$\bk$, whereas the corresponding Ricci rotation coefficients as $L_{ab}$.

Since the KS vector $\bk$ appears in many terms of the Christoffel symbols
$\Gamma^a_{bc}$ \eqref{KS:Christoffel}, the simplest component of the Ricci
tensor is the boost weight two component $R_{00} = R_{ab} k^a k^b$. After quite
involved calculations using $k_a k^a = k_{a,b} k^a = k^a_{\phantom{a},b} k_a = 0$
we end up with a remarkably simple result
\begin{equation}
  R_{00} = 2 \H k_{c;a} k^{a} k^{c}_{\phantom{c};b} k^{b}
    - \half (n-2) \left( \frac{\Omega_{,ab}}{\Omega}
    - \frac{3}{2} \frac{\Omega_{,a} \Omega_{,b}}{\Omega^2}  \right) k^{a} k^{b},
\end{equation}
for an arbitrary conformal factor $\Omega$. Therefore, for the conformal factor
of an (anti-)de Sitter background metric \eqref{KS:Omega(A)dS} obeying
\eqref{KS:bgEFEs} we obtain
\begin{equation}
  R_{00} = 2 \H k_{c;a} k^{a} k^{c}_{\phantom{c};b} k^{b} = 2 \H L_{i0} L_{i0}.
\end{equation}
It now follows from the Einstein field equations that $L_{i0} = 0$, i.e.\ $\bk$
is geodetic, if and only if $T_{00} = 0$.

\begin{proposition}
  \label{KS:proposition:geodetic_k}
  The Kerr--Schild  vector $\bk$ in the generalized Kerr--Schild metric
  \eqref{KS:GKSmetric} is geodetic if and only if the boost weight 2 component
  of the energy--momentum tensor \mbox{$T_{00}=T_{ab} k^a k^b$} vanishes.
\end{proposition}

Proposition \ref{KS:proposition:geodetic_k} implies that the Kerr--Schild vector
$\bk$ is geodetic not only in Einstein GKS spacetimes, where the energy--momentum
tensor is absent, but also in spacetimes with aligned matter content such as
aligned Maxwell field $F_{ab} k^a \propto k_b$ or aligned pure radiation
$T_{ab} \propto k_a k_b$.  

Moreover, if the Kerr--Schild vector $\bk$ is geodetic then we may assume,
without loss of generality, that it is also affinely parametrized since we are
still able to rescale $\bk$ by an appropriate scalar factor. Subsequently, this
factor can be included to the KS function $\mathcal{H}$ and hence the GKS form
of the original metric remains unchanged.
In order to preserve the normalization \eqref{intro:frame:constraints} of an
already chosen null frame, one has to rescale $\bk$ by preforming the boost
\eqref{intro:frame:boosts}
\begin{equation}
  \hat{\bk} = \lambda \bk, \qquad
  \hat{\bn} = \lambda^{-1} \bn.
\end{equation}
Recall that the Kerr--Schild vector $\bk$ is geodetic if $k_{a;b} k^b = L_{10} k_a$
and since we require $\hat{\bk}$ to be affinely parametrized, therefore,
$\hat{L}_{10} = 0$. From the transformation properties of the Ricci rotation
coefficient $L_{10}$ under the boost \eqref{intro:Riccicoeff:boosts}, it follows
\begin{equation}
  \D \lambda = - \lambda L_{10},
\end{equation}
which determines the necessary scalar factor $\lambda$. Finally, we denote
$\hat{\mathcal{H}} \equiv \lambda^{-2} \mathcal{H}$, therefore, the Kerr--Schild
term $\mathcal{H} k_a k_b$ transforms under the boost as a quantity with boost
weight 0, i.e.\ $\hat{\mathcal{H}} \hat{k}_a \hat{k}_b = \mathcal{H} k_a k_b$,
and the GKS form is not affected by this operation.

Thus, in the following sections, the geodetic Kerr--Schild vector $\bk$ is
assumed to be affinely parametrized. This will lead to a significant
simplification of the necessary calculations.

\subsection[Kerr--Schild congruence in the background spacetime]{Kerr--Schild congruence in the background\\spacetime}
\label{sec:KS:riccicoeff}
One may compare the geodesicity and the optical properties of the null
Kerr--Schild congruence $\bk$ in the full GKS spacetime and in the background
(A)dS spacetime. It will be shown that there is a close relation between the
congruences in both spacetimes. 

Quantities constructed from the background metric are easily obtained from
quantities constructed from the full GKS metric simply by setting $\mathcal{H}$
to zero. For instance, using the Christoffel symbols \eqref{KS:Christoffel}, it
is straightforward to show that
\begin{equation}
  \begin{split}
  k_{a;b} k^b &= k_{a,b} k^b = k_{a\overline{;}b} k^b, \\
  k^a_{\phantom{a};b} k^b &=  k^a_{\phantom{a},b} k^b + \frac{\Omega_{,b}}{\Omega} k^a k^b
    = k^a_{\phantom{a}\overline{;}b} k^b ,
  \end{split}
\end{equation}
where `` $\overline{;}$ '' in the expression $k_{a\overline{;}b}$ denotes the
covariant derivative with respect to the background (A)dS metric $\bar{g}_{ab}$.
Thus, one can immediately see that the Kerr--Schild vector $\bk$ is geodetic in
the full GKS metric if and only if it is geodetic in the (A)dS background
$\bar{g}_{ab}$.

The geometrical properties of the Kerr--Schild congruence $\bk$ in the full GKS
spacetime are encoded in the optical matrix $L_{ij}$ \eqref{intro:Riccicoeff}.
Following \cite{OrtaggioPravdaPravdova2008}, we define a null frame $\bar{\bn}$,
$\bk$, $\bm^{(i)}$ in the background spacetime $\bar{g}_{ab}$, where
\begin{equation}
  \bar{n}_a = n_a + \H k_a,
  \label{KS:background_n}
\end{equation}
and the remaining frame vectors $\bk$, $\bm^{(i)}$ are same as in the full
spacetime. This choice guarantees
\begin{equation}
  \bar{g}_{ab} = 2k_{(a} \bar{n}_{b)} + \delta_{ij} m^{(i)}_a m^{(j)}_b
\end{equation}
and allows us to express the optical matrix $\bar{L}_{ij}$ in the background
spacetime, which can be then compared with $L_{ij}$. Using
\eqref{KS:Christoffel}, it follows
\begin{equation}
  L_{ij}\equiv k_{a;b} m^{(i)a} m^{(j)b} =
  k_{a\overline{;} b} m^{(i)a} m^{(j)b} \equiv {\bar{L}_{ij}}.
\end{equation}
Therefore, the optical matrices of the Kerr--Schild congruence $\bk$ with respect
to the full GKS metric $g_{ab}$ and the (A)dS background metric $\bar{g}_{ab}$
are equal, i.e.\ the corresponding expansion, shear and twist scalars have the
same values in both spacetimes. Note that for $\bk$ being geodetic, $L_{ij}$ does
not depend on our particular choice \eqref{KS:background_n} since in such case
$L_{ij}$ is invariant under null rotations with $\bk$ fixed
\eqref{intro:Riccicoeff:nullrot}.

It should be emphasized that the index of the background frame covector
$\bar{n}_a$ \eqref{KS:background_n} is raised by the full metric as
$\bar n^a \equiv g^{ab} \bar n_b = n^a + \H k^a$, whereas the contraction with
the background metric gives $\bar g^{ab} \bar n_b = n^a - \H k^a$. Similarly as
for the vector $\bk$ \eqref{KS:raiselower_k}, the vector indices of $\bm^{(i)}$
may be raised and lowered by both metrics since
$m^a_{(i)} \equiv g^{ab} m^{(i)}_b = \bar g^{ab} m^{(i)}_b$.

One may also compare the remaining Ricci rotational coefficients with respect to
the full GKS spacetime and the background spacetime. Thus, we obtain
\begin{equation}
  \begin{split}
  &L_{i0} = \bar{L}_{i0}, \qquad
  L_{10} = \bar{L}_{10}, \qquad
  L_{1i} = \bar{L}_{1i} - \H \bar{L}_{i0}, \qquad
  L_{i1} = \bar{L}_{i1}, \phantom{\half} \\
  &L_{11} = \bar{L}_{11} - \D\H - \H \bar{L}_{10} + \H \frac{\Omega_{,a}}{\Omega} k^a, \qquad
  N_{i0} = \bar{N}_{i0}, \\
  &N_{i1} = \bar{N}_{i1} + 2 \H \bar{L}_{1i} - \H \bar{L}_{i1} + \H \bar{N}_{i0} - \H^2 \bar{L}_{i0}
    + \delta_i \H + \H \frac{\Omega_{,a}}{\Omega} m_{(i)}^a, \\
  &N_{ij} = \bar{N}_{ij} + \H \bar{L}_{ji} - \H \frac{\Omega_{,a}}{\Omega} k^a \delta_{ij}, \qquad
  \M{i}{j0} = \bM{i}{j0}, \\
  &\M{i}{jk} = \bM{i}{jk}, \qquad
  \M{i}{j1} = \bM{i}{j1} + \H \left( \bar{L}_{ij} - \bar{L}_{ji} \right) + \H \bM{i}{j0}.
  \end{split}
\end{equation}

Let us mention that for $\bk$ being geodetic and affinely parametrized
the following relations hold
\begin{equation}
  n_{a;b} k^b = \bar{n}_{a\bar{;}b} k^b, \qquad
  m^{(i)}_{a;b} k^b = m^{(i)}_{a\bar{;}b} k^b.
\end{equation}
In other words, the frame vectors $\bn$, $\bm^{(i)}$ are parallelly transported
along $\bk$ in the full GKS spacetime if and only if the frame vectors
$\bar{\bn}$, $\bm^{(i)}$ are parallelly transported along $\bk$ in the background
spacetime. This can help us in finding a parallelly propagated frame in the full
GKS spacetime, which can be a nontrivial task. Instead, it could be easier to
find such a frame in the background (A)dS spacetime and then use
\eqref{KS:background_n} relating these two frames.

In fact, we partially employ this procedure in section \ref{sec:KS:Kerr-(A)dS}
since the full GKS metric of the five dimensional Kerr--(A)dS spacetime is quite
complicated and non-diagonal, whereas the background (A)dS metric is diagonal
and the calculations are not so involved.

\section{Geodetic Kerr--Schild vector field}
\label{sec:KS:geodetic_k}

So far, we discussed the properties of the GKS metric \eqref{KS:GKSmetric} with
an arbitrary null Kerr--Schild vector field $\bk$ without any additional
assumptions. In the rest of this chapter, we will consider Einstein GKS
spacetimes and GKS spacetimes with aligned matter fields. Then it follows from
proposition \ref{KS:proposition:geodetic_k} that the Kerr--Schild vector $\bk$
is geodetic as discussed in section \ref{sec:KS:general_k}. We also assume an
affine parametrization of $\bk$. Using higher dimensional Newman--Penrose
formalism, we employ the Einstein field equations and analyze the conditions
imposed on the GKS ansatz.

Assuming geodesicity of $\bk$, we arrive at the convenient expressions for the
contracted Christoffel symbols frequently used in the following calculations
\begin{equation}
  \begin{split}
  \Gamma^a_{bc} k^b &= - \D\H k^a k_c + \half \frac{\Omega_{,c}}{\Omega} k^a
    + \half \frac{\Omega_{,b}}{\Omega} k^b \delta^a_c
    - \half \frac{\Omega_{,b}}{\Omega} \bar{g}^{ab} k_c, \\
  \Gamma^a_{bc} k_a &= \D\H k_b k_c + \half \frac{\Omega_{,c}}{\Omega} k_b
    + \half \frac{\Omega_{,b}}{\Omega} k_c - \half \frac{\Omega_{,a}}{\Omega} k^a \bar{g}_{bc}.
  \end{split}
\end{equation}

\subsection{Ricci tensor}
\label{sec:KS:Ricci}

Despite the simple form of the GKS metric \eqref{KS:GKSmetric} along with the
assumption that $\bk$ is geodetic, expressing the Ricci tensor is a quite
complicated task since hundreds of terms appear during the derivation. Hence,
we were able to perform this and some of the following tedious calculations only
by means of the computer algebra system {\sc Cadabra}
\cite{Peeters2006,Peeters2007}. Fortunately, after many operations, we obtain
the Ricci tensor in the compact form
\begin{equation}
  \begin{split}
  R_{ab} &= \left( \H k_a k_b \right)_{;cd} g^{cd} - \left( \H k^s k_a \right)_{;bs}
    - \left( \H k^s k_b \right)_{;as} + \frac{2 {\Lambda}}{n-2} \bar{g}_{ab} \\
    &\qquad - 2 \H \left( \D^2 \H + L_{ii} \D\H + 2 \H \omega^2 \right) k_a k_b,
  \end{split}
  \label{KS:Ricci}
\end{equation}
which for $\Lambda=0$ reduces to the result of \cite{OrtaggioPravdaPravdova2008}
where, at first sight, the sign before the last term is opposite. However, it can
be easily shown that both results are in full accordance, if one rewrites the
covariant derivatives in \eqref{KS:Ricci} in terms of the partial derivatives.

From \eqref{KS:Ricci} it follows that the Kerr--Schild vector $\bk$ is an
eigenvector of the Ricci tensor
\begin{equation}
  R_{ab} k^b = - \left[ \D^2 \H + (n - 2) \theta \D \H + 2 \H \omega^2
    - \frac{2 \Lambda}{n - 2} \right] k_a
  \label{KS:Ricci:eigenvector}
\end{equation}
and one can immediately see that the positive boost weight frame components
of the Ricci tensor vanish
\begin{equation}
  R_{00} = 0, \qquad R_{0i} = 0.
  \label{KS:Ricci:bw2&1}
\end{equation}
The non-vanishing frame components of the Ricci tensor can be straightforwardly
obtained from \eqref{KS:Ricci} by performing appropriate contractions with the
corresponding frame vectors
\begin{align}
  R_{01} &= - \D^{2}\H - (n - 2) \theta \D\H - 2\H \omega^{2}
    + \frac{2 \Lambda}{n - 2},
  \label{KS:R01} \\
  R_{ij} &= 2 \H L_{ik} L_{jk} - 2 \left( \D\H + (n-2) \theta \H \right) S_{ij}
    + \frac{2 {\Lambda}}{n - 2} \delta_{ij},
  \label{KS:Rij} \\
  R_{1i} &= - \delta_i (\D\H)
     + 2 L_{[i1]} \D\H
     + 2 L_{ij} \delta_j \H
     - S_{jj} \delta_i \H
     + 2 \H \bigg( \delta_j A_{ij} \notag \\
     &\qquad + A_{ij} \M{j}{kk}
     - A_{jk} \M{i}{jk}
     - L_{1i} S_{jj}
     + 3 L_{ij} L_{[1j]}
     + L_{ji} L_{(1j)} \bigg),
  \label{KS:R1i}\\
  R_{11} &= \delta_i (\delta_i\H)
     + \left( N_{ii} - 2 \H S_{ii} \right) \D\H
     + \left( 4 L_{1i} - 2 L_{i1} + \M{i}{jj} \right) \delta_i\H \notag \\
     &\qquad - S_{ii} \Delta \H
     + 2 \H \bigg( 2 \delta_i L_{[1i]} + 4 L_{1i} L_{[1i]} + L_{i1} L_{i1}
     - L_{11} S_{ii} \notag \\
     &\qquad + 2 L_{[1i]} \M{i}{jj}
     - 2 A_{ij} N_{ij} - 2 \H \omega^2 \bigg)
     + \frac{4 \H {\Lambda}}{n - 1},
  \label{KS:R11}
\end{align}
where the components are sorted by their boost weight and some of them were
further simplified using the Ricci identities \cite{OrtaggioPravdaPravdova2007}.

\subsection[Riemann tensor and algebraic type of the Weyl tensor]
{Riemann tensor and algebraic type of the Weyl\\tensor}
\label{sec:KS:Weyl}

In this section, we point out that GKS spacetimes \eqref{KS:GKSmetric} with
a geodetic Kerr--Schild vector $\bk$ are algebraically special and $\bk$ is a
multiple WAND of the Weyl tensor. First, we express the frame components of the
Riemann tensor. As in the case of the Ricci tensor, the positive boost weight
components of the Riemann tensor identically vanish
\begin{equation}
  R_{0i0j} = 0, \qquad R_{010i} = 0, \qquad R_{0ijk} = 0.
  \label{KS:Riemann:bw2&1}
\end{equation}
The non-vanishing frame components of the Riemann tensor sorted by their boost
weight read
\begin{align}
  R_{0101} &= \D^2\H - \frac{2 \Lambda}{(n - 2)(n - 1)},
  \label{KS:R0101} \\
  R_{01ij} &= - 2 A_{ij} \D\H + 4 \H S_{k[j} A_{i]k}, \phantom{\frac{1}{()}}
  \label{KS:R01ij} \\
  R_{0i1j} &= - L_{ij} \D\H + 2 \H A_{ik} L_{kj}
    + \frac{2 \Lambda}{(n - 2)(n - 1)} \delta_{ij},
  \label{KS:R0i1j} \\
  R_{ijkl} &= 4 \H \left( A_{ij} A_{kl} + A_{l[i} A_{j]k} + S_{l[i} S_{j]k} \right) \phantom{\frac{1}{()}} \notag \\
    &\qquad + \frac{2 \Lambda}{(n - 2)(n - 1)} \left( \delta_{ik} \delta_{jl}
    - \delta_{il} \delta_{jk} \right),
  \label{KS:Rijkl} \\
  R_{011i} &= - \delta_i \left( \D\H \right) + 2 L_{[i1]} \D\H + L_{ji} \delta_j\H
    + 2 \H \left( L_{1j} L_{ji} - L_{j1} S_{ij} \right), \phantom{\frac{1}{()}} \\
  R_{1ijk} &= 2 L_{[j|i} \delta_{|k]}\H + 2 A_{jk} \delta_i\H
    + 4 \H \bigg( \delta_{[k} S_{j]i} + \M{l}{[jk]} S_{il} - \M{l}{i[j} S_{k]l} \notag \\
    &\qquad + L_{1i} A_{jk} + L_{1[k} A_{j]i} \bigg), \\
  R_{1i1j} &= \delta_i ( \delta_j\H ) + \M{k}{(ij)} \delta_k\H + 4 L_{1(i} \delta_{j)}\H
    - 2 L_{(i|1} \delta_{j)}\H + N_{(ij)} \D\H \phantom{\frac{1}{()}} \notag \\
    &\qquad - S_{ij} \Delta\H
    + 2 \H \bigg( \delta_{(i} L_{1|j)} - \Delta S_{ij} - 2 L_{1(i} L_{j)1}
    + 2 L_{1i} L_{1j} \notag \\
    &\qquad - L_{k(i} N_{k|j)}
    + L_{1k} \M{k}{(ij)}
    - 2 \H L_{k(i} A_{j)k}
    - 2 \H A_{ik} A_{jk} \phantom{\frac{1}{()}} \notag \\
    &\qquad - L_{k(i} \M{k}{j)1}
    - L_{(i|k} \M{k}{j)1} \bigg).
  \label{KS:R1i1j}
\end{align}
Note that the boost weight zero components of the Riemann tensor $R_{0101}$,
$R_{01ij}$, $R_{0i1j}$ and $R_{ijkl}$ are given only by the Kerr--Schild
function $\H$, the optical matrix $L_{ij}$ and the cosmological constant
$\Lambda$. Actually, in section \ref{sec:KS:rdep}, this fact allows us to
explicitly determine the dependence of these components on an affine parameter
$r$ along the geodesics of the Kerr--Schild congruence $\bk$.

It is also convenient to express explicitly the frame components of the Weyl
tensor \eqref{intro:Weyl:definition} for a general Ricci tensor that will be
occasionally employed throughout the thesis, explicitly
\begin{align}
  C_{0i0j} &= R_{0i0j} - \frac{1}{n-2} R_{00} \delta_{ij}, \phantom{\frac{1}{(n-1)(n-2)}}
  \label{Weyl:C0i0j} \\
  C_{010i} &= R_{010i} + \frac{1}{n-2} R_{0i}, \phantom{\frac{1}{(n-1)(n-2)}}
  \label{Weyl:C010i} \\
  C_{0ijk} &= R_{0ijk} + \frac{1}{n-2} (R_{0k} \delta_{ij} - R_{0j} \delta_{ik}), \phantom{\frac{1}{(n-1)(n-2)}}
  \label{Weyl:C0ijk} \\
  C_{0101} &= R_{0101} + \frac{2}{n-2} R_{01} - \frac{1}{(n-1)(n-2)} R, 
  \label{Weyl:C0101} \\
  C_{01ij} &= R_{01ij}, \phantom{\frac{1}{(n-1)(n-2)}}
  \label{Weyl:C01ij} \\
  C_{0i1j} &= R_{0i1j} - \frac{1}{n-2} (R_{ij} + R_{01} \delta_{ij}) + \frac{1}{(n-1)(n-2)} R \delta_{ij},
  \label{Weyl:C0i1j} \\
  C_{ijkl} &= R_{ijkl} - \frac{2}{n-2} (R_{j[l} \delta_{k]i} - R_{i[l} \delta_{k]j})
    + \frac{2}{(n-1)(n-2)} R \delta_{i[k} \delta_{l]j},
  \label{Weyl:Cijkl} \\
  C_{011i} &= R_{011i} - \frac{1}{n-2} R_{1i}, \phantom{\frac{1}{(n-1)(n-2)}}
  \label{Weyl:C011i} \\
  C_{1ijk} &= R_{1ijk} + \frac{1}{n-2} (R_{1k} \delta_{ij} - R_{1j} \delta_{ik}), \phantom{\frac{1}{(n-1)(n-2)}}
  \label{Weyl:C1ijk} \\
  C_{1i1j} &= R_{1i1j} - \frac{1}{n-2} R_{11} \delta_{ij}. \phantom{\frac{1}{(n-1)(n-2)}}
  \label{Weyl:C1i1j}
\end{align}

Since the positive boost weight frame components of the Ricci tensor
\eqref{KS:Ricci:bw2&1} and the Riemann tensor \eqref{KS:Riemann:bw2&1}
identically vanish it means that this holds also for the corresponding
components of the Weyl tensor \eqref{Weyl:C0i0j}--\eqref{Weyl:C0ijk}, i.e.\
\begin{equation}
  C_{0i0j} = 0, \qquad C_{010i} = 0, \qquad C_{0ijk} = 0,
  \label{KS:Weyl:bw2&1}
\end{equation}
and therefore
\begin{proposition}
  \label{KS:proposition:Weyltypes}
  Generalized Kerr--Schild spacetimes \eqref{KS:GKSmetric} with a geodetic
  Kerr--Schild vector $\bk$ are algebraically special with $\bk$ being the
  multiple WAND.
\end{proposition}
In other words, GKS spacetimes \eqref{KS:GKSmetric} with a geodetic Kerr--Schild
vector $\bk$ are of Weyl type II or more special. Let us remind that the
Kerr--Schild vector $\bk$ is geodetic if and only if the boost weight 2
component of the energy--momentum tensor vanish $T_{00} = 0$ as follows from
proposition \ref{KS:proposition:geodetic_k}. Consequently, Einstein GKS
spacetimes and GKS spacetimes with an aligned matter field are algebraically
special.

It can be shown \cite{PravdaPravdovaOrtaggio2007} that in static spacetimes
admitting a WAND $\bl = (\ell_t, \ell_A)$ with the metric not depending on the
direction of time one may always construct a distinct WAND
$\bn = (-\ell_t, \ell_A)$ with the same order of alignment. Thus, static
spacetimes are compatible only with Weyl types G, I$_i$, D or O.

A similar statement also holds for stationary spacetimes with the metric
remaining unchanged under reflection symmetry, for instance, the symmetry
$t \rightarrow -t$, $\varphi_i \rightarrow -\varphi_i$ of the higher dimensional
Kerr--(A)dS metric \cite{GibbonsLuPagePope2004} expressed in the Boyer--Lindquist
coordinates. If $\ell$ is a WAND then, again due to the symmetry, there is a
distinct WAND $\bn$. Additional assumption has to be imposed, namely, one has to
require non-vanishing ``divergence scalar'' corresponding to the expansion scalar
$\theta$ in case of geodetic WAND. This ensures that both WANDs do not
represent the same null direction $\bl \neq - \bn$ and hence such spacetimes are
of Weyl types G, I$_i$, D or conformally flat. See
\cite{PravdaPravdovaOrtaggio2007} for further details.

Proposition \ref{KS:proposition:Weyltypes} along with the above statements for
static and stationary spacetimes lead to
\begin{corollary}
  \label{KS:proposition:Weyltypes2}
  Generalized Kerr--Schild spacetimes \eqref{KS:GKSmetric} with a geodetic
  Kerr--Schild vector $\bk$ which are
  \begin{enumerate}
    \item[(a)] either static
    \item[(b)] or stationary with reflection symmetry and non-vanishing expansion
  \end{enumerate}
  are of Weyl type D or conformally flat.
\end{corollary}
		     
Note that these results immediately imply that the Kerr--(A)dS metrics in
all dimensions \cite{GibbonsLuPagePope2004} are of Weyl type D, as was shown
previously in \cite{Hamamotoetal2006} by explicit calculation of the Weyl tensor.

\section[Brinkmann warp product of Kerr--Schild spacetimes]
{Brinkmann warp product of Kerr--Schild\\spacetimes}
\label{sec:KS:warp}

The Brinkmann warp product introduced in section \ref{sec:intro:warp} is a
convenient method for generating new $n$-dimensional solutions of the vacuum
Einstein field equations from known $(n-1)$-dimensional Ricci-flat or Einstein
metrics. Naturally, this warp product can be also applied to Einstein GKS
spacetimes. Thus, let us consider the seed metric of the GKS form
\eqref{KS:GKSmetric}
\begin{equation}
  \d \tilde{s}^2 = \bar{g}_{ab} \, \d x^a \, \d x^b - 2 \H k_a k_b \, \d x^a \, \d x^b.
  \label{KS:GKSwarped:seed}
\end{equation}
Then the warped metric $\d s^2$ is given by \eqref{intro:warp:warpedmetric}
\begin{equation}
  \d s^2 = \frac{1}{f} \, \d z^2 + f \bar{g}_{ab} \, \d x^a \, \d x^b
    - 2 f \H k_a k_b \, \d x^a \, \d x^b.
  \label{KS:GKSwarped}
\end{equation}
Since the seed background metric $\bar{g}_{ab}$ represents Einstein space of
Weyl type O, i.e.\ conformally flat, and moreover the warp product preserves
the Weyl type of algebraically special spacetimes, as mentioned in section
\ref{sec:intro:warp}, the new warped background metric
${f}^{-1} \, \d z^2 + f \bar{g}_{ab} \, \d x^a \, \d x^b$ describes necessarily
an (anti-)de Sitter or Minkowski spacetime. The remaining term
$2 f \H k_a k_b \, \d x^a \, \d x^b$ in \eqref{KS:GKSwarped} is again a multiple
of two null vectors since the original Kerr--Schild vector $\bk$ is obviously
null with respect to the new warped metric as well. Therefore, the warped metric
\eqref{KS:GKSwarped} is also an Einstein GKS metric and thus the Brinkmann warp
product preserves the GKS form.

If the $(n-1)$-dimensional seed background metric $\bar{g}_{ab}$ takes the
canonical form \eqref{KS:bgmetric}, \eqref{KS:flatmetric} with the conformal
factor $\tilde \Omega$ given by the cosmological constant of the seed
\eqref{intro:warp:seedricci} as
\begin{align}
  &\tilde \Omega_\text{AdS} = \frac{1}{(\lambda b + d^2) t^2}, \\
  &\tilde \Omega_\text{dS} = - \frac{1}{(\lambda b + d^2) {x_1}^2},
\end{align}
respectively, then the $n$-dimensional warped background metric can be also put
to the corresponding canonical form
\begin{align}
  f \bar{g}_{ab} \d x^a \d x^b + f^{-1} \d z^2
    &= f \tilde \Omega ( -\d t^2 + \d x_1^2 + \ldots + \d x_{n - 2}^2 ) + f^{-1} \d z^2 \notag \\
    &= \Omega ( - \d \hat t^2 + \d \hat x_1^2 + \ldots + \d \hat x_{n - 2}^2 + \d \hat z^2 ), 
  \label{KS:warpedbackground}
\end{align}
where $\Omega$ is defined exactly as in \eqref{KS:Omega(A)dS}, the warp factor
$f(z)$ is given by \eqref{intro:warp:factor} and the new and old coordinates
are related by an appropriate transformation depending on the signs of the Ricci
scalars $\tilde R$ and $R$ of the seed and the warped metric, respectively.
Recall that, as discussed in section \ref{sec:intro:warp}, not all combinations
of the signs of $\tilde R$ and $R$ are possible. In the trivial case
$\tilde R = 0$, $R = 0$ corresponding to the direct product, just one extra flat
dimension is added to the $(n - 1)$-dimensional Minkowski metric. In other cases,
the following coordinate transformations has to be performed in order to cast
the warped background metric to the canonical form

\bigskip
$\mathrm{AdS}_{n-1} \Rightarrow \mathrm{AdS}_{n}:$
\begin{equation}
  t = \hat{t}, \qquad
  x_1^2 = \hat x_1^2 + \hat z^2, \qquad
  x_{\itilde} = \hat x_{\itilde}, \qquad
  z = \frac{\sqrt{- \left( d^2 + \lambda b \right)} \hat z}{\lambda \hat x_1} + \frac{d}{\lambda},
  \label{KS:warp:AdStoAdS}
\end{equation}

$\mathrm{dS}_{n-1} \Rightarrow \mathrm{AdS}_{n}:$
\begin{equation}
  t^2 = \hat t^2 - \hat z^2, \qquad
  x_i = \hat x_i, \qquad
  z = \frac{\sqrt{d^2 + \lambda b} \, \hat t}{- \lambda \hat z} + \frac{d}{\lambda},
  \label{KS:warp:dStoAdS}
\end{equation}

$\mathrm{dS}_{n-1} \Rightarrow \mathrm{dS}_{n}:$
\begin{align}
  t^2 = \hat t^2 - \hat z^2, \qquad
  x_i = \hat x_i, \qquad
  z = \frac{\sqrt{d^2 + \lambda b} \, \hat z}{\lambda \hat t} + \frac{d}{\lambda},
  \label{KS:warp:dStodS}
\end{align}

$\mathrm{M}_{n-1} \Rightarrow \mathrm{AdS}_{n}:$
\begin{align}
  t = \hat t, \qquad
  x_i = \hat x_i, \qquad
  z = \frac{1}{\lambda \hat z} + \frac{d}{\lambda}.
  \label{KS:warp:MtoAdS}
\end{align}
We have to emphasize that in the above expressions
\eqref{KS:warp:AdStoAdS}--\eqref{KS:warp:MtoAdS}, exceptionally, the index $i$
goes from 1 to $n - 2$ and the index $\itilde$ ranges from 2 to $n - 2$.

\section{Einstein Kerr-Schild spacetimes}

In section \ref{sec:KS:geodetic_k}, we discussed the algebraic properties and
expressed the Ricci and Riemann tensors of the GKS metric with the only
additional assumption that the Kerr--Schild vector $\bk$ is geodetic. As follows
from proposition \ref{KS:proposition:geodetic_k}, this includes a wide class
of spacetimes, for instance, Einstein spaces or spacetimes with an aligned
matter field. In this section, we will restrict our analysis to the simplest
case, namely, Einstein GKS spacetimes. Thus, we employ the
vacuum Einstein field equations in order to study their implications for the GKS
metric.

In $n$ dimensions, the vacuum Einstein field equations can be rewritten as
\begin{equation}
  R_{ab} = \frac{2 \Lambda}{n-2} g_{ab}.
  \label{KS:vacuum:Ricci}
\end{equation}
Since the Kerr--Schild vector $\bk$ is geodetic in the case of Einstein GKS
spacetimes, we can substitute the Ricci tensor \eqref{KS:Ricci} to
\eqref{KS:vacuum:Ricci}. The term containing the cosmological constant $\Lambda$
on the right hand side of \eqref{KS:vacuum:Ricci} is multiplied by the full GKS
metric $g_{ab}$, whereas the term with the cosmological constant in the
expression of the Ricci tensor \eqref{KS:Ricci} is multiplied by the background
metric $\bar{g}_{ab}$. If we assume that both cosmological constants are equal,
the difference between these terms is proportional to the Kerr--Schild term
$2 \H k_a k_b$ and we arrive at the Einstein field equations for Einstein GKS
metrics in the form
\begin{equation}
  \begin{split}
  &\left( \H k_a k_b \right)_{;cd} g^{cd} - \left( \H k^s k_a \right)_{;bs}
    - \left( \H k^s k_b \right)_{;as} \\ 
  &\qquad - 2 \H \left( \D^2 \H + L_{ii} \D\H + 2 \H \omega^2 - \frac{2 \Lambda}{n-2} \right) k_a k_b = 0.
  \end{split}
  \label{KS:vacuum:EFEs}
\end{equation}
Using the frame components of the Ricci tensor
\eqref{KS:Ricci:bw2&1}--\eqref{KS:R11}, one may express the frame components of
the Einstein field equations \eqref{KS:vacuum:EFEs} as
\begin{align}
  &\D^{2}\H + (n - 2) \theta \D\H + 2\, \H \omega^{2} = 0, \phantom{\half}
  \label{KS:vacuum:EFE01} \\
  &2 \H L_{ik} L_{jk} - 2 \left( \D\H + (n-2) \theta \H \right) S_{ij} = 0, \phantom{\half}
  \label{KS:vacuum:EFEij} \\
  &\delta_i (\D\H)
    - 2 L_{[i1]} \D\H
    - 2 L_{ij} \delta_j \H
    + S_{jj} \delta_i \H
    - 2 \H \bigg( \delta_j A_{ij} \notag \\
    &\qquad + A_{ij} \M{j}{kk}
    - A_{jk} \M{i}{jk}
    - L_{1i} S_{jj}
    + 3 L_{ij} L_{[1j]}
    + L_{ji} L_{(1j)} \bigg) = 0,
  \label{KS:vacuum:EFE1i} \\
  &\delta_i (\delta_i\H)
    + \left( N_{ii} - 2 \H S_{ii} \right) \D\H
    + \left( 4 L_{1i} - 2 L_{i1} + \M{i}{jj} \right) \delta_i\H \notag \\
    &\qquad - S_{ii} \Delta \H
    + 2 \H \bigg( 2 \delta_i L_{[1i]} + 4 L_{1i} L_{[1i]} + L_{i1} L_{i1}
    - L_{11} S_{ii} \notag \\
    &\qquad + 2 L_{[1i]} \M{i}{jj}
    - 2 A_{ij} N_{ij} - 2 \H \omega^2 \bigg)
    + \frac{4 \H {\Lambda}}{n - 1} = 0.
  \label{KS:vacuum:EFE11}
\end{align}
Note that the cosmological constant appears only in the boost weight $-2$ frame
component \eqref{KS:vacuum:EFE11} corresponding to the contraction of the
Einstein field equations \eqref{KS:vacuum:EFEs} with two frame vectors $\bn$.
It may seem that the terms containing $\Lambda$ in \eqref{KS:vacuum:EFEs} and
\eqref{KS:vacuum:EFE11} are not in accordance, but the additional term with
$\Lambda$ appears when one rewrites the derivatives of the Ricci rotation
coefficients in \eqref{KS:vacuum:EFEs} by means of the Ricci identities
\cite{OrtaggioPravdaPravdova2007}.

Following \cite{OrtaggioPravdaPravdova2008}, we rewrite the trace of
\eqref{KS:vacuum:EFEij} as
\begin{equation}
  2 \H L_{ij} L_{ij} - 2 \left( \D\H + (n-2) \theta \H \right) (n-2) \theta = 0,
\end{equation}
using $\H^{-1} \D \H = \D \log\H$ and the decomposition of the optical matrix
$L_{ij}$ \eqref{intro:optical_matrix:decomposition}, as
\begin{align}
  (n - 2) \theta (\D \log\H) &= L_{ij} L_{ij} - (n - 2)^2 \theta^2 \nonumber \\
  &= \sigma^2 + \omega^2 - (n - 2)(n - 3) \theta^2.
  \label{KS:vacuum:EFEijTrace}
\end{align}
Obviously, $\H$ appears in \eqref{KS:vacuum:EFEijTrace} only if $\theta \neq 0$,
therefore, it is natural to study non-expanding GKS spacetimes, where $\theta=0$,
and expanding GKS spacetimes, where $\theta \neq 0$, separately. This will be
done in the following sections \ref{sec:KS:nonexpanding} and
\ref{sec:KS:expanding}, respectively.

The positive boost weight frame components of the Weyl tensor vanish identically,
as was shown in general for GKS spacetimes with a geodetic Kerr--Schild vector
field $\bk$ \eqref{KS:Weyl:bw2&1}.
Substituting the Ricci tensor of Einstein spaces \eqref{KS:vacuum:Ricci} to the
definition of the Weyl tensor \eqref{intro:Weyl:definition}, one obtains
\begin{equation}
  C_{abcd} = R_{abcd} - \frac{4 \Lambda}{(n - 1)(n - 2)} g_{a[c} g_{d]b}
  \label{KS:Weyl:Einsteinspaces}
\end{equation}
and the non-trivial frame components of the Weyl tensor
\eqref{Weyl:C0101}--\eqref{Weyl:C1i1j} for Einstein GKS spacetimes read
\begin{align}
  C_{0101} &= R_{0101} + \frac{2 \Lambda}{(n - 2)(n - 1)},
  \label{KS:C0101} \\
  C_{01ij} &= R_{01ij}, \phantom{\frac{1}{()}}
  \label{KS:C01ij} \\
  C_{0i1j} &= R_{0i1j} - \frac{2 \Lambda}{(n - 2)(n - 1)} \delta_{ij},
  \label{KS:C0i1j} \\
  C_{ijkl} &= R_{ijkl} - \frac{2 \Lambda}{(n - 2)(n - 1)} \left( \delta_{ik} \delta_{jl}
              - \delta_{il} \delta_{jk} \right),
  \label{KS:Cijkl} \\
  C_{011i} &= R_{011i}, \qquad
  C_{1ijk} = R_{1ijk}, \qquad
  C_{1i1j} = R_{1i1j}, \phantom{\frac{1}{()}}
  \label{KS:C1i1j}
\end{align}
where the frame components of the Riemann tensor are given in
\eqref{KS:R0101}--\eqref{KS:R1i1j}. Note that the terms containing cosmological
constant $\Lambda$ in \eqref{KS:C0101}--\eqref{KS:C1i1j} cancel the corresponding
terms in the Riemann tensor \eqref{KS:R0101}--\eqref{KS:R1i1j} and therefore
the cosmological constant actually does not enter the frame components of the
Weyl tensor.

\section{Non-expanding Kerr--Schild spacetimes}
\label{sec:KS:nonexpanding}

In this section, we will consider the simplest subclass of Einstein GKS
spacetimes where the null Kerr--Schild congruence $\bk$ is non-expanding
($\theta = 0$). Occasionally, we will also admit an additional aligned radiation
term in the Ricci tensor.

Substituting $\theta = 0$ to the trace \eqref{KS:vacuum:EFEijTrace} of one of
the frame component of the vacuum Einstein field equations, we immediately see
that the sum of squares of the shear $\sigma$ and twist $\omega$ scalars has to
vanish
\begin{equation}
  \sigma^2 + \omega^2 = 0
\end{equation}
and therefore $\sigma = \omega = 0$.
In other words, a non-expanding Kerr--Schild vector field $\bk$ is also
non-shearing and non-twisting. In fact, this means that non-expanding Einstein
GKS spacetimes belong to the Kundt class of solutions and the optical matrix
vanishes
\begin{equation}
  L_{ij} = 0.
  \label{KS:nonexpanding:Lij}
\end{equation}
Vanishing of the optical matrix along with the already applied assumption that
$\bk$ is geodetic and affinely parametrized, i.e.\ $L_{a0} = 0$, significantly
simplifies our calculations. For instance, the vacuum Einstein field equations
\eqref{KS:vacuum:EFE01}--\eqref{KS:vacuum:EFE11} reduce to
\begin{align}
  &\D^{2}\H = 0, \phantom{\half}
  \label{KS:nonexpanding:EFE01} \\
  &\delta_i (\D\H) - 2 L_{[i1]} \D\H = 0, \phantom{\half}
  \label{KS:nonexpanding:EFE1i} \\
  &\delta_i (\delta_i\H) + N_{ii} \D\H
  + \left( 4 L_{1i} - 2 L_{i1} + \M{i}{jj} \right) \delta_i\H \notag \\
  &\qquad + 2 \H \left( 2 \delta_i L_{[1i]} + 4 L_{1i} L_{[1i]} + L_{i1} L_{i1}
  + 2 L_{[1i]} \M{i}{jj} \right) + \frac{4 \H {\Lambda}}{n - 1} = 0.
  \label{KS:nonexpanding:EFE11}
\end{align}
The considerable simplification occurs also when we express the frame components
of the Weyl tensor \eqref{KS:C0101}--\eqref{KS:C1i1j} for the non-expanding case
\begin{align}
  C_{0i0j} &= 0, \qquad C_{010i} = 0, \qquad C_{0ijk} = 0, \phantom{\half} \\
  C_{0101} &= \D^2\H, \qquad C_{01ij} = 0, \qquad C_{0i1j} = 0, \phantom{\half} \\
  C_{ijkl} &= 0, \qquad
  C_{011i} = - \delta_i \left( \D\H \right) + 2 L_{[i1]} \D\H, \qquad
  C_{1ijk} = 0, \phantom{\half} \\
  C_{1i1j} &= \delta_i ( \delta_j\H ) + \M{k}{(ij)} \delta_k\H + 4 L_{1(i} \delta_{j)}\H
    - 2 L_{(i|1} \delta_{j)}\H + N_{(ij)} \D\H \phantom{\frac{1}{()}} \notag \\
    &\qquad + 2 \H \bigg( \delta_{(i} L_{1|j)} - 2 L_{1(i} L_{j)1}
    + 2 L_{1i} L_{1j} + L_{1k} \M{k}{(ij)} \bigg),
  \label{KS:nonexpandC1i1j}
\end{align}
where the only nontrivial boost weight 0 component $C_{0101}$ and boost weight
$-1$ component $C_{011i}$ vanish due to the Einstein field equations
\eqref{KS:nonexpanding:EFE01} and \eqref{KS:nonexpanding:EFE1i}. Thus, we can
conclude that

\begin{proposition}
  Einstein generalized Kerr--Schild spacetimes \eqref{KS:GKSmetric} with a
  non-expanding Kerr--Schild congruence $\bk$ are of Weyl type N with $\bk$ being
  the multiple WAND. Twist and shear of the Kerr--Schild congruence $\bk$
  necessarily vanish and therefore these solutions belong to the Kundt class of
  Weyl type N Einstein spacetimes.
  \label{KS:proposition:nonexpanding} 
\end{proposition}
Let us point out the relation of non-expanding GKS spacetimes with the VSI and
CSI classes of spacetimes defined and discussed in \cite{Coleyetal2002,
ColeyMilsonPravdaPravdova2004,ColeyHervikPelavas2005,
ColeyFusterHervikPelavas2006,ColeyFusterHervik2007}. It was shown that a
spacetime is VSI, i.e.\ all curvature invariants of all orders constructed from
the Riemann tensor and its covariant derivatives vanish, if and only if there
exists a non-expanding, non-shearing and non-twisting congruence of null
geodesics along which only the negative boost weight frame components of the
Riemann tensor are non-zero. In fact, the Kundt class of spacetimes of Weyl
types III, N, or O with the Ricci tensor of algebraic types III, N or O is
equivalent to the VSI class and all metrics within these classes are presented
in \cite{ColeyFusterHervikPelavas2006}. On the other hand, the CSI class is
defined as spacetimes for which all scalar invariants constructed from the
Riemann tensor and its covariant derivatives are constant and, as was
conjectured in \cite{ColeyHervikPelavas2005}, a CSI spacetime is either locally
homogeneous or belongs to the Kundt class.

Obviously, non-expanding Ricci-flat GKS metrics \eqref{KS:GKSmetric}, i.e.\
non-expanding KS metrics \eqref{KS:KSansatz} with the flat background, thus
belong to the VSI class of spacetimes since the Riemann tensor is given exactly
by the Weyl tensor \eqref{KS:Weyl:Einsteinspaces} which has only the boost
weight $-2$ components \eqref{KS:nonexpandC1i1j} along the Kerr--Schild vector
field $\bk$ representing a non-expanding, non-shearing and non-twisting
congruence as follows from proposition \ref{KS:proposition:nonexpanding}.

Unlike the Ricci-flat case, some of the boost weight zero components
\eqref{KS:R0101}--\eqref{KS:Rijkl} of the Riemann tensor of non-expanding
Einstein GKS spacetimes are proportional to the cosmological constant $\Lambda$.
Therefore, all curvature invariants either vanish or are constants depending on
$\Lambda$ and thus these spacetimes belong to the CSI class.

The arbitrariness of the choice of the frame vectors $\bn$, $\bm^{(i)}$ can be
used to show that, without loss of generality, one may set $L_{[1i]} = 0$ in the
case of Kundt spacetimes $L_{ij} = L_{i0} = 0$. Note that all the following
Lorentz transformations preserve both $L_{ij} = 0$ and $L_{i0} = 0$. Let us
assume that $\bk$ is affinely parametrized. First, we perform a boost
\eqref{intro:Riccicoeff:boosts} to break the affine parametrization, namely
$\hat L_{10} = \D \lambda$, in such an appropriate way that will be clear from
the final step. Next, we employ null rotations with $\bk$ fixed
\eqref{intro:Riccicoeff:nullrot} under which $L_{10}$ remains unchanged and
$L_{[1i]}$ transform as $\hat L_{[1i]} = L_{[1i]} - \half z_i L_{10}$. This
simply determines the functions $z_i$ to set all $\hat L_{[1i]}$ to zero.
Furthermore, we are still able to align the spacelike frame vectors $\bm^{(i)}$
to $L_{1i}$ by spatial rotations \eqref{intro:Riccicoeff:spins}, where
$\hat L_{1i} = X_{ij} L_{1j}$ with $X_{ij}$ being an orthogonal matrix, so that
$L_{1i}$ has just one component, let say $L_{12} = L_{21} \neq 0$,
$L_{1\itilde} = L_{\itilde 1} = 0$. Again, $L_{10}$ is not affected by this
operation and therefore we can finally perform the first step reversely, i.e.
$\lambda' = \lambda^{-1}$, to recover the affine parametrization of $\bk$. But
since $\hat L_{[1i]} = L_{[1i]} + \half \lambda'^{-1} \delta_i \lambda'$, we now
require $\delta_i \lambda' = 0$.

Note that the natural frame \eqref{KS:VSI:frame} of VSI metrics
\eqref{KS:VSI:metric} or even the natural frame of general Kundt metrics in the
canonical form \cite{PodolskyZofka2008} are examples of such frames with
$L_{[1i]} = 0$ since $\bl = \d u$ are constant one-forms.

In the case that $L_{[1i]} = 0$, the Einstein field equations
\eqref{KS:nonexpanding:EFE01}--\eqref{KS:nonexpanding:EFE11} further simplify to
\begin{align}
  &\D^{2}\H = 0, \qquad
  \delta_i (\D\H) = 0, \phantom{\half}
  \label{KS:nonexpanding:EFE1i:2} \\
  &\delta_i (\delta_i\H) + N_{ii} \D\H
    + \left( 2 L_{1i} + \M{i}{jj} \right) \delta_i\H
    + 2 \H L_{i1} L_{i1} + \frac{4 \H {\Lambda}}{n - 1} = 0.
  \label{KS:nonexpanding:EFE11:2}
\end{align}
One may integrate \eqref{KS:nonexpanding:EFE1i:2} to determine the $r$-dependence
of the Kerr--Schild function $\H$
\begin{equation}
  \H = f^{(0)} r + g^{(0)},
\end{equation}
where $f^{(0)}$ and $g^{(0)}$ are functions not depending on $r$ subject to
$\delta_i f^{(0)} = 0$ and it remains only to satisfy
\eqref{KS:nonexpanding:EFE11:2}. Recall that $r$ is an affine parameter along
the null geodesics $\bk$.

Note also that the above statement of proposition
\ref{KS:proposition:nonexpanding} remains valid if we admit an additional aligned
null radiation term in the Ricci tensor
\begin{equation}
  R_{ab} = \frac{2 \Lambda}{n-2} g_{ab} + \Phi k_a k_b.
  \label{KS:nullrad:Ricci}
\end{equation}
Since the Ricci tensor \eqref{KS:nullrad:Ricci} differs from the case of Einstein
spaces \eqref{KS:vacuum:Ricci} just in the frame component $R_{11}$, the
aligned null radiation term appears only on the right hand side of the frame
component \eqref{KS:nonexpanding:EFE11} of the Einstein field equations
\begin{equation}
  \begin{split}
  &\delta_i (\delta_i\H) + N_{ii} \D\H
  + \left( 4 L_{1i} - 2 L_{i1} + \M{i}{jj} \right) \delta_i\H \\
  &\qquad + 2 \H \left( 2 \delta_i L_{[1i]} + 4 L_{1i} L_{[1i]} + L_{i1} L_{i1}
  + 2 L_{[1i]} \M{i}{jj}  \right) + \frac{4 \H {\Lambda}}{n - 1} = \Phi
  \end{split}
  \label{KS:nullrad:EFE11}
\end{equation}
and as one can immediately see from \eqref{Weyl:C0i0j}--\eqref{Weyl:C1i1j}, only
the boost weight $-2$ frame components of the Weyl tensor $C_{1i1j}$ depend on
$R_{11}$. Therefore, the null radiation term does not affect the derivation of
proposition \ref{KS:proposition:nonexpanding}.

\subsection[\texorpdfstring{Examples of non-expanding Einstein generalized\\Kerr--Schild spacetimes}
{Examples of non-expanding Einstein generalized Kerr--Schild spacetimes}]
{Examples of non-expanding Einstein generalized\\Kerr--Schild spacetimes}
\label{sec:KS:nonexpanding:examples}

Let us recall the statement of proposition \ref{KS:proposition:nonexpanding}
that all non-expanding Einstein GKS spacetimes belong to the Kundt class. As
mentioned above, this also holds for non-expanding GKS spacetimes with null
radiation aligned with the Kerr--Schild vector $\bk$.
Geometrically, the Kundt class of solutions is defined as spacetimes admitting
a geodetic, non-expanding, non-shearing and non-twisting null congruence
generated by a null vector field that will be represented by the Kerr--Schild
vector $\bk$ in the case of Kundt GKS spacetimes.

In four dimensions, the Goldberg--Sachs theorem implies that Kundt spacetimes
possibly with a cosmological constant or aligned matter fields are algebraically
special, i.e.\ of Petrov type II or more special, and the geodetic
non-expanding, non-shearing and non-twisting null congruence corresponds with
the PND \cite{StephaniKramer2003}.

Analogically, it was shown in \cite{OrtaggioPravdaPravdova2007} that higher
dimensional Kundt spacetimes with the vanishing positive boost weight frame
components of the Ricci tensor, $R_{00} = R_{0i} = 0$, that admit cosmological
constant $\Lambda$ and aligned matter content, are of Weyl type II or more
special again with the geodetic null congruence being the WAND. In fact, $R_{00}$
vanishes identically for a Kundt metric and if $R_{0i} \neq 0$, then the spacetime
is of Weyl type I \cite{PodolskyZofka2008}.

The metric of general $n$-dimensional Kundt spacetimes can be expressed in the
canonical form \cite{ColeyFusterHervikPelavas2006,PodolskyZofka2008}
\begin{equation}
  \d s^2 = 2 \d u \left[ \d v + H(u,v,x^k) \, \d u + W_{i}(u,v,x^k) \, \d x^i \right]
    + g_{ij}(u,x^k) \, \d x^i \, \d x^j,
  \label{KS:Kundt:metric}
\end{equation}
where the coordinate $v$ corresponds to an affine parameter along the geodesics
of the non-expanding, non-shearing and non-twisting null congruence
$\bk = \partial_v$ and the transverse metric $g_{ij}$ does not depend on $v$.
It should be emphasized that we denote the transverse spatial coordinates as
$x^2, \ldots, x^{n-1}$ since in our convention $i$, $j$ range from 2 to $n - 1$
in contrast with \cite{ColeyFusterHervikPelavas2006}.

In general, Kundt spacetimes do not admit the GKS form \eqref{KS:GKSmetric}.
This follows directly from the fact that, without any conditions on the Ricci
tensor, the Kundt metrics are of Weyl type I or more special, whereas the GKS
metrics with geodetic $\bk$ are of Weyl type II or more special. Even in the
case of Einstein spaces, there exist, for instance, type III Einstein Kundt
spacetimes which are, by proposition \ref{KS:proposition:nonexpanding},
incompatible with the GKS form.

However, it can be shown that all Weyl type N VSI metrics written in
appropriate coordinates with a flat transverse space as
\cite{ColeyFusterHervikPelavas2006}
\begin{equation}
  \d s^2 = 2 \d u \left[ \d v + H(u,v,x^k) \, \d u + W_{i}(u,v,x^k) \, \d x^i \right]
    + \delta_{ij} \, \d x^i \, \d x^j,
  \label{KS:VSI:metric}
\end{equation}
where $H$ and $W_i$ satisfy certain conditions, admit the KS form
\eqref{KS:KSansatz}. In fact, the class of type N VSI metrics is equivalent to
the class of Weyl type N Ricci-flat Kundt spacetimes. Recall that we use the
different convention for indices of $W_i$ and $x^k$. Whereas in
\cite{ColeyFusterHervikPelavas2006} indices $i, j, \ldots$ ranges from $1$ to
$n - 2$, throughout the thesis we consistently use $i, j, \ldots$ running from
$2$ to $n - 1$.

Obviously, one may introduce a natural null frame in the VSI spacetime
\eqref{KS:VSI:metric}
\begin{equation}
  \begin{aligned}
  &\ell_a \, \d x^a = \d u, \qquad
  n_a \, \d x^a = \d v + H \, \d u + W_i \, \d x^i, \qquad
  m^{(i)}_a \, \d x^a = \d x^i, \\
  &\ell^a \, \partial_a = \partial_v, \qquad
  n^a \, \partial_a = \partial_u - H \, \partial_v, \qquad
  m_{(i)}^a \, \partial_a = \partial_i - W_i \, \partial_v.
  \end{aligned}
  \label{KS:VSI:frame}
\end{equation}
Then, it immediately follows from \eqref{KS:VSI:metric} and \eqref{KS:VSI:frame}
that
\begin{equation}
  L_{1i} = \half W_{i,v}, \qquad
  L_{[1i]} = 0, \qquad 
  L_{11} = H_{,v}
  \label{KS:VSI:Riccicoeff}
\end{equation}
and all other components of $L_{ab}$ are zero. Using the notation of
\cite{ColeyFusterHervikPelavas2006}, the VSI class can be divided into two
distinct subclasses with vanishing ($\epsilon=0$) and non-vanishing
($\epsilon=1$) quantity $L_{1i} L_{1i}$, respectively. The canonical choice
of the functions $W_i$ is
\begin{equation}
  W_{2,v} = -2 \frac{\epsilon}{x^2}, \qquad W_{\itilde,v} = 0.
  \label{KS:VSI:W:canonical_choice}
\end{equation}
One may express the constraints imposed on the undetermined functions $H$ and
$W_i$ following from the Einstein field equations and from the condition on the
form of the Weyl tensor. Namely, for Ricci-flat spacetimes of Weyl type N,
in the case $\varepsilon=0$, we obtain \cite{ColeyFusterHervikPelavas2006}
\begin{equation}
  \begin{split}
  &W_2 = 0, \qquad
  W_{\tilde{\imath}} = x^2 C_{\tilde{\imath}}(u)
    + x^{\tilde{\jmath}} B_{\tilde{\jmath}\tilde{\imath}}(u), \phantom{\half} \\
  &H = H^0(u,x^i), \qquad
  \Delta H^0 - \frac{1}{2} \sum C^2_{\tilde{\imath}}
    - 2 \sum_{\tilde{\imath}<\tilde{\jmath}} B^2_{\tilde{\imath}\tilde{\jmath}} = 0,
  \end{split}
  \label{KS:VSI:typeN:epsilon0:W}
\end{equation}
whereas, in the case $\varepsilon=1$, one gets 
\begin{equation}
  \begin{split}
  &W_2 = - \frac{{{2}} v}{x^2}, \qquad
  W_{\tilde{\imath}} = C_{\tilde{\imath}}(u)
    + x^{\tilde{\jmath}} B_{\tilde{\jmath}\tilde{\imath}}(u), \qquad
  H = \frac{v^2}{2(x^2)^2} + H^0(u,x^i), \\
  &x^2 \Delta \left(\frac{H^0}{x^2}\right)
    - \frac{1}{(x^2)^2} \sum W^2_{\tilde{\imath}}
    - 2 \sum_{\tilde{\imath}<\tilde{\jmath}} B^2_{\tilde{\imath}\tilde{\jmath}} = 0,
  \end{split}
  \label{KS:VSI:typeN:epsilon1:W}
\end{equation}
where $B_{[\tilde{\imath}\tilde{\jmath}]}=0$ in both cases and
$\tilde{\imath}$, $\tilde{\jmath} = 3, \dots, n - 1$. Similarly, one may also
find $H^0(u, v, x^k)$ such that the VSI metric \eqref{KS:VSI:metric} with
$H = H^0$ is flat \cite{ColeyFusterHervikPelavas2006}
\begin{align}
  \epsilon = 0: \qquad
  H^0_\text{flat} &= \frac{1}{2} x^1 x^{\itilde} (C_{\itilde,u}
    + B_{\itilde \jtilde} C_{\jtilde})
    + \frac{1}{2} B_{\itilde \tilde{k}} B_{\jtilde \tilde{k}} x^{\itilde} x^{\jtilde}
    + x^i F_i(u) \notag \\
  &\qquad + \frac{1}{8} \left( \sum C^2_{\itilde} (x^2)^2
    + \sum_{\itilde \leq \jtilde} C_{\itilde} C_{\jtilde} x^{\itilde} x^{\jtilde} \right), \\
    \epsilon = 1: \qquad
  H^0_\text{flat} &= \frac{1}{2} \sum W_m^2
    - \frac{1}{16} + x^1 F_0(u) + x^1 x^i F_{i}(u),
\end{align} 
where $F_0(u)$, $F_i(u)$ are arbitrary functions of $u$ and $W_i$ are given as
for type N \eqref{KS:VSI:typeN:epsilon0:W}, \eqref{KS:VSI:typeN:epsilon1:W}.
The one-form $\d u$ is associated with the geodetic null vector field $\partial_v$
and therefore all Weyl type N VSI metrics \eqref{KS:VSI:metric} can be written
in the KS form \eqref{KS:KSansatz} as
\begin{equation}
  \d s^2= \d s^2_\text{flat} + \left( H^{0} - H^{0}_\text{flat} \right) \d u^2.
\end{equation}
 
Since a metric of general higher dimensional type N Einstein Kundt spacetimes
has not been given explicitly in the literature yet, we cannot simply follow the
above procedure in the case of such metrics. This prevents us from answering the
question whether the implication stated in proposition
\ref{KS:proposition:nonexpanding} is valid also in the opposite direction,
i.e.\ whether there is an equivalency between the classes of type N Einstein
Kundt spacetimes and non-expanding Einstein GKS spacetimes.

However, we are still able to show, using the results of
\cite{OzsvathRobinsonRozga1985, BicakPodolsky1999}, that at least all
four-dimensional type N Einstein Kundt metrics admit the GKS form. The metric
of type N Kundt spacetimes admitting cosmological constant and possibly containing 
pure radiation can be expressed as \cite{OzsvathRobinsonRozga1985} 
\begin{equation}
  \begin{split}
  \d s^2 &= - 2 \frac{Q^2}{P^2} \, \d u \, \d v + \left( 2 k \frac{Q^2}{P^2} v^2
    - \frac{\left(Q^2\right)_{,u}}{P^2} v - \frac{Q}{P} H \right) \d u^2 \\
    &\qquad+ \frac{1}{P^2} \left( \d x^2 + \d y^2 \right),
  \end{split}
  \label{KS:Kundt:4D:metric}
\end{equation}
where 
\begin{equation}
  \begin{split}
  &P = 1 + \frac{\Lambda}{12} (x^2 + y^2), \\
  &k = \frac{\Lambda}{6} \alpha(u)^2 + \frac{1}{2} \left(\beta(u)^2
    + \gamma(u)^2 \right), \\
  &Q = \left(1 - \frac{\Lambda}{12} (x^2 + y^2) \right) \alpha(u)
    + \beta(u) x + \gamma(u) y,
  \end{split}
  \label{KS:Kundt:4D:k}
\end{equation}
with $\alpha(u)$, $\beta(u)$ and $\gamma(u)$ being arbitrary functions of the
coordinate $u$ and $H=H(x,y,u)$. These spacetimes are Einstein if
\begin{equation}
  P^2 (H_{,xx} + H_{,yy}) + \frac{2}{3} \Lambda H = 0,
  \label{KS:Kundt:4D:Einstein}
\end{equation}
which has a general solution \cite{BicakPodolsky1999}
\begin{equation}
  H = 2 f_{1,x} - \frac{\Lambda}{3 P} (x f_1 + y f_2),
  \label{KS:Kundt:4D:Einstein:solution}
\end{equation}
where the functions $f_1 = f_1(u,x,y)$ and $f_2 = f_2(u,x,y)$ are subject to
$f_{1,x} = f_{2,y}$,  $f_{1,y}=-f_{2,x}$. It can be shown that the Einstein
metrics \eqref{KS:Kundt:4D:metric}, \eqref{KS:Kundt:4D:Einstein:solution} are
conformally flat for
\begin{equation}
  H(x,y,u) = \frac{1}{P} \left(A \left(1 - \frac{\Lambda}{12} (x^2 + y^2) \right)
    + B x + C y \right),
\end{equation}
where $A(u)$, $B(u)$ and $C(u)$ are arbitrary functions of $u$. Therefore, all
four-dimensional type N Kundt metrics \eqref{KS:Kundt:4D:metric} differ from
the conformally flat case only by a factor of $\d u^2$ and thus such metrics
take the GKS form \eqref{KS:GKSmetric}.

\subsection{\texorpdfstring{Warped Einstein Kundt generalized Kerr--Schild\\spacetimes}
{Warped Einstein Kundt generalized Kerr--Schild spacetimes}}
\label{sec:KS:nonexpanding:warp}

In the previous section \ref{sec:KS:nonexpanding:examples}, we have presented
explicitly known Weyl type N Einstein Kundt metrics in order to show that they
can be cast to the GKS form. However, these metrics are either only
four-dimensional but admitting cosmological constant or arbitrary dimensional
but only Ricci-flat. In this section, we employ these metrics again in order to
construct examples of higher dimensional Einstein Kundt spacetimes belonging to
the GKS class with an almost arbitrary cosmological constant by means of the
Brinkmann warp product.

As discussed in sections \ref{sec:intro:warp} and \ref{sec:KS:warp}, the
Brinkmann warp product \eqref{intro:warp:warpedmetric} allows us to generate new
$n$-dimensional Einstein GKS metrics $\d s^2$ \eqref{KS:GKSwarped} from known
$(n-1)$-dimensional Einstein GKS seed metrics $\d \tilde s^2$ in the form
\eqref{KS:GKSwarped:seed}. The cosmological constant of the warped metric
$\d s^2$ is not completely arbitrary since its sign depends on the sign of the
cosmological constant of the seed metric $\d \tilde s^2$. The allowed
combinations of these signs were discussed in section \ref{sec:intro:warp}. Let
us also recall that the Brinkmann warp product preserves the Weyl type of
algebraically special spacetimes.

First, let us choose the $n$-dimensional type N Ricci-flat Kundt metric
\eqref{KS:VSI:metric}, \eqref{KS:VSI:typeN:epsilon0:W} and
\eqref{KS:VSI:typeN:epsilon1:W}, i.e.\ type N subclass of VSI spacetimes, as a
seed $\d \tilde s^2$. The sign of the Ricci scalar $R$ of the warped metric
$\d s^2$ may be zero or negative. Omitting the trivial case of the direct
product, we thus construct $(n+1)$-dimensional type N Einstein GKS metrics with
a negative cosmological constant. One may use \eqref{intro:warp:conformal-0} to
cast such metrics to the form conformal to the direct product
\begin{equation}
  \begin{split}
  \d s^2 &= \frac{1}{-\lambda \tilde z^2} \Big( 2 \, \d u \left[ \d v
    + H(u,v,x^k) \, \d u + W_{i}(u,v,x^k) \, \d x^i \right] \\
    &\qquad+ \delta_{ij} \, \d x^i \, \d x^j + \d \tilde z^2 \Big), \phantom{\half}
  \end{split}
  \label{KS:VSI:warp:metric:conformal}
\end{equation}
where $i, j = 2,\dots,n-1$. Performing the coordinate transformation
$v = - \lambda \tilde v \tilde z^2$, we easily put the above metric to the
canonical Kundt form \eqref{KS:Kundt:metric}
\begin{equation}
  \d s^2 = 2 \, \d u \left[ \d\tilde v + \tilde H \, \d u
    + \tilde W_{\tilde \imath} \, \d x^{\tilde \imath} \right]
    + \frac{1}{-\lambda \tilde z^2} \delta_{\tilde\imath \tilde\jmath} \, \d x^{\tilde\imath} \, \d x^{\tilde\jmath},
  \label{KS:VSI:warp:metric:Kundt}
\end{equation}
with $\itilde,\jtilde=2,\dots,n$ and
\begin{equation}
  \begin{split}
  &\tilde H = \frac{1}{-\lambda \tilde z^2} H(u,v,x^k), \qquad
  \d x^n = \d\tilde z, \\
  &\tilde W_i = \frac{1}{-\lambda \tilde z^2} W_i(u,v,x^k), \qquad
  \tilde W_n = \frac{2 \tilde v}{\tilde z}.
  \end{split}
  \label{KS:VSI:warp:metricfunction:Kundt}
\end{equation}

In fact, metrics \eqref{KS:VSI:warp:metric:conformal},
\eqref{KS:VSI:warp:metric:Kundt} were already discussed in
\cite{ColeyHervikPelavas2005,ColeyFusterHervik2007} in the context of CSI
spacetimes and supergravity.

Although one may apply the Brinkmann warp product multiple times to obtain
further and further solutions it does not lead to new results in this case.
Since the $(n+1)$-dimensional metric \eqref{KS:VSI:warp:metric:conformal} is of
the form
\begin{equation}
  \d s^2 = \frac{1}{- \lambda z^2} \left( \d z^2 + g^{(0)}_{ab} \, \d x^a \, \d x^b \right),
  \label{KS:VSI:warp:metric:conformal2}
\end{equation}
with $g^{(0)}_{ab} \, \d x^a \, \d x^b = 2 \, \d u \left[ \d v + H \, \d u
+ W_{i} \, \d x^i \right] + \delta_{ij} \, \d x^i \, \d x^j$ not depending on
the coordinate $z$. The second application of the warp product
\eqref{intro:warp:warpedmetric} gives
\begin{equation}
  \d s'^2 = \frac{\d w^2}{f(w)}
    + f(w) \left( \frac{1}{- \lambda z^2} \left( \d z^2
    + g^{(0)}_{ab} \, \d x^a \, \d x^b \right) \right),
  \label{KS:VSI:doublewarp:metric:conformal}
\end{equation}
where the function $f(w)$ is defined in \eqref{intro:warp:factor}. The seed
metric \eqref{KS:VSI:warp:metric:conformal2} has a negative cosmological
constant and therefore the only possibility is that the cosmological constant of
the warped metric \eqref{KS:VSI:doublewarp:metric:conformal} is negative as well.
Then the coordinate transformation \eqref{KS:warp:AdStoAdS}, where we replace
$x_1 \rightarrow z$, $z \rightarrow w$ and $\lambda \rightarrow \lambda'$, along
with the relation $\lambda = \lambda' b + d^2$ following from
\eqref{intro:warp:seedricci} yields
\begin{align}
  & \frac{f(w)}{- \lambda z^2} = \frac{1}{- \lambda' \tilde z^2},
  \label{KS:VSI:doublewarp:metric:conformal:1} \\
  & \frac{\d w^2}{f(w)} + \frac{f(w)}{- \lambda z^2} \d z^2 =
    \frac{1}{- \lambda' \tilde z^2} \left( \d \tilde w^2 + \d \tilde z^2 \right).
  \label{KS:VSI:doublewarp:metric:conformal:2}
\end{align}
Substituting \eqref{KS:VSI:doublewarp:metric:conformal:1} and 
\eqref{KS:VSI:doublewarp:metric:conformal:2} to
\eqref{KS:VSI:doublewarp:metric:conformal} immediately
leads to
\begin{equation}
  \d s'^2 = \frac{1}{- \lambda' \tilde z^2} \left( g^{(0)}_{ab }\, \d x^a \, \d x^b
    + \d \tilde w ^2 + \d \tilde z^2 \right)
\end{equation}
and using $v = - \lambda' \tilde v \tilde z^2$, we finally arrive at
\eqref{KS:VSI:warp:metric:Kundt} with \eqref{KS:VSI:warp:metricfunction:Kundt}
where moreover
\begin{equation}
  \begin{split}
  W_{(n+1)} = 0, \qquad
  \d x^{(n+1)} = \d \tilde w.
  \end{split}
\end{equation}
Therefore, we obtained the same class of $(n + 2)$-dimensional metrics as in the
case of applying the warp product on the subclass of $(n + 1)$-dimensional
metrics \eqref{KS:VSI:metric} with $W_n = 0$ and then swaps
$n \leftrightarrow (n + 1)$.

So far we have used only the type N Ricci-flat VSI metrics \eqref{KS:VSI:metric}
as a seed and thus we constructed higher dimensional type N Einstein Kundt GKS
metrics with a negative cosmological constant. However, one can also warp the
four-dimensional Einstein Kundt metrics \eqref{KS:Kundt:4D:metric}. In this case,
there are more possible combinations of the signs of the Ricci scalars $\tilde R$
and $R$ of the seed and warped metric, respectively. Recall that only the case
with both Ricci scalars being zero or negative is free from curvature or
parallelly propagated singularities at a point where $f(z)=0$.

Such five-dimensional warped metrics can be expressed, for instance, either in
the form conformal to a direct product using
\eqref{intro:warp:conformal++}--\eqref{intro:warp:conformal-+} as in the
previous case or directly in the GKS form using \eqref{KS:GKSwarped},
\eqref{KS:warpedbackground} and the coordinate transformations
\eqref{KS:warp:AdStoAdS}--\eqref{KS:warp:MtoAdS}. Here, we use the latter
approach.

First, we have to split the four-dimensional type N Kundt metric
\eqref{KS:Kundt:4D:metric} into the background (anti-)de Sitter or Minkowski
metric $\bar{g}_{ab}$ and the Kerr--Schild term $\H k_a k_b$. The background
metric can be obtained as a weak-field limit of \eqref{KS:Kundt:4D:metric}
\cite{GriffithsPodolsky2009}
\begin{equation}
  \d s^2 = -2 \frac{Q^2}{P^2} \, \d u \, \d v
    + 2 k \frac{Q^2}{P^2} v^2 \, \d u^2
    + \frac{1}{P^2} \left( \d x^2 + \d y^2 \right).
  \label{KS:Kundt:4D:metric:background}
\end{equation}
Therefore, the four-dimensional type N Kundt metric \eqref{KS:Kundt:4D:metric}
can be straightforwardly cast to the GKS form simply by reordering the terms
\begin{equation}
  \begin{aligned}
  \d s^2 &= -2 \frac{Q^2}{P^2} \, \d u \, \d v
    + 2 k \frac{Q^2}{P^2} v^2 \, \d u^2
    + \frac{1}{P^2} \left( \d x^2 + \d y^2 \right) \\
    &\qquad - \left( \frac{\left( Q^2 \right)_{,u}}{P^2} v + \frac{Q}{P} \H \right) \d u^2,
  \end{aligned}
\end{equation}
where the functions $P$, $Q$ and $k$ are given in \eqref{KS:Kundt:4D:k} and,
obviously, the last term corresponds to the Kerr-Schild term $2 \H k_a k_b$.
These seed metrics with a four-dimensional cosmological constant denoted as
$\tilde \Lambda$ can be split to several geometrically distinct classes
\cite{GriffithsPodolsky2009}. Depending on whether $\tilde \Lambda$ and $k$ is
positive, negative or vanishing, we will denote such possible subclasses as
KN($\tilde \Lambda^+$, $k^+$), KN($\tilde \Lambda^0$, $k^+$),
KN($\tilde \Lambda^0$, $k^0$), KN($\tilde \Lambda^{-}, k^+$),
KN($\tilde \Lambda^{-}, k^-$) and KN($\tilde \Lambda^{-}, k^0$), respectively.

Five-dimensional Ricci-flat metrics obtained from KN($\tilde \Lambda^+$, $k^+$)
KN($\tilde \Lambda^0$, $k^+$) and KN($\tilde \Lambda^0$, $k^0$) belong to the
VSI class and the warped metrics constructed using the Ricci-flat seeds from
KN($\tilde \Lambda^0$, $k^+$) and KN($\tilde \Lambda^0$, $k^0$) are already
contained in the class of warped VSI metrics \eqref{KS:VSI:warp:metric:conformal}.
Therefore, we restrict ourselves to the cases with non-vanishing cosmological
constants of the seed and warped metrics. Here, we do not present the warped
metrics explicitly, instead, we give the coordinate transformations putting the
corresponding background metric \eqref{KS:Kundt:4D:metric:background} to the
canonical form. Then, one may perform the warp product \eqref{KS:GKSwarped} and
employ \eqref{KS:warp:AdStoAdS}--\eqref{KS:warp:MtoAdS} to cast the warped
background metric \eqref{KS:warpedbackground} back to the canonical form.

\subsubsection{Generalized Kundt waves KN($\tilde \Lambda^{-}, k^+$)}

The Kundt metric \eqref{KS:Kundt:4D:metric} with the canonical choice
$\alpha = 0$, $\beta = \sqrt{2}$, $\gamma = 0$, where the functions $Q$ and $k$
are given by the functions $\alpha$, $\beta$ and $\gamma$ via
\eqref{KS:Kundt:4D:k}
\begin{equation}
  Q = {\sqrt{2}} x, \qquad
  k = 1,
  \label{KS:Kundt:4D:KN-+}
\end{equation}
represents generalized Kundt waves KN($\tilde \Lambda^-, k^+$). One may put the
anti-de Sitter background metric to the canonical form by means of the coordinate
transformation 
\begin{equation}
  \begin{aligned}
  u& = \frac{Y \mp \sqrt{T^2 - X^2 - Z^2}}{a}, \qquad & T&= \frac{a^2 \left( 2 - P \right)}{2xv}, \\
  v& = \pm \frac{a}{2 \sqrt{T^2 - X^2 - Z^2}},        & X&= \frac{a^2 P}{2xv}, \\
  x& = \pm \frac{2 a \sqrt{T^2 - X^2 - Z^2}}{X + T},  & Y&= \frac{a \left( 1 + 2uv \right)}{2v}, \\
  y& = \frac{2 a Z}{X + T},                           & Z&= \frac{ay}{2xv},
  \end{aligned}
\end{equation}
where $a = \sqrt{-3/\tilde \Lambda}$.

\subsubsection{Generalized \pp waves KN($\tilde \Lambda^{-}, k^-$)}

The subclass KN($\tilde \Lambda^{-}, k^-$) which generalizes \pp waves can be
described by the canonical choice $\alpha=1$, $\beta=0$, $\gamma=0$ leading to
\begin{equation}
  Q = 1 - \frac{\tilde \Lambda}{12} (x^2 + y^2), \qquad
  k = \frac{\tilde \Lambda}{6}.
  \label{KS:Kundt:4D:KN--}
\end{equation}
In this case, the anti-de Sitter background metric can be cast to the canonical
form using the coordinate transformation
\begin{equation}
  \begin{aligned}
  u& = \sqrt{2} \left( \pm \sqrt{X^2 + Y^2 + Z^2} - T \right), \qquad &T& = \sqrt{2} \frac{a^2 - uv}{2v}, \\
  v& = \pm \frac{a^2}{\sqrt{2} \sqrt{X^2 + Y^2 + Z^2}},               &X& = \frac{a^2 P}{\sqrt{2} Qv}, \\
  x& = \frac{2 a Z}{X \pm \sqrt{X^2 + Y^2 + Z^2}},                    &Y& = \frac{ax}{\sqrt{2}Qv}, \\
  y& = \frac{2 a Y}{X \pm \sqrt{X^2 + Y^2 + Z^2}},                    &Z& = \frac{ay}{\sqrt{2}Qv},
  \end{aligned}
\end{equation}
where again $a = \sqrt{-3/\tilde \Lambda}$.

\subsubsection{Generalized Siklos waves KN($\tilde \Lambda^{-}, k^0$)}

The last subclass with a negative cosmological constant $\tilde \Lambda$,
generalized Siklos waves KN($\tilde \Lambda^{-}, k^0$), is determined by the
canonical choice $\alpha=1$, $\beta = \sqrt{-\Lambda/3} \cos\theta$ and
$\gamma = \sqrt{-\Lambda/3} \sin\theta$. The coordinate transformation
\begin{equation}
  \begin{aligned}
  u& = \frac{-T^2 + X^2 + Y^2 + Z^2}{\sqrt{2} \left( T + Y \right)},
  \qquad &T& = \frac{1}{\sqrt{2} v} \left( a^2 - uv - \frac{ax}{Q} \right), \\
  v& = \frac{a^2}{\sqrt{2} \left( T + Y \right)},                    &X& = \frac{a^2 P}{\sqrt{2}Qv}, \\
  x& = \sqrt{2} a \frac{(T + Y)^2 - X^2 - Z^2}{(T + X + Y)^2 + Z^2}, &Y& = \frac{u}{\sqrt{2}} + \frac{ax}{\sqrt{2} Qv}, \\
  y& = 2 \sqrt{2} a \frac{Z(T + Y)}{(T + X + Y)^2 + Z^2},            &Z& = \frac{ay}{\sqrt{2} Qv},
  \end{aligned}
\end{equation}
with $a = \sqrt{-3/\tilde \Lambda}$, brings the background metric to the
canonical form. Note that, in the special subcase when $\theta$ is independent
of $u$, such metrics are equivalent to the Siklos metric
\cite{GriffithsPodolsky2009}
\begin{equation}
  \d s^2 = \frac{3}{- \Lambda x^2} \left( \d u \, \d r + H \, \d u^2 + \d x^2 + \d y^2 \right).
  \label{KS:Siklos:metric}
\end{equation}
Obviously, since the Siklos metric \eqref{KS:Siklos:metric} is conformal to
\pp waves, one may obtain the same five-dimensional metric either by warping
the Siklos metric \eqref{KS:Siklos:metric} and using \eqref{KS:warp:AdStoAdS}
or by warping \pp waves and using \eqref{KS:warp:MtoAdS}. Thus, one may perhaps
conjecture that appropriate seeds from KN($\tilde \Lambda^-$, $k^0$) and
KN($\tilde \Lambda^0$, $k^0$) may lead to the same warped metric.

\subsubsection{Generalized Kundt and \pp waves KN($\Lambda^+$, $k^+$)}

In the case with a positive cosmological constant $\tilde \Lambda$, there is
only one canonical subclass KN($\Lambda^+$, $k^+$). The canonical choice is
either $\alpha = 0$, $\beta = \sqrt{2}$, $\gamma = 0$ or $\alpha = 1$,
$\beta = 0$, $\gamma = 0$, where both choices are equivalent and correspond to
generalized Kundt or \pp waves. The functions $Q$ and $k$ are given in
\eqref{KS:Kundt:4D:KN-+} and \eqref{KS:Kundt:4D:KN--}, respectively.
The de Sitter background metric can be cast to the canonical form using the
transformation
\begin{equation}
  \begin{aligned}
  u& = \frac{X \mp \sqrt{T^2 - Y^2 - Z^2}}{a}, \qquad &T& = \frac{a^2 P}{2xv}, \\
  v& = \pm \frac{a}{2 \sqrt{T^2 - Y^2 - Z^2}},        &X& = \frac{a \left( 1 + 2uv \right)}{2v}, \\
  x& = \pm \frac{2 a \sqrt{T^2 - Y^2 - Z^2}}{T + Z},  &Y& = \frac{ay}{2xv}, \\
  y& = \frac{2 a Y}{T + Z},                           &Z& = \frac{a^2 \left( 2 - P \right)}{2xv},
  \end{aligned}
\end{equation}
in the case of generalized Kundt waves or
\begin{equation}
  \begin{aligned}
  u& = \sqrt{2} \left( Z \mp \sqrt{T^2 - X^2 - Y^2} \right), \qquad &T& = \frac{a^2 P}{\sqrt{2}Qv}, \\
  v& = \pm \frac{a^2}{\sqrt{2} \sqrt{T^2 - X^2 - Y^2}},             &X& = \frac{ax}{\sqrt{2}Qv}, \\
  x& = \frac{2 a X}{T \pm \sqrt{T^2 - X^2 - Y^2}},                  &Y& = \frac{ay}{\sqrt{2}Qv}, \\
  y& = \frac{2 a Y}{T \pm \sqrt{T^2 - X^2 - Y^2}},                  &Z& = \sqrt{2} \frac{a^2 + uv}{2v},
  \end{aligned}
\end{equation}
in the case of generalized \pp waves, respectively, where
$a = \sqrt{3/\tilde \Lambda}$.

\section{Expanding Kerr--Schild spacetimes}
\label{sec:KS:expanding}

In the previous section \ref{sec:KS:nonexpanding}, we have discussed the
consequences of the Einstein field equations for non-expanding ($\theta = 0$)
Einstein GKS spacetimes. In that case, the Kerr--Schild function $\mathcal{H}$
does not enter the trace \eqref{KS:vacuum:EFEijTrace} of the components
\eqref{KS:vacuum:EFEij} of the Einstein field equations, which then implies that
such spacetimes belong to the Kundt class. In the expanding case
($\theta \neq 0$), on the other hand, we can immediately express $\D \log\H$
from the trace \eqref{KS:vacuum:EFEijTrace} as
\begin{equation}
  \D \log\H = \frac{L_{ik} L_{ik}}{\theta (n-2)} - (n-2) \theta.
  \label{KS:vacuum:DlogH}
\end{equation}

\subsection{Optical constraint}
\label{sec:KS:optical_constraint}

Substituting \eqref{KS:vacuum:DlogH} back to \eqref{KS:vacuum:EFEij} eliminates
the Kerr--Schild function $\mathcal{H}$ and therefore we obtain purely
geometrical condition on the geodetic Kerr--Schild congruence $\bk$
\begin{equation}
  L_{ik} L_{jk} = \frac{L_{lk} L_{lk}}{(n-2) \theta} S_{ij},
  \label{KS:vacuum:optical_constraint}
\end{equation}
referred to as ``the optical constraint'' \cite{OrtaggioPravdaPravdova2008}.
Suppressing indices in matrix notation, the optical constraint
\eqref{KS:vacuum:optical_constraint} reads
\begin{equation}
  \mathbf{L} \mathbf{L}^T = \alpha \mathbf{S}.
  \label{KS:vacuum:optical_constraint:matrix}
\end{equation}
It can be easily shown that $\mathbf{L}$ satisfying the optical constraint is
a normal matrix since \eqref{KS:vacuum:optical_constraint:matrix} can be
rewritten as \cite{Reallinprep}
\begin{equation}
  \left( \mathbf{1} - 2 \alpha^{-1} \mathbf{L} \right) \left( \mathbf{1} - 2 \alpha^{-1} \mathbf{L}^T \right) =
  \mathbf{1},
  \label{KS:vacuum:optical_constraint:unitary}
\end{equation}
where, obviously, the matrix $\mathbf{1} - 2 \alpha^{-1} \mathbf{L}$ is unitary,
one may swap the terms on the left hand side of
\eqref{KS:vacuum:optical_constraint:unitary} to obtain an equivalent relation
which then leads to the optical constraint in the form
\begin{equation}
  \mathbf{L}^T \mathbf{L} = \alpha \mathbf{S}.
  \label{KS:vacuum:optical_constraint:matrix2}
\end{equation}
Therefore, comparing \eqref{KS:vacuum:optical_constraint:matrix} with
\eqref{KS:vacuum:optical_constraint:matrix2}, the optical matrix $\mathbf{L}$
commutes with its transpose and indeed it is a normal matrix.

A real matrix $\mathbf{M}$ is normal if and only if there is a real orthogonal
matrix $\mathbf{O}$ such that \cite{HornJohnson1990}
\begin{equation}
  \mathbf{O}^T \mathbf{M} \mathbf{O} =
    \begin{pmatrix} \mathcal{M}_1 & & & \\ & \mathcal{M}_2 & & \\ & & \ddots & \\ & & & \mathcal{M}_k
    \end{pmatrix},
  \label{KS:normal_matrix}
\end{equation}
where $\mathcal{M}_i$ is either a real $1 \times 1$ matrix or a real $2 \times 2$
matrix of the form
\begin{equation}
  \mathcal{M}_i = \begin{pmatrix} s_i & a_i \\
    - a_i & s_i \end{pmatrix}.
  \label{KS:normal_matrix:block}
\end{equation}

This implies that the optical matrix $L_{ij}$ can be put into the block-diagonal form
\eqref{KS:normal_matrix}, \eqref{KS:normal_matrix:block} by appropriate spins
\eqref{intro:frame:spins} of the frame. Furthermore, such a canonical frame is
compatible with parallel transport along the geodetic null congruence $\bk$
\cite{OrtaggioPravdaPravdova2010}.

The sparse structure of the optical matrix $L_{ij}$ considerably simplifies the
determination of its dependence on the affine parameter $r$ along null geodesics
$\bk$ from the Sachs equation \eqref{intro:Sachs_equation}.
Due to the block-diagonal form \eqref{KS:normal_matrix},
\eqref{KS:normal_matrix:block} of the optical matrix $L_{ij}$ in the canonical
frame, one may integrate \eqref{intro:Sachs_equation}. In the case of a block
corresponding to the $1 \times 1$ matrix, we get
\begin{equation}
  L_{(i)(i)} = \frac{1}{r + b^0_i},
  \label{KS:expanding:Lij:block:1x1}
\end{equation}
whereas, for a block consisting of the $2 \times 2$ matrix, the Sachs equation
implies
\begin{equation}
  \begin{split}
  L_{i,i + 1} &= -L_{i + 1, i} = \frac{(a^0_i)^2}{(r + b^0_i)^2 + (a^0_i)^2}, \\
  L_{(i)(i)} &= L_{(i + 1)(i + 1)} = \frac{r + b^0_i}{(r + b^0_i)^2 + (a^0_i)^2},
  \label{KS:expanding:Lij:block:2x2}
  \end{split}
\end{equation}
where $a^0$ and $b^0_i$ are arbitrary functions not depending on $r$. Putting
\eqref{KS:expanding:Lij:block:1x1} and \eqref{KS:expanding:Lij:block:2x2}
to the optical constraint \eqref{KS:vacuum:optical_constraint:matrix}, we obtain
\begin{equation}
  \alpha ( r + b^0_i) = 1,
\end{equation}
for all values of the index $i$. Therefore, all functions $b^0_i$ are equal and,
without loss of generality, we may set them to zero. Thus, we may conclude that
the optical matrix $L_{ij}$ of expanding Einstein GKS spacetimes in a canonical
frame takes the block-diagonal form
\begin{align}
  L_{ij} = \left(
  \begin{array}{cccc}
    \fbox{$\mathcal{L}_{(1)}$} & & & \\
    & \ddots & & \\
    & & \fbox{$\mathcal{L}_{(p)}$} & \\
    & & & \fbox{$
      \begin{array}{ccc}
	& & \\
	\ \ & \tilde{\mathcal{L}} \ \ & \\
	& &
      \end{array} $}
    \end{array} \right),
  \label{KS:expanding:Lij:blockdiagonal}
\end{align}
with $2 \times 2$ blocks $\mathcal{L}_{(1)}, \dots, \mathcal{L}_{(p)}$ of the
form
\begin{equation}
\mathcal{L}_{(\mu)} = \left(
  \begin{array}{cc}
    s_{(2\mu)} & A_{2\mu,2\mu+1} \\
    -A_{2\mu,2\mu+1} & s_{(2\mu)}
  \end{array}
  \right), \qquad \mu=1, \ldots, p,
  \label{KS:expanding:Lij:block}
\end{equation}
where the corresponding symmetric diagonal and anti-symmetric anti-diagonal
parts are given by
\begin{equation}
  s_{(2\mu)} = \frac{r}{r^2+(a^0_{(2\mu)})^2},
  \qquad A_{2\mu,2\mu+1} = \frac{a^0_{(2\mu)}}{r^2+(a^0_{(2\mu)})^2},
  \label{KS:expanding:rdep:s+A}
\end{equation}
respectively. The remaining block $\tilde{\mathcal{L}}$ is
$(n-2-2p) \times (n-2-2p)$ diagonal matrix
\begin{equation}
  \tilde{\mathcal{L}} = \frac{1}{r} \text{diag}(\underbrace{1, \ldots, 1}_{(m-2p)},
    \underbrace{0, \ldots, 0}_{(n-2-m)}),
  \label{KS:expanding:rdep:Ldiagonal}
\end{equation}
where $m$ and $n - 2$ denote the rank and dimension of the optical matrix
$L_{ij}$ and $p$ corresponds to the number of $2 \times 2$ blocks. Clearly,
these quantities are subject to the relation $0 \le 2p \le m \le n-2$.

Now, we are able to express the expansion, shear and twist scalars defined in
\eqref{intro:optical_scalars} as
\begin{align}
  &\theta = \frac{1}{n - 2} \left( 2 \sum_{\mu=1}^p \frac{r}{r^2 + (a^0_{(2\mu)})^2} + \frac{m-2p}{r} \right), \\
  \label{KS:expanding:rdep:expansion}
  &\sigma^2 = 2 \sum_{\mu=1}^p \frac{r^2}{(r^2 + (a^0_{(2\mu)})^2)^2} + \frac{m-2p}{r^2} - (n - 2) \theta^2, \\
  &\omega^2 = 2 \sum_{\mu=1}^p \frac{(a^0_{(2\mu)})^2}{(r^2 + (a^0_{(2\mu)})^2)^2}
\end{align}
and the quantity $L_{ij} L_{ij}$ reads
\begin{equation}
  L_{ik} L_{ik} = (n-2) \theta \frac{1}{r}.
\end{equation}
Using the above results, we can determine the $r$-dependence of the Kerr--Schild
function $\H$ by integrating \eqref{KS:vacuum:DlogH}
\begin{equation}
  \H = \frac{\H_0}{r^{m - 2p - 1}} \prod_{\mu=1}^p \frac{1}{r^2 + (a^0_{(2\mu)})^2},
  \label{KS:expanding:rdep:H}
\end{equation}
which is identical to the case with vanishing $\Lambda$ discussed in
\cite{OrtaggioPravdaPravdova2008}.

Note that $\H$ behaves as $\H \approx \frac{\H_0}{r^{m - 1}}$ for large $r$ and
therefore Einstein GKS spacetimes \eqref{KS:GKSmetric} are asymptotically
(anti-)de Sitter or Minkowski depending on the cosmological constant of the
background metric $\bar{g}_{ab}$ as one approaches the null infinity along a
null geodesic of the Kerr--Schild congruence $\bk$. The behaviour near the
origin $r = 0$ will be investigated in section \ref{sec:KS:singularities}.

\subsection{Algebraic type}

In section \ref{sec:KS:Weyl}, we have already shown that GKS spacetimes
\eqref{KS:GKSmetric} with a geodetic Kerr--Schild vector $\bk$ which include
Einstein spaces are of Weyl type II or more special as follows from proposition
\ref{KS:proposition:geodetic_k}. Moreover, non-expanding Einstein GKS spacetimes
are necessarily only of Weyl type N by proposition
\ref{KS:proposition:nonexpanding}.

Now, we generalize the argument given in \cite{OrtaggioPravdaPravdova2008} for
the Ricci-flat case to Einstein spaces in order to show that expanding Einstein
GKS spacetimes are incompatible with Weyl types III and N and therefore such
spacetimes are only of types II, D or conformally flat.

The boost weight zero components of the Weyl tensor vanish, by definition, for
Weyl types III and N. In particular, the vanishing frame components $C_{0i1j}$
of Einstein GKS spacetimes given by \eqref{KS:R0i1j} and \eqref{KS:C0i1j} imply 
\begin{equation}
  L_{ij} \D\H = 2 \H A_{ik} L_{kj}.
\end{equation}
Multiplying the above equation by $L_{lj}$ and using the optical constraint
\eqref{KS:vacuum:optical_constraint}, one obtains
\begin{equation}
  S_{il} \D\H = 2 \H A_{ik} S_{kl}.
  \label{KS:algtype:eqn1}
\end{equation}
Taking the trace of \eqref{KS:algtype:eqn1} and eliminating the constant factor
$(n-2)$ gives
\begin{equation}
  \theta \, \D\H =0.
  \label{KS:expanding:Weylcond}
\end{equation}
Next, we employ the Einstein field equations. Substituting $\D\H = 0$ to
\eqref{KS:vacuum:EFE01} leads to $\omega = 0$, therefore, the optical matrix is
symmetric $L_{ij} = S_{ij}$ and one may rewrite \eqref{KS:vacuum:EFEij} as
\begin{equation}
  S_{ik} S_{jk} = (n - 2) \theta S_{ij}.
  \label{KS:expanding:EFEij:non-twisting}
\end{equation}
In a frame of the eigenvectors, $S_{ij}$ takes the form
$S_{ij} = \text{diag}(s_{(2)}, s_{(3)}, \ldots, s_{(n-1)})$ and
\eqref{KS:expanding:EFEij:non-twisting} reduces to
$s^2_{(i)} = s_{(i)} \sum_j s_{(j)}$ which has the only solution
\begin{equation}
  L_{ij} = \text{diag}(s, 0, \ldots, 0).
  \label{KS:expanding:Lij:typeIII&N}
\end{equation}
Indeed, substituting $\D\H = 0$ along with \eqref{KS:expanding:Lij:typeIII&N}
to the Weyl tensor cancels the remaining non-vanishing boost weight zero
components $C_{01ij}$ \eqref{KS:R01ij}, \eqref{KS:C01ij} and $C_{ijkl}$
\eqref{KS:Rijkl}, \eqref{KS:Cijkl}.

The canonical form of the optical matrix $L_{ij}$ for Ricci-flat spacetimes of
type N and non-twisting subclass of type III was determined in
\cite{PravdaPravdovaColeyMilson2004} using the Bianchi identities and the fact
that $C_{abcd} = R_{abcd}$. The same result can be also obtained for Einstein
spaces since the additional terms in the Weyl tensor of such spacetimes
\eqref{KS:Weyl:Einsteinspaces} are proportional to the cosmological constant
$\Lambda$ and the metric tensor $g_{ab}$ which does not affect the Bianchi
identities, i.e.\ $C_{ab[cd;e]} = R_{ab[cd;e]}$. Therefore, in the case of
non-twisting Einstein spacetimes of types III and N, the canonical form of the
optical matrix is
\begin{equation}
  L_{ij} = \text{diag}(s, s, 0, \ldots, 0).
  \label{KS:expanding:Lij:typeIII&N:Einsteinspaces}
\end{equation}

The form of the optical matrix of expanding Einstein GKS spacetimes with
$\D\H = 0$ \eqref{KS:expanding:Lij:typeIII&N} is not compatible with the form
of the optical matrix for general non-twisting Einstein spacetimes of Weyl types
III and N \eqref{KS:expanding:Lij:typeIII&N:Einsteinspaces}. Consequently,
expanding Einstein GKS solutions of types III and N do not exist and we can
conclude that
      
\begin{proposition}
  Einstein generalized Kerr--Schild spacetimes \eqref{KS:GKSmetric} with an
  expanding geodetic Kerr--Schild congruence $\bk$ are of Weyl types II, D or
  conformally flat.
  \label{KS:proposition:expanding}
\end{proposition}
Note that the conformally flat case occurs only if we admit the cosmological
constants of the full and background spacetimes not to be equal, otherwise $\H$
has to vanish and $g_{ab}$ is given just by the background metric $\bar{g}_{ab}$.

Conversely, one can immediately see that $\D\H$ has to be non-vanishing in
expanding Einstein GKS spacetimes and thus the Kerr--Schild function $\H$ has
to depend on an affine parameter along the null geodesics $\bk$. If we compare
this result with the $r$-dependence of $\H$ \eqref{KS:expanding:rdep:H}, it is
obvious that $m \neq 1$. Therefore, in accordance with the Goldberg--Sachs
theorem, the optical matrix of expanding Einstein GKS spacetimes is necessarily
non-shearing in the case $n = 4$ since it consists either of one $2 \times 2$
block \eqref{KS:expanding:Lij:block} or of the unit matrix multiplied by
$r^{-1}$. The optical constraint can thus be essentially considered as a higher
dimensional generalization of the Goldberg--Sachs theorem for Einstein GKS
spacetimes which states that there is no shear in two dimensional planes spanned
by two spacelike frame vectors $\bm^{(i)}$, $\bm^{(j)}$ corresponding to the
$2 \times 2$ blocks in the optical matrix.

Now, we are able to summarize all possible algebraic types of Einstein GKS
spacetimes. Omitting more general types G and I, excluded by proposition
\ref{KS:proposition:Weyltypes}, and the trivial conformally flat case, the
allowed combinations of the Weyl types and values of the expansion scalar are
depicted in table \ref{tab:KS:Weyltypes}.

\begin{table}
  \caption{Weyl types compatible with Einstein generalized Kerr--Schild
  spacetimes depending on the values of the expansion scalar $\theta$.}
  \begin{center}
    \begin{tabular}{ccccc}
      \toprule
      & \multicolumn{4}{c}{Weyl type} \\
      \cmidrule(l){2-5}
      Expansion & II & D & III & N \\
      \midrule
      $\theta = 0$ & $\times$ & $\times$ & $\times$ & \checkmark \\
      $\theta \neq 0$ & \checkmark & \checkmark & $\times$ & $\times$ \\
      \bottomrule
    \end{tabular}
  \end{center}
  \label{tab:KS:Weyltypes}
\end{table}

Let us conclude that Einstein GKS spacetimes of Weyl type N admit only a
non-expanding Kerr--Schild congruence $\bk$ and belong to the Kundt class. Type
III is incompatible with the GKS ansatz in the case of Einstein spaces, whereas
types II and D imply that such spacetimes are expanding.

Note also that the higher dimensional Robinson--Trautman class contains only
solutions of Weyl type D \cite{PodolskyOrtaggio2006} and although there is an
intersection with type D Einstein GKS spacetimes, such as the
Schwarzschild--Tangherlini black hole, in general higher dimensional
Robinson--Trautman metrics do not admit the GKS form. However, it was shown in
\cite{PodolskyOrtaggio2006} that general Robinson--Trautman metric
\begin{equation}
  \d s^2 = \frac{r^2}{P^2} \gamma_{ij} \, \d x^i \, \d x^j - 2 \d u \, \d r - 2 H \d u^2
  \label{KS:RobinsonTrautman:metric}
\end{equation}
is conformally flat and Einstein if $\gamma_{ij}$ is of constant curvature which
then implies
\begin{equation}
  \gamma_{ij} = \delta_{ij}, \qquad
  P = a(u) + b_i(u) x^i + c(u) \delta_{ij} x^i x^j,
  \label{KS:RobinsonTrautman:(A)dS}
\end{equation}
where $a(u)$, $b_i(u)$ and $c(u)$ are arbitrary functions of the coordinate $u$.
Therefore, every Einstein Robinson--Trautman metric
\eqref{KS:RobinsonTrautman:metric} possibly admitting aligned null radiation in
the Ricci tensor with $\gamma_{ij}$ and $P$ of the form
\eqref{KS:RobinsonTrautman:(A)dS} differs from the corresponding (anti-)de
Sitter or Minkowski background by the factor
\begin{equation}
  2 \H k_a k_b \, \d x^a \, \d x^b =
    2 (H - H^0) \, \d u^2 =
    - \frac{\mu(u)}{r^{n - 3}} \, \d u^2.
\end{equation}
If the function $\mu(u)$ does not depend on $u$, such metrics describe static
(A)dS--Schwarzschild--Tangherlini black holes. Otherwise, a null radiation term
appears in the Ricci tensor and the metric corresponds to the Vaidya solution.

\subsection{\texorpdfstring{$r$-dependence of the Weyl tensor}{r-dependence of the Weyl tensor}}
\label{sec:KS:rdep}

The boost weight zero components of the Weyl tensor of Einstein GKS spacetimes
\eqref{KS:C0101}--\eqref{KS:Cijkl}, \eqref{KS:R0101}--\eqref{KS:Rijkl} are given
only in terms of the optical matrix $L_{ij}$, the Kerr--Schild function $\H$
and its first and second derivatives $\D\H$ and $\D^2\H$, respectively. This
now allows us to determine easily the $r$-dependence of these components of the
Weyl tensor using the $r$-dependence of the quantities $L_{ij}$ and $\H$ derived
already in section \ref{sec:KS:optical_constraint}. We will employ these results
later in order to discuss the presence of curvature singularities in expanding
Einstein GKS spacetimes in section \ref{sec:KS:singularities}.

It is convenient to adopt more compact notation for the boost weight zero
components of the Weyl tensor \cite{PravdaPravdovaOrtaggio2007}
\begin{equation}
  \Phi_{ij} \equiv C_{0i1j}, \qquad
  \Phi = C_{0101}, \qquad
  \Phi^S_{ij} = - \frac{1}{2} C_{ikjk}, \qquad
  \Phi^A_{ij} = \frac{1}{2} C_{01ij}.
  \label{KS:expanding:Phi}
\end{equation}
First, we express the derivatives of the function $\H$ from
\eqref{KS:expanding:rdep:H}
\begin{align}
  \D\H &= - \frac{\H_0}{r^{m - 2p - 2}} \left( \frac{m - 2p - 1}{r^2}
      + 2 \sum^p_{\mu = 1} \frac{1}{r^2 + (a^0_{(2 \mu)})^2} \right)
      \prod^p_{\nu = 1} \frac{1}{r^2 + (a^0_{(2 \nu)})^2},
  \label{KS:expanding:rdep:DH} \\
  \D^2\H &= \frac{\H_0}{r^{m - 2p - 3}} \Bigg( \frac{(m - 2p - 1) (m - 2p)}{r^4} \notag \\
      &\qquad + 2 \frac{2 m - 4p - 3}{r^2} \sum^p_{\mu = 1} \frac{1}{r^2 + (a^0_{(2 \mu)})^2}
      + 4 \sum^p_{\mu = 1} \frac{1}{(r^2 + (a^0_{(2 \mu)})^2)^2} \notag \\
      &\qquad + 4 \sum^p_{\mu = 1} \frac{1}{r^2 + (a^0_{(2 \mu)})^2} \sum^p_{\rho = 1}
      \frac{1}{r^2 + (a^0_{(2 \rho)})^2} \Bigg) \prod^p_{\nu = 1} \frac{1}{r^2 + (a^0_{(2 \nu)})^2}.
  \label{KS:expanding:rdep:DDH}
\end{align}
Substituting the $r$-dependence of the optical matrix $L_{ij}$
\eqref{KS:expanding:Lij:blockdiagonal}--\eqref{KS:expanding:rdep:Ldiagonal} and
the function $\H$ \eqref{KS:expanding:rdep:H} and its derivatives
\eqref{KS:expanding:rdep:DH}, \eqref{KS:expanding:rdep:DDH} to the expressions
for the corresponding boost weight zero components of the Weyl tensor
\eqref{KS:R0101}--\eqref{KS:Rijkl}, \eqref{KS:C0101}--\eqref{KS:Cijkl}, we
immediately obtain the $r$-dependence of $\Phi_{ij}$
\begin{align}
  \Phi_{2\mu, 2\mu} &= \Phi_{2\mu + 1,2\mu + 1} = - \D\H s_{(2\mu)} - 2 \H A^2_{2\mu, 2\mu + 1} \phantom{\half} \notag \\
    &= \frac{\H_0}{r^{m - 2p - 3}} \frac{1}{r^2 + (a^0_{(2\mu)})^2} \Bigg( \frac{m - 2p - 2}{r^2}
      + \frac{1}{r^2 + (a^0_{(2\mu)})^2} \notag \\
    &\qquad + 2 \sum^p_{\nu=1} \frac{1}{r^2 + (a^0_{(2\nu)})^2} \Bigg)
    \prod^p_{\rho=1} \frac{1}{r^2 + (a^0_{(2\rho)})^2},
  \label{KS:expanding:rdep:Phi:1} \\
  \Phi_{2\mu, 2\mu + 1} &= \Phi^A_{2\mu, 2\mu + 1} = - \D (\H A_{2\mu, 2\mu + 1}) \phantom{\half} \notag \\
    &= \frac{\H_0}{r^{m - 2p - 2}} \frac{a^0_{(2\mu)}}{r^2 + (a^0_{(2\mu)})^2} \Bigg( \frac{m - 2p - 1}{r^2}
      + \frac{2}{r^2 + (a^0_{(2\mu)})^2} \notag \\
    &\qquad + 2 \sum^p_{\nu=1} \frac{1}{r^2 + (a^0_{(2\nu)})^2} \Bigg)
    \prod^p_{\rho=1} \frac{1}{r^2 + (a^0_{(2\rho)})^2},
  \label{KS:expanding:rdep:Phi:2} \\
  \Phi_{\alpha \beta} &= - r^{-1} \delta_{\alpha \beta}, \qquad
  \Phi = \D^2\H. \phantom{\half}
  \label{KS:expanding:rdep:Phi:3}
\end{align}
Hence, $\Phi_{ij}$ reproduces the block diagonal structure of the optical matrix
$L_{ij}$, where $\mu, \nu, \ldots = 1, \dots, p$ number the $2 \times 2$ blocks,
whereas the elements of the diagonal block are indexed by $\alpha, \beta,
\ldots = 2p + 2, \dots, n-1$. Similarly, one may determine the
$r$-dependence of the remaining non-vanishing boost weight zero components $C_{ijkl}$
\begin{align}
  C_{2\mu, 2\mu+1, 2\mu, 2\mu+1} &= 2 \H \left( 3 A^2_{2\mu, 2\mu+1} - s^2_{(2\mu)} \right) \phantom{\half} \notag \\
    &= - 2 \frac{\H_0}{r^{m - 2p - 1}} \frac{r^2 - 3 (a^0_{(2\mu)})^2}{\left(r^2 + (a^0_{(2\mu)})^2\right)^2}
      \prod^p_{\nu = 1} \frac{1}{r^2 + (a^0_{(2 \nu)})^2}, \\
  C_{2\mu, 2\mu+1, 2\nu, 2\nu+1} &= 2 C_{2\mu, 2\nu, 2\mu+1, 2\nu+1} = -2 C_{2\mu, 2\nu+1, 2\mu+1, 2\nu} \phantom{\half} \notag \\
    &= 4 \H A_{2\mu, 2\mu+1} A_{2\nu, 2\nu+1} \phantom{\half} \notag \\
    &= 4 \frac{\H_0}{r^{m - 2p - 1}} \frac{a^0_{(2\mu)}}{r^2 + (a^0_{(2\mu)})^2}
      \frac{a^0_{(2\nu)}}{r^2 + (a^0_{(2\nu)})^2}
      \prod^p_{\rho = 1} \frac{1}{r^2 + (a^0_{(2 \rho)})^2}, \\ 
  C_{2\mu, 2\nu, 2\mu, 2\nu} &= C_{2\mu, 2\nu+1, 2\mu, 2\nu+1} = -2 \H s_{(2\mu)} s_{(2\nu)} \phantom{\half} \notag \\
    &= -2 \frac{\H_0}{r^{m - 2p - 3}} \frac{1}{r^2 + (a^0_{(2\mu)})^2}
      \frac{1}{r^2 + (a^0_{(2\nu)})^2}
      \prod^p_{\rho = 1} \frac{1}{r^2 + (a^0_{(2 \rho)})^2}, \\ 
  C_{(\alpha)(i)(\alpha)(i)} &= - 2 \H s_{(i)} r^{-1} \phantom{\half} \notag \\
     &= -2 \frac{\H_0}{r^{m - 2p - 1}} \frac{1}{r^2 + (a^0_{(i)})^2}
      \prod^p_{\mu = 1} \frac{1}{r^2 + (a^0_{(2 \mu)})^2}, 
\end{align}
where $\mu \neq \nu$.

\subsection{Singularities}
\label{sec:KS:singularities}

Let us briefly discuss curvature singularities of expanding Einstein GKS
metrics. It is obvious from the $r$-dependence of the Kerr--Schild function
$\H$ \eqref{KS:expanding:rdep:H} that it may diverge for $r \to 0$. Namely,
$\H$ and consequently the full GKS metric blows up at $r = 0$ in the
following cases:
\begin{itemize}
  \item in the ``generic'' case when neither $2p = m$ nor $2p = m - 1$, or
  \item in the special cases when $2p = m$ (for $m$ even) or $2p = m - 1$
    (for $m$ odd) if at least one of the functions $a^0_{(2\mu)}$, not depending
    on $r$, admits a real root at $x = x_0$.
\end{itemize}
One may express the Kretschmann scalar at the singular point to verify whether
there is a real curvature singularity. Omitting the trivial conformally flat
case, expanding Einstein GKS spacetimes are of Weyl types D or II as follows
from proposition \ref{KS:proposition:expanding}. Thus, the positive boost weight
components of the Weyl tensor and consequently the corresponding components of
the Riemann tensor vanish. Note that the negative boost weight components do not
have the appropriate counterparts in the following contractions. For instance,
the term $R_{011i} n_a k_b k_c m_d^{(i)}$ may give a non-zero contribution only
with the term $R_{100i} k^a n^b n^c m^d_{(i)}$. Therefore, the Kretschmann
scalar is determined only by the boost weight zero components of the Riemann
tensor
\begin{equation}
  R_{abcd} R^{abcd} = 4 \left( R_{0101} \right)^2 - 4 R_{01ij} R_{01ij}
    + 8 R_{0i1j} R_{0j1i} + R_{ijkl} R_{ijkl}.
  \label{KS:Kretschmann}
\end{equation}
Expressing the frame components of the Riemann tensor in terms of the Weyl
tensor from \eqref{KS:C0101}--\eqref{KS:Cijkl} and using the notation
\eqref{KS:expanding:Phi}, we can rewrite the Kretschmann scalar as
\begin{equation}
  \begin{aligned}
  R_{abcd} R^{abcd} &= 4 \Phi^2 + 8 \Phi^S_{ij} \Phi^S_{ij} - 24 \Phi^A_{ij} \Phi^A_{ij} + C_{ijkl} C_{ijkl} \\
    &\qquad + \frac{8 n}{(n-1)(n-2)^2} \, \Lambda^2.
  \end{aligned}
  \label{KS:Kretschmann:Phi}
\end{equation}
The only additional term in comparison with the Ricci-flat case
\cite{OrtaggioPravdaPravdova2008} is the last constant term proportional to
$\Lambda^2$ which clearly cannot influence the divergence of the Kretschmann
scalar and thus the presence of singularities.

Let us discuss behavior of \eqref{KS:Kretschmann:Phi} for $r \to 0$ in the above
mentioned singular cases. The Kretschmann scalar consists of a sum of squares,
except the third term which is negative, and, as we will see, it is sufficient
to compare only the first and the third term.

In the ``generic'' case $2p \neq m$, $2p \neq m - 1$, one may determine the
behaviour of the following quantities near $r = 0$ from the $r$-dependence of
$\H$ \eqref{KS:expanding:rdep:H} and $A_{2\mu, 2\mu+1}$
\eqref{KS:expanding:rdep:s+A}
\begin{equation}
  \begin{aligned}
  &\H \sim r^{-(m - 2p - 1)}, \qquad
  \D\H \sim r^{-(m - 2p)}, \qquad
  \D^2\H \sim r^{-(m - 2p + 1)}, \\
  &A_{2\mu, 2\mu+1} \sim 1 \quad \text{if $a^0_{(2\mu)} \neq 0$}, \qquad
  A_{2\mu, 2\mu+1} = 0 \quad \text{if $a^0_{(2\mu)} = 0$ at $x = x_0$}.
  \end{aligned}
  \label{KS:expanding:rdep:various}
\end{equation}
Substituting \eqref{KS:expanding:rdep:various} to \eqref{KS:expanding:rdep:Phi:2} and
\eqref{KS:expanding:rdep:Phi:3}, we obtain that $\Phi \sim r^{-(m - 2p + 1)}$
and either $\Phi^A_{2\mu, 2\mu+1} \sim r^{-(m - 2p)}$ for $a^0_{(2\mu)} \neq 0$
or $\Phi^A_{2\mu, 2\mu+1} = 0$ for $a^0_{(2\mu)} = 0$. Therefore, the first term
dominates over the third term in \eqref{KS:Kretschmann:Phi} and the Kretschmann
scalar diverges. Thus, in the ``generic'' case, a curvature singularity is
always located at $r = 0$.

Note that this case also includes all non-twisting expanding Einstein GKS
solutions, where $p = 0$, such as the higher dimensional
(A)dS--Schwarzschild--Tangherlini black holes. A five-dimensional example of
these metrics will be presented in section \ref{sec:KS:Kerr-(A)dS} as a static
limit of the (A)dS--Kerr metric \eqref{KS:Kerr-(A)dS:5D:metric} with $m = 3$,
$p = 0$, where the corresponding optical metric $L_{ij}$ and the function $\H$
will be also given explicitly.

Similarly, one may analyze the special cases $2p = m$ (for $m$ even),
$2p = m - 1$ (for $m$ odd) where some of the functions $a^0_{(2\mu)}$ have real
roots. Let $q$ $(q \geq 1)$ denotes the number of such vanishing $a^0_{(2\mu)}$.
Now, from \eqref{KS:expanding:rdep:s+A} it follows that if $a^0_{(2\mu)}$ has no
root then $A_{2\mu, 2\mu+1} \sim 1$ near $r = 0$, whereas if $a^0_{(2\mu)}$
admits a root $x = x_0$ then $A_{2\mu, 2\mu+1} = 0$ at this point. The
Kerr--Schild function $\H$ \eqref{KS:expanding:rdep:H} and its derivatives
behave for $2p = m$ as
\begin{equation}
  \H \sim r^{-2q + 1} \, , \qquad
  \D\H \sim r^{-2q} \, , \qquad
  \D^2\H \sim r^{-2q - 1} \, ,
\end{equation}
therefore, $\Phi \sim r^{-2q - 1}$ and
$\Phi^A_{ij} \Phi^A_{ij} \sim 2 (p - q) r^{-4q}$.
In the case $2p = m - 1$ one gets
\begin{equation}
  \H \sim r^{-2q} \, , \qquad
  \D\H \sim r^{-2q - 1} \, , \qquad
  \D^2\H \sim r^{-2q - 2}
\end{equation}
and consequently $\Phi \sim r^{-2q - 2}$ and
$\Phi^A_{ij} \Phi^A_{ij} \sim 2 (p - q) r^{-2(2q + 1)}$.
Therefore, in both special cases $2p = m$ and $2p = m - 1$, the first term in
\eqref{KS:Kretschmann:Phi} dominates again over the third term and if any
$a^0_{(2\mu)}$ has a real root at $x=x_0$ then a curvature singularity is
located at $r=0$, $x=x_0$. Note that this corresponds, for instance, to the
well-known ring shaped singularity of the Kerr black hole.

As will be shown in section \ref{sec:KS:Kerr-(A)dS}, these special cases are
represented, e.g., by the five-dimensional Kerr--(A)dS metric
\eqref{KS:Kerr-(A)dS:5D:metric}, where the optical matrix $L_{ij}$ of rank
$m = 3$ has, in the rotating case, one $2 \times 2$ block, i.e.\ $p = 1$. The
function $a^0_{(2)}$ is given by the spins $a$, $b$ as $(a^0_{(2)})^2 =
a^2 \cos^2\theta + b^2 \sin^2\theta$. If one of the spins is zero, $a^0_{(2)}$
admits a root and, indeed, this corresponds to the special case ($2p = m - 1$,
$m$ odd) with a curvature singularity located at $r = 0$. In the case with both
spins being non-zero, this metric belongs neither to the ``generic'' case nor to
the special cases since $a^0_{(2)}$ never vanishes and therefore no singularity
is present at $r = 0$.

\subsection[\texorpdfstring{Example of expanding Einstein generalized Kerr--Schild\\spacetime: Kerr--(A)dS}
{Example of expanding Einstein generalized Kerr--Schild spacetime: Kerr--(A)dS}]
{Example of expanding Einstein generalized\\Kerr--Schild spacetime: Kerr--(A)dS}
\label{sec:KS:Kerr-(A)dS}

In this section, we compare our results obtained for general expanding Einstein
GKS spacetimes with an explicit physically interesting example, namely, the
higher dimensional Kerr--(A)dS metric. Considering such metrics in five
dimensions, we find a parallelly transported frame which allows us to express
the optical matrix $L_{ij}$ in the block-diagonal form. Subsequently, we compare
this optical matrix and the Kerr--Schild function $\H$ with the $r$-dependence
of the corresponding quantities of general expanding Einstein GKS spacetimes
derived in the previous sections and discuss the presence of curvature
singularities.

The higher dimensional Kerr--(A)dS metric is an example of an expanding Einstein
GKS spacetime describing a black hole rotating in $\lfloor (n-1)/2 \rfloor$
independent planes with a possible cosmological constant $\Lambda$. In this
sense, it is a generalization of the Myers--Perry black hole, which can be
obtained by taking the limit $\Lambda \rightarrow 0$. The Kerr--(A)dS metric in
arbitrary dimension was derived in \cite{GibbonsLuPagePope2004} in the GKS form
\eqref{KS:GKSmetric} using the spheroidal coordinates consisting of the radial
coordinate $r$, time coordinate $t$, $\lfloor (n-1)/2 \rfloor$ azimuthal angular
coordinates $\phi_i$ and $\lfloor n/2 \rfloor$ coordinates $\mu_i$ subject to
\begin{equation}
  \sum^{\lfloor n/2 \rfloor}_{i=1} \mu_i^2 = 1.
  \label{KS:Kerr-(A)dS:mu}
\end{equation}
The background metric $\bar g_{ab}$,
Kerr--Schild vector $\bk$ and function $\H$ in $n = 2k + 1$ dimensions are given
by
\begin{align}
  \bar g_{ab} \, \d x^a \, \d x^b &= - W (1 - \lambda r^2) \, \d t^2
    + F \, \d r^2
    + \sum^k_{i=1} \frac{r^2 + a_i^2}{1 + \lambda a_i^2} (\d \mu_i^2 + \mu_i^2 \, \d \phi_i^2) \notag \\
    &\qquad + \frac{\lambda}{W (1 - \lambda r^2)} \left( \sum^k_{i=1} \frac{(r^2 + a_i^2) \mu_i \, \d \mu_i}{1 + \lambda a_i^2} \right)^2, \\
  \label{KS:Kerr-(A)dS:metric:1}
  k_a \, \d x^a &= W \, \d t + F \, \d r
    - \sum^k_{i=1} \frac{a_i \mu_i^2}{1 + \lambda a_i^2} \, \d \phi_i, \\
  \H &= - \frac{M}{\sum^k_{i=1} \frac{\mu_i^2}{r^2 + a_i^2} \prod^k_{j=1} (r^2 + a_j^2)},
\end{align}
whereas in $n = 2k$ dimensions
\begin{align}
   \bar g_{ab} \, \d x^a \, \d x^b &= - W (1 - \lambda r^2) \, \d t^2
    + F \, \d r^2
    + \sum^k_{i=1} \frac{r^2 + a_i^2}{1 + \lambda a_i^2} \, \d \mu_i^2
    + \sum^{k-1}_{i=1} \frac{r^2 + a_i^2}{1 + \lambda a_i^2} \mu_i^2 \, \d \phi_i^2 \notag \\
    &\qquad + \frac{\lambda}{W (1 - \lambda r^2)} \left( \sum^{k}_{i=1} \frac{(r^2 + a_i^2) \mu_i \, \d \mu_i}{1 + \lambda a_i^2} \right)^2, \\
  k_a \, \d x^a &= W \, \d t + F \, \d r
    - \sum^{k-1}_{i=1} \frac{a_i \mu_i^2}{1 + \lambda a_i^2} \, \d \phi_i, \\
  \H &= - \frac{M}{r \sum^k_{i=1} \frac{\mu_i^2}{r^2 + a_i^2} \prod^{k-1}_{j=1} (r^2 + a_j^2)},
  \label{KS:Kerr-(A)dS:metric:2}
\end{align}
where the functions $W$ and $F$ are defined as
\begin{equation}
  W = \sum^k_{i=1} \frac{\mu_i^2}{1 + \lambda a_i^2}, \qquad
  F = \frac{r^2}{1 - \lambda r^2} \sum^k_{i=1} \frac{\mu_i^2}{r^2 + a_i^2}
\end{equation}
and $\lambda$ is related to the cosmological constant $\Lambda$ via
\begin{equation}
  \lambda = \frac{2\Lambda}{(n-1)(n-2)}.
  \label{KS:lambda}
\end{equation}

In five dimensions, one may choose the coordinates $\mu_i$ as
$\mu_1 = \sin \theta$, $\mu_2 = \cos \theta$ that clearly satisfy
\eqref{KS:Kerr-(A)dS:mu} and denote the azimuthal coordinates as $\phi_1 = \phi$,
$\phi_2 = \psi$ and the rotation parameters as $a_1 = a$, $a_2 = b$. The
background metric $\bar{g}_{ab}$, Kerr--Schild vector $\bk$ and function $\H$
are then given by
\begin{align}
  \bar g_{ab} \, \d x^a \, \d x^b &= - \frac{(1-\lambda r^2) \Delta}{(1 + \lambda a^2)(1 + \lambda b^2)} \, \d t^2
    + \frac{r^2 \rho^2}{(1 - \lambda r^2)(r^2 + a^2)(r^2 + b^2)} \, \d r^2 \notag \\
    &\qquad + \frac{\rho^2}{\Delta} \, \d \theta^2 
    + \frac{(r^2 + a^2) \sin^2\theta}{1 + \lambda a^2} \, \d \phi^2
    + \frac{(r^2 + b^2) \cos^2\theta}{1 + \lambda b^2} \, \d \psi^2,
  \label{KS:Kerr-(A)dS:5D:metric} \\
  k_a \, \d x^a  &= \frac{\Delta}{(1 + \lambda a^2)(1 + \lambda b^2)} \, \d t
    + \frac{r^2 \rho^2}{(1 - \lambda r^2)(r^2 + a^2)(r^2 + b^2)} \, \d r \notag \\
    &\qquad- \frac{a \sin^2\theta}{1 + \lambda a^2} \, \d \phi
    - \frac{b \cos^2\theta}{1 + \lambda b^2} \, \d \psi,
  \label{KS:Kerr-(A)dS:5D:k} \\
  \mathcal{H} &= - \frac{M}{\rho^2},
  \label{KS:Kerr-(A)dS:5D:H}
\end{align}
where the angular coordinate ranges are as usual $\theta \in \langle0,
\pi\rangle$, $\phi \in \langle 0, 2 \pi )$, $\psi \in \langle 0, 2 \pi)$ and the
functions $\rho$, $\Delta$ and $\nu$ are defined as
\begin{equation}
  \rho^2 = r^2 + \nu^2, \qquad
  \Delta = 1 + \lambda \nu^2, \qquad
  \nu = \sqrt{a^2 \cos^2\theta + b^2 \sin^2\theta}.
\end{equation}
In agreement with propositions \ref{KS:proposition:geodetic_k} and
\ref{KS:proposition:Weyltypes}, it can be shown by a straightforward calculation
that the Kerr--Schild vector $\bk$ \eqref{KS:Kerr-(A)dS:5D:k} is the geodetic
multiple WAND. Moreover, $\bk$ is already scaled to be affinely parametrized.

Now we construct a parallelly transported null frame, with $\bk$ being one of
the vectors, such that the optical matrix $L_{ij}$ takes the block-diagonal form
\eqref{KS:expanding:Lij:blockdiagonal}. However, first of all we find an
arbitrary null frame satisfying the constraints \eqref{intro:frame:constraints}
and then we can transform it to the desired form using null rotations and spins
which preserve $\bk$.

Note that, unlike the full Kerr--(A)dS metric, the background metric
$\bar{g}_{ab}$ \eqref{KS:Kerr-(A)dS:5D:metric} is diagonal and much more simpler.
Therefore, it is easier to construct a null frame in the background spacetime
$\bk$, $\bar\bn$, $\bm^{(i)}$ which is related to the frame in the full spacetime
$\bk$, $\bn$, $\bm^{(i)}$ by \eqref{KS:background_n}. One may immediately express
the inverse background metric as
\begin{equation}
  \begin{aligned}
  \bar g^{ab} \frac{\partial}{\partial x_a} \frac{\partial}{\partial x_b} &=
    - \frac{(1 + \lambda a^2)(1 + \lambda b^2)}{(1-\lambda r^2) \Delta} \left( \frac{\partial}{\partial t} \right)^2
    + \frac{\Delta}{\rho^2} \left( \frac{\partial}{\partial \theta} \right)^2 \\
    &\qquad + \frac{1 + \lambda a^2}{(r^2 + a^2) \sin^2\theta} \left( \frac{\partial}{\partial \phi} \right)^2
    + \frac{1 + \lambda b^2}{(r^2 + b^2) \cos^2\theta} \left( \frac{\partial}{\partial \psi} \right)^2 \\
    &\qquad + \frac{(1 - \lambda r^2)(r^2 + a^2)(r^2 + b^2)}{r^2 \rho^2} \left( \frac{\partial}{\partial r} \right)^2.
  \end{aligned}
  \label{KS:Kerr-(A)dS:5D:inversemetric}
\end{equation}
Since $\bar g^{ab}$ is diagonal we can easily choose the covector $\bar\bn$
same as $\bk$ with the only difference that we change the sign of the ``$\d t$''
component to ensure that $\bar\bn$ is also null and then we multiply it by an
appropriate factor to satisfy the frame normalization
$\bar g^{ab} k_a \bar n_b = 1$
\begin{equation}
  \begin{aligned}
  \bar n_a \d x^a &= - \frac{1 - \lambda r^2}{2} \, \d t
    + \frac{(1 + \lambda a^2)(1 + \lambda b^2) r^2 \rho^2}{2 \Delta (r^2 + a^2) (r^2 + b^2)} \, \d r \\
    &\qquad - \frac{a (1 + \lambda b^2)(1 - \lambda r^2) \sin^2\theta}{2 \Delta} \, \d \phi \\
    &\qquad - \frac{b (1 + \lambda a^2)(1 - \lambda r^2) \cos^2\theta}{2 \Delta} \, \d \psi.
  \end{aligned}
\end{equation}
Consequently, the corresponding frame vector $\bn$ in the full spacetime is
given by \eqref{KS:background_n} as
\begin{equation}
  \begin{aligned}
  n_a \d x^a &= \left( \frac{M \Delta}{\rho^2 (1 + \lambda a^2)(1 + \lambda b^2)}
      - \frac{1 - \lambda r^2}{2} \right) \d t \\
    &\qquad + \frac{r^2}{(r^2 + a^2) (r^2 + b^2)} \bigg( \frac{(1 + \lambda a^2)(1 + \lambda b^2) \rho^2}{2 \Delta} \\
      &\qquad + \frac{M}{(1 - \lambda r^2)} \bigg) \, \d r 
    - a \sin^2\theta \bigg( \frac{(1 + \lambda b^2)(1 - \lambda r^2)}{2 \Delta} \\
      &\qquad + \frac{M}{\rho^2 (1 + \lambda a^2)} \bigg) \, \d \phi 
    - b \cos^2\theta \bigg( \frac{(1 + \lambda a^2)(1 - \lambda r^2)}{2 \Delta} \\
      &\qquad + \frac{M}{\rho^2 (1 + \lambda b^2)} \bigg) \, \d \psi.
  \end{aligned}
  \label{KS:Kerr-(A)dS:5D:n}
\end{equation}

It remains to determine three spacelike frame vectors. A general spacelike
vector is of the form $m_a \, \d x^a = A \, \d t + B \,\d r + C \, \d\theta
+ E \, \d\phi + F \, \d\psi$. The orthogonality conditions
$\bar{g}^{ab} k_a m_b = \bar{g}^{ab} \bar{n}_a m_b = 0$ imply that $A=0$ and
$B = E \frac{a}{r^2 + a^2} + F \frac{b}{r^2 + b^2}$. Then from
$\bar{g}^{ab} m_a m_b = 1$ it follows that
\begin{multline}
  C^2 \frac{\Delta}{\rho^2}
  + E^2 \left[ \frac{1 + \lambda a^2 \cos^2 \theta}{\rho^2 \sin^2 \theta} + \frac{b^2}{r^2 \rho^2} \right]
  + 2 E F \left[ \frac{ab(1 - \lambda r^2)}{r^2 \rho^2} \right] \\
  + F^2 \left[ \frac{1 + \lambda b^2 \sin^2 \theta}{\rho^2 \cos^2 \theta} + \frac{a^2}{r^2 \rho^2} \right] = 1.
\end{multline}
Obviously, the simplest solution is $C = \rho/\sqrt{\Delta}$, $E = F = 0$
which determines the first of the spacelike frame vectors $\bm^{(2)}$
\begin{equation}
  m^{(2)}_a \, \d x^a = \frac{\rho}{\sqrt{\Delta}} \, \d\theta.
  \label{KS:Kerr-(A)dS:5D:m2}
\end{equation}
Expressing $\bar{g}^{ab} m^{(2)}_a m_b = 0$, one obtains $C = 0$ and therefore
the ``$\d \theta$'' component of two remaining spacelike vectors has to vanish,
i.e.
\begin{equation}
  m^{(\kappa)}_a \d x^a = \left( E_\kappa \frac{a}{r^2 + a^2}
    + F_\kappa \frac{b}{r^2 + b^2} \right) \d r
    + E_\kappa \, \d\phi
    + F_\kappa \, \d\psi, \qquad \kappa = 3,4.
\end{equation}
The four unknown functions $E_\kappa$, $F_\kappa$ have to satisfy three
equations following from the remaining constraints
$\bar{g}^{ab} m^{(3)}_a m^{(4)}_b = 0$, $\bar{g}^{ab} m^{(3)}_a m^{(3)}_b =
\bar{g}^{ab} m^{(4)}_a m^{(4)}_b = 1$, which are rather complicated to solve at
this moment. Since there is an arbitrariness, we can try to find $\bm^{(4)}$ to
be parallelly transported along $\bk$, i.e.\ to satisfy the
condition $m^{(4)}_{a;b} k^a = 0$. Only the contraction
$m^{(4)}_{a;b} k^a m_{(2)}^b$ is sufficiently simple to express and we obtain
$E_2 = F_2 \frac{b \sin^2 \theta}{a \cos^2 \theta}$ relating the functions $E_2$,
$F_2$. Putting this expression to the normalization
$\bar{g}^{ab} m^{(4)}_a m^{(4)}_b = 1$ yields
$F_2 = \frac{a r \cos^2 \theta}{\nu}$ and therefore
\begin{equation}
  \begin{aligned}
  m^{(4)}_a \, \d x^a &= \frac{abr}{\nu} \left( \frac{\sin^2 \theta}{r^2 + a^2}
    + \frac{\cos^2 \theta}{r^2 + b^2} \right) \d r \\
    &\qquad + \frac{b r \sin^2 \theta}{\nu} \, \d\phi + \frac{a r \cos^2 \theta}{\nu} \, \d\psi.
  \end{aligned}
  \label{KS:Kerr-(A)dS:5D:m4}
\end{equation}
The remaining unknown functions $E_1$, $F_1$ can be determined from
$\bar{g}^{ab} m^{(4)}_a m^{(3)}_b = 0$ leading to $E_1 = - F_1 \frac{a}{b}$ and
subsequently $\bar{g}^{ab} m^{(3)}_a m^{(3)}_b = 1$ implies
$F_1 = \frac{b \rho \sin\theta \cos\theta}{\sqrt{\Delta} \nu}$. Finally, we
thus obtain
\begin{equation}
  m^{(3)}_a \, \d x^a  = \frac{\rho \sin\theta \cos\theta}{\sqrt{\Delta} \nu} \left[
    \left( \frac{b^2}{r^2 + b^2} - \frac{a^2}{r^2 + a^2} \right) \d r
    - a \, \d\phi + b \, \d\psi \right].
  \label{KS:Kerr-(A)dS:5D:m3}
\end{equation}

So far we have constructed the null frame consisting of $\bk$
\eqref{KS:Kerr-(A)dS:5D:k}, $\bn$ \eqref{KS:Kerr-(A)dS:5D:n},
$\bm^{(2)}$ \eqref{KS:Kerr-(A)dS:5D:m2}, $\bm^{(3)}$ \eqref{KS:Kerr-(A)dS:5D:m3}
and $\bm^{(4)}$ \eqref{KS:Kerr-(A)dS:5D:m4} which can be expressed in
the contravariant form as
\begin{equation}
  \begin{aligned}
  \bk &= - \frac{1}{1 - \lambda r^2} \frac{\partial}{\partial t}
    + \frac{\partial}{\partial r}
    - \frac{a}{r^2 + a^2} \frac{\partial}{\partial \varphi}
    - \frac{b}{r^2 + b^2} \frac{\partial}{\partial \psi}, \\
  \bn &= \left( \frac{1}{2} \frac{(1 + \lambda a^2)(1 + \lambda b^2)(1 - \lambda r^2)}{\Delta}
    - \frac{M}{\rho^2} \right) \bk \\
    &\qquad + \frac{(1 + \lambda a^2)(1 + \lambda b^2)}{\Delta} \frac{\partial}{\partial t}, \\
  \bm^{(2)} &= \frac{\sqrt{\Delta}}{\rho} \frac{\partial}{\partial \theta}, \\
  \bm^{(3)} &= \frac{\rho \sin \theta \cos \theta}{\sqrt{\Delta} \nu }
    \bigg[ \frac{(b^2 - a^2)(1 - \lambda r^2)}{\rho^2} \frac{\partial}{\partial r}
    - \frac{a(1 + \lambda a^2)}{(r^2 + a^2) \sin^2 \theta} \frac{\partial}{\partial \varphi} \\
    &\qquad + \frac{b(1 + \lambda b^2)}{(r^2 + b^2) \cos^2 \theta} \frac{\partial}{\partial \psi} \bigg], \\
  \bm^{(4)} &= \frac{a b r}{\nu} \left[ \frac{1 - \lambda r^2}{r^2} \frac{\partial}{\partial r}
    + \frac{1 + \lambda a^2}{a (r^2 + a^2)} \frac{\partial}{\partial \varphi}
    + \frac{1 + \lambda b^2}{b (r^2 + b^2)} \frac{\partial}{\partial \psi} \right].
  \end{aligned}
  \label{KS:Kerr-(A)dS:5D:frame}
\end{equation}
One may show that although this frame is not parallelly transported along the
geodetic Kerr--Schild vector $\bk$, it is already adapted to the block-diagonal
form of the optical matrix $L_{ij}$, where the $2 \times 2$ block corresponds
to the plane spanned by the frame vectors $\bm^{(2)}$ and $\bm^{(3)}$.

Recall that the condition that a frame is parallelly transported along a
geodetic ($L_{i0} = 0$) and affinely parametrized ($L_{10} = 0$) null congruence
$\bk$ can be written using Ricci rotation coefficients as
\begin{equation}
  N_{i0} = \M{i}{j0} = 0.
\end{equation}
Considering our chosen frame \eqref{KS:Kerr-(A)dS:5D:frame}, we have to
transform away the following non-zero coefficients
\begin{equation}
  \begin{aligned}
  N_{20} &= \frac{\lambda (a^2 - b^2) \sin\theta \cos\theta}{\rho \sqrt{\Delta}}, \qquad
  N_{30} = - \frac{\lambda (a^2 - b^2) r \sin\theta \cos\theta}{\rho \nu \sqrt{\Delta}}, \\
  N_{40} &= \frac{\lambda a b}{\nu}, \qquad
  \M{2}{30} = \frac{\nu}{\rho^2}, \qquad
  \M{3}{20} = - \frac{\nu}{\rho^2},
  \end{aligned}
\end{equation}
by appropriate Lorentz transformations. First, it is convenient to perform spins
to set $\M{i}{j0}$ to zero since this simplifies the equations determining the
parameters $z_i$ of an appropriate null rotation around $\bk$ setting $N_{i0}$
to zero.

Note that null rotations with $\bk$ fixed and spins in any plane spanned by
$\bm^{(i)}$ and $\bm^{(j)}$ corresponding to a $2 \times 2$ block in the
canonical form of the optical matrix $L_{ij}$ preserve the block-diagonal form
of this matrix.

Therefore, we assume the transformation matrix of appropriate spins
\eqref{intro:Riccicoeff:spins} in the form
\begin{equation}
  X_{ij} = 
  \begin{pmatrix}
    \cos \varepsilon(r) & -\sin \varepsilon(r) & 0\\
    \sin \varepsilon(r) & \cos \varepsilon(r) & 0 \\
    0 & 0 & 1
  \end{pmatrix}.
\end{equation}
Requiring $\hM{i}{j0} = 0$ and using the orthogonality
$\mathbf{X}^{-1} = \mathbf{X}^T$ of the transformation matrix, it follows from
\eqref{intro:Riccicoeff:spins} that
\begin{equation}
  \D \mathbf{X}  = - \mathbf{X} \mathbf{M}.
\end{equation}
This is satisfied if $\D \varepsilon(r) \equiv \frac{\d \varepsilon(r)}{\d r} =
\frac{\nu}{r^2 + \nu^2}$ which has a solution $\varepsilon(r) =
\tan^{-1} \frac{r}{\nu}$ and therefore
\begin{equation}
  X_{ij} = 
  \begin{pmatrix}
    \frac{\nu}{\rho} & -\frac{r}{\rho} & 0\\
    \frac{r}{\rho} & \frac{\nu}{\rho} & 0 \\
    0 & 0 & 1
  \end{pmatrix}.
  \label{KS:Kerr-(A)dS:5D:Xij}
\end{equation}

Thus, the spin represented by the transformation matrix
\eqref{KS:Kerr-(A)dS:5D:Xij} ensures that all $\M{i}{j0}$ vanish and $N_{i0}$
transform as
\begin{equation}
  \begin{aligned}
  \hat{N}_{20} &= \frac{\nu}{\rho} N_{20} - \frac{r}{\rho} N_{30}
    = \frac{\lambda (a^2 - b^2) \sin \theta \cos \theta}{\sqrt{\Delta} \nu}, \\
  \hat{N}_{30} &= \frac{r}{\rho} N_{20} + \frac{\nu}{\rho} N_{30}  = 0, \qquad
  \hat{N}_{40} = N_{40} = \frac{\lambda a b}{\nu}
  \end{aligned}
  \label{KS:Kerr-(A)dS:5D:Ni0:spins}
\end{equation}
and now we set the remaining $\hat{N}_{20}$, $\hat{N}_{40}$ to zero using an
appropriate null rotation with $\bk$ fixed \eqref{intro:Riccicoeff:nullrot}
which leads to
\begin{equation}
  \hat{\hat{N}}_{i0} = \hat{N}_{i0} + \D z_i.
  \label{KS:Kerr-(A)dS:5D:nullrot}
\end{equation}
Requiring $\hat{\hat{N}}_{i0} = 0$ and substituting
\eqref{KS:Kerr-(A)dS:5D:Ni0:spins} to \eqref{KS:Kerr-(A)dS:5D:nullrot}, we can
integrate over $r$ to obtain the functions $z_i$
\begin{equation}
  z_2 = - \frac{\lambda (a^2 - b^2) r \sin \theta \cos \theta}{\sqrt{\Delta} \nu}, \qquad
  z_4 = - \frac{\lambda a b r}{\nu}.
\end{equation}
Therefore, we have found a parallelly transported frame $\bk$, $\hat{\hat{\bn}}$,
$\hat{\hat{\bm}}^{(2)}$, $\hat{\hat{\bm}}^{(3)}$, $\hat{\hat{\bm}}^{(4)}$ which
is related to the null frame $\bk$, $\bn$, $\bm^{(2)}$, $\bm^{(3)}$, $\bm^{(4)}$
\eqref{KS:Kerr-(A)dS:5D:frame} by
\begin{equation}
  \begin{aligned}
  \hat{\hat{\bn}} &= \bm + z_2 \left( \frac{\nu}{\rho} \bm^{(2)} - \frac{r}{\rho} \bm^{(3)} \right)
    + z_4 \bm^{(4)} \\
    &\qquad - \half \frac{\lambda^2 r^2}{\nu^2}
      \left[ a^2 b^2 + \frac{(a^2 - b^2)^2 \sin^2 \theta \cos^2 \theta}{\Delta} \right] \bk, \\
  \hat{\hat{\bm}}^{(2)} &= \frac{\nu}{\rho} \bm^{(2)} - \frac{r}{\rho} \bm^{(3)} - z_2 \bk, \\
  \hat{\hat{\bm}}^{(3)} &= \frac{r}{\rho} \bm^{(2)} + \frac{\nu}{\rho} \bm^{(3)}, \qquad
  \hat{\hat{\bm}}^{(4)} = \bm^{(4)} - z_4 \bk.
  \end{aligned}
  \label{KS:Kerr-(A)dS:5D:ptframe}
\end{equation}

The optical matrix $L_{ij}$ of the five-dimensional Kerr--(A)dS metric
\eqref{KS:Kerr-(A)dS:5D:metric}--\eqref{KS:Kerr-(A)dS:5D:H} can be
straightforwardly expressed in the parallelly transported frame
\eqref{KS:Kerr-(A)dS:5D:ptframe} as
\begin{equation}
  L_{ij} = \begin{pmatrix} \frac{r}{\rho^2} & \frac{\nu}{\rho^2} & \phantom{a}0\phantom{a} \\
    -\frac{\nu}{\rho^2} & \frac{r}{\rho^2} & 0 \\
    0 & 0 & \frac{1}{r}
  \end{pmatrix},
  \label{KS:Kerr-(A)dS:5D:L}
\end{equation}
which is obviously of rank $m = 3$ and contains one 2 $\times$ 2 block, i.e.\
$p = 1$, where
\begin{equation}
  s_{(2)} = \frac{r}{r^2 + \nu^2}, \qquad
  A_{2, 3} = \frac{\nu}{r^2 + \nu^2}.
\end{equation}
One may compare this particular optical matrix $L_{ij}$
\eqref{KS:Kerr-(A)dS:5D:L} and the Kerr--Schild function $\mathcal{H}$
\eqref{KS:Kerr-(A)dS:5D:H} of the five-dimensional Kerr--(A)dS metric with the
corresponding quantities $L_{ij}$ \eqref{KS:expanding:Lij:blockdiagonal}
\begin{equation}
  L_{ij} = \begin{pmatrix} s_{(2)} & A_{2,3} & \phantom{ab}0\phantom{ab} \\
    - A_{2,3} & s_{(2)} & 0 \\
    0 & 0 & \frac{1}{r}
    \end{pmatrix}
  \label{KS:Kerr-(A)dS:5D:L:general}
\end{equation}
and $\mathcal{H}$ \eqref{KS:expanding:rdep:H}
\begin{equation}
  \mathcal{H} = \mathcal{H}_0 \frac{1}{r^2 + (a^0_{(2)})^2}
  \label{KS:Kerr-(A)dS:5D:H:general}
\end{equation}
of general expanding Einstein GKS spacetimes with the same parameters $n=5$,
$m=3$, $p=1$, where
\begin{equation}
  s_{(2)} = \frac{r}{r^2 + (a^0_{(2)})^2}, \qquad
  A_{2, 3} = \frac{a^0_{(2)}}{r^2 + (a^0_{(2)})^2}.
\end{equation}
The presented optical matrices \eqref{KS:Kerr-(A)dS:5D:L},
\eqref{KS:Kerr-(A)dS:5D:L:general} and the functions $\H$
\eqref{KS:Kerr-(A)dS:5D:H}, \eqref{KS:Kerr-(A)dS:5D:H:general} are in accordance
and obviously
\begin{equation}
  a^0_{(2)} = \nu \equiv \sqrt{a^2 \cos^2 \theta + b^2 \sin^2 \theta}, \qquad
  \mathcal{H}_0 = - M.
  \label{KS:Kerr-(A)dS:5D:a}
\end{equation}

Now we employ the results of section \ref{sec:KS:singularities} to investigate
the presence of curvature singularities at $r = 0$ in five-dimensional
Kerr--(A)dS spacetimes. According to number of non-zero rotation parameters $a$,
$b$ our discussion can be split into three distinct cases.

In the first case where both rotation parameters are non-zero $a \neq 0$,
$b \neq 0$, the optical matrix $L_{ij}$ has one $2 \times 2$ block, i.e.\
$p = 1$, and therefore $2p = m - 1$. This corresponds to the special cases
in section \ref{sec:KS:singularities} when a curvature singularity is
presented if $a^0_{(2)}$ \eqref{KS:Kerr-(A)dS:5D:a} admits a real root. Since
$a^2 \cos^2 \theta + b^2 \sin^2 \theta$ never vanish there is no curvature
singularity .

If we set one of the rotation parameters to zero, for instance $a \neq 0$,
$b = 0$, then the optical matrix $L_{ij}$ has still one $2 \times 2$ block and
therefore $2p = m - 1$. However, now $a^0_{(2)} = a \cos \theta$ has a real
root at $\theta = \frac{\pi}{2}$ and this corresponds to the ring-shaped
singularity known from the four-dimensional Kerr solution.

Whereas in the previous cases the presence of a curvature singularity depends
on the behaviour of the function $a^0_{(2)}$, in the last static case
$a = b = 0$, the Kerr--(A)dS metric reduces to the
(A)dS--Schwarzschild--Tangherlini black hole where the optical matrix $L_{ij}$
is obviously diagonal, i.e.\ $p = 0$. Since neither $2p = m - 1$ nor $2p = m$,
this corresponds to the ``generic'' case in section \ref{sec:KS:singularities}
where a curvature singularity is located at $r = 0$.

\subsection[\texorpdfstring{Expanding generalized Kerr--Schild spacetimes with null\\radiation}
{Expanding generalized Kerr--Schild spacetimes with null radiation}]
{Expanding generalized Kerr--Schild spacetimes with null radiation}
\label{sec:KS:expanding:nullrad}

In this section, we present two explicit examples of expanding GKS spacetimes
with a null radiation term in the Ricci tensor which is aligned with the
Kerr--Schild vector $\bk$ as in \eqref{KS:nullrad:Ricci}. Namely, the Kinnersley
photon rocket and the Vaidya shining star. Both solutions belong to the
Robinson--Trautman class of spacetimes admitting an expanding, non-shearing and
non-twisting geodetic null congruence.

First, let us point out the differences between four and higher dimensional
cases. In four dimensions, the Robinson--Trautman class contains solutions of
all algebraically special Petrov types II, D, III and N, see, e.g.,
\cite{StephaniKramer2003}. On the other hand, in higher dimensions, the form of
the optical matrix of spacetimes with non-zero expansion and vanishing shear and
twist $L_{ij} = \theta \delta_{ij}$ excludes Weyl types III and N. It follows
directly from the fact that all type N and non-twisting type III Einstein
spacetimes have the optical matrix of rank 2 which, in an appropriate frame,
takes the form $L_{ij} = \text{diag}(s,s,0,\ldots,0)$
\cite{OrtaggioPravdaPravdova2010} and therefore, omitting the Kundt class, these
types III and N spacetimes are non-shearing only in dimension $n = 4$. Indeed,
it was shown in \cite{PodolskyOrtaggio2006} that the higher dimensional
Robinson--Trautman class is not so rich as in the four-dimensional case and
contains only solutions of Weyl type D.

\subsubsection{Kinnersley photon rocket}

The Kinnersley photon rocket \cite{Kinnersley1969} is a four-dimensional
solution of the Einstein field equations with a null radiation term on the
right-hand side representing arbitrarily moving test particle in the Minkowski
background due to the back-reaction caused by emitted photons. The metric of
such spacetime admitting the KS form \eqref{KS:KSansatz} can be generalized by
adding a cosmological constant $\Lambda$ and is of Petrov type D
\cite{GriffithsPodolsky2009}. This suggests that it could have a higher
dimensional analogue.

Such a solution in arbitrary dimension was found as a special case in
\cite{GursesSarioglu2002}, where Kerr--Schild metrics with an acceleration due
to null radiation were studied in the context of the Einstein--Maxwell theory.
Subsequently, a cosmological constant $\Lambda$ was taken into account in
\cite{GursesSarioglu2004} and enters the metric via the Kerr--Schild function
$\H$
\begin{equation}
  \H = - \frac{m(u)}{r^{n-3}} - \frac{\Lambda}{(n-2)(n-1)} r^2, \\
\end{equation}
in the KS form \eqref{KS:KSansatz} with the flat background metric expressed in
the Cartesian-like coordinates as
\begin{equation}
  \eta_{ab} \, \d x^a \, \,d x^b = \frac{r^2}{P^2} \delta_{ij} \, \d x^i \, \d x^j
    - 2 \d u \, \d r
    - \left( 1 - 2r\left( \log P \right)_{,u} \right) \d u^2,
\end{equation}
where
\begin{equation}
  P = \left( \dot z^0 - \dot z^1 \right) - \delta_{ij} \dot z^j x^i
    + \frac{1}{4} \left( \dot z^0 + \dot z^1 \right) \delta_{ij} x^i x^j
\end{equation}
and the Kerr--Schild vector $\bk$ is given by
\begin{equation}
  k^a \, \partial_a = \partial_r.
\end{equation}
The limit $m = 0$ represents an (anti-)de Sitter spacetime. Therefore, the
metric can be cast to the GKS form \eqref{KS:GKSmetric} since we can split
$\H = \H_m + \H_\Lambda$ to the parts containing only $m(u)$ or $\Lambda$,
respectively, and then denote $\bar{g}_{ab} = \eta_{ab} - 2 \H_\Lambda k_a k_b$
which is obviously an (anti-)de Sitter background metric and where $\H_m k_a k_b$
now corresponds to the Kerr--Schild term.

The Ricci tensor takes the form \eqref{KS:nullrad:Ricci} with a null radiation
term aligned with the Kerr--Schild vector $\bk$, where $\Phi$ is given by
\begin{equation}
  \Phi = \frac{n - 2}{r^{n - 2}} \left( (n - 1) m \left( \log P \right)_{,u} - m_{,u} \right).
\end{equation}

As already mentioned above, this higher dimensional solution is of the Weyl type
D with an expanding, non-twisting and non-shearing null congruence corresponding
to the Kerr--Schild vector $\bk$. An arbitrary motion of a photon rocket can be
prescribed by an appropriate choice of the functions $z^a(u)$ defining a
timelike worldline in the flat background metric with $u$ being the proper time.
Such various trajectories and corresponding radiation patterns of emitted
photons and mass loss of the rocket were studied in \cite{Podolsky2010}.

Note that, for a rocket at rest, photons are radiated isotropically and the
Kinnersley metric reduces to the spherically symmetric Vaidya solution.

\subsubsection{Vaidya shining star}

The four-dimensional Vaidya solution originally found in \cite{Vaidya1943} is a
non-static generalization of the Schwarzschild metric representing a spherically
symmetric star losing mass which is carried away by emitted radiation. The
higher dimensional Vaidya metric obtained in the Ricci-flat case
\cite{IyerVishveshwara1989} and subsequently in the case of Einstein spaces
\cite{PatelDesai1997} can be transformed to the KS form \eqref{KS:KSansatz}
with the flat background metric as
\begin{equation}
  \begin{aligned}
  \d s^2 &= - \d t^2 + \d r^2 + r^2 \, \d \Omega^2_{(n-2)} \\
    &\qquad + 2 \left[ \frac{m(u)}{r^{n-3}}
    + \frac{\Lambda r^2}{(n-2)(n-1)} \right] \left( \d t - \d r \right)^2
  \end{aligned}
  \label{KS:Vaidya}
\end{equation}
and differs from the Schwarzschild--Tangherlini metric \eqref{KS:AdS-S-T:KS} in
the way that the mass $m(u)$ may vary and depends on the retarded time
$u = t - r$. This leads to the null radiation term
\begin{equation}
  \Phi = \frac{(n - 2)}{r^{n - 2}} m_{,u},
\end{equation}
appearing in the Ricci tensor of the form \eqref{KS:nullrad:Ricci}. The metric
\eqref{KS:Vaidya} can be transformed to the GKS form \eqref{KS:GKSmetric} with
an (A)dS background by introducing a new time coordinate $t'$
\begin{equation}
  t' = t + \int \frac{\lambda r^2}{1 - \lambda r^2} \, \d r,
\end{equation}
where $\lambda$ is given by the cosmological constant $\Lambda$ via
\eqref{KS:lambda}, or explicitly for a positive and negative cosmological
constant
\begin{equation}
  t' = \begin{cases}
    t - r + \frac{1}{\sqrt{\lambda}} \tanh^{-1} \sqrt{\lambda} r& \quad \lambda > 0, \\
    t - r + \frac{1}{\sqrt{-\lambda}} \tan^{-1} \sqrt{-\lambda} r& \quad \lambda < 0,
  \end{cases}
\end{equation}
respectively. Therefore, the cosmological constant $\Lambda$ moves from the
Kerr--Schild function $\H$ to the background metric and one thus obtains
\begin{equation}
  \begin{aligned}
  \d s^2 &= - (1 - \lambda r^2) \, \d t'^2
    + \frac{\d r^2}{1 - \lambda r^2}
    + r^2 \, \d \Omega^2_{(n-2)} \\
    &\qquad + \frac{2 m(u)}{r^{n-3}} \left( \d t' - \frac{\d r}{1 - \lambda r^2} \right)^2,
  \end{aligned}
\end{equation}
where the first three terms represent the corresponding (anti-)de Sitter
background metric in the spherical coordinates.

Note that some other generalizations of the Vaidya metric are known.
Four-dimensional charged Vaidya metrics in the electro-vacuum case were
investigated in \cite{IbohalKapil2010}. It was shown there that the
four-dimensional Reissner--Nordstr\"om--Vaidya metric is of Petrov type D and
the four-dimensional Kerr--Newman--Vaidya solution is of Petrov type II unlike
the Kerr--Newman black hole which is of type D. Vaidya solutions were studied
also in the Einstein--Maxwell theory with sources, for instance, a four-current
proportional to the Kerr--Schild vector $\bk$ was considered in
\cite{BhattVaidya1991}.

\subsection[\texorpdfstring{Warped expanding Einstein generalized Kerr--Schild\\spacetimes}
{Warped expanding Einstein generalized Kerr--Schild spacetimes}]
{Warped expanding Einstein generalized\\Kerr--Schild spacetimes}
\label{sec:KS:expanding:warp}

In section \ref{sec:KS:nonexpanding:warp}, we employed the Brinkmann warp
product, presented in section \ref{sec:intro:warp}, in order to generate
examples of higher dimensional non-expanding Einstein GKS metrics from known
four-dimensional Einstein Kundt metrics and higher dimensional VSI metrics. In
this way we obtained solutions with one extra dimension and this also allowed us
to introduce cosmological constant to Ricci-flat metrics.

Naturally, one may use this method also in the case of expanding Einstein GKS
spacetimes. We already know that the Brinkmann warp product
\eqref{intro:warp:warpedmetric} preserves the Weyl type of algebraically special
spacetimes. Furthermore, as we have shown in section \ref{sec:KS:warp}, it
preserves the GKS form as well. Therefore, starting with an $(n-1)$-dimensional
expanding Einstein GKS seed metric, which is of Weyl type II or D by proposition
\ref{KS:proposition:expanding}, we construct an $n$-dimensional Einstein GKS
metric of the same Weyl type as the seed. This implies that the warped metric
has non-zero expansion.

More precisely, the optical matrices of general seed and warped metrics in
parallelly propagated frames satisfy \cite{OrtaggioPravdaPravdova2010}
\begin{equation}
  L_{ij} = \tilde L_{ij}, \qquad
  L_{i, n-1} = L_{n-1, j} = L_{n-1, n-1} = 0,
\end{equation}
where $i, j = 2, \ldots, n - 2$. Obviously, the optical matrix $L_{ij}$ of the
warped metric is degenerate and the corresponding optical scalars read
\begin{equation}
  \theta = \frac{n-3}{n-2} \tilde \theta, \qquad
  \sigma^2 = \tilde \sigma^2 + \frac{n-3}{n-2} \tilde \theta^2, \qquad
  \omega^2 = \tilde \omega^2.
\end{equation}
The zeros in the last row and column of the optical matrix $L_{ij}$ lead to the
fact that an expanding warped metric is shearing even if the seed metric is
shear-free. Thus, for instance, metrics belonging to the Robinson--Trautman
class do not remain within this class after application of the Brinkmann warp
product.

The simplest example of a warped expanding GKS metric is an anti-de Sitter
spacetime which has a non-trivial Kerr--Schild term with expanding $\bk$ if it
is rewritten in the KS form as in \eqref{KS:AdS:KS}. It can be obtained by using
\eqref{intro:warp:kk-0}, i.e.\ the case $\tilde R = 0$, $R < 0$, and taking the
Minkowski spacetime as a seed $\d \tilde s^2$. Then the warp product leads to
the metric expressed in the form
\begin{equation}
  \d s^2 = \d y^2 + e^{2\sqrt{-\lambda} y} \, \eta_{ab} \, \d x^a \, \d x^b.
\end{equation}
In fact, these coordinates cover only a part of the complete AdS spacetime and
a similar five-dimensional metric was already considered to solve the hierarchy
problem in the Randall--Sundrum braneworld scenario \cite{RandallSundrum1999a,
RandallSundrum1999b}
\begin{equation}
  \d s^2 = \d y^2 + e^{-2\sqrt{-\lambda} |y|} \, \eta_{ab} \, \d x^a \, \d x^b,
  \label{KS:Randall-Sundrum}
\end{equation}
where the absolute value causes a jump discontinuity in the first derivatives of
the metric leading subsequently to a delta function in the second derivatives.
Effectively, there appears a term proportional to $\delta(y)$ in the Ricci tensor
corresponding to an energy--momentum tensor lying on the brane $y=0$.



Of course, one may choose any Einstein spacetime as a seed in
\eqref{KS:Randall-Sundrum} instead of the flat Minkowski metric $\eta_{ab}$.
For instance, substituting the Schwarzschild metric as a slice $\d \tilde s^2$,
we obtain the Chamblin--Hawking--Reall black hole on a brane
\begin{equation}
  \d s^2 = \d y^2 + e^{-2\sqrt{-\lambda} |y|} \, \d \tilde s^2,
\end{equation}
discussed in \cite{ChamblinHawkingReall1999}.

In general, choosing expanding Einstein GKS metrics representing black holes as
a seed, one obtains black string solutions. It can be easily seen directly from
the form of the warp product with a flat extra dimension
\eqref{intro:warp:kk++}--\eqref{intro:warp:kk-+} that an $(n-1)$-dimensional
slice $\d \tilde s^2$ of the full $n$-dimensional metric $\d s^2$ is multiplied
by a conformal factor depending on the extra dimension coordinate $y$ and
therefore these black strings are in general non-homogeneous. They are
homogeneous only in the case of direct product $\tilde R = 0, R = 0$.

Thus, one may generate various black string solutions by warping appropriate
black holes, for instance, taking the Kerr--(A)dS metric as a seed, we get
spinning black string. Nevertheless, let us remind that these warped spacetimes
suffer from naked singularity at points where the warp factor vanishes, i.e.
$f(z) = 0$, except the cases $\tilde R = 0, R = 0$ and $\tilde R < 0, R < 0$
when $f(z)$ does not admit a root.

Here we present only an example of the latter case, i.e.\ $n$-dimensional black
string with a negative cosmological constant $\Lambda$ sliced by an
$(n - 1)$-dimensional Kerr--AdS metric
\eqref{KS:Kerr-(A)dS:metric:1}--\eqref{KS:Kerr-(A)dS:metric:2}. Using the form
conformal to a direct product \eqref{intro:warp:conformal--}, the warped metric
can be expressed in the GKS form \eqref{KS:GKSmetric} with the background metric
$\bar{g}_{ab}$, Kerr--Schild vector $\bk$ and function $\H$ given by
\begin{align}
  \bar g_{ab} \, \d x^a \, \d x^b &= \frac{1}{\cos^2 (\sqrt{-\lambda} x)} \Bigg[ \d x^2
    - W (1 - \lambda r^2) \, \d t^2
    + F \, \d r^2 \notag \\
    &\qquad + \sum^{\lfloor (n-1)/2 \rfloor}_{i=1} \frac{r^2 + a_i^2}{1 + \lambda a_i^2} \, \d \mu_i^2
    + \sum^{\lfloor n/2 - 1 \rfloor}_{i=1} \frac{r^2 + a_i^2}{1 + \lambda a_i^2} \mu_i^2 \, \d \phi_i^2 \notag \\
    &\qquad + \frac{\lambda}{W (1 - \lambda r^2)} \left( \sum^{\lfloor (n-1)/2 \rfloor}_{i=1} \frac{(r^2 + a_i^2) \mu_i \, \d \mu_i}{1 + \lambda a_i^2} \right)^2 \Bigg], \\
  \label{KS:Kerr-AdS:warped:1}
  k_a \, \d x^a &= \frac{1}{\cos (\sqrt{-\lambda} x)} \Bigg[ W \, \d t
    + F \, \d r
    - \sum^{\lfloor n/2 - 1 \rfloor}_{i=1} \frac{a_i \mu_i^2}{1 + \lambda a_i^2} \, \d \phi_i \Bigg], \\
  \H &= \begin{cases}
    - \frac{M}{r \sum^{\lfloor (n-1)/2 \rfloor}_{i=1} \frac{\mu_i^2}{r^2 + a_i^2} \prod^{\lfloor n/2 -1 \rfloor}_{j=1} (r^2 + a_j^2)}
      & \text{if $n$ is even,} \\
    - \frac{M}{\sum^{\lfloor (n-1)/2 \rfloor}_{i=1} \frac{\mu_i^2}{r^2 + a_i^2} \prod^{\lfloor (n-1)/2 \rfloor}_{j=1} (r^2 + a_j^2)}
      & \text{if $n$ is odd,}
  \end{cases}
  \label{KS:Kerr-AdS:warped:2}
\end{align}
where the functions $W$ and $F$ are defined as
\begin{equation}
  W = \sum^{\lfloor (n-1)/2 \rfloor}_{i=1} \frac{\mu_i^2}{1 + \lambda a_i^2}, \qquad
  F = \frac{r^2}{1 - \lambda r^2} \sum^{\lfloor (n-1)/2 \rfloor}_{i=1} \frac{\mu_i^2}{r^2 + a_i^2},
\end{equation}
$\lambda < 0$ is related to the cosmological constant $\Lambda$ via
\eqref{KS:lambda} and the coordinate $x$ is subject to
$- \frac{\pi}{2} < \sqrt{-\lambda} x < \frac{\pi}{2}$. In the case $n = 5$, such
metrics have been already constructed in \cite{OrtaggioPravdaPravdova2010b}.

Let us remark that if we set $\lambda = 0$, the black string metric
\eqref{KS:Kerr-AdS:warped:1}--\eqref{KS:Kerr-AdS:warped:2} reduces to the
direct product of an $(n - 1)$-dimensional Myers--Perry black hole with a flat
extra dimension since, in this limit, the form of a warped metric
\eqref{intro:warp:conformal--} for $\tilde R < 0, R < 0$ smoothly transforms to
the form \eqref{intro:warp:conformal00} for $\tilde R = 0, R = 0$ and again a
naked singularity is not present.

\section[Other generalizations of the Kerr--Schild ansatz]{Other generalizations of the Kerr--Schild\\ansatz}

Apart from the GKS metric \eqref{KS:GKSmetric}, there are other possible
generalizations of the original Kerr--Schild ansatz \eqref{KS:KSansatz} that
also lead to exact solutions. One of such examples, referred to as the extended
Kerr--Schild ansatz, was proposed in \cite{AlievCiftci2008}. In fact, it was
shown that the CCLP solution \cite{ChongCveticLuPope2005} representing charged
rotating black holes in five-dimensional minimal gauged supergravity previously
found by the trial and error method can be put into this form. Besides an
(anti-)de Sitter background metric and the null Kerr--Schild congruence
appearing in the GKS form, this ansatz also consists of an additional spacelike
vector field and will be studied in more detail in chapter \ref{sec:xKS}.

Another extension of the Kerr--Schild ansatz was also discovered by rewriting an
already known solution. If one performs a Wick rotation of the higher
dimensional Kerr--NUT--(A)dS spacetime \cite{ChenLuPope2006} in order to change
the signature $(1, n-1)$ of the metric either to the signature $(k, k+1)$ in odd
dimension $n = 2k + 1$ or to the signature $(k,k)$ in even dimension $n = 2k$,
respectively, then the resulting metric can be cast to the multi-Kerr--Schild
form \cite{ChenLu2007}
\begin{equation}
  g_{ab} = \bar{g}_{ab} - 2 \sum^k_{\mu=1} \H_{\mu} k^{(\mu)}_a k^{(\mu)}_b.
\end{equation}
In odd dimensions, the (A)dS background metric $\bar{g}_{ab}$, the Kerr--Schild
vectors $\bk^{(\mu)}$ and the functions $\H_\mu$ of the Kerr--NUT--(A)dS
spacetime are given by
\begin{align}
  \bar{g}_{ab} \, \d x^a \, \d x^b &=  \frac{W}{\prod^n_{i=1} \Xi_i} \, \d\tau^2
    + \sum_{\mu=1}^{n} \frac{U_\mu}{\bar X_\mu} \, \d x_\mu^2
    - \sum_{i=1}^n \frac{\gamma_i}{\Xi_i \prod_{j \neq i} (a_i^2 - a_j^2)} \, \d\varphi_i^2, \\
  k^{(\mu)}_a \, \d x^a &= \frac{W}{1 + \lambda x_\mu^2} \frac{\d\tau}{\prod^k_{i=1} \Xi_i}
    - \frac{U_{\mu} \, \d x_{\mu}}{\bar X_{\mu}} \notag \\
    &\qquad - \sum_{i = 1}^k \frac{a_i \gamma_i \, \d\varphi_i}{(a_i^2 - x_\mu^2) \Xi_i \prod_{j \neq i} (a_i^2 - a_j^2)},\\
  \H_\mu &= - \frac{b_\mu}{U_\mu}, \qquad
  \bar X_\mu = \frac{(1 + \lambda x_\mu^2)}{x_\mu^2} \prod_{i-1}^k (a_i^2 - x_\mu^2),
\end{align}
whereas, in even dimensions, they can be expressed as
\begin{align}
  \bar{g}_{ab} \, \d x^a \, \d x^b &=  \frac{W}{\prod^{k-1}_{i=1} \Xi_i} \, \d\tau^2
    + \sum_{\mu=1}^{k} \frac{U_\mu}{\bar X_\mu } \, \d x_\mu^2
    - \sum_{i=1}^{k-1} \frac{\gamma_i}{a_i^2 \Xi_i \prod_{j \neq i} (a_i^2 - a_j^2)} \, \d\varphi_i^2, \\
  k^{(\mu)}_a \, \d x^a &= \frac{W}{1 + \lambda x_\mu^2} \frac{\d\tau}{\prod^{k-1}_{i=1} \Xi_i}
    - \frac{U_{\mu} \, \d x_{\mu}}{\bar X_\mu} \notag \\
    &\qquad - \sum_{i=1}^{k-1} \frac{\gamma_i \, \d\varphi_i}{(a_i^2 - x_\mu^2) a_i \Xi_i \prod_{j \neq i} (a_i^2 - a_j^2)},\\
  \H_\mu &= - \frac{b_\mu x_\mu}{U_\mu}, \qquad
  \bar X_\mu = - (1 + \lambda x_\mu^2) \prod_{i=1}^{k-1}(a_i^2-x_\mu^2),
\end{align}
where $\lambda$ is given as in \eqref{KS:lambda}, $b_\mu$ corresponds to the
mass and NUT parameters and $\Xi_i$, $\gamma_i$, $W$, $U_\mu$ are defined as
\begin{equation}
  \begin{aligned}
    \Xi_i &= 1 + \lambda a_i^2,& \qquad
    \gamma_i &= \prod_{\nu=1}^k (a_i^2 - x_\nu^2), \\
    W &= \prod_{\nu=1}^k (1 + \lambda x_\nu^2),& \qquad
    U_\mu &= \prod_{\nu \neq \mu} (x_\nu^2 - x_\mu^2).
  \end{aligned}
\end{equation}
Each of the scalar functions $\H_{\mu}$ contains just one of the mass or NUT
parameters which appears in $\H_{\mu}$ linearly. Therefore, the NUT parameters
enter the metric of this form in an analogous way as the mass parameter. The
vectors $\bk^{(\mu)}$ are linearly independent, mutually orthogonal and tangent
to the corresponding affinely parametrized geodetic null congruences.

The last presented example of possible generalizations of the Kerr--Schild
ansatz has appeared recently in \cite{Wu2011} where the higher dimensional
charged rotating Kaluza--Klein AdS black hole has been obtained in the form
\begin{equation}
  g_{ab} = H^{\frac{1}{D-2}} \left( \bar{g}_{ab} + \frac{2m}{UH} k_a k_b \right), \qquad
  A_a = \frac{2ms}{UH} k_a, \qquad
  \Phi = - \frac{\ln(H)}{D-2},
\end{equation}
as a solution of the Einstein--Maxwell--dilaton theory with the Lagrangian
\begin{align}
  \mathcal{L} &= \sqrt{-g} \bigg( R - \frac{1}{4} (n-1)(n-2) (\partial\Phi)^2
    -\frac{1}{4} e^{-(n-1) \Phi} \mathcal{F}^2 \notag \\
  &\qquad + g^2 (n - 1) \left( (n - 3) e^{\Phi} + e^{-(n - 3) \Phi} \right) \bigg).
  \label{KS:KKAdS}
\end{align}
The background metric $\bar{g}_{ab}$ represents again an AdS spacetime, but now
$\bk$ is a timelike vector field
\begin{align}
  \bar{g}_{ab} \, \d x^a \, \d x^b &= - \left( 1 + g^2 r^2 \right) W \, \d\bar{t}^2
    + \sum_{i=1}^{N + \epsilon} \frac{r^2 + a_i^2}{\chi_i} \, \d\mu_i^2
    + \sum_{i=1}^N \frac{r^2 + a_i^2}{\chi_i} \mu_i^2 \, \d\bar{\phi}_i^2 \notag \\
    &\qquad + F \, \d r^2 
    - \frac{g^2}{\left(1 + g^2 r^2 \right) W}
      \left( \sum_{i=1}^{N + \epsilon} \frac{r^2 + a_i^2}{\chi_i} \mu_i \, \d\mu_i \right)^2, \\
  k_a \, \d x^a &= c W\, \d\bar{t}
    +\sqrt{f(r)} F \, \d r
    - \sum_{i=1}^N \frac{a_i \sqrt{\Xi_i}}{\chi_i} \mu_i^2 \, \d\bar{\phi}_i,
\end{align}
where the functions $H$, $U$, $W$, $F$, $f(r)$, $\Xi_i$ and $\chi_i$ are defined as
\begin{equation}
  \begin{aligned}
  H &= 1 + \frac{2m s^2}{U}, \qquad
  U = r^\epsilon \sum_{i=1}^{N + \epsilon} \frac{\mu_i^2}{r^2 + a_i^2} \prod_{j=1}^N \left(r^2 + a_j^2 \right), \\
  F &= \frac{r^2}{1 + g^2 r^2} \sum_{i=1}^{N + \epsilon} \frac{\mu_i^2}{r^2 + a_i^2}, \qquad
  f(r) = c^2 - s^2 \left(1 + g^2 r^2 \right), \\
  W &= \sum_{i=1}^{N + \epsilon} \frac{\mu_i^2}{\chi_i}, \qquad
  \Xi_i = c^2 - s^2 \chi_i, \qquad
  \chi_i = 1 - g^2 a_i^2
  \end{aligned}
\end{equation}
and $c \equiv \cosh\delta$, $s \equiv \sinh\delta$ with $\delta$ being a charge.
The timelike vector field $\bk$ is geodetic and its norm with respect to the
background metric depends on the charge $\bar{g}_{ab} k^a k^b = - s^2$.
Note that in the uncharged case, the metric \eqref{KS:KKAdS} reduces to the
GKS form \eqref{KS:GKSmetric} and the vector $\bk$ becomes null.

\chapter{Extended Kerr--Schild spacetimes}
\label{sec:xKS}

The aim of this chapter is to investigate general properties of an extension of
the original Kerr--Schild ansatz, referred to as the extended Kerr--Schild
ansatz, in a similar way as it is performed for GKS spacetimes in chapter
\ref{sec:KS}. Later, we may find the results of such analysis useful in finding
new exact solutions in this form. In fact, this extension has been already
studied in \cite{EttKastor2010} using the method of perturbative expansion. It
has been shown that the vacuum field equations truncate beyond a certain low
order in an expansion around the background metric similarly as in the case of
GKS spacetimes. Here we employ the Newman--Penrose formalism which allows us to
formulate some statements about geodesicity and optical properties of the null
congruence and Weyl types of extended Kerr--Schild spacetimes. However, these
results have not been published yet since the analysis is not completed and some
aspects can be further investigated.

We define the extended Kerr--Schild ansatz (xKS) as a metric of the form
\begin{equation}
  g_{ab} = \bar{g}_{ab} - 2 \H k_a k_b - 2 \K k_{(a} m_{b)},
  \label{xKS:metric}
\end{equation}
where the background metric $\bar{g}_{ab}$ represents an (anti-)de Sitter or
Minkowski spacetime, $\H$ and $\K$ are scalar functions, $\bk$ is a null vector
and $\bm$ is a spacelike unit vector both with respect to the full metric
\begin{equation}
  k^a k_a \equiv g_{ab} k^a k^b = 0, \qquad
  k^a m_a \equiv g_{ab} k^a m^b = 0, \qquad
  m^a m_a \equiv g_{ab} m^a m^b = 1.
\end{equation}
From the form of the xKS metric \eqref{xKS:metric} it then follows that the same
holds also with respect to the background metric
\begin{equation}
  \bar{g}_{ab} k^a k^b = 0, \qquad
  \bar{g}_{ab} k^a m^b = 0, \qquad
  \bar{g}_{ab} m^a m^b = 1
\end{equation}
and the inverse metric is simply given by
\begin{equation}
  g^{ab} = \bar{g}^{ab} + \left( 2 \H - \K^2 \right) k^a k^b + 2 \K k^{(a} m^{b)}.
  \label{xKS:inversemetric}
\end{equation}
Note that our definition of the xKS ansatz \eqref{xKS:metric} slightly differs
from those ones in \cite{AlievCiftci2008,EttKastor2010} by the factors since we
follow the notation of the GKS form \eqref{KS:GKSmetric} and also by the fact
that we consider the spacelike vector $\bm$ to be normalized to unity since we
will identify it with one of the frame vectors.

First, let us provide a motivation for studying xKS spacetimes. As will be shown
in section \ref{sec:xKS:geodetic}, one of the reasons why to consider xKS ansatz
\eqref{xKS:metric} is that such metrics cover more general algebraic types than
the GKS metrics \eqref{KS:GKSmetric}. Recall the results of chapter \ref{sec:KS}
that GKS spacetimes with a geodetic Kerr--Schild vector $\bk$ without any
further assumptions are of Weyl type II or more special. Expanding Einstein GKS
spacetimes are compatible only with types II or D unless conformally
flat and non-expanding Einstein GKS spacetimes are of type N and belong to the
Kundt class. Therefore, expanding Einstein xKS spacetimes could include black
hole solutions of a more general Weyl type than II, for instance black rings
\cite{EmparanReall2001} that are of type I$_i$ \cite{PravdaPravdova2005}.
Non-expanding Einstein xKS spacetimes could contain Kundt metrics of a more
general type than N.

Furthermore, unlike the static charged black hole, rotating charged black hole
as an exact solution of higher dimensional Einstein--Maxwell theory is unknown.
The four-dimensional Kerr--Newman black hole can be cast to the KS form
\eqref{KS:KSansatz} with the background metric $\eta_{ab}$, the Kerr--Schild
vector $\bk$ and the function $\H$ given by
\begin{equation}
  \begin{aligned}
  &\eta_{ab} \, \d x^a \, \d x^b = -\d t^2 + \d x^2 + \d y^2 + \d z^2, \phantom{\half} \\
  &k_a \, \d x^a = \d t
    + \frac{rx+ay}{r^2 + a^2} \, \d x
    + \frac{ry-ax}{r^2 + a^2} \, \d y
    + \frac{z}{r} \, \d z, \\
  &\mathcal{H} = - \frac{r^2}{r^4 + a^2 z^2} \left( Mr - \frac{Q^2}{2} \right),
  \end{aligned}
\end{equation}
where $M$, $Q$ are mass and charge of the black hole, respectively, the
coordinate $r$ satisfy
\begin{equation} 
  \frac{x^2+y^2}{r^2 + a^2} + \frac{z^2}{r^2} = 1
\end{equation}
and the vector potential is proportional to the Kerr--Schild vector $\bk$
\begin{equation} 
  A = \frac{Qr^3}{r^4 + a^2z^2} \bk.
\end{equation}
However, the attempt to generalize this ansatz to higher dimensions using the KS
form of the Myers--Perry black hole has failed \cite{MyersPerry1986}. It has
turned out that a vector potential proportional to $\bk$ cannot simultaneously
satisfy the corresponding Einstein and Maxwell field equations.
An open question is whether the xKS ansatz may resolve these problems.

Another significant reason is that some of the known exact solutions can be cast
to the xKS form. For instance, VSI metrics are examples of such xKS spacetimes
as will be pointed out in section \ref{sec:xKS:Kundt}. Some expanding xKS
spacetimes representing black holes are also known, namely CCLP solution
\cite{ChongCveticLuPope2005} discussed in section \ref{sec:xKS:CCLP}. In fact,
the investigation of the CCLP metric has led to the xKS ansatz introduced in
\cite{AlievCiftci2008}.

Lastly, the xKS metric seems to be still sufficiently simple to be treated
analytically. As in the case of GKS spacetimes in chapter \ref{sec:KS}, we
assume that the background metric $\bar{g}_{ab}$ representing an (anti-)de
Sitter or Minkowski spacetime takes the conformally flat form
\eqref{KS:bgmetric} with a conformal factor \eqref{KS:Omega(A)dS} and with the
flat Minkowski metric in Cartesian coordinates \eqref{KS:flatmetric}. Note that
throughout this chapter we do not discuss the limit $\K = 0$ when the xKS ansatz
\eqref{xKS:metric} reduces to the GKS metric \eqref{KS:GKSmetric} studied in
chapter \ref{sec:KS}.

\section{General Kerr--Schild vector field}
\label{sec:xKS:general}

As in the case of the GKS ansatz, it can be seen directly from the form of the
xKS metric \eqref{xKS:metric} and its inverse \eqref{xKS:inversemetric} that the
index of the Kerr--Schild vector $\bk$ can be raised and lowered by both the
full metric $g_{ab}$ and the background metric $\bar{g}_{ab}$. The form of the
xKS ansatz also implies that it is convenient to employ the higher dimensional
Newman--Penrose formalism reviewed in section \ref{sec:intro:NP} and identify
the null and spacelike vectors $\bk$, $\bm$ appearing in the xKS metric
\eqref{xKS:metric} with the vectors $\bl$, $\bm^{(2)}$ of the null frame
\eqref{intro:frame:constraints}, respectively. The corresponding Ricci rotation
coefficients will be denoted as $L_{ab}$ and $M_{ab} \equiv \M{2}{ab}$.

Despite the fact that the xKS metric \eqref{xKS:metric} has only few additional
terms with respect to the GKS metric \eqref{KS:GKSmetric}, the Christoffel
symbols contain several times more terms and it is convenient or in fact
necessary to use the computer algebra system \textsc{Cadabra} \cite{Peeters2006,
Peeters2007}. Although the calculations are much more involved than in the case
of GKS spacetimes the boost weight 2 component of the Ricci tensor
$R_{00} = R_{ab} k^a k^b$ contains again many vanishing contractions with the
null Kerr--Schild vector $\bk$ leading to the simple expression
\begin{equation}
  \begin{aligned}
  R_{00} &= 2 \H L_{i0} L_{i0}
  	- \half \K^2 L_{\itilde 0} L_{\itilde 0}
	+ 2 \K L_{i(i} L_{2)0}
	+ \K L_{\itilde 0} M_{\itilde 0}
	+ 2 \D\K L_{20} \\
	&\qquad + \K \D L_{20}
  	- \half (n - 2) \left( \frac{\Omega_{,ab}}{\Omega}
  	- \frac{3}{2} \frac{\Omega_{,a} \Omega_{,b}}{\Omega^2} \right) k^a k^b,
  \end{aligned}
  \label{xKS:R00:general}
\end{equation}
where apart from the indices $i, j, \ldots = 2, \ldots, n - 1$ employed
throughout the thesis, we define new indices $\itilde, \jtilde, \ldots = 3,
\ldots, n - 1$ denoted by tilde such that the vector $\bm$ is excluded in the
notation $\bm^{(\itilde)}$.

Since the conformal factor $\Omega$ of an (anti-)de Sitter background satisfies
\eqref{KS:bgEFEs} or $\Omega = 1$ in the case of Minkowski background, the last
term in \eqref{xKS:R00:general} is identically zero. Moreover, the component
$R_{00}$ completely vanishes if $L_{i0} = 0$ and therefore
\begin{proposition}
  \label{xKS:proposition:geodetic_k}
  If the Kerr--Schild vector $\bk$ in the extended Kerr--Schild metric
  \eqref{xKS:metric} is geodetic then the boost weight 2 component of the
  energy--momentum tensor $T_{00} = T_{ab} k^a k^b$ vanishes.
\end{proposition}
Note that if $\K = 0$, the xKS metric \eqref{xKS:metric} reduces to the GKS form
\eqref{KS:GKSmetric} and the implication of proposition
\ref{xKS:proposition:geodetic_k} also holds in the opposite direction as follows
from proposition \ref{KS:proposition:geodetic_k}. However, in the case of xKS
spacetimes, we obtain only a sufficient condition and it is possible to set
$R_{00}$ to zero also for non-geodetic $\bk$ by a special choice of $\H$, $\K$
and $\bm$.

Inspired by the relation \eqref{xKS:CCLP:k_m} valid in CCLP spacetimes, we can
formulate an additional covariant condition in order to restrict the geometry
of the vectors $\bk$ and $\bm$ so that the implication of proposition
\ref{xKS:proposition:geodetic_k} becomes an equivalence. It turns out that such
an convenient condition is $\mathcal{L}_{\bk} m_a \propto m_a$,
$\mathcal{L}_{\bm} k_a = 0$ since then
$L_{20} = 0$, $M_{\itilde 0} = - L_{2 \itilde} = - L_{\itilde 2}$ as will be
shown in section \ref{sec:xKS:relation_k_m} and therefore all terms apart from
the first two in \eqref{xKS:R00:general} vanish
\begin{equation}
  R_{00} = \left( 2\H - \half \K^2 \right) L_{\itilde 0} L_{\itilde 0},
  \label{xKS:R00:special}
\end{equation}
which ensures that the implication holds in both directions and we can conclude
that 
\begin{corollary}
  \label{xKS:proposition:geodetic_k:2}
  In the special case when
  $\mathcal{L}_{\bk} m_a \propto m_a$ and $\mathcal{L}_{\bm} k_a = 0$, the
  Kerr--Schild vector $\bk$ in the extended Kerr--Schild metric
  \eqref{xKS:metric} is geodetic if and only if the boost weight 2 component of
  the energy--momentum tensor $T_{00} = T_{ab} k^a k^b$ vanishes.
\end{corollary}

\subsection{Kerr--Schild congruence in the background}

Following section \ref{sec:KS:riccicoeff}, the full metric $g_{ab}$ can be
expressed in the null frame \eqref{intro:frame:constraints} simply as
\begin{equation}
  g_{ab} = 2 k_{(a} n_{b)} + m_a m_b
    + \delta_{\tilde{\imath}\tilde{\jmath}} m_a^{(\tilde{\imath})} m_b^{(\tilde{\jmath})}
\end{equation}
and can be compared with the xKS form \eqref{xKS:metric}. Obviously, if one
chooses
\begin{equation}
  \bar{n}_a = n_a + \H k_a + \K m_a
  \label{xKS:background_n}
\end{equation}
then $\bar{\bn}$, $\bk$, $\bm$, $\bm^{(\tilde{\imath})}$ form a null frame in
the background metric $\bar{g}_{ab}$. Although the indices of the vectors $\bk$
and $\bm^{(\itilde)}$ can be raised and lowered by both the full metric $g_{ab}$
and the background metric $\bar{g}_{ab}$ one has to operate with the vectors
$\bar{\bn}$ and $\bm$ carefully since
\begin{equation}
  \bar{n}^a \equiv g^{ab} \bar{n}_b = n^a + \H k^a + \K m^a, \qquad
  m^a \equiv g^{ab} m_a,
\end{equation}
whereas
\begin{equation}
  \bar{g}^{ab} \bar{n}_a = n^b - \H k^b, \qquad
  \bar{g}^{ab} m_a = m^b - \K k^b.
  \label{xKS:m:raised_in_bg}
\end{equation}
Similarly as for GKS spacetimes in section \ref{sec:KS:riccicoeff}, the
covariant derivative compatible with the background metric $\bar{g}_{ab}$ can
be easily obtained from the covariant derivative compatible with the full metric
$g_{ab}$ by setting $\H = \K = 0$. Thus, we can compare the Ricci rotation
coefficients constructed in the full spacetime using the frame $\bk$, $\bn$,
$\bm$, $\bm^{(\itilde)}$ with those ones in the background spacetime denoted by
barred letters and expressed in terms of the frame $\bk$, $\bar{\bn}$, $\bm$,
$\bm^{(\itilde)}$
\begin{align}
  L_{i0} &= \bar L_{i0}, \qquad
  L_{10} = \bar L_{10}, \qquad
  L_{ij} = \bar L_{ij} - \K \delta_{2[i} \bar L_{j]0}, \phantom{\half}
  \label{xKS:Riccicoeff:1} \\
  L_{1i} &= \bar L_{1i}
    - \H \bar L_{i0}
    - \half \K (\bar L_{2i} + \bar M_{i0})
    + \half \K \left( \bar L_{10}
    - \frac{\D\K}{\K}
    + \frac{\D \Omega}{\Omega} \right) \delta_{2i}, \\
  L_{i1} &= \bar L_{i1}
    - \half \K (\bar L_{2i} + \bar M_{i0})
    - \half \K \left( \bar L_{10}
    + \frac{\D\K}{\K}
    - \frac{\D \Omega}{\Omega} \right) \delta_{2i}, \\
  L_{11} &= \bar L_{11}
    - \H \bar L_{10}
    - \D\H
    + \H \frac{\D\Omega}{\Omega}
    - \K (L_{21} - N_{20}), \\
  M_{\itilde 0} &= \bar M_{\itilde 0} + \half \K \bar L_{\itilde 0}, \qquad
  M_{\itilde \jtilde} = \bar M_{\itilde \jtilde}
    + \K \bar S_{\itilde \jtilde}
    - \half \K \frac{\D\Omega}{\Omega} \delta_{\itilde \jtilde}, \\
  M_{\itilde 2} &= \bar M_{\itilde 2}
    + \K (\bar L_{2 \itilde} + \bar M_{\itilde 0}), \phantom{\half} \\
  M_{\itilde 1} &= \bar M_{\itilde 1}
    - \H \bar L_{\itilde 2}
    + (\H - \half \K) (\bar L_{2 \itilde} + \bar M_{\itilde 0})
    - \half \H \K \bar L_{\itilde 0}
    + \K \bar L_{(1 \itilde)} \notag \\
    &\qquad - \half \K \bar M_{\itilde 2}
    + \half \delta_{\itilde} \K, \\
  N_{i0} &= \bar N_{i0}
    - \half \K (\bar L_{2i} + \bar M_{i0})
    - \half \left( \K \bar L_{10}
    + \D\K \right) \delta_{2i}, \\
  N_{21} &= \bar N_{21}
    + \delta_2 \H
    + 2 \H \bar L_{12}
    - \H \K \bar L_{22}
    - \H^2 \bar L_{20}
    + \H \bar N_{20}
    - \H \bar L_{21} \phantom{\half} \notag \\
    &\qquad - \K \bar N_{22}
    + \half \K (\H + \half \K^2) \frac{\D\Omega}{\Omega}
    - \half \K \frac{\T\Omega}{\Omega}
    + \frac{3}{2} \K^2 \frac{\delta_2 \Omega}{\Omega}
    + \H \K \bar L_{10} \notag \\
    &\qquad+ \K^2 (\bar L_{21} - \bar N_{20})
    - \T \K, \phantom{\half} \\
  N_{i1} &= \bar N_{i1}
    + \delta_i \H
    + 2 \H \bar L_{1i}
    - \H \K (\bar L_{2i} + \bar M_{i0})
    - \H^2 \bar L_{i0}
    + \H \bar N_{i0}
    - \H \bar L_{i1} \phantom{\half} \notag \\
    &\qquad - \K \bar M_{i1} 
    - \K \bar N_{2i}
    + \half \K^2 (1 - \K) (\bar L_{2i} + \bar M_{i0}), \\
  N_{i2} &= \bar N_{i2}
    + \H \bar L_{2i}
    - \half \H \K \bar L_{i0}
    + \half \delta_i \K
    + \K \bar L_{[1i]}
    - \half \K^2 (\bar L_{2i} + \bar M_{i0}) \notag \\
    &\qquad - \half \K \bar M_{i2}
    + \K \bar N_{i0}, \\
  N_{22} &= \bar N_{22}
    - \half \K \frac{\delta_2 \Omega}{\Omega}
    + \H \bar L_{22}
    - (\H + \half \K^2) \frac{\D\Omega}{\Omega}
    - \K (\bar L_{21} - \bar N_{20}), \\
  N_{2i} &= \bar N_{2i}
    + \H \bar L_{i2}
    + \half \H \K \bar L_{i0}
    - \half \delta_i \K
    - \K \bar L_{(1i)}
    + \half \K^2 (\bar L_{2i} + \bar M_{i0}) \notag \\
    &\qquad + \half \K \bar M_{i2}, \\
  N_{ij} &= \bar N_{ij}
    + \H \bar L_{ji}
    - \half \K \bar M_{ij}
    + \half \K \bar M_{ji}
    - \H \frac{\D\Omega}{\Omega} \delta_{ij}
    - \half \K \frac{\delta_2 \Omega}{\Omega} \delta_{ij}, \\
  \M{\itilde}{\jtilde 0} &= \bM{\itilde}{\jtilde 0}, \qquad
  \M{\itilde}{\jtilde k} = \bM{\itilde}{\jtilde k}
    + \K \left( \bar L_{[ij]} + \bM{\itilde}{\jtilde 0} \right) \delta_{2k}, \\
  \M{\itilde}{\jtilde 1} &= \bM{\itilde}{\jtilde 1}
    + 2 \H \bar L_{[\itilde \jtilde]}
    + \H \bM{\itilde}{\jtilde 0}
    + \K \bar M_{[\itilde \jtilde]}. \phantom{\half}
  \label{xKS:Riccicoeff:2}
\end{align}
From \eqref{xKS:Riccicoeff:1} it follows that the Kerr--Schild vector $\bk$ is
geodetic with respect to the full spacetime $g_{ab}$ if and only if it is
geodetic with respect to the background spacetime $\bar{g}_{ab}$. Moreover, an
affine parametrization in the former spacetime corresponds to the affine
parametrization in the latter spacetime and vice versa. Note also that a
geodetic $\bk$ has the same optical properties in the full and background
spacetimes since the corresponding optical matrices are equal
$L_{ij} = \bar L_{ij}$.

\subsection{\texorpdfstring{Relation of the vector fields $\bk$ and $\bm$}{Relation of the vector fields k and m}}
\label{sec:xKS:relation_k_m}

Motivated by the observation that the congruences $\bk$ and $\hat{\bm}$ of the
CCLP black hole obey \eqref{xKS:CCLP:k_m}, let us study the consequences
of the relation
\begin{equation}
  (m_{a;b} - m_{b;a}) k^b = - \frac{\D\zeta}{\zeta} m_a, \qquad
  (k_{a;b} - k_{b;a}) m^b = 0,
  \label{xKS:relation_k_m}
\end{equation}
for general xKS spacetimes. The contractions of the first equation in
\eqref{xKS:relation_k_m} with the vectors $\bn$, $\bm$ and $\bm^{(\itilde)}$
give
\begin{equation}
  L_{21} - N_{20} = 0, \qquad
  L_{22} = - \frac{\D\zeta}{\zeta}, \qquad
  L_{2 \itilde} + M_{\itilde 0} = 0,
  \label{xKS:relation_k_m:Riccicoeff1}
\end{equation}
respectively. Similarly, the contractions of the second equation in
\eqref{xKS:relation_k_m} with $\bk$, $\bar{\bn}$ and  $\bm^{(\itilde)}$ lead to
\begin{equation}
  L_{20} = 0, \qquad
  L_{[12]} = 0, \qquad
  L_{[2 \itilde]} = 0,
  \label{xKS:relation_k_m:Riccicoeff2}
\end{equation}
respectively. We can also express the relations
\eqref{xKS:relation_k_m:Riccicoeff1} and \eqref{xKS:relation_k_m:Riccicoeff2}
for the Ricci rotation coefficients in the background spacetime using
\eqref{xKS:Riccicoeff:1}--\eqref{xKS:Riccicoeff:2}
\begin{equation}
  \begin{aligned}
  &\bar L_{20} = L_{20}, \qquad
  \bar L_{21} - \bar N_{20} = L_{21} - N_{20} - \half \K \frac{\D\Omega}{\Omega}, \\
  &\bar L_{22} = L_{22}, \qquad
  \bar L_{[12]} = L_{[12]} + \half \H L_{20} - \half \K L_{10}, \\
  &\bar L_{[2 \itilde]} = L_{[2 \itilde]} + \half \K L_{\itilde 0}, \qquad
  \bar L_{2 \itilde} + \bar M_{\itilde 0} = L_{2 \itilde} + M_{\itilde 0}.
  \label{xKS:relation_k_m:Riccicoeff:bg}
  \end{aligned}
\end{equation}

Obviously, since $\bk$ is geodetic and the conformal factor reads $\Omega = 1$
for the CCLP metric with the Minkowski background, from
\eqref{xKS:relation_k_m:Riccicoeff:bg} it follows that \eqref{xKS:relation_k_m}
also holds in the background metric where the covariant derivatives ``;'' reduce
to the ones ``$\bar ;$'' compatible with the background metric.
Then substituting $\hat m_a = \zeta m_a$ where $\zeta$ is the
norm of $\hat{\bm}$, we obtain \eqref{xKS:CCLP:k_m}. In other words,
\eqref{xKS:relation_k_m} is indeed equivalent to \eqref{xKS:CCLP:k_m} for the
CCLP black hole.

Let us point out that the relation \eqref{xKS:relation_k_m} can be expressed in
terms of the Lie derivative in the full spacetime $g_{ab}$. Note that if
$X^a \omega_a = 0$, then the Lie derivative of a one-form $\omega$ along a
vector field $X$ in coordinate notation reads
\begin{equation}
  \begin{aligned}
  (\mathcal{L}_{X} \omega)_a = \omega_{a;b} X^b + X^b_{\phantom{b};a} \omega_b
    = (\omega_{a;b} - \omega_{b;a}) X^b
  \end{aligned}
\end{equation}
and therefore \eqref{xKS:relation_k_m} is equivalent to
\begin{equation}
  \mathcal{L}_{\bk} m_a \propto m_a, \qquad
  \mathcal{L}_{\boldsymbol{m}} k_a = 0.
  \label{xKS:relation_k_m:lie}
\end{equation}

Note also that the Lie bracket of the vectors $\bk$ and $\bm$ in terms of the Ricci
rotation coefficients reads
\begin{equation}
  [\bk, \bm]^a = L_{20} \, n^a
    + (L_{12} + N_{20}) k^a
    + (L_{i2} - M_{i0}) m^a_{(i)}.
\end{equation}
If $\bk$ and $\bm$ satisfy the relation \eqref{xKS:relation_k_m} then
\begin{equation}
  [\bk, \bm]^a =  2 L_{12} \, k^a
    + L_{22} \, m^a
    + 2 L_{2\itilde} \, m^a_{(\itilde)}
    \label{xKS:relation_k_m:liebracket}
\end{equation}
and therefore the vector fields $\bk$ and $\bm$ are surface-forming provided
that $L_{2 \itilde} = 0$. 

To conclude this section, let us point out the compatibility of the relation
\eqref{xKS:relation_k_m} with the form of the optical matrix
\eqref{KS:expanding:Lij:blockdiagonal} satisfying the optical constraint
\eqref{KS:vacuum:optical_constraint} since it holds also for the CCLP black
hole as it is shown in section \ref{sec:xKS:CCLP}. Comparing
$L_{[2\itilde]} = 0$ \eqref{xKS:relation_k_m:Riccicoeff2} with the optical
matrix \eqref{KS:expanding:Lij:blockdiagonal}, it follows that the vector $\bm$
must not lie in any plane spanned by two spacelike frame vectors corresponding
to a $2 \times 2$ block with non-vanishing twist of the null geodetic congruence
$\bk$ and therefore, omitting the degenerate case $L_{22} = 0$, $\bm$ lies in 
a $1 \times 1$ block of the optical matrix, i.e.\ $L_{22} = \frac{1}{r}$.
Then one may integrate \eqref{xKS:relation_k_m:Riccicoeff1} to obtain
$\zeta = \frac{C}{r}$, where $C$ is an arbitrary function not depending on the
affine parameter $r$ along null geodesics $\bk$. In the case of the CCLP metric,
the function $C$ corresponds to $\nu$.

\section{Geodetic Kerr--Schild vector field}
\label{sec:xKS:geodetic}

From now on, we assume the null Kerr--Schild vector field $\bk$ to be geodetic.
Proposition \ref{xKS:proposition:geodetic_k} then implies that $T_{00} = 0$
which holds not only if the energy--momentum tensor is absent, i.e.\ in Einstein
spaces, but also for spacetimes with aligned matter content such as aligned
Maxwell field $F_{ab} k^b \propto k_a$ or aligned null radiation
$T_{ab} \propto k_a k_b$. As commented in section \ref{sec:KS:general_k},
without loss of generality, we also assume that $\bk$ is affinely parametrized
which further simplifies the following calculations. Note that in the case of
GKS spacetimes, i.e.\ if $\K = 0$, the assumption of Einstein spaces or
spacetimes with aligned matter fields is a sufficient condition for the
Kerr--Schild vector $\bk$ to be geodetic as stated in proposition
\ref{KS:proposition:geodetic_k}.

\subsection{Ricci tensor}

Now we express the frame components of the Ricci tensor for xKS spacetimes
\eqref{xKS:metric} with the Kerr--Schild vector $\bk$ being geodetic and
affinely parametrized. Apart from $R_{00}$, these components are much more
complicated then for GKS spacetimes and $R_{0i}$ no longer vanishes identically.
In addition, we present only the components $R_{01}$ and $R_{ij}$ which have
been crucial in the analysis of GKS spacetimes in chapter \ref{sec:KS}
\begin{align}
  R_{00} &= 0, \phantom{\half}
  \label{xKS:R00} \\
  R_{0i} &= - \half \left( \D^2\K + L_{jj} \D\K + 2 \K \omega^2 \right) \delta_{2i}
    + 2 \K S_{2j} S_{ij}
    - 2 \K L_{2j} L_{ij}
    - S_{2i} \D\K \notag \\
    & \qquad + \half \K \D \! \left( L_{2i} - M_{i0} \right)
    + \left( \D\K + \half \K L_{jj} \right) \left( L_{2i} - M_{i0} \right) \notag \\
    & \qquad + \half \K \left( L_{ij} - \M{i}{j0} \right) \left( L_{2j} - M_{j0} \right),
  \label{xKS:R0i} \\
  R_{01} &= - \D^2\H
    - L_{ii} \D\H
    - 2 \H \omega^2
    + \half \K \D^2\K
    + \half (\D\K)^2
    - \half \K \left( L_{2i} + M_{i0} \right) \M{i}{jj} \notag \\
    &\qquad + \K^{-1} \D \! \left( \K^2 N_{20} \right)
    - \half \delta_i \! \left( \K L_{2i} + \K M_{i0} \right)
    + \half \left( \K L_{ii} - M_{ii} \right)\D\K
    - A_{2i} \delta_i \K \notag \\
    &\qquad + \half \D \! \left( \K^2 L_{22} \right)
    + \K \left( L_{i1} S_{2i} - L_{1i} A_{2i} - A_{ij} M_{ij} + L_{ii} N_{20} + M_{i0} N_{i0} \right) \notag \\
    &\qquad - \half \delta_2 \D\K
    - \K^2 \left( L_{2i} + M_{i0} \right) A_{2i}
    + \half \K^2 L_{22} L_{ii}
    + \frac{2 \Lambda}{n-2},
  \label{xKS:R01} \\
  R_{ij} &= - 2 S_{ij} \D\H
    + 2 \H L_{ik} L_{jk}
    - 2 \H L_{kk} S_{ij}
    - \D \! \left( \K M_{(ij)} \right)
    + 2 \K S_{ij} \D\K
    - \delta_2 \left( \K S_{ij} \right) \phantom{\half} \notag \\
    &\qquad - \half \left( L_{j2} + M_{j0} \right) \delta_i\K
    - \half \left( L_{i2} + M_{i0} \right) \delta_j\K
    + \K \bigg[ \left( L_{21} + N_{20} \right) S_{ij} \notag \\
    &\qquad - \half \left( L_{j2} + M_{j0} \right) L_{1i}
    - \half \left( L_{i2} + M_{i0} \right) L_{1j}
    + \left( A_{2j} + M_{j0} \right) L_{i1} \notag \\
    &\qquad + \left( A_{2i} + M_{i0} \right) L_{j1}
    + \half \left( L_{2i} - M_{i0} \right) N_{j0}
    + \half \left( L_{2j} - M_{j0} \right) N_{i0}
    - S_{ij} M_{kk} \notag \\
    &\qquad + 2 A_{(i|k} M_{j)k}
    + L_{ik} M_{[jk]}
    + L_{jk} M_{[ik]}
    - L_{kk} M_{(ij)}
    - 2 S_{k(i} \M{k}{j)2}
    + M_{(ik)} \M{j}{k0} \phantom{\half} \notag \\
    &\qquad + M_{(jk)} \M{i}{k0} \bigg]
    + \K^2 \bigg[ A_{2i} A_{2j}
    + \half \left( L_{2i} + M_{i0} \right) \left( L_{2j} + M_{j0} \right)
    - 2 L_{(i|2} M_{j)0} \notag \\
    &\qquad - L_{ik} S_{jk}
    - L_{ki} A_{jk}
    + L_{kk} S_{ij} \bigg]
    + \bigg[ 2 \K A_{2j} \D\K
    - \half \delta_j \D\K
    + L_{j1} \D\K \notag \\
    &\qquad - \half \K \left( L_{j2} - M_{j0} \right) \D\K
    + L_{jk} \delta_k \K
    - \half L_{kk} \delta_{j} \K
    - \half \K L_{kk} L_{1j}
    + \half \K L_{1k} L_{jk} \notag \\
    &\qquad + \K L_{[k1]} \M{k}{j0}
    + \half \K L_{kk} L_{j1}
    - \half \K L_{k1} \left( 2 L_{ik} - L_{ki} \right)
    + \K A_{jk} \M{k}{ll} \notag \\
    &\qquad - \K L_{kl} \M{j}{[kl]}
    - \K A_{jk} N_{k0}
    - \K \D L_{[1j]}
    + \K \delta_k A_{jk}
    + \K^2 A_{2j} L_{kk}
    + \K^2 A_{k2} L_{jk} \phantom{\half} \notag \\
    &\qquad - \K^2 \half L_{2k} L_{kj}
    - \half \K^2 L_{k2} \M{k}{j0}
    + \half \K^2 L_{kj} M_{k0}
    - \half \K^2 \D L_{j2} \bigg] \delta_{2i} \notag \\
    &\qquad + \bigg[ 2 \K A_{2i} \D\K
    - \half \delta_i \D\K
    + L_{i1} \D\K
    - \half \K \left( L_{i2} - M_{i0} \right) \D\K
    + L_{ik} \delta_k \K \notag \\
    &\qquad - \half L_{kk} \delta_{i} \K
    - \half \K L_{kk} L_{1i}
    + \half \K L_{1k} L_{ik}
    + \K L_{[k1]} \M{k}{i0}
    + \half \K L_{kk} L_{i1} \notag \\
    &\qquad - \half \K L_{k1} \left( 2 L_{ik} - L_{ki} \right)
    + \K A_{ik} \M{k}{ll}
    - \K L_{kl} \M{i}{[kl]}
    - \K A_{ik} N_{k0}
    - \K \D L_{[1i]} \notag \\
    &\qquad + \K \delta_k A_{ik}
    + \K^2 A_{2i} L_{kk}
    + \K^2 A_{k2} L_{ik}
    - \K^2 \half L_{2k} L_{ki}
    - \half \K^2 L_{k2} \M{k}{i0} \notag \\
    &\qquad + \half \K^2 L_{ki} M_{k0}
    - \half \K^2 \D L_{i2} \bigg] \delta_{2j}
    + \bigg[ \half (\D\K)^2 - \K^2 \omega^2 \bigg] \delta_{2i} \delta_{2j} \notag \\
    &\qquad + \frac{2 \Lambda}{n-2} \delta_{ij}.
    \label{xKS:Rij}
\end{align}
In the case of Einstein GKS spacetimes, the constraints following from the
Einstein field equations involving the components $R_{ij}$ imply that
non-expanding Einstein GKS spacetimes belong to the Kundt class and that the
optical matrices of expanding Einstein GKS spacetimes satisfy the optical
constraint \eqref{KS:vacuum:optical_constraint}. Moreover, $R_{01}$ has been
also employed to show that expanding Einstein GKS spacetimes are not compatible
with Weyl types III and N.

Unfortunately, in the case of xKS spacetimes these components of the Ricci
tensor are rather complicated and it seems that we cannot straightforwardly
obtain similar results. However, if one considers a restrictive assumption on
the geometry of the vectors $\bk$ and $\bm$, for instance, such as
\eqref{xKS:relation_k_m:lie}, then \eqref{xKS:relation_k_m:Riccicoeff1} and
\eqref{xKS:relation_k_m:Riccicoeff2} obviously lead to the simplification of
the Ricci tensor components and this analysis is left for future work.

\subsection[Riemann tensor and algebraic type of the Weyl tensor]{Riemann tensor and algebraic type of the Weyl\\tensor}

The frame components of the Riemann tensor for xKS spacetimes with a geodetic
Kerr--Schild vector field $\bk$ are also more complex than those ones of GKS
spacetimes. Therefore, we list only the non-negative boost weight components
which are be employed in the following sections
\begin{align}
  R_{0i0j} &= 0, \phantom{\half}
  \label{xKS:R0i0j} \\
  R_{010i} &= \frac{1}{2} \D^2\K \, \delta_{2i}
    + \frac{1}{2} \D\! \left( \K L_{2i} + \K M_{i0} \right)
    + \frac{1}{2} M_{i0} \D\K \notag \\
    &\qquad - \frac{1}{2} \K \left( L_{2j} + M_{j0} \right) \M{i}{j0},
  \label{xKS:R010i} \\
  R_{0ijk} &= L_{i[j} \delta_{k]2} \, \D\K
    - \left( L_{[jk]} \D\K + 2 \K S_{l[j} A_{k]l} \right) \delta_{2i}
    - 2 \K A_{il} L_{l[j} \delta_{k]2} \phantom{\half} \notag \\
    & \qquad - \K \left( L_{2i} + M_{i0} \right) L_{[jk]}
    + \K L_{2[j} L_{k]i}
    - \K M_{[j|0} L_{i|k]}, \phantom{\half}
  \label{xKS:R0ijk} \\
  R_{0101} &= \D^2\H
    - \frac{1}{4} (\D\K)^2
    - \K \left( L_{2i} + M_{i0} \right) N_{i0}
    + \D\! \left( \K L_{21} - \K N_{20} \right) \notag \\
    &\qquad - N_{20} \D\K
    - \frac{1}{2} \K L_{2 2} \D\K
    - \frac{1}{4} \K^2 \left( L_{2i} + M_{i0} \right) \left( L_{2i} + M_{i0} \right) \notag \\
    &\qquad - \frac{2 \Lambda}{(n - 2)(n - 1)}, \\
  R_{01ij} &= - 2 A_{ij} \D\H
    - 4 \H S_{k[i} A_{j]k}
    - M_{[ij]} \D\K
    - M_{[i|0} \, \delta_{j]}\K
    + \delta_{[i}\D\K \, \delta_{j]2} \phantom{\half} \notag \\
    &\qquad + \K L_{2[i} L_{1|j]}
    + \K L_{2[i} L_{j]1}
    - 2 \K A_{ij} \left( L_{21} - N_{20} \right)
    - \K L_{k[i} M_{j]k} \phantom{\half} \notag \\
    &\qquad - \K \left( L_{2k} + M_{k0} \right) \M{k}{[ij]}
    + \K L_{k[i} M_{k|j]}
    - \frac{1}{2} \K \, \delta_j \! \left( L_{2i} + M_{i0} \right) \notag \\
    &\qquad + \frac{1}{2} \K \, \delta_i \! \left( L_{2j} + M_{j0} \right)
    + \K^2 L_{2[i} M_{j]0}
    - L_{k[i} \delta_{j]2} \, \delta_k\K
    + \K L_{2[i} \delta_{j]2} \, \D\K \notag \\
    &\qquad - 2 \K L_{[1k]} L_{k[i} \delta_{j]2}, \phantom{\half} \\ 
  R_{0i1j} &= - \D\H L_{ij}
    + 2 \H A_{ik} L_{kj}
    - \half \delta_j\D\K \, \delta_{2i}
    + \frac{1}{4} (\D\K)^2 \delta_{2i} \delta_{2j}
    + \half L_{2j} \delta_{i}\K \notag \\
    &\qquad + \half \left( 2 L_{(i|1} \delta_{2|j)} - M_{ij}
    + \K \left( L_{(ij)} + A_{2i} \delta_{2j} - S_{j2} \delta_{2i} \right)
    \right) \D\K \notag \\
    &\qquad + \half L_{kj} \delta_k\K \delta_{2i}
    + \frac{1}{4} \K \left( L_{2i} + M_{i0} \right) \D\K \delta_{2j}
    + \frac{1}{4} \K \left( L_{2j} + M_{j0} \right) \D\K \delta_{2i} \notag \\
    &\qquad - \half \delta_j \! \left( \K L_{2i} + \K M_{i0} \right)
    + \half \K \left( L_{2i} + M_{i0} \right) L_{j1}
    + \half \K \left( L_{2j} + M_{j0} \right) L_{i1} \notag \\
    &\qquad - \half \K \left( L_{2k} + M_{k0} \right) \M{k}{ij}
    - \K \left( L_{21} - N_{20} \right) L_{ij}
    - \K L_{(1i)} L_{2j} \notag \\
    &\qquad + \K L_{kj} M_{[ik]}
    + \K L_{21} S_{ij}
    - \K L_{k1} A_{ik} \delta_{2j}
    + \half \K^2 L_{22} S_{ij} \notag \\
    &\qquad - \frac{1}{4} \K^2 \left( L_{2j} - M_{j0} \right) \left( L_{2i} + M_{i0} \right)
    + \K^2 \left( L_{2k} + M_{k0} \right) A_{k(i} \delta_{j)2} \notag \\
    &\qquad + \half \K \left( L_{1k} L_{kj} - L_{k1} L_{jk} \right) \delta_{2i}
    + \frac{2 \Lambda}{(n - 2)(n - 1)} \delta_{ij}.
  \label{xKS:R0i1j}
\end{align}
Unlike for the GKS spacetimes, the boost weight 1 components $R_{010i}$
\eqref{xKS:R010i} and $R_{0ijk}$ \eqref{xKS:R0ijk} of xKS spacetimes do not
vanish identically. However, since the boost weight 2 components $R_{0i0j}$
\eqref{xKS:R0i0j} and $R_{00}$ \eqref{xKS:R00} of the Riemann and Ricci tensor,
respectively, are identically zero, from \eqref{Weyl:C0i0j} it follows that the
same also holds for the Weyl tensor
\begin{equation}
  C_{0i0j} = 0
\end{equation}
and therefore
\begin{proposition}
  \label{xKS:proposition:Weyltype}
  Extended Kerr--Schild spacetimes \eqref{xKS:metric} with a geodetic
  Kerr--Schild vector $\bk$ are of the Weyl type I or more special with $\bk$
  being the WAND.
\end{proposition}
This proposition confirms one of our motivations for considering the xKS form of
metrics that such spacetimes with a geodetic $\bk$ may cover more general
algebraic types than Einstein GKS spacetimes.

Note also that if we assume that the vectors $\bk$ and $\bm$ obey
\eqref{xKS:relation_k_m:lie} the frame components of the Riemann tensor
are dramatically simplified as in the case of the Ricci tensor.

\section{Kundt extended Kerr--Schild spacetimes}
\label{sec:xKS:Kundt}

In this section, we study xKS spacetimes with a non-expanding, non-shearing and
non-twisting geodetic null congruence of integral curves of the Kerr--Schild
vector $\bk$, i.e.\ the subclass of Kundt metrics admitting the xKS form
\eqref{xKS:metric}. First, substituting $L_{ij} = 0$, we rewrite the components
of the Ricci tensor \eqref{xKS:R00}--\eqref{xKS:Rij} of general xKS spacetimes
with a geodetic $\bk$ for the Kundt case
\begin{align}
  R_{00} &= 0, \phantom{\half} \\
  R_{0i} &= - \half \D^2\K \delta_{2i}
    - \half \K \D M_{i0}
    - M_{i0} \D\K
    + \half \K M_{j0} \M{i}{j0},
  \label{xKS:Kundt:R0i} \\
  R_{01} &= - \D^2\H
    + \half \K \D^2\K
    + \half (\D\K)^2
    - \half \delta_2 \D\K
    + \K^{-1} \D \! \left( \K^2 N_{20} \right) \notag \\
    &\qquad - \half M_{ii} \D\K
    - \half \delta_i \! \left( \K M_{i0} \right)
    + \K M_{i0} N_{i0}
    - \half \K M_{i0} \M{i}{jj}
    + \frac{2\Lambda}{n-2},
  \label{xKS:Kundt:R01} \\
  R_{22} &= \K M_{k2} M_{k0}
    - \delta_2 \D\K
    + 2 \D ( \K L_{21} )
    - 2 \K L_{[k1]} M_{k0}
    - 2 \K \D L_{(12)} \notag \phantom{\half} \\
    &\qquad + \half (\D\K)^2
    + \frac{2 \Lambda}{n-2},
  \label{xKS:Kundt:R22} \\
  R_{\itilde 2} &= - \half \delta_{\itilde} \D\K
    - \half \D \! \left( \K M_{\itilde 2} \right)
    - \half M_{\itilde 0} \delta_2\K
    + L_{\itilde 1} \D\K
    + \half \K M_{\itilde 0} \D\K \notag \\
    &\qquad - \half \K \bigg[
    \left( L_{12} - 2 L_{21} + N_{20} \right) M_{\itilde 0}
    - M_{\itilde k} M_{k0}
    - M_{k\itilde} M_{k0} \notag \\
    &\qquad + M_{k2} \M{k}{\itilde 0} \bigg]
    + \K L_{[k1]} \M{k}{\itilde 0}
    - \K \D L_{[1\itilde]},
  \label{xKS:Kundt:RI2} \\
  R_{\itilde \jtilde} &= - \D \! \left( \K M_{(\itilde \jtilde)} \right)
    - M_{(\itilde|0} \delta_{\jtilde)} \K
    - \K \bigg[ \half \left( L_{1\jtilde} - 2 L_{\jtilde 1} + N_{\jtilde 0} \right) M_{\itilde 0} \notag \\
    &\qquad + \half \left( L_{1\itilde} - 2 L_{\itilde 1} + N_{\itilde 0} \right) M_{\jtilde 0}
    + M_{(\itilde|k} \M{k}{\jtilde)0}
    + M_{k(\itilde} \M{k}{\jtilde)0} \bigg] \notag \\
    &\qquad + \half \K^2 M_{\itilde 0} M_{\jtilde 0}
    + \frac{2 \Lambda}{n-2} \delta_{\itilde \jtilde}.
  \label{xKS:Kundt:RIJ}
\end{align}
Similarly, the components of the Riemann tensor
\eqref{xKS:R0i0j}--\eqref{xKS:R0i1j} reduce to
\begin{align}
  R_{0i0j} &= 0, \qquad
  R_{0ijk} = 0, \phantom{\half}
  \label{xKS:Kundt:R0ijk} \\
  R_{010i} &= \half \D^2\K \, \delta_{2i}
    + \half \K \D M_{i0}
    + M_{i0} \D\K
    - \half \K \M{i}{j0} M_{j0},
  \label{xKS:Kundt:R010i} \\
  R_{0101} &= \D^2\H
    - \frac{1}{4} (\D\K)^2
    - \K M_{i0} N_{i0}
    + \D\! \left( \K L_{21} - \K N_{20} \right)
    - N_{20} \D\K \nonumber \\*
    &\qquad - \frac{1}{4} \K^2 M_{i0} M_{i0}
    - \frac{2 \Lambda}{(n - 2)(n - 1)},
  \label{xKS:Kundt:R0101} \\
  R_{01\itilde 2} &= \half \delta_{\itilde}\D\K
    - \half \delta_2 \! \left( \K M_{\itilde 0} \right)
    - \half M_{\itilde 2} \D\K
    - \K M_{k0} \M{k}{[\itilde 2]},
  \label{xKS:Kundt:R01I2} \\
  R_{01\itilde\jtilde} &= \delta_{[\itilde}\! \left( \K M_{\jtilde]0} \right)
    - M_{[\itilde\jtilde]} \D\K
    - \K M_{k0} \M{k}{[\itilde\jtilde]},
  \label{xKS:Kundt:R01IJ} \phantom{\half} \\
  R_{0212} &= - \half \delta_2\D\K
    + \frac{1}{4} (\D\K)^2
    + L_{21} \D\K
    + \half \K M_{k0} M_{k2}
    + \frac{2 \Lambda}{(n - 2)(n - 1)}, \\
  R_{021\itilde} &= - \half \delta_{\itilde} \D\K
    + \frac{1}{4} \left( 2 L_{\itilde 1} + \K M_{\itilde 0} \right) \D\K
    + \half \K M_{\itilde 0} L_{21}
    + \half \K M_{k0} M_{k\itilde}, \\
  R_{0\itilde 12} &= \frac{1}{4} \left( 2 L_{\itilde 1}
    - 2 M_{\itilde 2} + \K M_{\itilde 0} \right) \D\K
    - \half \delta_2 \! \left( \K M_{\itilde 0} \right) \notag \\
    &\qquad + \half \K \left( M_{\itilde 0} L_{21}
    - M_{k0} \M{k}{\itilde 2} \right), \\
  R_{0\itilde 1 \jtilde} &= - \half M_{\itilde\jtilde} \D\K
    - \half \delta_{\jtilde} \! \left( \K M_{\itilde 0} \right)
    + \K L_{(\itilde|1} M_{\jtilde)0}
    - \half \K M_{k0} \M{k}{\itilde \jtilde} \notag \\
    &\qquad + \frac{1}{4} \K^2 M_{\itilde 0} M_{\jtilde 0}
    + \frac{2 \Lambda}{(n - 2)(n - 1)} \delta_{\itilde\jtilde}.
\end{align}
As has been shown in section \ref{sec:xKS:general}, a general xKS metric
\eqref{xKS:metric} with a geodetic Kerr--Schild vector $\bk$ is of Weyl type I
or more special. In the case of Kundt xKS spacetimes, the only non-trivial boost
weight 1 frame components of the Riemann tensor satisfy $R_{010i} = -R_{0i}$ as
one can directly see from \eqref{xKS:Kundt:R0i} and \eqref{xKS:Kundt:R010i}.
Putting this relation to \eqref{Weyl:C0i0j} and $R_{0ijk} = 0$
\eqref{xKS:Kundt:R0ijk} to \eqref{Weyl:C010i}, we obtain the boost weight 1
components of the Weyl tensor
\begin{equation}
  C_{0ijk} = \frac{1}{n - 2}(R_{0k} \delta_{ij} - R_{0j} \delta_{ik}), \qquad
  C_{010i} = \frac{3-n}{n-2} R_{0i}
  \label{xKS:Kundt:Weyl:bw0}
\end{equation}
and thus Kundt xKS spacetimes are of Weyl type II or more special if and only if
$T_{0i} = 0$. Note that this statement holds even for general Kundt metrics
\cite{OrtaggioPravdaPravdova2007} not necessarily of the xKS form. Assuming
$T_{0i} = 0$, the Einstein field equations $R_{02} = 0$ and $R_{0\itilde} = 0$,
where $R_{0i}$ is given by \eqref{xKS:Kundt:R0i}, can be written as
\begin{align}
  &\D^2\K = \K M_{\jtilde0} M_{\jtilde0},
  \label{xKS:Kundt:G02} \\
  &\D \! \left( \K^2 M_{\itilde 0} \right) = \K^2 M_{\jtilde0} \M{\itilde}{\jtilde0},
  \label{xKS:Kundt:G0I}
\end{align}
respectively. The trivial solution $\K = 0$ corresponds to the GKS limit when
$T_{0i} = 0$ is a necessary condition in order to satisfy the Einstein field
equations since in this case the components $R_{0i}$ identically vanish as was
shown in section \ref{sec:KS:Ricci}. The Ricci rotation coefficients $\M{i}{ja}$
are antisymmetric in the indices $i$ and $j$, therefore by multiplying
\eqref{xKS:Kundt:G0I} with $2 \K^2 M_{\itilde 0}$ we eliminate the term on the
right-hand side and the remaining term can be combined as
\begin{equation}
  0 = 2 \D \! \left( \K^2 M_{\itilde 0} \right) \K^2 M_{\itilde 0}
  = \D \! \left( \K^4 M_{\itilde 0} M_{\itilde 0} \right).
  \label{xKS:Kundt:GOI:rewritten}
\end{equation}
Consequently, $\K^4 M_{\itilde 0}M_{\itilde 0} = (c^0)^2$, where the function
$c^0$ does not depend on the affine parameter $r$ along the null geodesics of
the Kerr--Schild congruence $\bk$. Obviously, since we assume $\K$ to be
non-zero, $c^0$ vanishes if and only if all $M_{i0}$ vanish. Substituting the
result of \eqref{xKS:Kundt:GOI:rewritten} to \eqref{xKS:Kundt:G02}, we obtain
a differential equation $\K^3 \D^2\K = (c^0)^2$ determining the $r$-dependence
of the function $\K$ which has two distinct branches of solutions
\begin{align}
  \K &= \sgn d^0 \sqrt{\big(d^0 (r + b^0) \big)^2 + \frac{(c^0)^2}{(d^0)^2}}
    & \text{if $c^0 \neq 0$},
  \label{xKS:Kundt:K:rdep:c<>0} \\
  \K &= f^0 r + e^0 & \text{if $c^0 = 0$},
  \label{xKS:Kundt:K:rdep:c=0}
\end{align}
where $b^0$, $d^0$, $e^0$ and $f^0$ are arbitrary functions not depending on
$r$. Finally, we can conclude that
\begin{proposition}
  \label{xKS:proposition:Kundt}
  For Kundt extended Kerr--Schild spacetimes with $\K \neq 0$ solving the
  Einstein field equations the following statements are equivalent
  \begin{itemize}
    \item[(i)] the boost weight 1 components
      $T_{0i} \equiv T_{ab} k^a m^b_{(i)} = 0$ of the energy--momentum tensor
      vanish,
    \item[(ii)] the spacetime is of Weyl type II or more special,
    \item[(iii)] the function $\K$ takes the form \eqref{xKS:Kundt:K:rdep:c<>0} or
      \eqref{xKS:Kundt:K:rdep:c=0}.
  \end{itemize}
  Moreover, $\K$ is a linear function \eqref{xKS:Kundt:K:rdep:c=0} of an affine
  parameter $r$ along the null geodesics of the Kerr--Schild congruence $\bk$
  if and only if $M_{\itilde 0} = 0$.
\end{proposition}

Note that in Kundt spacetimes the relation between the vectors $\bk$ and $\bm$
\eqref{xKS:relation_k_m} holds if and only if
\begin{equation}
  L_{21} = N_{20}, \qquad
  D \zeta = 0, \qquad
  M_{\itilde 0} = 0, \qquad
  L_{[12]} = 0.
  \label{xKS:Kundt:relation_k_m}
\end{equation}
Obviously, \eqref{xKS:Kundt:relation_k_m} is not satisfied for $\K$ of the form
\eqref{xKS:Kundt:K:rdep:c<>0}.

\subsection{Explicit example}

Now we present explicit examples of Ricci-flat Kundt xKS spacetimes, namely
the class of spacetimes with vanishing scalar invariants \eqref{KS:VSI:metric}.
Such VSI metrics can be written as
\begin{equation}
  \d s^2 = 2 \d u \, \d r + \delta_{ij} \, \d x^i \, \d x^j
    + 2 H(u, r, x^k) \, \d u^2
    + 2 W_i(u, r, x^k) \, \d u \, \d x^i,
  \label{xKS:VSI:metric}
\end{equation}
which better exhibits its xKS form \eqref{xKS:metric} with the flat background
metric described by the first two terms in \eqref{xKS:VSI:metric}. Obviously,
one may identify the null one-form $\d u$ with the Kerr--Schild vector $\bk$
\begin{equation}
  k_a \, \d x^a = \d u.
\end{equation}
Therefore, the function $\H$ appearing in the xKS ansatz \eqref{xKS:metric} is
given by
\begin{equation}
  \H = - H(u, r, x^k)
\end{equation}
and the remaining term $\K \bm = - W_i(u, r, x^k) \, \d x^i$ can be split to the
function $\K$ and the vector $\bm$ as
\begin{equation}
  \K = - \sqrt{W_i W_i}, \qquad
  \bm = \frac{W_i \, \d x^i}{\sqrt{W_j W_j}},
  \label{xKS:VSI:K}
\end{equation}
so that $\bm$ is a unit vector. It turns out that all VSI metrics
\eqref{xKS:VSI:metric} admit the xKS form \eqref{xKS:metric}.

Recall that xKS spacetimes with a geodetic Kerr--Schild vector field $\bk$ are
of Weyl type I or more special as follows from proposition
\ref{xKS:proposition:geodetic_k}. Furthermore, proposition
\ref{xKS:proposition:Kundt} implies that Kundt xKS spacetimes with $R_{0i} = 0$
are of Weyl type II or more special. On the other hand, it is known
\cite{ColeyMilsonPravdaPravdova2004} that VSI spacetimes are of Weyl types III,
N or O with the Ricci tensor of type III or more special, i.e.\ $R_{00} =
R_{0i} = R_{01} = R_{ij} = 0$, and all VSI metrics with the Ricci tensor of
types N and O have been given explicitly in \cite{ColeyFusterHervikPelavas2006}.
As has been already mentioned in section \ref{sec:KS:nonexpanding:examples}, the
VSI class can be divided into two distinct subclasses with vanishing and
non-vanishing quantity $L_{1i} L_{1i}$ denoted as $\epsilon = 0$ and
$\epsilon = 1$, respectively. The canonical choices of the functions $W_i$ in
these subclasses are given in \eqref{KS:VSI:W:canonical_choice}. The functions
$W_i$ and $H$ can be further constrained employing the Einstein field equations
and assuming an appropriate special algebraic type of the spacetime. Note also
that for VSI spacetimes the statements (i) and (ii) of proposition
\ref{xKS:proposition:Kundt} are clearly satisfied and therefore the function
$\K$ takes one of the forms \eqref{xKS:Kundt:K:rdep:c<>0} or
\eqref{xKS:Kundt:K:rdep:c=0} depending on the functions $W_i(u, r, x^k)$ which
differ in the subclasses $\epsilon = 0$ and $\epsilon = 1$.

\subsubsection{Case $\epsilon = 0$}
In the subclass $\epsilon = 0$ of VSI spacetimes, the functions $W_i(u, r, x^k)$
are given by \cite{ColeyFusterHervikPelavas2006}
\begin{equation}
  W_2 = 0, \qquad
  W_{\itilde} = W_{\itilde}^0(u, x^k),
\end{equation}
where the superscript 0 denotes that the quantity is independent on the
coordinate $r$ corresponding to an affine parameter along the geodesics of the
Kerr--Schild congruence $\bk$. Then from \eqref{xKS:VSI:K} it follows that $\K$
is of the form \eqref{xKS:Kundt:K:rdep:c=0} with $f^0 = 0$ and does not depend
on $r$
\begin{equation}
  \K = - \sqrt{W_{\itilde}^0 W_{\itilde}^0} = e^0.
\end{equation}
Therefore $c^0 = 0$ and thus all the Ricci rotation coefficients $M_{i0}$
vanish. Moreover, if also $N_{20} = 0$, then the vector $\bm$ is parallelly
transported along the null geodesics of the congruence $\bk$. On the other hand,
if $N_{20}$ is non-vanishing, it can be transformed to zero by a null rotation
with $\bk$ fixed \eqref{intro:Riccicoeff:nullrot} setting $\D z_2 = - N_{20}$,
however, this Lorentz transformation changes the vector $\bm$.

\subsubsection{General case $\epsilon = 1$}

In the subclass $\epsilon = 1$, the functions $W_i(u, r, x^k)$ read
\cite{ColeyFusterHervikPelavas2006}
\begin{equation}
  W_2 = - \frac{2}{x^2} r, \qquad
  W_{\itilde} = W_{\itilde}^0.
  \label{xKS:VSI:W:e=1}
\end{equation}
If at least one of $W_{\itilde}$ is non-zero, the function $\K$
\eqref{xKS:VSI:K} takes the form \eqref{xKS:Kundt:K:rdep:c<>0}
\begin{equation}
  \K = - \sqrt{\frac{4}{(x^2)^2} r^2 + W_{\itilde}^0 W_{\itilde}^0}.
  \label{xKS:VSI:W:e=1:K}
\end{equation}
Comparing \eqref{xKS:VSI:W:e=1:K} with \eqref{xKS:Kundt:K:rdep:c<>0}, it
immediately follows that $b^0 = 0$,
$d^0 = - \frac{2}{x^2}$ and
$(c^0)^2 = \frac{(x^2)^2}{4} W^0_{\itilde} W^0_{\itilde}$
. Since $M_{\itilde 0} M_{\itilde 0} = \K^{-4} (c^0)^2 \neq 0$
the vector $\bm$ is not parallelly transported along $\bk$.

\subsubsection{Special case $\epsilon = 1$}

In the special case of the subclass $\epsilon = 1$ of VSI spacetimes where all
$W^0_{\itilde}$ in \eqref{xKS:VSI:W:e=1} vanish, the function $\K$ is of the
form \eqref{xKS:Kundt:K:rdep:c=0} with $e^0 = 0$
\begin{equation}
  \K = - \frac{2}{|x^2|} r = f^0 r.
\end{equation}
As in the case $\epsilon = 0$, $c^0$ and consequently $M_{\itilde0}$ vanish and
if $N_{20}$ is zero, the vector $\bm$ is parallelly transported along $\bk$.

\subsection{\texorpdfstring{Not all vacuum higher dimensional \pp waves belong to the class of Ricci-flat xKS spacetimes}
{Not all vacuum higher dimensional pp-waves belong to the class of Ricci-flat xKS spacetimes}}

In the previous section, we have shown that all VSI metrics admit the xKS form
\eqref{xKS:metric}. The question is whether also all vacuum \pp waves belong
to the class of xKS spacetimes.
As mentioned in section \ref{sec:QG:Einstein:comparison}, higher dimensional
\pp waves are of Weyl type II or more special and Ricci-flat if not supported by
an appropriate matter field. It is also known that vacuum \pp waves of type III
or more special belong to the VSI class. Therefore, it remains to investigate
the situation of the Weyl type II \pp waves.

All Ricci-flat \pp wave metrics can be written in the form
\begin{equation}
  \d s^2 = 2 \d u \left[ \d v + H(u, x^k) \, \d u + W_{i}(u, x^k) \, \d x^i \right]
    + g_{ij}(u,x^k) \, \d x^i \, \d x^j
  \label{xKS:Kundt:ppwave}
\end{equation}
and it does not seem that for general transversal metric $g_{ij}$, \pp wave
metrics \eqref{xKS:Kundt:ppwave} can be cast to the xKS form. However, one may
use the result of \cite{PodolskyZofka2008} that the transversal Riemannian
metric $g_{ij}$ of vacuum \pp waves is Ricci-flat and the statement that the
transversal Riemannian space is locally homogeneous in the case of CSI Kundt
metrics as has been shown in \cite{ColeyHervikPelavas2005}. Since a Ricci-flat
locally homogeneous Riemannian space is flat \cite{PruferTricerriVanhecke1996},
we can conclude that Ricci-flat CSI \pp wave metrics can be written in the form
\eqref{xKS:Kundt:ppwave} with a flat transversal space, i.e.\
$g_{ij} = \delta_{ij}$, which is obviously in the xKS form with the Minkowski
background. However, note that there could exist non-CSI Ricci-flat \pp waves
and therefore we cannot decide whether all higher dimensional Ricci-flat
\pp waves belong to the class of xKS spacetimes.

Recall also that type N Ricci-flat VSI spacetimes and consequently vacuum type N
\pp waves can be cast to the KS form \eqref{KS:KSansatz} as has been shown in
\cite{OrtaggioPravdaPravdova2008} and as also immediately follows from the
results of section \ref{sec:KS:nonexpanding:examples}. All the mentioned facts
about higher dimensional vacuum \pp waves are summarized in table
\ref{tab:xKS:ppwaves}.
\begin{table}
  \caption{Properties of higher dimensional Ricci-flat \pp waves.}
  \begin{center}
    \begin{tabular}{cccc}
      \toprule
      Weyl type & KS & xKS & VSI \\
      \midrule
      N   & \checkmark & \checkmark & \checkmark \\
      III & $\times$   & \checkmark & \checkmark \\
      II  & $\times$   & only CSI   & $\times$   \\
      \bottomrule
    \end{tabular}
  \end{center}
  \label{tab:xKS:ppwaves}
\end{table}
Finally, note that in four dimensions, all vacuum \pp wave metrics are only of
Weyl type N, belong to the VSI class and take the KS form.

\section[Expanding extended Kerr--Schild spacetimes]{Expanding extended Kerr--Schild\\spacetimes}
\label{sec:xKS:CCLP}

In this section, we construct a null frame in the CCLP spacetime
\cite{ChongCveticLuPope2005} which then allows us to express the optical matrix,
show that it obeys the optical constraint and determine the algebraic type. The
CCLP solution represents a charged rotating black hole in five-dimensional
minimal gauged supergravity or equivalently in the
Einstein--Maxwell--Chern--Simons theory with a negative cosmological constant
$\Lambda$ and the Chern--Simons coefficient $\chi = 1$ described by the field
equations
\begin{align}
  &R^a_b -2 (F_{bc} F^{ac} - \frac{1}{6} \delta^a_b F_{cd} F^{cd})
    + \frac{2}{3 \Lambda} \delta^a_b = 0, \\
  &\nabla_b F^{ab} + \frac{\chi}{2\sqrt{3}\sqrt{-g}} \epsilon^{abcde} F_{bc} F_{de} = 0
\end{align}
and it is an example of expanding xKS spacetime. In fact, the xKS ansatz has
been proposed in \cite{AlievCiftci2008} where it has been shown that the CCLP
black hole can be cast to the form
\begin{equation}
  g_{ab} = \bar g_{ab} - 2 \H k_a k_b - 2 \hat \K k_{(a} \hat m_{b)},
  \label{xKS:CCLP:original}
\end{equation}
where we put hat over the $\hat{\K}$ and $\hat{\bm}$ since unlike our definition
of the xKS ansatz \eqref{xKS:metric}, the vector $\hat{\bm}$ is not normalized
to unity.

In the case $\Lambda = 0$, the CCLP metric can be rewritten in terms of spheroidal
coordinates in the form \eqref{xKS:CCLP:original} with the flat background metric
and the vectors $\bk$, $\hat{\bm}$ given by \cite{AlievCiftci2008}
\begin{align}
  \bar{g}_{ab} \, \d x^a \, \d x^b &= - \d t^2
    - 2 \d r \left( \d t - a \sin^2 \theta \, \d \phi - b \cos^2 \theta \, \d \psi \right)
    + \rho^2 \, \d \theta^2 \notag \phantom{\half} \\
    &\qquad + \left( r^2 + a^2 \right) \sin^2 \theta \, \d \phi^2
    + \left( r^2 + b^2 \right) \cos^2 \theta \, \d \psi^2, \\
  k_a \, \d x^a &= - \d t + a \sin^2 \theta \, \d \phi + b \cos^2 \theta \, \d \psi, \phantom{\half}
  \label{xKS:CCLP:k} \\
  \hat m_a \, \d x^a &= b \sin^2 \theta \, \d \phi + a \cos^2 \theta \, \d \psi. \phantom{\half}
  \label{xKS:CCLP:mhat}
\end{align}
The functions $\H$, $\hat{\K}$ and the one-form gauge potential proportional to
the Kerr--Schild vector $\bk$ then read
\begin{equation}
  \H = - \frac{M}{\rho^2} + \frac{Q^2}{2 \rho^4}, \qquad
  \hat \K = - \frac{Q}{\rho^2}, \qquad
  A = - \frac{\sqrt{3} Q}{2 \rho} \bk,
\end{equation}
where $r$ is the spheroidal radial coordinate, the angular coordinates have
usual ranges $\phi \in \langle 0, 2 \pi)$, $\psi \in \langle 0, 2 \pi)$,
$\theta \in \langle 0, \pi \rangle$ and
$\rho^2 = r^2 + a^2 \cos^2 \theta + b^2 \sin^2 \theta$. One may show that the
vectors $\bk$ \eqref{xKS:CCLP:k} and $\hat{\bm}$ \eqref{xKS:CCLP:mhat} in the
background spacetime satisfy
\cite{EttKastor2010}
\begin{equation}
  (\hat m_{a\bar{;}b} - \hat m_{b\bar{;}a}) k^b = 0, \qquad
  (k_{a\bar{;}b} - k_{b\bar{;}a}) \hat m^b = 0,
  \label{xKS:CCLP:k_m}
\end{equation}
where ``$\bar{;}$'' denotes the covariant derivative compatible with the
background metric $\bar{g}_{ab}$ and as follows from \eqref{xKS:m:raised_in_bg}
the contractions can be performed by both the full and background metrics since
$\bk$ is geodetic. It is also convenient to define $\nu^2 \equiv \rho^2 - r^2$.

Since the metric is in the form \eqref{xKS:CCLP:original} we put it into our
definition of the xKS form \eqref{xKS:metric} with a unit vector $\bm$ by
rescaling the vector $\hat{\bm}$ and including its norm to the function
$\hat{\K}$. Thus, the norm of the vector $\hat \bm$ is
$|\hat \bm| = \frac{\nu}{r}$, therefore the normalized $\bm$ reads
\begin{equation}
  m_a \, \d x^a = \frac{b r \sin^2 \theta}{\nu} \d \phi
    + \frac{a r \cos^2 \theta}{\nu} \d \psi
  \label{xKS:CCLP:m}
\end{equation}
and since the scalar functions are related via $\K = \hat \K |\hat \bm|$ one
gets
\begin{equation}
  \K = - \frac{Q \nu}{r \rho^2}.
  \label{xKS:CCLP:K}
\end{equation}
Obviously, one may identify $\bk$ and $\bm$ with the null and spacelike frame
vectors $\bl$ and $\bm^{(2)}$, respectively. The remaining vectors $\bn$,
$\bm^{(3)}$ and $\bm^{(4)}$ can be obtained by solving the constraints
\eqref{intro:frame:constraints}. Similarly as for the five-dimensional
Kerr--(anti-)de Sitter metric in section \ref{sec:KS:Kerr-(A)dS}, it is easier
to find a null frame in the background spacetime first
\begin{align}
  m^{(3)}_a \, \d x^a &= \frac{\rho}{\nu} \cot\theta \left( \d t - b \, \d\psi \right), 
  \label{xKS:CCLP:frame:m3} \\
  m^{(4)}_a \, \d x^a &= \rho \, \d \theta, \phantom{\half}
  \label{xKS:CCLP:frame:m4} \\
  \bar n_a \, \d x^a &= \half \frac{\rho^2 - r^2 \sin^2 \theta}{\nu^2 \sin^2 \theta} \, \d t
    + \d r
    + \half a \frac{\rho^2 - r^2 \sin^2\theta}{\nu^2} \, \d\phi \notag \\
    &\qquad - \half b \cot^2\theta \frac{\rho^2 + r^2 \sin^2\theta}{\nu^2} \, \d\psi,
\end{align}
from which we then construct the frame in the full spacetime using
\eqref{xKS:background_n}. Therefore, the vectors $\bm^{(3)}$ and $\bm^{(4)}$ are
same in both frames and $\bn$ is given by
\begin{equation}
\begin{aligned}
  n_a \, \d x^a &= \left( \half \frac{\rho^2 - r^2 \sin^2 \theta}{\nu^2 \sin^2 \theta}
    - \frac{2 M \rho^2 - Q^2}{2 \rho^4} \right) \d t
    + \d r \\
    &\qquad + \left( \half a \frac{\rho^2 - r^2 \sin^2\theta}{\nu^2}
    + \left( a \frac{2 M \rho^2 - Q^2}{2 \rho^4} + b \frac{Q}{\rho^2} \right) \sin^2\theta \right) \d\phi \\
    &\qquad - \left( \half b \frac{\rho^2 + r^2 \sin^2\theta}{\nu^2 \sin^2\theta}
    - \left( b \frac{2 M \rho^2 - Q^2}{2 \rho^4} + a \frac{Q}{\rho^2} \right) \right) \cos^2\theta \, \d\psi.
  \end{aligned}
  \label{xKS:CCLP:frame:n}
\end{equation}
Straightforwardly, one may also obtain the contravariant components of the frame
vectors \eqref{xKS:CCLP:k}, \eqref{xKS:CCLP:m}, \eqref{xKS:CCLP:frame:m3},
\eqref{xKS:CCLP:frame:m4} and \eqref{xKS:CCLP:frame:n} 
\begin{align}
  k^a \, \partial_a &= \partial_r, \phantom{\half} \\
  m_{(2)}^a \, \partial_a &= \frac{ab}{r \nu} \, \partial_t
    - \frac{Q \nu^2 + a b \rho^2}{r \rho^2 \nu} \, \partial_r
    + \frac{b}{r \nu} \, \partial_\phi
    + \frac{a}{r \nu} \, \partial_\psi, \\    
  m_{(3)}^a \, \partial_a &= \frac{\sin\theta \cos\theta}{\rho \nu} \bigg[ (a^2 - b^2) \, \partial_t
    - \frac{(r^2 + a^2)}{\sin^2\theta} \, \partial_r \notag \\
    &\qquad + \frac{a}{\sin^2\theta} \, \partial_\phi
    - \frac{b}{\cos^2\theta} \, \partial_\psi \bigg], \\
  m_{(4)}^a \, \partial_a &= \frac{1}{\rho} \, \partial_\theta, \\
  n^a \, \partial_a &= - \frac{b^2}{\nu^2} \, \partial_t
    - \left( \frac{\rho^2 - (r^2 + 2 b^2) \sin^2\theta}{2 \nu^2 \sin^2\theta}
      + \frac{M}{\rho^2} - \frac{Q^2}{2 \rho^4} \right) \partial_r \notag \\
    &\qquad + \frac{a \cot^2\theta}{\nu^2}\,  \partial_\phi
    - \frac{b}{\nu^2} \, \partial_\psi.
  \label{xKS:CCLP:frame}
\end{align}

Having established the frame, we can easily calculate the Ricci rotation
coefficients \eqref{intro:Riccicoeff}. It can be shown that $L_{a0} = 0$ and
therefore $\bk$ is geodetic and affinely parametrized. The frame
\eqref{xKS:CCLP:frame} is not parallelly transported along the geodesics of the
Kerr--Schild congruence $\bk$ since some of the independent components $N_{i0}$
and $\M{i}{j0}$ are non-vanishing
\begin{align}
  N_{20} &= - Q \frac{\nu}{\rho^4}, \qquad
  N_{30} = \frac{r \cot\theta}{\rho \nu}, \qquad
  N_{40} = - \frac{\cot\theta}{\rho},
  \label{xKS:CCLP:frame:Na0} \\
  \M{2}{30} &= 0, \qquad
  \M{2}{40} = 0, \qquad
  \M{3}{40} = - \frac{\nu}{\rho^2}.
\end{align}
Note that we cannot set this frame to be parallelly transported using
appropriate spins and null rotations with $\bk$ fixed as in the case of the
Kerr--(anti-)de Sitter black hole in section \ref{sec:KS:Kerr-(A)dS} since we
are not able to transform away the component $N_{20}$ unless the identification
of the vector $\bm$ appearing in the xKS metric \eqref{xKS:metric} with the
frame vector $\bm^{(2)}$ is relaxed.

The optical matrix $L_{ij}$ takes the block-diagonal form
\begin{equation}
  L_{ij} = \begin{pmatrix} \frac{1}{r} & 0 & 0 \\
    0 & \frac{r}{\rho^2} & \frac{\nu}{\rho^2} \\
    0 & -\frac{\nu}{\rho^2} & \frac{r}{\rho^2}
  \end{pmatrix}
  \label{xKS:CCLP:Lij}
\end{equation}
and it immediately follows that $L_{ij}$ is a normal matrix. Moreover,
interestingly, the optical matrix $L_{ij}$ \eqref{xKS:CCLP:Lij} of the CCLP
black hole satisfies the optical constraint \eqref{KS:vacuum:optical_constraint}.
Note that, in section \ref{sec:KS:optical_constraint}, it has been pointed out
that in a certain sense the optical constraint can be considered as a
possible generalization of the Goldberg--Sachs theorem to higher dimensions
restricted to Einstein GKS spacetimes.

As has been mentioned in section \ref{sec:xKS:relation_k_m}, the vectors $\bk$
and $\bm$ of the CCLP metric satisfy the relation \eqref{xKS:relation_k_m} and
therefore \eqref{xKS:relation_k_m:Riccicoeff1} and
\eqref{xKS:relation_k_m:Riccicoeff2} also hold. It can be seen directly from
\eqref{xKS:CCLP:Lij} that $L_{22} = \frac{1}{r}$, $L_{2 \itilde} = 0$ and using
\eqref{xKS:relation_k_m:Riccicoeff1} from \eqref{xKS:CCLP:frame:Na0} it follows
that $L_{12} = - Q \frac{\nu}{\rho^4}$. Substituting these Ricci rotation
coefficients to \eqref{xKS:relation_k_m:liebracket}, we obtain
\begin{equation}
  [\bk, \bm]^a = -2 Q \frac{\nu}{\rho^4} k^a + \frac{1}{r} m^a
\end{equation}
and therefore the vector fields $\bk$ and $\bm$ are surface-forming.
Note also that $L_{22} = \frac{1}{r}$ following from the optical matrix
\eqref{xKS:CCLP:Lij} is in accordance with \eqref{xKS:relation_k_m:Riccicoeff1}
since $\zeta = |\hat{\bm}| = \frac{\nu}{r}$ implies that
$\frac{\D \zeta}{\zeta} = - \frac{1}{r}$.

Employing the frame \eqref{xKS:CCLP:frame}, we can express frame components of
the Weyl tensor and determine the algebraic type. In accordance with proposition
\ref{xKS:proposition:Weyltype}, the boost weight 2 components of the Weyl tensor
vanish
\begin{equation}
  C_{0i0j} = 0
\end{equation}
and therefore the spacetime is of Weyl type I or more special. Due to the
tracelessness of the Weyl tensor, the boost weight 1 components $C_{010i}$ are
not independent \eqref{intro:Weyl:symmetries} and thus vanishing of the
components $C_{0ijk}$ is sufficient condition for a spacetime to be of Weyl type
II or more special. In the case of the CCLP black hole, one gets the non-zero
components
\begin{equation}
  \begin{aligned}
  C_{0234} &= - C_{0243} = - \frac{4 Q \nu^2}{\rho^6}, \qquad \\
  C_{0323} &= - C_{0332} = C_{0424} = - C_{0442} = - \frac{2 Q r \nu}{\rho^6}, \\
  C_{0324} &= - C_{0342} = - C_{0423} = C_{0432} = - \frac{2 Q \nu^2}{\rho^6},
  \end{aligned}
\end{equation}
which suggests that the spacetime is of Weyl type I.

In section \ref{sec:KS:Weyl}, we have mentioned the results of
\cite{PravdaPravdovaOrtaggio2007} that stationary spacetimes with the metric
remaining unchanged under reflection symmetry and with non-vanishing expansion
are of Weyl types G, I$_i$, D or conformally flat. The CCLP spacetimes obey
these conditions along with the reflection symmetry $t \rightarrow -t$,
$\phi \rightarrow -\phi$, $\psi \rightarrow -\psi$ of the metric in
Boyer--Lindquist-type coordinates given in \cite{ChongCveticLuPope2005}.
Therefore we can conclude that in general the CCLP solution is of Weyl type
I$_i$ with $\bk$ being the WAND unless a distinct multiple WAND exists. However,
the form of the xKS metric \eqref{xKS:metric} suggests that $\bk$ should have
the highest order of alignment. Then only either in the uncharged case when the
metric corresponds to the five-dimensional Myers--Perry black hole or in the
non-rotating limit $\nu = 0$, the spacetime is of more special Weyl type D and
the metric reduces to the GKS form since $\K$ \eqref{xKS:CCLP:K} vanishes.

\chapter{Quadratic gravity}
\label{sec:QuadGr}

In this chapter, we mainly present our results published in
\cite{MalekPravda2011}, which will be occasionally extended. First of all,
let us provide a motivation for considering generalizations of the Einstein
theory referred to as quadratic gravity with a general Lagrangian containing all
possible polynomial curvature invariants up to the second order in the Riemann
tensor and summarize some basic properties of such theories.

As it is known, the Einstein--Hilbert action \eqref{intro:Eistein-Hilbert:action}
is non-renormalizable \cite{tHooftVeltman1974}. In perturbative quantum gravity,
corrections have to be added and, demanding coordinate invariance, these
corrections should consist of various curvature invariants. If one includes
curvature squared terms $\alpha R^2 + \beta R_{ab} R^{ab}$ to the
four-dimensional Einstein--Hilbert action, the theory becomes renormalizable
\cite{Stelle1977,Stelle1978}. Note that such terms give the most general action
up to the second order in curvature since the term $R_{abcd} R^{abcd}$ can be
rewritten using the squared Ricci tensor, squared curvature scalar and the
Gauss--Bonnet term which is topological invariant in four dimensions and thus
does not contribute to the dynamics. Unfortunately, the field equations then
contain the fourth derivatives of the metric and unitarity is lost due to the
introduced massive ghost-like graviton.

In string theory, the ghost freedom of low-energy effective action requires that
the quadratic corrections to the Einstein--Hilbert term, which is the lowest
order term in the Regge slope expansion of strings, are of the Gauss--Bonnet
form \cite{Zwiebach1985}. Therefore, the action up to the second order consists
of dimensionally extended Euler densities which correspond to the first three
terms of the Lovelock Lagrangian yielding the field equations of the second
order in derivatives of the metric. However, the forms of the corrections depend
on the type of string theory \cite{BentoBertolami1995} and in higher orders
they are no longer given only by Euler densities.

Recently, in four dimensions \cite{LuPope2011} and later in arbitrary dimension
\cite{Deseretal2011}, it has been shown that despite non-zero $\alpha$ and
$\beta$ producing ghosts, the massive spin-0 mode can be eliminated and the
massive spin-2 mode becomes massless by an appropriate choice of the parameters
of the theory denoted as the critical point. This fact has led to a current
growing interest in such theories \cite{AlishahihaFareghbal2011,
GulluGursesSismanTekin2011,Pang2011,PorratiRoberts2011,LiuLuLuo2011}.

We will consider a general action of quadratic gravity which can be rearranged
to a more convenient form \cite{DeserTekin2002}
\begin{equation}
  \begin{aligned}
  \mathcal{S} &= \int \d^n x \, \sqrt{-g} \bigg( \frac{1}{\kappa} \left( R - 2 \Lambda_0 \right)
    + \alpha R^2 + \beta R_{ab} R^{ab} \\
    &\qquad + \gamma \left(R_{abcd} R^{abcd} - 4 R_{ab} R^{ab} + R^2 \right) \bigg),
  \end{aligned}
  \label{QG:action}
\end{equation} 
where the term multiplied by $\kappa^{-1}$ is the well-known Einstein--Hilbert
term leading to the Einstein field equations. The last term multiplied by
$\gamma$ is the Gauss--Bonnet term and these both terms appear as the first
three terms in the Lovelock theory \cite{Lovelock1971} with the Lagrangian
\begin{equation}
  \mathcal{L} = \sqrt{-g} \ \sum^p_{k=0} \alpha_k \mathcal{L}_k,
\end{equation}
which consists of a linear combination of dimensionally extended
$2k$-dimensional Euler densities
\begin{equation}
  \mathcal{L}_k = \frac{1}{2^k} \delta^{a_1 b_1 \dots a_k b_k}_{c_1 d_1 \dots c_k d_k}
  \prod_{m=1}^k R_{a_m b_m}^{\phantom{a_m b_m} c_m d_m},
\end{equation}
where $\delta^{a_1 b_1 \dots a_k b_k}_{c_1 d_1 \dots c_k d_k} =
\frac{1}{k!} \delta_{\lbrack c_1}^{a_1} \delta_{d_1}^{b_1} \cdots \delta_{c_k}^{a_k} \delta_{d_k]}^{b_k}$
is the totally anti-symmetric generalized Kronecker delta. In $n$ dimensions,
$2k$-dimensional Euler densities, where $2k \geq n$, are topological invariants
or vanish identically and thus do not contribute to the field equations.
Since these equations are quasi-linear of the second order in derivatives of the
metric, Lovelock theories are natural generalizations of the Einstein gravity to
higher dimensions, in contrast to general quadratic gravity \eqref{QG:action}
where, due to non-zero $\alpha$ and $\beta$, the field equations are of the
fourth order.

Varying the action \eqref{QG:action} with respect to the metric leads to the
source-free field equations of quadratic gravity \cite{GulluTekin2009}
\begin{equation}
  \begin{aligned}
  & \frac{1}{\kappa} \left( R_{ab} - \frac{1}{2} R g_{ab} + \Lambda_0 g_{ab} \right)
  + 2 \alpha R \left( R_{ab} - \frac{1}{4} R g_{ab} \right) \\
  &\qquad + \left( 2 \alpha + \beta \right)\left( g_{ab} \Box - \nabla_a \nabla_b \right) R
  + 2 \gamma \bigg( R R_{ab} - 2 R_{acbd} R^{cd} \\
  &\qquad + R_{acde} R_{b}^{\phantom{b}cde} - 2 R_{ac} R_{b}^{\phantom{b}c}
  - \frac{1}{4} g_{ab} \left( R_{cdef} R^{cdef} - 4 R_{cd} R^{cd} + R^2 \right) \bigg) \\
  &\qquad + \beta \Box \left( R_{ab} - \frac{1}{2} R g_{ab} \right)
  + 2 \beta \left( R_{acbd} - \frac{1}{4} g_{ab} R_{cd} \right) R^{cd} = 0.
  \end{aligned}
  \label{QG:fieldeqns}
\end{equation}
As in the Gauss--Bonnet theory, quadratic gravity in dimension $n > 4$ admits,
in general, two distinct maximally symmetric vacua with the corresponding
cosmological constants $\Lambda$ given by
\begin{equation}
  \frac{\Lambda - \Lambda_0}{2 \kappa} + \Lambda^2 \bigg( \frac{(n-4)}{(n-2)^2} (n \alpha + \beta)
    + \frac{(n-3) (n-4)}{(n-2) (n-1)} \gamma \bigg) = 0.
\end{equation}

In general relativity, Birkhoff's theorem states that a spherically symmetric
solution of the Einstein field equations in vacuum is locally isometric to the
Schwarzschild solution, consequently, spherically pulsating objects cannot emit
gravitational waves. It has been shown in \cite{Zegers2005} that Birkhoff's
theorem is valid also in the Lovelock theories, namely, that solutions of the
source-free Lovelock field equations with spherical, planar or hyperbolic
symmetry are locally isometric to the corresponding static black hole. Although
the fact that the field equations are of the second order is crucial in the
proof of the Birkhoff's theorem, as has been discussed recently in
\cite{OlivaRay2011}, this theorem can be extended also to higher derivative
theories with the field equations of the fourth order if the traced field
equations are of the second order, i.e.\ the massive spin-0 mode is not present
in the linearized field equations of the theory and if the field equations for
spherically, plane or hyperbolically symmetric spacetimes reduce to the second
order. However, note that in the case of quadratic gravity \eqref{QG:action} the
extended Birkhoff's theorem requires $\alpha = \beta = 0$ and thus the action is
given only by the Gauss--Bonnet term.

Let us also point out that at least certain subclasses of quadratic gravity
possess well-posed initial value formulation. For instance, the Cauchy problem
in the case of Einstein--Gauss--Bonnet gravity, i.e.\ $\alpha = \beta = 0$,
$\gamma \neq 0$, was studied in \cite{ChoquetBruhat1988}. In four dimensions,
effectively with $\gamma = 0$, it has been shown that the Cauchy problem can be
solved for initial data if $\beta \neq 0$ \cite{TeyssandierTourrenc1983}.

As has been already mentioned, although quadratic gravity at the linearized
level around any of two vacua describes massless and massive spin-2 and massive
spin-0 modes, by an appropriate choice of the parameters of the theory, one may
eliminate the massive scalar mode and subsequently ensure that the massive
spin-2 mode becomes massless. Following \cite{Deseretal2011}, the trace of the
linearized field equations for the metric fluctuations
$h_{ab} = \bar{g}_{ab} + g_{ab}$ around an (A)dS vacuum $\bar{g}_{ab}$ reads 
\begin{equation}
  \left[ (4 (n-1) \alpha + n \beta) \bar{\Box}
    - (n-2) \left( \frac{1}{\kappa} + 4 \epsilon \Lambda \right) \right] R^L = 0,
\end{equation}
where $R^L$ is the linearized Ricci scalar and $\epsilon$ is defined in
\eqref{QG:fieldeqns:epsilon}. Setting
\begin{equation}
  4(n-1) \alpha + n \beta = 0,
  \label{QG:critical_point:1}
\end{equation}
we get rid of the massive spin-0 mode provided that
$\kappa \neq - 4 \epsilon \Lambda$ and then, choosing the gauge
$\bar{\nabla}^a h_{ab} = \bar{\nabla}_b h$, the linearized field equations
reduce to \cite{Deseretal2011}
\begin{equation*}
  \left( \bar{\Box} - \frac{4 \Lambda}{(n-1)(n-2)} - M^2 \right)
  \left( \bar{\Box} - \frac{4 \Lambda}{(n-1)(n-2)} \right) h_{ab} = 0.
\end{equation*}
Therefore, vanishing $M^2$ defines the critical point only with the massless
spin-2 excitation
\begin{equation}
  M^2 \equiv - \frac{4\Lambda}{\beta} \left( \frac{1}{4\Lambda\kappa} 
  + \frac{n\alpha}{n-2} + \frac{\beta}{n-2}
  + \frac{(n-3)(n-4)}{(n-1)(n-2)} \gamma \right) = 0.
  \label{QG:critical_point:2}
\end{equation}
Note that the parameters cannot be tuned so that both distinct (A)dS vacua
become simultaneously critical. However, as has been also shown in
\cite{Deseretal2011}, one may still employ the remaining arbitrariness of the
parameters to obtain a theory with only one unique critical (A)dS vacuum. The
cosmological constant of such (A)dS vacuum is then given by
\begin{equation}
  \Lambda = \Lambda_0 = \frac{(n - 1)(n - 2)}{8(n - 3) \kappa \gamma}
\end{equation}
and the action reduces to the form of the Einstein--Weyl gravity
\begin{equation}
  \mathcal{S} = \int \d^n x \, \sqrt{-g} \bigg( \frac{1}{\kappa} \left( R - 2 \Lambda_0 \right)
    + \gamma C_{abcd} C^{abcd} \bigg)
  \label{QG:action:Weyl}
\end{equation}
with only one additional parameter $\gamma$ besides Einstein's constant $\kappa$.
Although at first sight it could seem that all type III and N solutions of the
Einstein gravity are also solutions of the Einstein--Weyl gravity since the last
term in the action \eqref{QG:action:Weyl} vanishes for these Weyl types such
reasoning is incorrect. In the following sections, we will thus study type III
and N solutions of the field equations of general quadratic gravity
\eqref{QG:fieldeqns} in detail. It will be shown, for instance, that apart from
a subclass of type III Einstein spacetimes that are solutions of quadratic
gravity with arbitrary parameters of theory, there exist a subclass of type III
Einstein spacetimes that do not satisfy the field equations if the Gauss--Bonnet
term is present in the action \eqref{QG:action}, i.e.\ $\gamma \neq 0$.


The field equations of quadratic gravity \eqref{QG:fieldeqns} are very complex
and a direct approach to finding exact solutions seems to be hopeless, therefore,
so far the known solutions of quadratic gravity has been obtained mostly by means
of an appropriate ansatz for the metric inspired by the form of known solutions
in the Einstein theory. Apart from the recently found AdS-wave solution of
general quadratic gravity in arbitrary dimension \cite{GulluGursesSismanTekin2011}
using the Kerr--Schild ansatz, at least to the author's knowledge, previously
known exact solutions of quadratic gravity belong only to one of two subclasses
with either $\gamma = 0$ or $\alpha = \beta = 0$.

In the former case $\gamma = 0$, few exact solutions are known. Four-dimensional
plane wave spacetimes have been analyzed in \cite{Madsen1990}. It has turned out
that such solutions with null radiation in a theory with $\beta = 0$ are the same
as in the Einstein gravity, whereas if $\beta \neq 0$, an additional condition is
imposed on the null radiation term. Note that this is in accordance with our
more general results in section \ref{sec:QG:nullradiation}. A perturbative
solution of the field equations has been used in \cite{NetoAguiarAraujo2003} to
compare gravitational waves in the linearized quadratic gravity with those in the
linearized Einstein gravity. It has been shown that the corrections to amplitude
depend on the parameter $\beta$ and the angular frequency of the wave but not
on $\alpha$. Charged black holes were also studied within this subclass of
theories. It has been pointed out in \cite{EconomouLousto1994} that the
four-dimensional Reissner--Nordstr\"om solution of the Einstein field equations
is a solution of quadratic gravity if $\beta = 0$.

In the case $\alpha = \beta = 0$ corresponding to the Einstein--Gauss--Bonnet
gravity, much more exact solutions have been found. The spherically symmetric
solutions in this theory consist of two branches, asymptotically Schwarzschild
black holes with a positive mass parameter and asymptotically Schwarzschild--AdS
spacetimes with a negative one \cite{BoulwareDeser1985}. Both are included in
the more general solution found in \cite{Anabalonetal2009} using the
Kerr--Schild ansatz with an (anti-)de Sitter background metric in the spheroidal
coordinates. However, this solution does not represent rotating black hole
solutions which have been so far studied only numerically \cite{BrihayeRadu2008,
BrihayeKleihausKunzRadu2010,Brihaye2011} or in the limit of small angular
momentum \cite{KimCai2007}.

In the following section, we employ a different strategy to find a solution. In
section \ref{sec:QG:Einstein}, we determine under which conditions known
solutions of the source-free Einstein theory solve also the field equations of
quadratic gravity and in section \ref{sec:QG:nullradiation} we find explicit
solutions of reduced field equations assuming a special form of the Ricci tensor.

\section{Einstein spacetimes}
\label{sec:QG:Einstein}

One of the approaches which may lead to the simplification of the very complex
field equations of quadratic gravity \eqref{QG:fieldeqns} is to assume a
special form of the Ricci tensor. Let us first study the simplest case of such
a form, i.e. Einstein spacetimes
\begin{equation}
  R_{ab} = \frac{2\Lambda}{n-2} g_{ab},
  \label{QG:Einstein:Ricci}
\end{equation}
as exact solutions to quadratic gravity. 

One may substitute the expression of the Riemann tensor in terms of the Weyl
tensor \eqref{intro:Weyl:definition}
\begin{equation}
  R_{abcd} = C_{abcd} + \frac{2}{n-2} (g_{a[c} R_{d]b} - g_{b[c} R_{d]a})
    - \frac{2}{(n-1)(n-2)} R g_{a[c} g_{d]b},
\end{equation}
along with the Ricci tensor \eqref{QG:Einstein:Ricci} and the corresponding
scalar curvature $R = \frac{2n}{n-2}\Lambda$ to the field equations
\eqref{QG:fieldeqns}. Since the Weyl tensor is traceless its contractions with
the metric vanish and \eqref{QG:fieldeqns} reduces to the simple form
\begin{equation}
  \mathcal{B} g_{ab} - \gamma \left( C_a^{\phantom{a}cde} C_{bcde}
    - \frac{1}{4} g_{ab} C^{cdef} C_{cdef} \right) = 0,
  \label{QG:Einstein:fieldeqns}
\end{equation}
with the constant factor $\mathcal{B}$ given only by the effective cosmological
constant $\Lambda$, parameters of theory $\alpha$, $\beta$, $\gamma$, $\kappa$,
$\Lambda_0$ and dimension of spacetime $n$ as
\begin{equation}
  \mathcal{B} = \frac{\Lambda - \Lambda_0}{2 \kappa} + \epsilon \Lambda^2,
  \label{QG:fieldeqns:B}
\end{equation}
where
\begin{equation}
    \epsilon = \frac{(n-4)}{(n-2)^2} (n \alpha + \beta)
      + \frac{(n-3) (n-4)}{(n-2) (n-1)} \gamma. 
  \label{QG:fieldeqns:epsilon}
\end{equation}

Obviously, in the special case $\gamma = 0$, i.e.\ when the Gauss--Bonnet term
is not present in the action \eqref{QG:action}, all Einstein spacetimes
\eqref{QG:Einstein:Ricci} with a cosmological constant $\Lambda$ solve the field
equations of quadratic gravity \eqref{QG:fieldeqns} provided that
$\mathcal{B} = 0$.

The condition $\mathcal{B} = 0$ appears several times in this chapter and, in
fact, it is a quadratic equation determining the effective cosmological constant
$\Lambda$ in terms of the parameters of particular theory $\alpha$, $\beta$,
$\gamma$, $\kappa$, $\Lambda_0$. In four dimensions, $\epsilon = 0$ and then
\eqref{QG:fieldeqns:B} implies that $\mathcal{B} = 0$ admits only one root
$\Lambda = \Lambda_0$. In dimension $n > 4$, there are two possible roots
\begin{equation}
  \Lambda = - \frac{1}{4 \kappa \epsilon} \left( 1 \pm \sqrt{1 + 8 \kappa \epsilon \Lambda_0} \right),
  \label{QG:effective_Lambda}
\end{equation}
where the Einstein constant $\kappa$ in our convention is assumed to be
positive. In the case that $1 + 8 \kappa \epsilon \Lambda_0 = 0$, the unique
solution is $\Lambda = 2 \Lambda_0$. If $\mathcal{B} = 0$ admits two roots, the
following combinations of their signs depending on the parameters of the theory
are possible:
\begin{itemize}
  \item if $\Lambda_0 < 0$ then either
    \begin{enumerate}
      \item $0 < 4 \kappa \epsilon < - \frac{1}{2 \Lambda_0}$ when both roots
	$\Lambda$ are negative, or
      \item $4 \kappa \epsilon < 0$ when one root is positive and the other is
	negative.
  \end{enumerate}
  \item If $\Lambda_0 = 0$ then one of the roots is zero $\Lambda = 0$ and the
    sign of second root $\Lambda= - \frac{1}{2 \kappa \epsilon}$ is either
    positive or negative depending on the sign of $\epsilon$.
  \item If $\Lambda_0 > 0$ then either
    \begin{enumerate}
      \item $0 > 4 \kappa \epsilon > - \frac{1}{2 \Lambda_0}$ when both roots
	are positive, or
      \item $4 \kappa \epsilon > 0$ when one root is positive and the other is
	negative.
    \end{enumerate}
\end{itemize}
Finally, if $1 + 8 \kappa \epsilon \Lambda_0 < 0$ then $\mathcal{B} = 0$ does
not admit any real root. Obviously, the same discussion also applies to the
possible signs of cosmological constants of two conformally flat vacua of
quadratic gravity since in that case the Weyl tensor vanishes.

Note also that in four dimensions the Gauss--Bonnet term is purely topological
and does not contribute to the field equations \eqref{QG:fieldeqns}. The
assumption that the spacetime is Einstein leads to \eqref{QG:Einstein:fieldeqns}
effectively with $\gamma = 0$ and thus $\mathcal{B} = 0$. For $n = 4$, from
\eqref{QG:fieldeqns:epsilon} it follows that $\epsilon = 0$ and then
\eqref{QG:fieldeqns:B} implies that the cosmological constants $\Lambda$ and
$\Lambda_0$ have to be equal. In other words, all four-dimensional Einstein
spaces with $\Lambda = \Lambda_0$ are solutions of the field equations of
quadratic gravity \eqref{QG:fieldeqns}. In fact, it has been already pointed out
in \cite{LuPope2011}, where the four-dimensional action of quadratic gravity has
been studied, that the corresponding field equations reduce for Einstein spaces
to the Einstein field equations. At the level of the Weyl tensor, this can be
seen as a consequence of the identity $C_a^{\phantom{a}cde} C_{bcde} =
\frac{1}{4} g_{ab} C^{cdef} C_{cdef}$ which holds in four dimensions but is not
valid without additional restrictions in dimension $n>4$ \cite{Lovelock1970}.

In the rest of this chapter, we study the conditions under which
\eqref{QG:Einstein:fieldeqns} can be satisfied in a general case $\gamma \neq 0$
in arbitrary dimension. Thus we will look for various classes of spacetimes
satisfying $C_a^{\phantom{a}cde} C_{bcde} = \frac{1}{4} g_{ab} C^{cdef} C_{cdef}$.

\subsection{Type N Einstein spacetimes} 

Now we show that the relation
$C_a^{\phantom{a}cde} C_{bcde} = \frac{1}{4} g_{ab} C^{cdef} C_{cdef}$ holds
for Weyl type N spacetimes since both sides vanish. The Weyl tensor expressed in
the frame \eqref{intro:frame:constraints} has only boost weight $-2$ components
in the decomposition \eqref{intro:Weyl:decomposition}, i.e.\
\begin{equation}
  C_{abcd} = 4 \Omega'_{ij} \ell_{\{ a} {m^i}_b \ell_c {m^j}_{d\}},
  \label{QG:Weyl:typeN}
\end{equation}
where we adopt more compact notation $\Omega'_{ij} \equiv C_{1i1j}$ from
\cite{DurkeePravdaPravdovaReall2010}. Note that the null vector $\bl$ is a
multiple WAND and $\Omega'_{ij}$ is symmetric and traceless. Obviously, the
following contractions of the Weyl tensor appearing in the field equations of
quadratic gravity for Einstein spacetimes \eqref{QG:Einstein:fieldeqns} vanish
\begin{equation}
  C_a^{\phantom{a}cde} C_{bcde} = C^{cdef} C_{cdef} = 0
\end{equation}
and thus we are left with the algebraic constraint $\mathcal{B} = 0$ which,
similarly as in the case of (A)dS vacua, prescribes two possible effective
cosmological constants $\Lambda$ of the solution for given parameters $\alpha$,
$\beta$, $\gamma$, $\kappa$, $\Lambda_0$. Therefore,

\begin{proposition}
  All Weyl type N Einstein spacetimes \eqref{QG:Einstein:Ricci} in arbitrary
  dimension with appropriately chosen effective cosmological constant $\Lambda$
  \eqref{QG:effective_Lambda} are exact solutions of the vacuum field equations
  of quadratic gravity \eqref{QG:fieldeqns}.
  \label{QG:proposition:typeNEinstein}
\end{proposition}

Note that in the special case of Ricci-flat spacetimes $\Lambda = 0$ in the
Gauss--Bonnet gravity $\alpha = \beta = 0$ with vanishing cosmological constant
$\Lambda_0 = 0$ this result has been already pointed out in
\cite{PravdovaPravda2008}.

We can also relate the result of proposition \ref{QG:proposition:typeNEinstein}
with chapter \ref{sec:KS} using the statement of proposition
\ref{KS:proposition:nonexpanding}. It follows that all non-expanding Einstein GKS
spacetimes \eqref{KS:GKSmetric} with a cosmological constant
\eqref{QG:effective_Lambda} are consequently solutions of the field equations of
quadratic gravity.

Let us briefly overview known type N Einstein spacetimes in higher dimensions.
The multiple WAND of such spacetimes is always geodetic. This was shown in
\cite{PravdaPravdovaColeyMilson2004} for the Ricci-flat case where the Bianchi
identity (B.9) in \cite{PravdaPravdovaColeyMilson2004} leads to
$C_{1i1[j} L_{k]0} = 0$ which then implies that the WAND $\ell$ is geodetic.
Although it is obvious from \eqref{KS:C0101}, \eqref{KS:C0i1j} and
\eqref{KS:Cijkl} that the Riemann tensor of type N Einstein spacetimes not only
has $R_{1i1j}$ components as in the Ricci-flat case but also $R_{0101}$,
$R_{0i1j}$ and $R_{ijkl}$ are non-zero. Still, the Bianchi identity (B.9) in
\cite{PravdaPravdovaColeyMilson2004} reduces to the equation
$C_{1i1[j} L_{k]0} = 0$ as well and one may follow the same steps as in
\cite{PravdaPravdovaColeyMilson2004} to show immediately that the WAND is
geodetic. Moreover, without loss of generality, one may assume the geodetic WAND
is affinely parametrized.

In an appropriately chosen frame the optical matrix $L_{ij}$ of type N Einstein
spacetimes consists of just one block $2 \times 2$
\cite{OrtaggioPravdaPravdova2010}
\begin{equation}
  L_{ij} = \left( \begin{array}{c|c}
    \begin{matrix} s & a \\ -a & s \end{matrix} & \hspace{14pt} \mathbf{0} \hspace{14pt} \\ \hline
    \phantom{\bigg[} \mathbf{0} \phantom{\bigg[} & \mathbf{0}
  \end{array} \right).
\end{equation}
From the definition of the optical scalars \eqref{intro:optical_scalars} follows
that the expansion, shear and twist are given by
\begin{equation}
  \theta = \frac{2}{n - 2} s, \qquad 
  \sigma^2 = 2 \frac{n - 4}{n - 2} s^2, \qquad
  \omega^2 = 2 a^2,
  \label{QG:Einstein:typeN:optical_scalars}
\end{equation}
respectively. Type N Einstein spacetimes can be thus further classified
according to the optical properties of the multiple WAND.

\subsubsection{Type N Einstein Kundt spacetimes}

The Kundt class is defined as spacetimes admitting a non-expanding $\theta = 0$,
non-shearing $\sigma^2 = 0$ and non-twisting $\omega^2 = 0$ geodetic null
congruence $\bl$, in other words, the optical matrix vanishes, $L_{ij} = 0$.
Spacetimes belonging to this class can be described by a metric of the form
\cite{ColeyHervikPelavas2005, PodolskyZofka2008}
\begin{equation}
  \d s^2 = 2 \d u \left[ \d v + H(u,v,x^k) \, \d u + W_i(u,v,x^k) \, \d x^i \right]
    + g_{ij}(u,x^k) \, \d x^i \, \d x^j.
  \label{QG:Kundt:metric}
\end{equation}
In general, higher dimensional Einstein Kundt metrics are of Weyl type II or
more special \cite{OrtaggioPravdaPravdova2007} and the functions $H$, $W_i$ and
$g_{ij}$ have not been expressed explicitly in the literature for the Weyl type
N yet. However, the situation differs for Ricci-flat Kundt metrics of types N
and III where one can set \cite{ColeyFusterHervikPelavas2006}
\begin{equation}
  g_{ij}(u,x^k) = \delta_{ij}
  \label{QG:Kundt:VSI}
\end{equation}
and the corresponding functions $W_{i}$ and $H$ for Weyl type N are given in
\eqref{KS:VSI:typeN:epsilon0:W} and \eqref{KS:VSI:typeN:epsilon1:W}. In four
dimensions, all type N Kundt metrics are completely known
\cite{GriffithsPodolsky2009} and given by \eqref{KS:Kundt:4D:metric},
\eqref{KS:Kundt:4D:k} and \eqref{KS:Kundt:4D:Einstein:solution}.

As we have already mentioned, non-expanding Einstein GKS spacetimes
\eqref{KS:GKSmetric} discussed in section \ref{sec:KS:nonexpanding} solve the
field equations of quadratic gravity since they belong to this class of
spacetimes.

Further details and examples of type N Kundt metrics can be found, for instance,
in section \ref{sec:KS:nonexpanding:examples} and in
\cite{ColeyHervikPelavas2005, ColeyFusterHervik2007}. In section
\ref{sec:KS:nonexpanding:warp}, the Brinkmann warp product is used to generate
new Einstein type N Kundt metrics from the known ones.

\subsubsection{Expanding, non-twisting type N Einstein spacetimes}

In four dimensions, expanding $\theta \neq 0$ and non-twisting $\omega^2 = 0$
type N Einstein metrics are necessarily shear-free due to the Goldberg--Sachs
theorem. This fact follows also immediately from 
\eqref{QG:Einstein:typeN:optical_scalars}. Therefore, such metrics belong to the
Robinson--Trautman class and are completely known \cite{GarciaDiazPlebanski1981},
see also \cite{GriffithsPodolsky2009} and references therein.

In contrast, for type N Einstein spacetimes in dimension $n > 4$, it follows
from \eqref{QG:Einstein:typeN:optical_scalars} that non-vanishing expansion
$\theta \neq 0$ implies non-zero shear $\sigma^2 > 0$. Higher dimensional
metrics belonging to this class of spacetimes can be constructed by warping
four-dimensional type N Einstein Robinson--Trautman metrics
\cite{GarciaDiazPlebanski1981}
\begin{equation}
  \begin{aligned}
  \d\tilde s^2 &= -2 \psi \, \d u \, \d r
    + 2 r^2 (\d x^2 + \d y^2)
    - 2 r (2 r f_1 + \epsilon x) \, \d u \, \d x \\
    &\qquad - 2 r (2 r f_2 + \epsilon y) \, \d u \, \d y
    + 2 (\psi B + A) \, \d u^2,
  \end{aligned}
\end{equation}
with 
\begin{equation}
\begin{aligned}
  A &= \frac{1}{4} \epsilon^2 (x^2 + y^2)
    + \epsilon (f_1 x + f_2 y) r
    + (f_1^2 + f_2^2) r^2, \\
  B &= - \frac{1}{2} \epsilon
    - r \partial_x f_1
    + \frac{1}{6} \tilde \Lambda r^2 \psi, \\
  \psi &= 1 + \half \epsilon (x^2 + y^2),
  \end{aligned}
\end{equation}
where $\epsilon=\pm 1$ or $0$, $\tilde \Lambda$ is a four-dimensional
cosmological constant and the functions $f_1=f_1(x,y)$  and $f_2=f_2(x,y)$ are
subject to
\begin{equation}
  \partial_x f_1 = \partial_y f_2, \quad \partial_y f_1 = - \partial_x f_2.
\end{equation}
Then the five-dimensional metric takes one of the forms
\eqref{intro:warp:conformal++}, \eqref{intro:warp:conformal00},
\eqref{intro:warp:conformal--}--\eqref{intro:warp:conformal-+} depending on the
signs of $\Lambda$ and $\tilde \Lambda$, where $\lambda = \frac{\Lambda}{6}$ and
the five-dimensional cosmological constant $\Lambda$ obeys
$|\Lambda| = 2 |\tilde \Lambda|$. Whereas, in the case $\Lambda = 0$,
$\tilde \Lambda > 0$, the metric is given by \eqref{intro:warp:conformal-0} with
$\tilde \Lambda = 3$.

\subsubsection{Twisting type N Einstein spacetimes}

Very few four-dimensional exact solutions of Einstein gravity within this class
are known. This includes the Ricci-flat Hauser metric \cite{Hauser1974} and the
Leroy metric \cite{Leroy1970} for a negative cosmological constant, see also
\cite{StephaniKramer2003}.

As in the previous case, higher dimensional solutions in this class can be
constructed from four-dimensional twisting solutions using the Brinkmann warp
product. An example of such warped spacetimes can be obtained using the Leroy
metric in the form given in \cite{OrtaggioPravdaPravdova2010} as a seed. Since
the cosmological constant of the seed metric is negative, there is just one
possibility of the sign of the cosmological constant of the warped metric and
thus we employ \eqref{intro:warp:conformal--}. After the substitution
$\tilde z = \sqrt{-\lambda} z$, the five-dimensional metric reads
\begin{equation}
  \begin{split}
    \d \tilde{s}^2 &=  \frac{1}{- \Lambda y^2 \cos^{2} \tilde z} \bigg[
    \frac{2}{3} (\d x + y^3 \, \d u) \big[6 y \, \d r + y^3 (1 - r^2) \, \d u \\
    &\qquad + (13 - r^2) \, \d x + 12 r \, \d y \big]
    + 3 (r^2 + 1) (\d x^2 + \d y^2)
    + 6 \d \tilde z^2 \bigg],
  \end{split}
\end{equation}
where $\Lambda = 6 \lambda < 0$ is a five-dimensional cosmological constant.

\subsection{Type III Einstein spacetimes}
\label{sec:QG:Einstein:typeIII}

In general, for type III Einstein spacetimes the term
$C_a^{\phantom{a}cde} C_{bcde} - \frac{1}{4} g_{ab} C^{cdef} C_{cdef}$ in
\eqref{QG:Einstein:fieldeqns} does not vanish as in the case of Weyl type N.
Using the compact notation \cite{DurkeePravdaPravdovaReall2010}
\begin{equation}
  \Psi_i' \equiv C_{101i}, \qquad
  \Psi_{ijk}' \equiv C_{1ijk}, \qquad
  \Omega_{ij}' \equiv C_{1i1j},
\end{equation}
where 
\begin{equation}
  \Psi'_{ijk} = - \Psi'_{ikj}, \qquad \Psi'_{[ijk]}=0, \qquad \Psi'_i = \Psi'_{kik},
  \label{QG:Psi:symmetries}
\end{equation}
the Weyl tensor of type III can be expressed from
\eqref{intro:Weyl:decomposition} as
\begin{equation}
  C_{abcd} = 8 \Psi'_i  \ell_{\{a} n_b \ell_c m^i_{d\}}
    + 4 \Psi'_{ijk} \ell_{\{a} m^i_b m^j_c m^k_{d \}}
    + 4 \Omega'_{ij} \ell_{\{a} m^i_b \ell_c m^j_{d\}}.
\end{equation}
Obviously, $C^{cdef} C_{cdef}$ vanishes and if we define $\tilde \Psi$ as
\begin{equation}
  \tilde \Psi \equiv \half \Psi'_{ijk} \Psi'_{ijk} - \Psi'_i \Psi'_i,
  \label{QG:Einstein:typeIII:tildePsi}
\end{equation}
then
\begin{equation}
  C_a^{\phantom{a}cde} C_{bcde} = \tilde \Psi \ell_a \ell_b.
  \label{QG:Einstein:typeIII:Weylcontraction}
\end{equation}
Therefore, in the case of Weyl type III, one may further simplify the field
equations of quadratic gravity for Einstein spaces \eqref{QG:Einstein:fieldeqns}
to the form
\begin{equation}
  \mathcal{B} g_{ab} - \gamma \tilde \Psi \ell_a \ell_b = 0.
  \label{QG:Einstein:fieldeqns:typeIII}
\end{equation}
The trace of \eqref{QG:Einstein:fieldeqns:typeIII} implies $\mathcal{B} = 0$,
which again determines two possible effective cosmological constants $\Lambda$
for the given parameters of theory, and subsequently it remains to satisfy
$\tilde \Psi = 0$.

\begin{proposition}
  Weyl type III Einstein spacetimes with an effective cosmological constant
  $\Lambda$ subject to \eqref{QG:effective_Lambda} are exact solutions of the
  vacuum field equations of quadratic gravity \eqref{QG:fieldeqns} if and only
  if $\tilde \Psi = 0$.
  \label{QG:proposition:typeIIIEinstein}
\end{proposition}

From \eqref{QG:Psi:symmetries} it follows that in four dimensions
$\tilde \Psi = 0$. This is in agreement with the already mentioned statement
that all four-dimensional Einstein spacetimes with an appropriate cosmological
constant are solutions of quadratic gravity since in this case effectively
$\gamma = 0$ in \eqref{QG:Einstein:fieldeqns}.

A wide class of higher dimensional Einstein spacetimes with $\tilde \Psi = 0$
can be obtained by using the Brinkmann warp product. In order to show that we
start with the following observation. The components of the Weyl tensor of the
seed $\d\tilde s^2$ and warped metric $\d s^2$ \eqref{intro:warp:warpedmetric}
expressed in coordinates $x^a = (z,x^\mu)$ are related by
\cite{OrtaggioPravdaPravdova2010b}
\begin{equation}
  C_{\mu \nu \rho \sigma} = f {\tilde C}_{\mu \nu \rho \sigma}, \qquad
  C_{z \mu \nu \rho} = C_{z \mu z \nu} = 0.
\end{equation}
Raising the indices by the corresponding metrics, it then follows that the
contractions $C_a^{\phantom{a}cde} C_{bcde}$ are given by
\begin{equation}
  C_\mu^{\phantom{\mu}  \nu \rho \sigma} C_{\tau \nu \rho \sigma} =
    \frac{1}{f} {\tilde C}_\mu^{\phantom{\mu}  \nu \rho \sigma} {\tilde C}_{\tau \nu \rho \sigma},
\end{equation}
with all $z$-components being zero.

Let us also recall that the Brinkmann warp product preserves the Weyl type of
algebraically special spacetimes. Therefore, if one takes an arbitrary
four-dimensional type III Einstein metric as a seed, for which
${\tilde C}_\mu^{\phantom{\mu}  \nu \rho \sigma} {\tilde C}_{\tau \nu \rho \sigma} = 0$
holds identically, then the warped metric represents a type III Einstein
spacetime with $\tilde \Psi = 0$ and thus, by proposition
\ref{QG:proposition:typeIIIEinstein}, it is also an exact solution of the field
equations of quadratic gravity \eqref{QG:fieldeqns} provided that the effective
cosmological constant $\Lambda$ satisfies \eqref{QG:effective_Lambda}. A few
examples of such twisting and non-twisting type III Einstein spacetimes obtained
by the Brinkmann warp product are given in \cite{OrtaggioPravdaPravdova2010}.

It should be emphasized that in contrast with the type N case, there exist type
III Einstein spacetimes which are not solutions of quadratic gravity. For
instance, $\tilde \Psi$ is clearly non-vanishing for the type III(a) subclass of
type III spacetimes characterized by $\Psi'_i = 0$ \cite{Coleyetal2004}. Type
III(a) Kundt spacetimes with null radiation given in
\cite{ColeyFusterHervikPelavas2006} contain type III(a) Ricci-flat metrics
\eqref{QG:Kundt:metric}, \eqref{QG:Kundt:VSI} with a covariantly constant null
vector, where
\begin{gather}
  H = H(u, x^k), \qquad
  W_2 = 0, \qquad
  W_{\itilde} = W_{\itilde}(u, x^k), \phantom{\half} \notag \\
  W^{\itilde}_{\phantom{\itilde},\itilde 2} = 0, \qquad
  W^{\itilde}_{\phantom{\itilde},\itilde \jtilde} = \Delta W_{\jtilde}, \phantom{\half} \\
  \Delta H - \frac{1}{4} (W_{\itilde,\jtilde} - W_{\jtilde,\itilde})(W^{\itilde,\jtilde} - W^{\jtilde,\itilde})
    - W^{\itilde}_{\phantom{\itilde},\itilde u} = 0. \notag
\end{gather}
Here the indices $\itilde$, $\jtilde$ run from 3 to $n-1$ and necessarily at
least one of the components $C_{1k \jtilde \itilde} = \half (W_{\itilde,\jtilde}
- W_{\jtilde,\itilde})_{,k}$ has to be non-vanishing, otherwise the metric
reduces to Weyl type N. An explicit five-dimensional example of such pp-wave
metric is given by \cite{OrtaggioPravdaPravdova2008}
\begin{gather}
  W_2 = 0, \qquad
  W_3 = h(u) x^2 x^4, \qquad
  W_4 = h(u) x^2 x^3,\\
  H = H_0 = h(u)^2 \left[ \frac{1}{24} \left( \left( x^3 \right)^4 + \left( x^4 \right)^4 \right)
    + h^0(x^2,x^3,x^4) \right],
\end{gather}
where $h^0(x^2,x^3,x^4)$ is subject to $\Delta h^0 = 0$. In fact, all type III
Ricci-flat pp-waves belong to the type III(a) subclass since the existence of
the covariantly constant null vector $\bl$ implies $C_{abcd} \ell^a =0$ and thus
$\Psi'_i$ vanishes. Therefore, type III Ricci-flat pp-waves are not solutions of
quadratic gravity.  

Motivated by the above results we naturally introduce two new subclasses of the
principal Weyl type III, namely type III(A) characterized by $\tilde \Psi \not=0$
and type III(B) defined by $\tilde \Psi=0$. Obviously, type III(a) is a subclass
of type III(A) since, as already mentioned,
$\tilde \Psi = \half \Psi'_{ijk} \Psi'_{ijk} \neq 0$ in this case.

\subsection{Comparison with other classes of spacetimes}
\label{sec:QG:Einstein:comparison}

It is of interest to compare the set of exact solutions of quadratic gravity
(QG) with other overlapping classes of spacetimes. Namely, spacetimes with
vanishing curvature invariants (VSI) \cite{ColeyMilsonPravdaPravdova2004},
spacetimes with constant curvature invariants (CSI)
\cite{ColeyHervikPelavas2005}, Kundt subclass of CSI (KCSI), \pp waves which
will be denoted as ppN, ppIII, etc., depending on a particular Weyl type,
and universal metrics (U) for which quantum corrections, i.e.\ all symmetric
covariantly conserved tensors of rank 2 constructed from the metric, Riemann
tensor and its covariant derivatives, are a multiple of the metric
\cite{ColeyGibbonsHervikPope2008,ColeyHervik2011}. Einstein or Ricci-flat
subclasses of these sets will be indicated by the appropriate subscript, for
instance, QG$_{\text{E}}$ and QG$_{\text{RF}}$.

The class of \pp waves is defined geometrically as spacetimes admitting a
covariantly constant null vector field, say $\bl$, and thus $\ell_{a;b} = 0$. It
then follows from the definition of the Riemann tensor that
$R_{abcd} \ell^a = 0$. The contraction with respect to the second and fourth
index yields $R_{0b} = 0$, where we identify $\bl$ with the corresponding null
frame vector \eqref{intro:frame:constraints}. On the other hand, in the case of
Einstein spaces, the frame component of the Ricci tensor $R_{01}$ is
proportional to the cosmological constant $R_{01} = \frac{2 \Lambda}{n - 2}$ and
therefore Einstein \pp waves do not admit non-vanishing $\Lambda$, i.e.\
pp$_\text{E}$ = pp$_\text{RF}$. In the Ricci-flat case, the Weyl tensor is given
exactly by the Riemann tensor and thus for pp$_\text{RF}$ we obtain
$C_{abcd} \ell^a = 0$. In four dimensions this corresponds to the Bel criterion
ensuring that the spacetime is of type N, whereas, in higher dimensions, it
implies that the spacetime is of Weyl type II or more special. 

We will consider $n>4$ since in four dimensions all pp$_{\text{RF}}$ are of type
N and as discussed above all Einstein spacetimes with $\Lambda = \Lambda_0$
belong to QG which leads to a considerable simplification.

From the definition of U, it is obvious that U $\subset$ QG since the action of
quadratic gravity \eqref{QG:action} contains quantum corrections only up to the
second order in curvature. The results of \cite{ColeyMilsonPravdaPravdova2004}
that VSI spacetimes are of Weyl type III or more special and admit a congruence
of non-expanding, non-shearing and non-twisting null geodesics and thus belong
to the Kundt class imply VSI $\subset$ KCSI. All pp-waves from the set
ppN$_{\text{RF}}$ $\cup$ ppIII$_{\text{RF}}$ belong to VSI, whereas, as was shown
above, ppIII$_{\text{RF}}$ $\cap$ QG is $\emptyset$ and therefore
VSI$_{\text{RF}}$ $\nsubseteq$ QG and pp$_{\text{RF}}$ $\nsubseteq$ QG.
It also holds that pp$_{\text{RF}}$ $\nsubseteq$ VSI since ppII$_{\text{RF}}$
solutions exist in higher dimensions.

Recently, it was conjectured in \cite{ColeyHervik2011} that U $\subset$ KCSI.
Obviously, ppIII$_{\text{RF}}$ are examples of spacetimes which are VSI and thus
KCSI but not U. However, note that QG$_{\text{E}}$ $\nsubseteq$ CSI since
examples of QG$_{\text{E}}$ metrics with non-vanishing expansion mentioned in
this section have in general non-trivial curvature invariants
\cite{OrtaggioPravdaPravdova2010}.

\section{Spacetimes with aligned null radiation}
\label{sec:QG:nullradiation}

One may attempt to find a wider class of solutions of quadratic gravity
considering more general form of the Ricci tensor than for Einstein spacetimes
\eqref{QG:Einstein:Ricci} but still sufficiently simple to considerably reduce
the field equations \eqref{QG:fieldeqns}. Therefore, we will assume that the
Ricci tensor contains an additional aligned null radiation term
\begin{equation}
  R_{ab} = \frac{2\Lambda}{n-2} g_{ab} + \Phi \ell_a \ell_b.
  \label{QG:nullradiation:Ricci}
\end{equation}
Then the contracted Bianchi identities $\nabla^a R_{ab} = \frac{1}{2} \nabla_b R$
imply that the null radiation term has to be covariantly conserved
$(\Phi \ell^a \ell^b)_{;a} = 0$, which one may rewrite using the derivatives
\eqref{intro:frame:derivatives} and the optical scalars
\eqref{intro:optical_scalars:ell} as
\begin{equation} 
  \left[ \D \Phi + \Phi (n-2) \theta \right] \ell_a + \Phi \ell_{a;b} \ell^b = 0.
  \label{QG:nullradiation:conservation}
\end{equation}
We identify $\bl$ with the corresponding null frame vector
\eqref{intro:frame:constraints} so that the contraction of
\eqref{QG:nullradiation:conservation} with the vector $\bm^{(i)}$ implies
$L_{i0} = 0$ in terms of the Ricci rotation coefficients
\eqref{intro:Riccicoeff}. Therefore, $\bl$ has to be geodetic and, without loss
of generality, we choose $\bl$ to be affinely parametrized. Consequently, the
contraction of \eqref{QG:nullradiation:conservation} with the frame vector $\bn$
yields
\begin{equation}
  \D \Phi = - (n-2) \theta \Phi.
  \label{QG:nullradiation:Phi}
\end{equation}

Now, following the same steps as for Einstein spaces in section
\ref{sec:QG:Einstein}, we express the field equations \eqref{QG:fieldeqns} in
terms of the Weyl tensor and the Ricci tensor \eqref{QG:nullradiation:Ricci} and
simplify them using the tracelessness of the Weyl tensor and the fact that $\bl$
is a null vector
\begin{equation}
  \begin{aligned}
  &(\beta \Box + \mathcal{A} ) ( \Phi \ell_a \ell_b )
    - 2 \mathcal{B} g_{ab}
    + 2 \gamma \left( C_a^{\phantom{a}cde} C_{bcde}
    - \frac{1}{4} g_{ab} C^{cdef} C_{cdef} \right) \\
    &\qquad + 2 \Phi \left( \beta - 2 \frac{n - 4}{n - 2} \gamma \right) C_{acbd} \ell^c \ell^d = 0,
  \end{aligned}
  \label{QG:fieldeqns:nullradiation}
\end{equation}
where $\mathcal{A}$ is defined as
\begin{equation}
  \mathcal{A} = \frac{1}{\kappa}
  + 4 \Lambda \bigg( \frac{n \alpha}{n-2}
  + \frac{\beta}{n-1}
  + \frac{(n-3)(n-4)}{(n-2)(n-1)} \gamma \bigg)
  \label{QG:fieldeqns:A}
\end{equation}
and $\mathcal{B}$ is given by \eqref{QG:fieldeqns:B}. If the Weyl tensor is of
type III or more special then the last term in \eqref{QG:fieldeqns:nullradiation}
vanishes and one may rewrite \eqref{QG:fieldeqns:nullradiation} using
$\tilde{\Psi}$ \eqref{QG:Einstein:typeIII:Weylcontraction} as
\begin{equation}
  (\beta \Box + \mathcal{A} ) ( \Phi \ell_a \ell_b )
    - 2 \mathcal{B} g_{ab}
    + 2 \gamma \tilde{\Psi} \ell_a \ell_b = 0.
  \label{QG:fieldeqns:nullradiation:typeIII}
\end{equation}

From now on, we restrict ourselves to spacetimes with $\tilde{\Psi} = 0$. The
trace of \eqref{QG:fieldeqns:nullradiation:typeIII} yields $\mathcal{B}=0$ which
again determines two possible effective cosmological constants $\Lambda$ via
\eqref{QG:effective_Lambda}. The remaining part of
\eqref{QG:fieldeqns:nullradiation:typeIII} reads
\begin{equation}
  (\beta \Box + \mathcal{A} ) ( \Phi \ell_a \ell_b ) = 0.
  \label{QG:fieldeqns:nullradiation:typeN}
\end{equation}
Using the notion of the subclasses III(A) and III(B) of the Weyl type III
defined at the end of section \ref{sec:QG:Einstein:typeIII}, we arrive to

\begin{proposition}
  \label{QG:proposition:nullrad}
  All spacetimes of Weyl types III(B), N and O with the Ricci tensor of the
  form \eqref{QG:nullradiation:Ricci} are vacuum solutions of quadratic gravity
  provided that $\mathcal{B} = 0$ and the null radiation term
  $\Phi \ell_a \ell_b$ satisfies \eqref{QG:fieldeqns:nullradiation:typeN}.
\end{proposition}
It should be emphasized that these spacetimes with a null radiation term
in the Ricci tensor, i.e.\ solutions of the non-vacuum Einstein field equations,
are solutions of the vacuum field equations of quadratic gravity without any
matter field terms in the action.

Let us briefly comment the special case $\beta = 0$. Then it follows that both
$\mathcal A$ \eqref{QG:fieldeqns:A} and $\mathcal B$ \eqref{QG:fieldeqns:B} have
to vanish and from these relations, one may eliminate the parameter $\gamma$ to
obtain
\begin{equation}
  \frac{8 n \alpha \kappa}{(n - 2)^2} \Lambda^2 - \Lambda + 2 \Lambda_0 = 0.
  \label{QG:fieldeqns:nullradiation:beta=0}
\end{equation}
Therefore, the effective cosmological constant $\Lambda$ is determined only by
the parameters $\alpha$, $\kappa$ and $\Lambda_0$. If the constraint on $\Lambda$
\eqref{QG:fieldeqns:nullradiation:beta=0} admits a real solution then the
remaining parameter $\gamma$ is subject to $\mathcal{A} = 0$ or
$\mathcal{B} = 0$. In other words, for special values of the parameters $\alpha$,
$\gamma$, $\kappa$ and $\Lambda_0$ of a theory with $\beta = 0$, all spacetimes
of Weyl types III(B), N and O with the Ricci tensor of the form
\eqref{QG:nullradiation:Ricci} with an arbitrary $\Phi$ and an effective
cosmological constant $\Lambda$ given by
\eqref{QG:fieldeqns:nullradiation:beta=0} are exact solutions of quadratic
gravity. However, none of such spacetimes with null radiation satisfies the
field equations of quadratic gravity if $\mathcal{A} \neq 0$, which occurs, for
instance, in the case $\Lambda = 0$. For the pure Gauss--Bonnet gravity
$\alpha = \beta = 0$, this implies that if $\Lambda = 2\Lambda_0$ following from
\eqref{QG:fieldeqns:nullradiation:beta=0} and simultaneously
\begin{equation}
  \gamma = - \frac{(n-2)(n-1)}{(n-3)(n-4)} \frac{1}{8\Lambda_0 \kappa},
  \label{QG:fieldeqns:nullradiation:GB:gamma}
\end{equation}
then both $\mathcal{A}$, $\mathcal{B}$ vanish and $\Phi$ can be arbitrary,
otherwise $\Phi$ has to be zero in order to satisfy the field equations.

Since we are interested in solutions of quadratic gravity with arbitrary
parameters we assume $\beta \neq 0$ in the rest of this chapter. The
contraction of \eqref{QG:fieldeqns:nullradiation:typeN} with the frame vectors
$\bl$ and $\bn$ gives
\begin{equation}
  \Phi L_{ij} L_{ij} = \Phi [(n-2) \theta^2 + \sigma^2 + \omega^2] = 0,
\end{equation}
where the optical matrix $L_{ij}$ is defined in \eqref{intro:Riccicoeff} and
the scalars $\theta$, $\sigma$ and $\omega$ \eqref{intro:optical_scalars}
correspond to expansion, shear, and twist, respectively. This implies that the
geodetic vector $\bl$ is non-expanding, $\theta=0$, non-shearing, $\sigma=0$,
and non-twisting, $\omega = 0$, and thus the optical matrix vanishes
$L_{ij} = 0$. Then it immediately follows from \eqref{QG:nullradiation:Phi}
that $\mathrm{D} \Phi = 0$, i.e.\ $\Phi$ does not depend on an affine parameter
along the null geodesics $\bl$.

\begin{proposition}
  \label{QG:proposition:nullrad:Kundt}
  All Weyl type III(B), N or conformally flat solutions of the source-free field
  equations of quadratic gravity \eqref{QG:fieldeqns} with $\beta \neq 0$ and
  the Ricci tensor of the form \eqref{QG:nullradiation:Ricci} belong to the
  Kundt class.
\end{proposition}
Note that in general, not only in quadratic gravity, conformally flat spacetimes
with null radiation belong to the Kundt class as follows directly from the
Bianchi identities.

Contracting \eqref{QG:fieldeqns:nullradiation:typeN} twice with the frame vector
$\bn$ and substituting $L_{a0} = 0$, i.e.\ $\bl$ being geodetic and affinely
parametrized, $L_{ij} = 0$ and $\D \Phi = 0$, we obtain the remaining non-trivial
frame component of \eqref{QG:fieldeqns:nullradiation:typeN}
\begin{equation}
  \begin{aligned}
  &\delta_i \delta_i \Phi
    + \left(2 L_{1i} + 4 L_{[1i]} + \M{i}{jj} \right) \delta_i \Phi
    + 2 \Phi \bigg( 2 \delta_i L_{[1i]} + L_{i1} L_{i1} \\
  &\qquad+ 4 L_{1i} L_{[1i]} + 2 L_{[1i]} \M{i}{jj} \bigg)
    + \frac{4 \Lambda \Phi}{n-2} 
    + \mathcal{A} \beta^{-1} \Phi = 0,
  \end{aligned}
  \label{QG:fieldeqns:nullradiation:typeN:comp11}
\end{equation}
where we also employed the Ricci identities \cite{OrtaggioPravdaPravdova2007}
and the commutators of the derivatives along the frame vectors
\eqref{intro:commutators}. The Ricci rotation coefficients $L_{1i}$, $L_{i1}$
and $\M{i}{jk}$ are defined in \eqref{intro:Riccicoeff}.

It has been shown in section \ref{sec:KS:nonexpanding} that one may always set
the frame in Kundt spacetimes so that $L_{[1i]} = 0$, $L_{12} \neq 0$,
$L_{1\itilde} = 0$. For Kundt metrics in the canonical form
\eqref{QG:Kundt:metric}, the natural frame with the geodetic WAND
$\ell_a \d x^a = \d u$ as the null frame vector $\bl$ implies that
the antisymmetric part of $L_{1i}$ vanishes
\begin{equation}
  L_{[1i]} = \ell_{[a;b]} n^a m^b_{(i)} =
    \ell_{[a,b]} n^a m^b_{(i)} - \Gamma^c_{[ab]} \ell_c n^a m^b_{(i)} = 0,
    \label{QG:Kundt:L1i}
\end{equation}
since $\ell_{a,b} = 0$ in these coordinates and the Christoffel symbols are
symmetric in the lower indices. However, $L_{\itilde 0}$ are in general
non-vanishing in this frame. Let us emphasize that this result does not
depend on the choice of the frame vectors $\bn$ and $\bm^{(i)}$ and it can be
used for further simplification of
\eqref{QG:fieldeqns:nullradiation:typeN:comp11}. Moreover, one may rewrite
$\delta_i \delta_i \Phi$ in terms of the d'Alembert operator to get
\begin{equation}
  \Box \Phi
  + 4 L_{1i} \delta_i \Phi
  + 2 L_{1i} L_{1i} \Phi 
  + \frac{4 \Lambda \Phi}{n-2} 
  + \mathcal{A} \beta^{-1} \Phi = 0.
  \label{QG:fieldeqns:nullradiation:typeN:comp11:dalembert}
\end{equation}

Note that, as argued under proposition \ref{KS:proposition:nonexpanding},
non-expanding GKS spacetimes \eqref{KS:GKSmetric} with the Ricci tensor of the
form \eqref{QG:nullradiation:Ricci} belong to the Kundt class of Weyl type N.
Therefore, such spacetimes with an appropriate effective cosmological constant
$\Lambda$ solve the field equations of quadratic gravity provided that the null
radiation term $\Phi \ell_a \ell_b$ satisfies
\eqref{QG:fieldeqns:nullradiation:typeN} or equivalently
\eqref{QG:fieldeqns:nullradiation:typeN:comp11:dalembert} in the case
$L_{[1i]} = 0$.

Interestingly, if we assume that the function $\H$ of non-expanding GKS
spacetimes with aligned null radiation is independent on an affine parameter
along null geodesics of the Kerr--Schild congruence $\bk \equiv \bl$, i.e.\
$\D\H = 0$, and set $L_{[1i]} = 0$, the corresponding Einstein field equations
\eqref{KS:nonexpanding:EFE01}, \eqref{KS:nonexpanding:EFE1i},
\eqref{KS:nullrad:EFE11} reduce to
\begin{equation}
  \Box \mathcal{H}
    + 4 L_{1i} \delta_i \mathcal{H}
    + 2 L_{1i} L_{1i} \mathcal{H}
    + \frac{4 \Lambda \mathcal{H}}{n-1} = \Phi.
  \label{QG:GKS:nullradiation:EFEs}
\end{equation}
Let us point out the similarity of
\eqref{QG:fieldeqns:nullradiation:typeN:comp11:dalembert} and
\eqref{QG:GKS:nullradiation:EFEs} which effectively decouples these two equations
and thus allows us to solve for $\H$ and $\Phi$ independently.

\subsection{Explicit solutions of Weyl type N}
\label{sec:QG:nullradiation:examples}

In this section, we present a few examples of type N solutions of the field
equations of quadratic gravity with the Ricci tensor of the form
\eqref{QG:nullradiation:Ricci}. We solve the cases with a vanishing and
non-vanishing effective cosmological constant $\Lambda$ separately since the
form of all higher dimensional type N Kundt metrics with $\Lambda = 0$ and
aligned null radiation is explicitly known \cite{ColeyFusterHervikPelavas2006},
whereas, in the case $\Lambda \neq 0$ we employ at least a particular example
of such Kundt metrics known as the Siklos metric.

\subsubsection{Case $\Lambda=0$}

Type N Kundt metrics with aligned null radiation and the vanishing cosmological
constant $\Lambda$ which belong to a subclass of VSI spacetimes, admit the
form \eqref{QG:Kundt:metric}, \eqref{QG:Kundt:VSI}. As in the Ricci-flat case
discussed in section \ref{sec:KS:nonexpanding:examples}, these metrics can be
split into two subclasses with vanishing ($\epsilon=0$) and non-vanishing
($\epsilon=1$) quantity $L_{1i} L_{1i}$. One may choose the same null frame as
in \eqref{KS:VSI:frame} and, consequently, the Ricci rotation coefficients
$L_{1i}$, $L_{i1}$ and $L_{11}$ are given as in \eqref{KS:VSI:Riccicoeff}.

The constraints on the undetermined metric functions $W_i$ and $H$ imposed by
the form of the Weyl tensor \eqref{QG:Weyl:typeN} and the Ricci tensor
\eqref{QG:nullradiation:Ricci} differ from the Ricci-flat case since the null
radiation term $\Phi$ now occurs \cite{ColeyFusterHervikPelavas2006}
\begin{equation}
  \begin{split}
  &W_2 = 0, \qquad
  W_{\tilde{\imath}} = x^2 C_{\tilde{\imath}}(u)
    + x^{\tilde{\jmath}} B_{\tilde{\jmath}\tilde{\imath}}(u), \phantom{\half} \\
  &H = H^0(u,x^i), \qquad
  \Delta H^0 - \frac{1}{2} \sum C^2_{\tilde{\imath}}
    - 2 \sum_{\tilde{\imath}<\tilde{\jmath}} B^2_{\tilde{\imath}\tilde{\jmath}} + \Phi = 0
  \end{split}
  \label{QG:VSI:EFEs:0}
\end{equation}
in the case $\epsilon=0$ and
\begin{equation}
  \begin{split}
  &W_2 = - \frac{{{2}} v}{x^2}, \qquad
  W_{\tilde{\imath}} = C_{\tilde{\imath}}(u)
    + x^{\tilde{\jmath}} B_{\tilde{\jmath}\tilde{\imath}}(u), \qquad
  H = \frac{v^2}{2(x^2)^2} + H^0(u,x^i), \\
  &x^2 \Delta \left(\frac{H^0}{x^2}\right)
    - \frac{1}{(x^2)^2} \sum W^2_{\tilde{\imath}}
    - 2 \sum_{\tilde{\imath}<\tilde{\jmath}} B^2_{\tilde{\imath}\tilde{\jmath}} + \Phi = 0
  \end{split}
  \label{QG:VSI:EFEs:1}
\end{equation}
in the case $\epsilon=1$, respectively. Where the indices $\tilde{\imath}$,
$\tilde{\jmath}$, \dots range from 3 to $n - 1$ and
$B_{[\tilde{\imath}\tilde{\jmath}]}=0$ in both cases.

Now we determine the form of the function $\Phi$. Obviously,
$\Phi_{,v} = \D \Phi= 0$ and thus $\Phi$ does not depend on the coordinate $v$.
In the Einstein gravity, no further conditions are imposed on $\Phi$, whereas
in quadratic gravity theory with $\beta \neq 0$, $\Phi$ still has to satisfy
\eqref{QG:fieldeqns:nullradiation:typeN} or equivalently
\eqref{QG:fieldeqns:nullradiation:typeN:comp11:dalembert} since $L_{[1i]} = 0$
in our chosen frame. The latter one is simpler to express and leads directly to
\begin{equation}
  \Phi_{,ii} - \frac{2 \epsilon}{x^2} \Phi_{,2}
    + \frac{2 \epsilon}{(x^2)^2} \Phi + (\kappa\beta)^{-1} \Phi = 0.
  \label{QG:examples:VSI:Phi}
\end{equation}
In the case $\epsilon = 0$, \eqref{QG:examples:VSI:Phi} reads
\begin{equation}
  \Delta \Phi + (\kappa \beta)^{-1} \Phi = 0.
  \label{QG:examples:VSI:Phi:0}
\end{equation}
Substituting \eqref{QG:examples:VSI:Phi:0} into \eqref{QG:VSI:EFEs:0}, we obtain 
\begin{equation}
  \Delta H^0_{\text{vac}} - \frac{1}{2} \sum C^2_{\tilde{\imath}}
    - 2 \sum_{\tilde{\imath}<\tilde{\jmath}} B^2_{\tilde{\imath}\tilde{\jmath}} = 0,
  \label{QG:VSI:vacuumEFEs:0}
\end{equation}
where we define $H^0_{\text{vac}} = H^0 - \kappa \beta \Phi$. Thus,
$H^0_{\text{vac}}$ corresponds to a vacuum VSI solution of the Einstein gravity,
i.e.\ it obeys \eqref{QG:VSI:EFEs:0} with $\Phi=0$.

Similarly, in the case $\epsilon = 1$, \eqref{QG:examples:VSI:Phi} reads
\begin{equation}
  x^2 \Delta \left( \frac{\Phi}{x^2} \right) + (\kappa \beta)^{-1} \Phi = 0.
  \label{QG:examples:VSI:Phi:1}
\end{equation}
Putting \eqref{QG:examples:VSI:Phi:1} to \eqref{QG:VSI:EFEs:1} and denoting
$H^0_{\text{vac}} = H^0 - \kappa \beta \Phi$ gives the Einstein field equations
for Ricci-flat type N VSI metrics, i.e.\ \eqref{QG:VSI:EFEs:1} with $\Phi = 0$
for $H^0_{\text{vac}}$
\begin{equation}
  x^2 \Delta \left(\frac{H^0_{\mathrm{vac}}}{x^2}\right)
    - \frac{1}{(x^2)^2} \sum W^2_{\tilde{\imath}}
    - 2 \sum_{\tilde{\imath}<\tilde{\jmath}} B^2_{\tilde{\imath}\tilde{\jmath}} = 0.
  \label{QG:VSI:vacuumEFEs:1}
\end{equation}

In other words, these steps may be performed backwards. We take an arbitrary
vacuum type N VSI metric \eqref{QG:Kundt:metric}, \eqref{QG:Kundt:VSI} with
$H^0_{\text{vac}}$, $W_i$ and $\Phi = 0$, i.e.\ a solution of the Einstein field
equations \eqref{QG:VSI:vacuumEFEs:0} or \eqref{QG:VSI:vacuumEFEs:1}, and
independently find $\Phi$ solving the corresponding equation
\eqref{QG:examples:VSI:Phi:0} or \eqref{QG:examples:VSI:Phi:1}, respectively.
Therefore, we finally arrive at a solution of the field equations of quadratic
gravity \eqref{QG:fieldeqns} represented by the metric \eqref{QG:Kundt:metric},
\eqref{QG:Kundt:VSI} with $\Phi \neq 0$,
$H^0 = H^0_{\mathrm{vac}} + \kappa \beta \Phi$ and $W_i$ unchanged.

Note that in this case where we assume $\Lambda = 0$ the condition
$\mathcal{B} = 0$ \eqref{QG:fieldeqns:B} determining the effective cosmological
constant $\Lambda$ in terms of the parameters $\alpha$, $\beta$, $\gamma$,
$\kappa$ and $\Lambda_0$ implies $\Lambda_0 = 0$ and therefore we are not able
to satisfy the criticality condition \eqref{QG:critical_point:1},
\eqref{QG:critical_point:2} by tuning the remaining parameters.

\subsubsection{Case $\Lambda \not=0$}

One may perform a similar procedure also in the case of a non-vanishing
effective cosmological constant $\Lambda$. Unlike in the previous case
$\Lambda = 0$, to our knowledge a general form of the metric describing all type
N Kundt spacetimes with $\Lambda \neq 0$ and aligned null radiation is not
explicitly known. Therefore, we consider the $n$-dimensional Siklos metric
\cite{ChamblinGibbons2000}
\begin{equation}
  \d s^2 = \frac{1}{-\lambda z^2} \left[ 2 \d u \, \d v
    + 2 H(u, v, x^k) \, \d u^2
    + \delta_{ij} \, \d x^i \, \d x^j \right],
  \label{QG:Siklos:metric}
\end{equation}
where
\begin{equation}
  \lambda = \frac{2 \Lambda}{(n-1)(n-2)},
  \qquad z = x^{n-1},
\end{equation}
as an example belonging to this class of spacetimes.

The Siklos metric is conformally related to \pp waves. However, note that the
terms inside the parentheses correspond to the metric which describes all
\pp waves only in four dimensions. In higher dimensions, this metric does not
represent the \pp wave class completely and not even the type N \pp wave
subclass.

Obviously, the metric \eqref{QG:Siklos:metric} takes the GKS form
\eqref{KS:GKSmetric} since the first and third terms in the parentheses
represent an anti-de Sitter background and $\d u$ corresponds to the null
vector. Then proposition \ref{KS:proposition:nonexpanding} ensures that this
metric is indeed of Weyl type N even if we admit an aligned null radiation term
in the Ricci tensor as commented under this proposition.

We need to split the Kerr--Schild term $\H k_a k_b \, \d x^a \, \d x^b
= \frac{H}{\lambda z^2} \, \d u^2$ into the vector $\bk$ and the function $\H$
so that $\bk$ is affinely parametrized. The simplest way how to do that is to
employ the coordinate transformation $v = - \lambda \tilde{v} z^2$ to put
\eqref{QG:Siklos:metric} to the canonical Kundt form
\begin{equation}
  \d s^2 = 2 \d u \left[ \d \tilde{v}
    - \frac{H}{\lambda z^2} \, \d u
    + \frac{2\tilde{v}}{z} \, \d z \right]
    - \frac{1}{\lambda z^2} \delta_{\itilde\jtilde} \, \d x^{\itilde} \, \d x^{\jtilde}.
  \label{QG:Siklos:canonicalKundt}
\end{equation}
In this form, $\d u$ corresponds to the congruence of non-expanding, non-shearing
and non-twisting affinely parametrized null geodesics. Therefore,
$\bk = k_a \d x^a = \d u$ and $\mathcal{H} = \frac{1}{\lambda z^2} H$.
Furthermore, if we identify the Kerr--Schild vector $\bk$ with the frame vector
$\bl$, then from \eqref{QG:Kundt:L1i} follows that $L_{[1i]} = 0$. One may
complete the frame by a natural choice of the remaining vectors
\begin{equation}
  \begin{aligned}
  \ell_a \, \d x^a &= \d u, \qquad
  n_a \, \d x^a = \frac{1}{- \lambda z^2} \, \d v + \frac{H}{- \lambda z^2} \, \d u, \qquad
  m^{(i)}_a \, \d x^a = \d x^i, \\
  \ell^a \, \partial_a &= - \lambda z^2 \, \partial_v, \qquad
  n^a \, \partial_a = \partial_u - H \, \partial_v, \qquad
  m_{(i)}^a \, \partial_a = \partial_i.
  \end{aligned}
  \label{QG:Siklos:frame}
\end{equation}

For the simplicity, we assume that $\D\H = 0$, i.e.\ $\H$ is independent on an
affine parameter along the null geodesics $\bk$. On the other hand, using the
frame \eqref{QG:Siklos:frame}, it follows that $\D\H = - \lambda z^2 \H_{,v}$
and thus $\H$ and consequently $H$ does not depend on the coordinate $v$.
Therefore, we can use the Einstein field equations for non-expanding GKS
spacetimes with aligned null radiation \eqref{QG:GKS:nullradiation:EFEs} which
for the Siklos metric \eqref{QG:Siklos:metric} leads to
\begin{equation}
  \Delta H - \frac{n-2}{z} H_{,z} = \Phi.
  \label{QG:Siklos:EFEs}
\end{equation}
The condition on $\Phi$ \eqref{QG:fieldeqns:nullradiation:typeN:comp11:dalembert}
for the metric \eqref{QG:Siklos:metric} reads
\begin{equation}
  \Delta\!\left( - \lambda z^2 \Phi \right) - \frac{n-2}{z} \left( - \lambda z^2 \Phi \right)_{,z}
    - \frac{\mathcal{C}}{z^2} ( - \lambda z^2 \Phi ) = 0,
  \label{QG:Siklos:Phi}
\end{equation}
where we defined
\begin{equation}
  \mathcal{C} \equiv 2 + \frac{\mathcal{A}}{\beta \lambda}
    = \frac{2}{\beta} \left( \frac{1}{2 \lambda \kappa} + (n-1)(n \alpha + \beta) + (n-3)(n-4) \gamma \right).
\end{equation}
Now we eliminate $\Phi$ from \eqref{QG:Siklos:EFEs} by combining 
\eqref{QG:Siklos:EFEs} with \eqref{QG:Siklos:Phi} and denoting
$H^{\text{vac}} = H - \mathcal{C}^{-1} z^2 \Phi$. This leads to the vacuum
($\Phi = 0$) equation \eqref{QG:Siklos:EFEs} for $H^{\text{vac}}$
\begin{equation}
  \Delta H^{\rm vac} - \frac{n-2}{z} H^{\rm vac}_{,z} = 0.
  \label{QG:Siklos:EFEs:vacuum}
\end{equation}

Therefore, we can take an arbitrary higher dimensional Siklos metric
\eqref{QG:Siklos:metric} obeying the vacuum Einstein field equations
\eqref{QG:Siklos:EFEs:vacuum} and find a solution $\Phi$ of
\eqref{QG:Siklos:Phi}. Then the metric \eqref{QG:Siklos:metric} with
$H = H^{\text{vac}} + \mathcal{C}^{-1} z^2 \Phi$, where the obtained $\Phi$
enters also the corresponding Ricci tensor \eqref{QG:nullradiation:Ricci},
satisfies the field equations of quadratic gravity \eqref{QG:fieldeqns}.

Unfortunately, this procedure cannot be performed for quadratic gravity theories
with the parameters at critical points \eqref{QG:critical_point:1},
\eqref{QG:critical_point:2}. Since $\mathcal{C} = 0$ in these cases, we are not
able to solve for $\Phi$ and $H$ independently. Instead, one has to find $\Phi$
obeying \eqref{QG:Siklos:Phi} and then solve \eqref{QG:Siklos:EFEs} for $\H$
with the given $\Phi$.

Let us point out how the solution \eqref{QG:Siklos:metric},
\eqref{QG:Siklos:EFEs}, \eqref{QG:Siklos:Phi} is related to the AdS-wave
solution that has been found in \cite{GulluGursesSismanTekin2011} by direct
substitution of the Siklos metric to the field equations of quadratic gravity
\eqref{QG:fieldeqns}. After necessary long calculations of the Riemann tensor
and its various contractions, the authors of \cite{GulluGursesSismanTekin2011}
have obtained the constraint for the function $H$ which can be equivalently
expressed from \eqref{QG:Siklos:EFEs} and \eqref{QG:Siklos:Phi} eliminating
$\Phi$
\begin{equation}
  \left( \Delta - \frac{n - 2}{z} \, \partial_z - \frac{\mathcal{C}}{z^2} \right) \left[ z^2
  \left( \Delta - \frac{n - 2}{z} \, \partial_z \right) H \right] = 0.
\end{equation}
In \cite{GulluGursesSismanTekin2011}, it has been further integrated to get a
particular solution.

\chapter{Conclusions and outlook}

In chapter \ref{sec:KS}, we have investigated GKS spacetimes \eqref{KS:GKSmetric}
with an (anti-)de Sitter background in arbitrary dimension. It has turned out that the
Kerr--Schild vector $\bk$ is geodetic if and only if the boost weight zero
component $T_{00}$ of the energy--momentum tensor vanishes as stated in
proposition \ref{KS:proposition:geodetic_k}. It has been shown that the vector field $\bk$
is geodetic in the background spacetime if and only if it is geodetic in the full spacetime
and the same also holds for the affine parametrization.
For GKS spacetimes with a geodetic Kerr--Schild vector $\bk$ including Einstein
spaces and spacetimes containing matter fields aligned with $\bk$ such as
aligned Maxwell field or aligned null radiation, we have given the explicit form
of the Ricci tensor and shown that the optical properties of the Kerr--Schild
congruence $\bk$ encoded in the optical matrix $L_{ij}$ in the full spacetime
are same as those in the background spacetime. If $\bk$ is geodetic then $T_{0i}$
as well as the positive boost weight components of the Weyl tensor vanish and
thus such GKS spacetimes are algebraically special, i.e.\ of Weyl type II or more special,
see proposition \ref{KS:proposition:Weyltypes}.

In section \ref{sec:KS:nonexpanding}, we have focused on non-expanding Einstein
GKS spacetimes. The Einstein field equations imply that these spacetimes
belong to the Kundt class and are only of Weyl type N, see proposition
\ref{KS:proposition:nonexpanding}. It has been pointed out that the same statement
also holds if we admit null radiation term in the Ricci tensor and that
the Kerr--Schild function $\H$ of non-expanding Einstein spacetimes is a linear
function of the affine parameter $r$ along the null geodesics $\bk$.
We have also presented some known examples of non-expanding Einstein GKS spacetimes
in section \ref{sec:KS:nonexpanding:examples} and constructed some new ones in section
\ref{sec:KS:nonexpanding:warp} using the Brinkmann warp product.

Expanding Einstein GKS spacetimes have been discussed in section \ref{sec:KS:expanding}.
The compatible Weyl types are II or D with $\bk$ being the multiple WAND as stated in
proposition \ref{KS:proposition:expanding} and the corresponding optical matrix $L_{ij}$ satisfies
the optical constraint \eqref{KS:vacuum:optical_constraint} implying that $L_{ij}$ is 
a normal matrix and in an appropriate frame takes the block diagonal form consisting
of $2 \times 2$ and identical $1 \times 1$ blocks and zeros. This sparse form has allowed us to integrate
the Sachs equation and explicitly express the $r$-dependence of the optical matrix $L_{ij}$
and subsequently of the Kerr--Schild function $\H$ and boost weight zero components of
the Weyl tensor. The $2 \times 2$ blocks in the optical matrix correspond to planes
spanned by pairs of the spacelike frame vectors in which the geodetic congruence $\bk$
is twisting. Only if $L_{ij}$ is non-degenerate and does not contain any
$2 \times 2$ block or if in even dimensions $L_{ij}$ contains only identical $2 \times 2$ blocks
then the congruence $\bk$ is non-shearing. It has been also shown that the rank of $L_{ij}$ is at least 2.
Therefore, in four dimensions, the optical matrix consists of one $2 \times 2$ block or
two identical $1 \times 1$ blocks which is in accordance with the Goldberg--Sachs theorem.
Expressing the Kretschmann scalar, we have discussed presence of curvature singularities
at the origin $r = 0$ in section \ref{sec:KS:singularities}. It has turned out that there are
three possible cases depending on the form of the optical matrix $L_{ij}$.
Namely, either no singularity is present or there is a point or Kerr-like singularity.
In section \ref{sec:KS:Kerr-(A)dS}, the analysis of expanding Einstein GKS spacetimes
has been compared with the explicit example of the five-dimensional Kerr--(anti-)de Sitter
black hole. We have established the null frame parallelly transported along the geodetic Kerr--Schild
vector $\bk$ and expressed the optical matrix. Using the results of section \ref{sec:KS:singularities},
we have discussed the presence of curvature singularities depending on the two rotation parameters
of the black hole.

In future work, the analysis of the GKS spacetimes could be extended to higher
order theories of gravity such as the Gauss--Bonnet or more general Lovelock
theories, some basic analysis in this field has been already done in \cite{EttKastor2011} and
the particular case of the GKS metrics in five-dimensional Gauss--Bonnet gravity
has been studied in \cite{Anabalonetal2009,AnabalonDeruelleTempoTroncoso2010}.
It may be also useful to employ the GKS ansatz to investigate Weyl type D solutions of quadratic gravity
and thus extend the work started in chapter \ref{sec:QuadGr}.

Higher dimensional GKS metrics could be also studied in the context of the Einstein--Maxwell theory.
As has been mentioned above, the generalization of the four-dimensional Kerr--Newman black hole to higher dimensions
using its KS form with the vector potential proportional to the Kerr--Schild vector $\bk$ has failed.
However, if one assumes the vector potential given by a linear combination of the Kerr--Schild vector $\bk$
and a spacelike vector $\bm$, i.e.\ $A_a = \alpha k_a + \beta m_a$, where $\alpha$ and $\beta$ are scalar functions
then it can be shown that this choice is compatible with $T_{00} = T_{0i} = 0$ if the relations
\eqref{xKS:relation_k_m:Riccicoeff1} appearing in the analysis of xKS spacetimes hold.
This suggests that a solution in the GKS form with this vector potential could be found
if $\bm$ corresponds to a $1 \times 1$ block in the optical matrix and thus does not lie in
any plane in which $\bk$ is twisting.


In chapter \ref{sec:xKS}, we have studied xKS spacetimes \eqref{xKS:metric},
i.e.\ an extension of the GKS ansatz where, in addition to the null Kerr–
Schild vector $\bk$, a spacelike vector field $\bm$ appears in the metric.
Unlike for GKS spacetimes, in general we have obtained only the necessary condition
for the component $T_{00}$ under which the Kerr--Schild vector $\bk$ is geodetic, see
proposition \ref{xKS:proposition:geodetic_k}. However, if one appropriately restricts the
geometry of the vectors $\bk$ and $\bm$ this condition becomes sufficient, see corollary
\ref{xKS:proposition:geodetic_k:2}.
As for GKS spacetimes, the Kerr--Schild vector field $\bk$ is geodetic and affinely
parametrized in the background spacetime if it is geodetic and affinely parametrized
in the full spacetime and vice versa. Similarly, if $\bk$ is geodetic the optical matrices
in both spacetimes are identical.
In contrast with the GKS case, it has been shown that xKS metrics with a geodetic
Kerr--Schild vector $\bk$ are of Weyl type I or more special, as stated in proposition \ref{xKS:proposition:Weyltype},
and the components $R_{0i}$ of the Ricci tensor do not vanish identically.

In section \ref{sec:xKS:Kundt}, we have restricted ourselves to Kundt xKS spacetimes
and determined the $r$-dependence of the scalar function $\K$ in the case of Weyl type
II when the components $T_{0i}$ vanish, see proposition \ref{xKS:proposition:Kundt}.
It has turned out that all VSI spacetimes
belong to the xKS class of solutions and explicit examples with the corresponding
forms of the function $\K$ have been given.
Higher dimensional \pp waves are of Weyl type II or more special and type III and N Ricci-flat \pp waves
belong to the VSI and consequently to the xKS class.
It has been also shown that type II CSI Ricci-flat \pp waves admit the xKS form.
However, it is not clear whether all Ricci-flat \pp waves can be cast to the xKS form.

An important example of an expanding xKS spacetime, namely
the charged rotating CCLP black hole in five-dimensional minimal gauged supergravity,
has been studied in section \ref{sec:xKS:CCLP}.
In the case with the flat background, we have established a null frame and expressed the optical
matrix which interestingly satisfies the optical constraint \eqref{KS:vacuum:optical_constraint}.
Non-vanishing boost weight 1 components of the Weyl tensor suggest that the CCLP black hole
is in general of Weyl type I$_i$. In the uncharged case which corresponds to the Myers--Perry black hole
and in the non-rotating limit, the metric reduces to the GKS form and is of Weyl type D.

Although the necessary calculations are more involved than for GKS spacetimes,
it is obvious that the Ricci and Riemann tensors of the xKS metrics dramatically simplify
if one assumes that the relations \eqref{xKS:relation_k_m} restricting the geometry of the vectors $\bk$
and $\bm$ hold. The analysis of these simplified expressions is left for future work.



In chapter \ref{sec:QuadGr}, we have studied exact solutions of the field equations of quadratic gravity.
In fact, we have determined under which conditions certain exact solutions of the Einstein theory solve also
the source-free field equations of quadratic gravity.
In the case of Einstein spacetimes, it has turned out that only the Gauss--Bonnet term in the action
of quadratic gravity imposes an additional conditions on the Weyl types of a solution.
If the Gauss--Bonnet term is not present, i.e.\ $\gamma = 0$, or in four dimensions when the Gauss--Bonnet term effectively
vanishes then Einstein spaces of all Weyl types with an appropriate effective cosmological constant $\Lambda$
given by $\mathcal{B} = 0$ \eqref{QG:fieldeqns:B} are solutions of quadratic gravity.
Otherwise, restricting ourselves to types III and more special, Einstein spacetimes only of Weyl types N or III(B)
with an effective cosmological constant $\Lambda$ given by $\mathcal{B} = 0$
solve the field equations of quadratic gravity as stated in propositions \ref{QG:proposition:typeNEinstein}
and \ref{QG:proposition:typeIIIEinstein}.
The subclass III(B) of type III is defined via vanishing quantity $\tilde \Psi$ \eqref{QG:Einstein:typeIII:tildePsi}
constructed from the Weyl tensor.
Note that type III Einstein spaces belonging to the subclass denoted as III(A) are not solutions of quadratic gravity.
Moreover, the class of type III(A) spacetimes contains the class of III(a) spacetimes and
therefore type III(a) Ricci-flat Kundt metrics including all type III Ricci-flat \pp waves
do not solve the field equations of quadratic gravity. 
We have also referred to known examples of higher dimensional type N metrics solving quadratic gravity
and constructed some examples of type III solutions using the fact that the Brinkmann warp product
of four-dimensional type III Einstein spacetimes leads to the five-dimensional type III(B) Einstein spacetimes.

In section \ref{sec:QG:nullradiation}, we have found a wider class of solutions of quadratic gravity considering
the Ricci tensor with an additional null radiation term aligned with a WAND.
Restricting to the subclasses of type III with vanishing quantity $\tilde \Psi$,
i.e.\ types III(B), N and O, we have obtained the algebraic condition $\mathcal{B} = 0$ determining the
effective cosmological constant $\Lambda$ and also the constraint for the null radiation term
which implies that the only allowed metrics belong to the Kundt class,
see propositions \ref{QG:proposition:nullrad} and \ref{QG:proposition:nullrad:Kundt}.

Recall that GKS spacetimes with null radiation belong to the type N Kundt class
and thus solve the field equations of quadratic gravity with an appropriate
$\Lambda$ given by $\mathcal{B} = 0$.
In section \ref{sec:QG:nullradiation:examples}, we have given explicit examples
of such metrics with $\Lambda = 0$ and $\Lambda \neq 0$.
Due to the similarity of the corresponding equations for the null radiation term $\Phi$
and for the Kerr--Schild function $\H$, we have been able to express
two completely independent equations for $\Phi$ and $\H_{\text{vac}}$, where $\H_{\text{vac}}$
corresponds to a Ricci-flat or Einstein part of the solution.
However, at critical points of quadratic gravity the equations for $\Phi$ and $\H$ cannot be 
separated in this way.

In future work, type D solutions of quadratic gravity could be also studied,
however, the significant simplification as in the type III cases does not
occur and a non-trivial form of the Ricci tensor could be necessary. 
Therefore, it could be convenient to employ, for instance, the GKS ansatz.
Note also that the field equations of quadratic gravity \eqref{QG:fieldeqns:nullradiation:typeIII}
for spacetimes of Weyl type III and the Ricci tensor with aligned null radiation
can be also solved with the assumption $\tilde \Psi \neq 0$,
i.e.\ for spacetimes of type III(A).
Therefore, unlike III(A) Einstein spaces which do not solve the field equations of quadratic gravity,
type III(A) solutions of quadratic gravity with null radiation in the Ricci tensor could exist.

\cleardoublepage
\phantomsection
\addcontentsline{toc}{chapter}{Bibliography}
\bibliography{bibtex,mypapers}
\bibliographystyle{utcaps}

\cleardoublepage
\phantomsection
\addcontentsline{toc}{chapter}{List of Tables}
\listoftables

\end{document}